\documentclass[useAMS]{mn2e}
\usepackage{times}
\usepackage{rotate}
\usepackage[pdftex]{graphicx}
\usepackage{graphics}
\usepackage{psfig}
\usepackage{lscape}
\input{epsf}
\usepackage[pdftex]{graphicx}
\newif\ifAMStwofonts
\AMStwofontstrue

\def\gsim{\mathrel{\hbox{\rlap{\hbox{\lower4pt\hbox{$\sim$}}}\hbox{$>$}}}}
\def\lsim{\mathrel{\hbox{\rlap{\hbox{\lower4pt\hbox{$\sim$}}}\hbox{$<$}}}}

\def\asca{{\it ASCA}}
\def\chandra{{\it Chandra}}
\def\xmm{{\it XMM-Newton}}

\def\rbr{{$r_{\rm br}$}}

\def\delchi{{$\Delta \chi^{2}$}}

\title[XMM observations of iron lines]{An XMM-Newton survey of broad iron lines in Seyfert galaxies}
\author[K. Nandra et al.]{K. Nandra$^{1}$, P. O'Neill$^{1}$\thanks{present address: School of Computing and Mathematics, Charles Sturt University, Locked Bag 588, Wagga Wagga 2678, Australia}, I.M. George$^{2,3}$, J.N. Reeves$^{3,4}$\thanks{present address: Astrophysics Group, School of Physics and Geographical Science, Keele University, Staffordshire, ST5 5BG, UK}  \\
$^1$Astrophysics Group, Imperial College London, Blackett Laboratory,
Prince Consort Road, London SW7 2AW, UK \\ 
$^2$Department of Physics, University of Maryland, Baltimore County, 1000 Hilltop Circle, Baltimore, MD 21250, USA \\
$^3$Laboratory for High Energy Astrophysics, NASA/Goddard Space Flight Center, Code 660, Greenbelt, MD 20771, USA \\
$^4$Department of Physics \& Astronomy, Johns Hopkins University, 3400 N. Charles Street, Baltimore, MD 21218, USA
}
\date{}

\begin{document}

\maketitle
\label{firstpage}

\begin{abstract}
We present an analysis of the X-ray spectra of a sample of 37 observations of 26 Seyfert galaxies observed by \xmm\ in order to characterize their iron K$\alpha$ emission. All objects show evidence for iron line emission in the 6-7 keV band. A narrow ``core" at 6.4 keV is seen almost universally in the spectra, and we model this using a neutral Compton reflection component, assumed to be associated with distant, optically thick material such as the molecular torus. Once this, and absorption by a zone of ionized gas in the line-of-sight is accounted for, less than half of the sample observations show an acceptable fit.  Approximately 2/3 of the sample shows evidence for further, broadened emission in the iron K-band. When modeled with a Gaussian, the inferred energy is close to that expected for neutral iron, with a slight redshift, with an average velocity width of $\sim 0.1$c. The mean parameters are consistent with previous {\it ASCA} results and support the idea that the broad components can be associated with the accretion disk.  Before proceeding to that conclusion, we test an alternative model comprising a blend of 3-4 narrow, unshifted emission lines (including the 6.4 keV core), together with 1-2 zones of highly ionized gas in the line-of-sight. Around 1/3 of the objects are not adequately fit by this model, and in general better fits are obtained with a relativistic disk line model, which has fewer free parameters. Nonetheless we find that absorption by ionized gas affects the spectrum above 2.5 keV in approximately half the sample. There is evidence for multiple ionized zones in at least 3 objects, but in all those cases a blurred reflector is required in addition to the complex absorption.  We also identify a number of narrow emission and absorption features around the Fe complex, and the significance and interpretation of these lines is discussed. After accounting for these additional complexities, we determine the typical parameters for the broad reflection. The emission is found to come, on average, from a characteristic radius  $\sim 15 r_{\rm g}$ and the average disk inclination is $\sim 40^{\circ}$. The broad reflection is on average significantly weaker, by a factor $\sim 2$,  than that expected from a flat disk illuminated by a point source. Notwithstanding these average proties, the objects exhibit a significant and wide range of reflection parameters. We find that 35~per cent of the sample observations can be explained solely with narrow line components, with no evidence for broadened emission at all. A further 25~per cent show evidence for significant broad emission, but at a characteristic radius relatively far from the black hole. The remaining $\sim 45$~per cent are best fit with a relativistically blurred reflection model. In 12/37 observations the characteristic emission radius is constrained to be $<50 r_{\rm g}$, where the gravitational redshift is measurable. For at least this subsample, our observations verify the potential for X-ray spectroscopy to diagnose the strong gravity regime of supermassive black holes.  
\end{abstract}


\section{INTRODUCTION}
\label{Sec:Introduction}

Fabian et al. (1989) predicted that line emission from accretion disks in active galactic nuclei (AGN) and X-ray binaries should be broadened by Doppler motions and relativistic effects close to the central black hole. These predictions were made contemporaneously with the discovery that iron K$\alpha$ emission, most likely from the accretion disk (e.g. George, Nandra \& Fabian 1990), was being found to be extremely common in AGN X-ray spectra obtained by the Ginga satellite (e.g. Pounds et al. 1990; Nandra \& Pounds 1994). Ginga's resolution was insufficient to search definitively for the relativistic broadening, so it was left to the first X-ray CCD detectors, flown aboard the ASCA satellite (Tanaka, Inoue \& Holt 1994), to test Fabian et al.'s prediction. Clear evidence for line broadening in excess of the instrumental resolution was found in early spectra (e.g. Mushotzky et al. 1995), but the most convincing example of an accretion disk line came with a long observation of the type 1 Seyfert galaxy MCG-6-30-15 (Tanaka et al. 1995). This showed a very broad, redshifted profile in spectacular agreement with the accretion disk models, and very difficult to model in any other way (Fabian et al. 1995). 

Subsequent work on a sample of Seyfert galaxies by Nandra et al. (1997; hereafter N97) showed statistical evidence for broadening in a large number ($\sim 75$~per cent) of objects. N97 demonstrated that a skewed, accretion disk line, rather than a symmetric profile, was the preferred interpretation for the broadening. The average properties of the line were found to be similar to MCG-6-30-15, but clear differences were found in the profiles observed in different objects. The 
general picture derived from the N97 sample was confirmed with other high signal-to-noise ratio ASCA spectra such as those of NGC 3516 (Nandra et al. 1999) and IC4329A (Done, Madejski \& Zycki 2000). 

N97 also concluded, based on statistical evidence, that narrow emission around 6.4 keV was relatively weak in the ASCA spectra, with a mean equivalent width for such lines of only 30 eV. Newer data with better resolution and signal-to-noise ratio has shown that this may not be the case, with Chandra grating spectra  showing a nearly-ubiquitous narrow component with typical equvalent width $\sim 100$~eV (e.g. Yaqoob et al. 2001; Yaqoob \& Padhmanaban 2004). 

X-ray spectra from \xmm\ have also revealed these narrow line cores (e.g. Reeves et al. 2001; Pounds et al. 2001) although at the resolution of the EPIC instrument is difficult to obtain strong constraints on their origin. In the disk-line archetype, MCG-6-30-15, both XMM and Chandra show that an apparently narrow core may in fact be significantly broadened (e.g. Fabian et al. 2002; Lee et al. 2002) and plausibly arises in the outer accretion disk. In most other cases, however, these narrow emission lines probably arise in more distant material,  most likely the putative molecular torus of AGN unification schemes (Krolik \& Kallman 1987; Awaki et al. 1991; Ghisellini, Haardt \& Matt 1994; Krolik, Madau \& Zycki 1994).  The optical broad line region is another possible site (BLR; e.g. Holt et al. 1980), although overall this seems to be disfavoured (Yaqoob \& Padhmanhaban 2004; Nandra 2006). 

Due to its relatively small effective area the Chandra HETG data are only sensitive to narrow features in AGN spectra, but the \xmm\ data are also able to constrain the broad component of the lines. Indeed, the higher throughput of the EPIC camera should result in better information, compared to \asca,  regarding the contribution of a relativistic accretion disk. Thus far, this increase in signal-to-noise ratio has not necessarily produced a corresponding increase in our understanding of the disk line phenomenon. Individual cases have been found in which no broadening seems to be required (e.g. Gondoin et al. 2001, 2003;  Pounds et al. 2003; Bianchi et al. 2004), or where alternative, physically plausible explanations can account for the broad features (e.g. Reeves et al. 2004). On the other hand, there are numerous examples where a relativistic accretion disk line is the preferred interpretation (e.g. Reeves et a. 2001; Gondoin et al. 2002; Longinotti et al. 2003), including the seemingly cast--iron case of MCG-6-30-15 (Fabian et al. 2002; Vaughan \& Fabian 2004). Relativistic lines have also now been reported in high redshift AGN (Comastri, Brusa \& Civano 2004; Brusa, Gilli \& Comastri 2005; Streblyanska et al. 2005). Reviews of the broad line pheonemenon have been published by, e.g. Fabian et al. (2000), Reynolds \& Nowak (2003), Liedahl \& Torres (2005) and Miller (2007). 

Overall, it is currently unclear whether or not broadening of the iron K$\alpha$ lines is more or less common than found with ASCA, nor can we be sure whether the evidence for relativistic disk emission is more or less secure. It is our intention to investigate these issues with our sample of Seyfert galaxies. This involves essentially repeating the analysis of N97 with a similar, but larger sample, with higher quality data and new insights into the nature of the spectra.  With this sample, our intention is to address the key issues: 1) How common is line broadening in Seyferts; 2) Where present, can the broadening be accounted for by a relativistic accretion disk; 3) how robust and unique is the accretion disk interpretation and 4) if there is robust evidence, what are the constraints on the innermost regions of accreting supermassive black hole systems. In \S 2 we describe our observations and data reduction. Simple fits are presented in \S 3. These reveal complexities in the spectra and we attempt to account for these without recourse to relativistic effects in \S 4. Accretion disk line models are presented, and compared to the competing interpretations in \S 5. In \S 6 we explore the effects of additional spectral complexities on the line parameters. \S 7 discusses the ensemble properties of the broadened emission for the sample as a whole. In \S 8, we discuss our results, in \S 9 future prospects for iron line studies are reviewed, and our conclusions are given in \S 10.  

\begin{table*}
\centering
\caption{\xmm\ Seyfert sample.
Col.(1): Common name of object;
Col.(2,3,4): Position and Redshift  from NASA/IPAC Extragalactic Database (NED);
Col.(5): Galactic absorption column from Elvis, Wilkes \& Lockman (1989); Stark et al. (1992) or Dickey \& Lockman (1990); Col.(6): References for previously-published \xmm\ spectra: (1) Vaughan et al. 2004; 
(2) Matt et al. 2006; (3) Immler et al. 2003 (4) Schurch et al. 2006;  (5) Dasgupta \& Rao 2006;
(6) Dewangan et al. 2003; (7) Balestra et al. 2004; (8) Awaki et al. 2006; (9) Braito et al. 2007; 
(10) Turner et al. 2002; (11) Bianchi et al. 2004; (12) Iwasawa et al. 2004; (13) Turner et al. 2005;
(14) Blustin et al. 2002; (15) Reeves et al. 2004; (16) Uttley et al. 2004;
(17)  Ogle et al. 2004; (18) Pounds et al. 2004; (19) Ponti et al. 2006; (20) Schurch et al. 2003;
(21) Schurch et al. 2004 (22) Mason et al. 2003; (23) Pounds et al. 2003a; (24) Turner et al. 2006; (25) Miller et al. 2006; (26) Vaughan et al. 2005; (27) Steenbrugge et al. 2003; (28) Reynolds et al. 2004;
(29) Wilms et al. 2001; (30) Fabian et al. 2002; (31) Vaughan \& Fabian 2004;   (32) Gondoin et al. 2001; 
(33) Matt et al. 2001; (34) Dewangan \& Griffiths 2005; (35) Pounds et al. 2003b; (36) Pounds et al. 2001; (37) Dadina et al. 2005; (38) Bianchi et al. 2003; (39) Starling et al. 2005; (40) Arevalo et al. 2006; (41) Blustin et al. 2003 (42);  Blustin et al. (2007)
\label{tab:sample}}
\begin{center}
\begin{tabular}{lccccc}
\hline
Name & RA & DEC & Redshift & $N_{\rm H}$ (Gal) & References \\
 & (J2000) & (J2000) & & $10^{20}$~cm$^{-2}$  & \\
(1) & (2) & (3) & (4) & (5) & (6) \\
\hline
	
NGC 526A 	& 01 23 54.4 & $-35$ 03 56 & 0.019 & 2.2 & ...\\
Mrk 590 		& 02 14 33.6 & $-00$ 46 00 & 0.035 & 4.1 & ... \\
Ark 120 		& 05 16 11.4 & $-00$ 09 00 & 0.033 & 12.6 & 1 \\
NGC 2110 	& 05 52 11.4 & $-07$ 27 23 & 0.007 & 18.3 & ...  \\
MCG+8-11-11 	& 05 54 53.6 & $+46$ 26 2 & 0.020 & 20.9 & 2 \\
Mrk 6 		& 06 52 12.3 & $+74$ 25 38 & 0.019 & 6.4 & 3,4 \\
Mrk 110 		& 09 25 12.9 & $+52$ 17 11 & 0.036 & 1.5 &  5 \\
NGC 2992 	& 09 45 42.0 & $-14$ 19 35 & 0.008 & 5.3 & ... \\
MCG-5-23-16	& 09 47 40.1 & $-30$ 56 56 & 0.008 & 8.4 & 6,7,8,9  \\
NGC 3516    	& 11 06 47.5 & $+72$ 34 07 & 0.009 & 2.9  & 10,11, 12, 13  \\
NGC 3783    	& 11 39 01.7 & $-37$ 44 19  & 0.010 & 8.5   & 14, 15 \\
HE 1143-1810  & 11 45 40.5 & $-18$ 27 16 & 0.033 & 3.5 & ... \\
NGC 4051    	& 12 03 09.6 & $+44$ 31 53  & 0.002 & 1.3  & 16, 17, 18, 19 \\
NGC 4151    	& 12 10 32.6 & $+39$ 24 21  & 0.003 & 2.1  & 20, 21 \\
Mrk 766     	& 12 18 26.5 & $+29$ 48 46  & 0.013 & 1.6  & 22, 23, 24, 25 \\
NGC 4395 	& 12 25 48.9 & $+33$ 32 48 & 0.001 & 1.4 & 26  \\
NGC 4593    	& 12 39 39.4 & $-05$ 20 39  & 0.009 & 2.0 & 27, 28 \\
MCG-6-30-15 	& 13 35 53.8 & $-34$ 17 44  & 0.008 & 4.1  & 29, 30, 31  \\
IC 4329A    	& 13 49 19.2 & $-30$ 18 34  & 0.016 & 10.4 & 11, 32 \\
NGC 5506 	& 14 13 14.8 & $-03$ 12 26 & 0.007 & 3.8 & 11, 33, 34 \\
NGC 5548    	& 14 17 59.5 & $+25$ 08 12  & 0.017 & 1.7  & 11, 35 \\
Mrk 509     	& 20 44 09.7 & $-10$ 43 25  & 0.034 & 4.2  & 36, 37 \\
NGC 7213 	& 22 09 16.2 & $-47$ 10 00 & 0.006 & 2.1 & 11, 38, 39  \\
NGC 7314 	& 22 35 46.2 & $-26$ 03 01 & 0.005 & 1.5 & 34 \\
Ark 564 		& 22 42 39.3 & $+29$ 43 32 & 0.025  & 6.4 & 40  \\
NGC 7469    	& 23 00 44.4 & $+08$ 36 17  & 0.017 & 4.8  & 41, 42  \\
\hline
\end{tabular}
\end{center}
\end{table*}

\begin{table*}
\centering
\caption{\xmm\ EPIC-pn observations.
Col.(1): Name and Observation number
Col.(2): \xmm\ Observation ID;
Col.(3): \xmm\ revolution;
Col.(4+5): Observation date and time (UT);
Col.(6): Duration of observation before screening (ks);
Col.(7): Exposure time after screening and accounting for detector deadtime;
Col.(8): EPIC-pn read-out mode. FF=Full Frame, LW=Large Window, SW=Small Window;
Col.(9): Filter in place during observation; 
Col.(10): Event grades used in the analysis; 
Col.(11): EPIC-pn net source count rate in 2.0-10.0 keV band.
\label{tab:obs}}
\begin{center}
\begin{tabular}{llrllrrllll}
\hline

Name & OBSID & Rev. & Date & Time & Duration & Exposure & Mode & Filter & Grades & Count rate\\
 (1) & (2) & (3) & (4) & (5) & (6) & (7) & (8) & (9) & (10) & (11) \\
\hline
\hline
NGC 526A         & 0150940101 & 647      & 2003-06-21 & 23:18:50 & 47.9          & 37.0          & LW & Thin1 & 0--4 & $2.113 \pm 0.008 $  \\
Mrk 590          & 0201020201 & 837      & 2004-07-04 & 08:36:22 & 225.3         & 51.9          & SW & Thin1 & 0--4 & $0.711 \pm 0.004 $  \\
Ark 120          & 0147190101 & 679      & 2003-08-24 & 05:41:48 & 112.1         & 57.4          & SW & Thin1 & 0--4 & $4.243 \pm 0.009 $  \\
NGC 2110         & 0145670101 & 593      & 2003-03-05 & 18:06:37 & 59.5          & 34.0          & LW & Thin1 & 0--4 & $2.183 \pm 0.008 $  \\
MCG+8-11-11      & 0201930201 & 794      & 2004-04-09 & 19:53:36 & 38.5          & 20.2          & SW & Medium & 0--4 & $4.597 \pm 0.015 $  \\
Mrk 6    & 0144230101 & 619      & 2003-04-26 & 14:43:25 & 59.0          & 31.6          & FF & Medium & 0--4 & $1.354 \pm 0.007 $  \\
Mrk 110          & 0201130501 & 904      & 2004-11-15 & 05:10:49 & 47.4          & 32.5          & SW & Thin1 & 0--4 & $3.209 \pm 0.010 $  \\
NGC 2992         & 0147920301 & 630      & 2003-05-19 & 13:04:56 & 28.9          & 21.8          & FF & Medium & 0 & $5.264 \pm 0.016 $  \\
MCG-5-23-16(1)   & 0112830301 & 261      & 2001-05-13 & 11:56:19 & 76.9          & 9.3   & FF & Medium & 0 & $4.954 \pm 0.023 $  \\
MCG-5-23-16(2)   & 0112830401 & 363      & 2001-12-01 & 22:19:11 & 24.9          & 19.6          & FF & Medium & 0 & $4.286 \pm 0.015 $  \\
NGC 3516(1)      & 0107460601 & 245      & 2001-04-10 & 11:14:24 & 130.2         & 37.9          & SW & Thin1 & 0--4 & $2.158 \pm 0.008 $  \\
NGC 3516(2)      & 0107460701 & 352      & 2001-11-09 & 23:12:51 & 130.0         & 81.5          & SW & Thin1 & 0--4 & $1.369 \pm 0.004 $  \\
NGC 3783(1)      & 0112210101 & 193      & 2000-12-28 & 17:57:06 & 40.4          & 25.8          & SW & Medium & 0--4 & $5.385 \pm 0.014 $  \\
NGC 3783(2)      & 0112210201 & 371      & 2001-12-17 & 19:34:20 & 413.5         & 173.3         & SW & Medium & 0--4 & $4.921 \pm 0.005 $  \\
HE 1143-1810     & 0201130201 & 824      & 2004-06-08 & 23:24:58 & 34.1          & 21.7          & SW & Thin1 & 0--4 & $3.112 \pm 0.012 $  \\
NGC 4051         & 0109141401 & 263      & 2001-05-16 & 11:52:05 & 122.0         & 41.2          & SW & Medium & 0--4 & $2.572 \pm 0.008 $  \\
NGC 4151(1)      & 0112310101 & 190      & 2000-12-21 & 17:07:48 & 33.0          & 20.6          & SW & Medium & 0--4 & $3.533 \pm 0.013 $  \\
NGC 4151(2)      & 0112830501 & 190      & 2000-12-22 & 03:43:38 & 85.4          & 68.6          & FF & Medium & 0--4 & $3.533 \pm 0.007 $  \\
NGC 4151(3)      & 0143500101 & 633      & 2003-05-25 & 01:43:21 & 56.9          & 31.8          & SW & Medium & 0--4 & $19.080 \pm 0.025 $  \\
Mrk 766(1)       & 0096020101 & 82       & 2000-05-20 & 10:50:12 & 58.8          & 24.3          & SW & Medium & 0--4 & $1.735 \pm 0.008 $  \\
Mrk 766(2)       & 0109141301 & 265      & 2001-05-20 & 08:55:37 & 129.9         & 72.4          & SW & Medium & 0--4 & $2.784 \pm 0.006 $  \\
NGC 4395         & 0142830101 & 728      & 2003-11-30 & 03:40:59 & 113.4         & 90.4          & FF & Medium & 0--4 & $0.506 \pm 0.002 $  \\
NGC 4593         & 0059830101 & 465      & 2002-06-23 & 13:08:48 & 87.2          & 50.6          & SW & Medium & 0--4 & $4.328 \pm 0.009 $  \\
MCG-6-30-15(1)   & 0111570101 & 108      & 2000-07-11 & 07:20:08 & 112.7         & 64.0          & SW & Medium & 0--4 & $3.502 \pm 0.007 $  \\
MCG-6-30-15(2)   & 0029740101 & 301      & 2001-07-31 & 15:39:18 & 349.3         & 204.5         & SW & Medium & 0--4 & $4.678 \pm 0.005 $  \\
IC 4329A(1)      & 0101040401 & 210      & 2001-01-31 & 15:09:01 & 13.9          & 9.3   & FF & Medium & 0--4 & $7.201 \pm 0.028 $  \\
IC 4329A(2)      & 0147440101 & 670      & 2003-08-06 & 06:56:27 & 136.0         & 68.4          & SW & Thin1 & 0 & $10.100 \pm 0.012 $  \\
NGC 5506(1)      & 0013140101 & 211      & 2001-02-02 & 23:20:29 & 20.0          & 13.7          & LW & Medium & 0--4 & $5.407 \pm 0.020 $  \\
NGC 5506(2)      & 0013140201 & 382      & 2002-01-09 & 18:20:15 & 13.8          & 9.9   & LW & Medium & 0--4 & $9.129 \pm 0.030 $  \\
NGC 5548(1)      & 0109960101 & 191      & 2000-12-24 & 22:34:30 & 26.1          & 15.9          & SW & Medium & 0--4 & $3.342 \pm 0.015 $  \\
NGC 5548(2)      & 0089960301 & 290      & 2001-07-09 & 16:08:04 & 135.0         & 66.2          & SW & Thin1 & 0--4 & $4.484 \pm 0.008 $  \\
Mrk 509          & 0130720101 & 161      & 2000-10-25 & 04:26:16 & 31.6          & 16.6          & SW & Thin1 & 0--4 & $3.187 \pm 0.014 $  \\
NGC 7213         & 0111810101 & 269      & 2001-05-29 & 00:18:06 & 49.6          & 29.3          & SW & Medium & 0--4 & $2.328 \pm 0.009 $  \\
NGC 7314         & 0111790101 & 256      & 2001-05-02 & 09:56:21 & 44.7          & 25.6          & SW & Medium & 0--4 & $4.292 \pm 0.013 $  \\
Ark 564          & 0206400101 & 930      & 2005-01-05 & 19:47:37 & 101.8         & 67.4          & SW & Medium & 0--4 & $2.095 \pm 0.006 $  \\
NGC 7469(1)      & 0112170101 & 192      & 2000-12-26 & 04:07:46 & 43.6          & 28.3          & SW & Medium & 0--4 & $2.937 \pm 0.010 $  \\
NGC 7469(2)      & 0207090101 & 912      & 2004-11-30 & 21:12:21 & 164.1         & 112.2         & SW & Medium & 0--4 & $3.200 \pm 0.005 $  \\
\hline
\end{tabular}
\end{center}
\end{table*}

\section{OBSERVATIONS AND DATA REDUCTION}

\subsection{Sample selection}

Our main aim is to characterize the iron K$\alpha$ emission in a uniform and systematic way in bright, local AGN. Our sample is chosen  from the observations available in the \xmm\ public archive, as of 1 January 2006, cross-correlated with the Veron \& Veron (2001) AGN catalogue. The analysis is restricted to nearby objects, imposing a cutoff of $z<0.05$. We excluded Seyfert 2 galaxies for this study as in these objects it is usually the case that the central engine is heavily obscured in the X-ray (Awaki et al. 1991). Unlike N97 we do include lightly obscured, intermediate Seyfert classifications (Sy 1.8 and 1.9) and also objects which are classically Seyfert 2 galaxies (i.e. without any broad emission at H$\alpha$) which nonetheless have broad emission lines detected in the near-IR (e.g. Blanco, Ward \& Wright 1990). We exclude one case of a Seyfert 1.8, NGC 1365, which has been shown to have a highly complex spectrum which at least at times is Compton thick (Risaliti et al. 2005), severely complicating analysis of the iron line. We also exclude radio loud objects (e.g. Blazars/BL Lac type objects, broad-line radio galaxies), and central cluster galaxies even if they are nominally classified as AGN.

Finally, as we wish to characterise the line properties in detail and thus require high signal-to-noise ratio in the hard X-ray band, we impose the condition that after background and other screening discussed below, that there are a minimum of 30,000 net counts in the EPIC pn spectrum in the 2-10 keV band.  

The final sample consists of 37 observations of 26 sources. Basic information about the sample is given in Table~\ref{tab:sample}. 

\subsection{Observations}

The observations described here  were taken with the EPIC-pn CCD camera 
(Struder et al. 2001) on board \xmm.  Details of the 37 observations used in the present analysis are given in Table~\ref{tab:obs} including the CCD mode, the filter used and the total net source count rates in the 2-10 keV band. Where several \xmm\ exposures of a give source have been taken, we combine data which were taken in the same or consective orbits, except in a single case, NGC 4151, where the CCD mode changed. 

\subsection{Data reduction}

\begin{figure}
\includegraphics[angle=0,width=90mm]{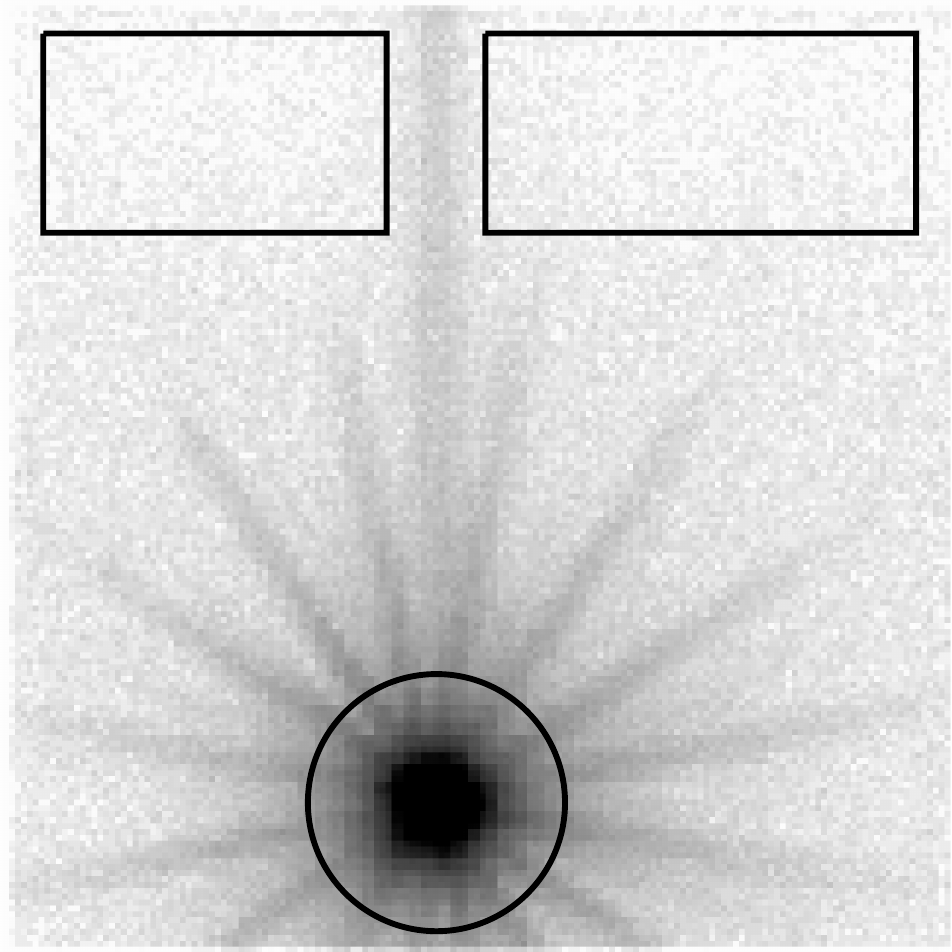}
\caption{Example of a a standard  source and  background region  geometry used  to to
extract events. The case shown is NGC 3783(2).}
\label{fig:ngc3783regs}
\end{figure}

\begin{figure}
\includegraphics[angle=0,width=90mm]{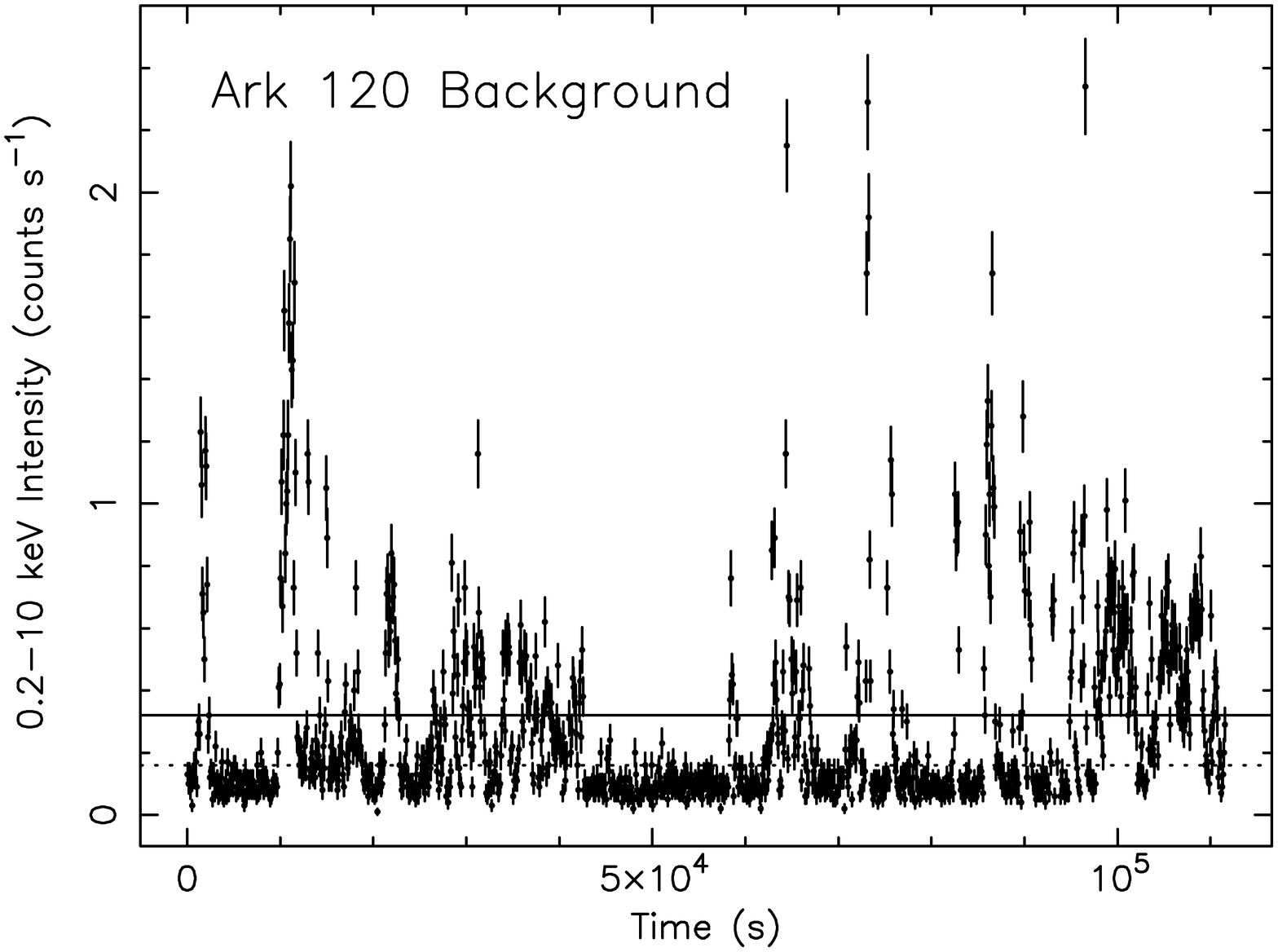}
\caption{An example background light curve illustrating our screening criteria.
The
data below  the lower horizontal  line have zero excess  variance, and
these represent the lower envelope  of the background light curve. The
upper horizontal line is at twice  this level and is adopted to define the GTIs used for the 
analysis of the source.}
\label{fig:akn120_bgd}
\end{figure}

The data were reduced as homogenously as possible. We used \textsc{XMMSAS} version 6.5.0 and calibration files as  of April 2006.   First, each  ODF was  processed using  the task  \textsc{epchain} to yield  a single events  list per  exposure for  the primary  chip (CCD number 4).  Source  and  background  events  were then  extracted  using  detector co-ordinates.  The nominal  arrangement of the    source    and   background    regions    is   illustrated   in Fig.~\ref{fig:ngc3783regs}.   A standardised source region was defined in detector co-ordinates as a circle with  a
radius of 700 detector  pixels (35\arcsec) at the nominal source position.  EPIC-pn data are collected in a variety of different windowing modes: Small Window (SW), Large Window (LW) or  Full Frame(FF), depending on the area of the detector being read out. The majority of the data (26/37 observations) were taken in SW mode. 
A standard background region was therefore defined which comprises  two rectangles located  at edge of the  SW
region  away  from  the  nominal  on-axis source  location.   The  two background regions are separated so as to avoid the out-of-time events strip, as seen  in Fig.~\ref{fig:ngc3783regs}.  Note that these background regions
were used regardless of the actual mode because the SW area is a subset of the LW and FF areas.  
The actual positioning of  the  source in  the  SW  area varied  somewhat  between
observations. The radius  of the source region was  decreased to avoid
the edge of the chip if  required. In cases where the source was located
closer to  the background regions than the nominal position, the size of the
background regions was  reduced. Each image was inspected for interloping
sources.   These  were  excised   from  the  background  region  using
additional circular  regions. In  the case of  the source  region, the
radius was reduced  to avoid other sources if  required.  The relative
areas of  the source and  background regions was determined  using the
{\tt xselect}  task,  and  the   nominal  ratio  between  the  source  and background areas is 3.7.

The events were filtered to include only those with \textsc{XMMSAS} quality flag of zero. Generally we used event files 
comprising of single-  and double-pixel events (i.e., CCD event patterns 0--4). However in 4 observations (Table~\ref{tab:obs}) significant pile-up was identified using the \textsc{XMMSAS}  task \textsc{epatplot}. In these cases single-pixel events (CCD event pattern 0) only were used, hence significantly reducing pileup effects (Ballet 1999).

A background light  curve was then extracted in  the 0.2--10~keV range using time bins of  100~s, and requiring  each bin  to be  fully  exposed.  We  then determined a background count rate limit, below which there is no evidence for intrinsic background variability beyond that expected from Poisson statistics. We achieved this by calculating the count rate at which the excess variance of the background lightcurve (as described in O'Neill et al. 2005) is equal to zero. We adopted  twice this limit  as  the  maximum  allowed  background  count rate  for  the observation, to allow for modest background variations. A typical example of the background screening limit is shown in Fig.~\ref{fig:akn120_bgd}.
The start  and  stop times  of  each 100~s  bin in  the filtered  background light  curve  were  then adopted  as  the set  of low-background good time intervals (GTIs).

Source  and   background  spectra   were  then  extracted   using  the low-background time intervals. The appropriate redistribution matrix and ancillary response  files  were generated  using  the  \textsc{XMMSAS}  tasks \textsc{rmfgen}  and \textsc{arfgen}, respectively. The redistribution  matrix was used to determine full-width-at-half-maximum  (FWHM) of  the detector  response  at each spectral channel.   The source spectral channels were  then grouped so that each  channel in  the grouped  spectrum had a  width of  at least $\sim$0.5 times  the FWHM and contained  no fewer than  20 counts, so that the error bars were approximately Gaussian. 

The subsequent background-subtracted spectra for each source  were fitted using \textsc{XSPEC} v11.3.2. As our main interest is in the iron K$\alpha$ region of the spectrum, we restrict our analysis to the 2.5-10 keV energy range.  Strong calibration-related residuals are invariably observed at $1.8-2.5$ keV, near the mirror Au M-shell edge and detector Si edge. This motivates our choice of lower energy bound of 2.5 keV. 

\subsection{Spectral models}
\label{sec:models}

As the quality of X-ray spectral data have improved, so has the complexity of the models used to fit them. There are currently a huge array of model components available in  \textsc{XSPEC}, and in many studies these are mixed and matched until a good fit to the data is obtained. It is not clear whether this yields any insight into the central regions of AGN, because with a sufficient number of model components and fit attempts one can almost always find an adequate parameterization of any given spectrum. Rather than adopt such an ad hoc approach, we have chosen to fit the spectra systematically, with models motivated by current observations and ideas about the structure of the central regions of AGN. 

In the standard model, the central black hole is fed by an accretion disk which emits radiation predominantly in the optical-UV (e.g. Malkan 1983). Hotter gas, often termed the ``corona'', Compton upscatters lower energy photons into the X-ray band having an approximate power-law form (e.g. Sunyaev \& Titarchuk 1980; Haardt \& Maraschi 1991, 1993), possibly with a thermal high energy cutoff (e.g. Gondek et al. 1996). For simplicity we adopt a power law continuum throughout. More complex continua are possible in certain geometries (e.g. Petrucci et al. 2000), but in general these would not be expected to affect strongly the properties of relatively narrow features, or induce false ones.. 

If the X-rays intersect the disk, or indeed any other optically thick material, they induce fluorescence and are Compton backscattered, resulting in a so-called ``Compton reflection'' spectrum (Guilbert \& Rees 1988; Lightman \& White 1988; George \& Fabian 1991; Matt et al. 1991). The most prominent emission line in this spectrum is iron K$\alpha$, the primary subject of the present study. If this reflection component arises close to the central black hole, it is distorted and blurred by Doppler and gravitational effects, giving rise to a characteristic line profile  (Fabian et al. 1989; Stella 1990; Laor 1991). X-ray reflection may also occur at greater distances. In particular, a likely site for is the molecular torus envisioned in orientation-dependent unification schemes (Antonucci \& Miller 1985). This can give rise to appreciable narrow line emission at iron K$\alpha$, as well as a Compton reflection continuum (Awaki et al. 1991; Ghisselini, Haardt \& Matt 1994; Krolik, Madau \& Zycki 1994).

Finally, the whole of the emission is seen through whatever obscuring gas exists in the line-of-sight to the X-ray source. When Seyfert 2 galaxies are excluded, this is usually due to partially ionized gas: the so--called "warm absorber" (Halpern 1984; Reynolds 1997; George et al. 1998b). Via high resolution observations we now know that this gas takes the form of an outflowing wind, with column density $\sim 10^{21-23}$~cm$^{-2}$ and a wide range of ionization states (e.g. Crenshaw, Kraemer \& George 2003). 

Thus, the minimum set of models required to describe AGN X-ray spectra are, in addition to the power law continuum, a reflection model, which may be blurred by relativistic effects or not, and an ionized absorber model. Here we describe the versions of these models we employ in this paper. 

\subsubsection{Compton  reflection}

The narrow cores seen in the line profiles indicates the presence of neutral gas at large distances, most plausibly identified with the torus. If this is the case, the narrow lines should be associated with a Compton reflection component and therefore to be self-consistent, the strength of the line and Compton reflecton continuum should be tied together. 

To model neutral reflection our basis is the model of Magdziarz \& Zdziarski (1995; known as 
\textsc{pexrav}) in  \textsc{xspec}  which charactertizes reflection in a slab geometry. This model is characterized by a number of parameters, namely the incident power law index, $\Gamma$, an upper exponential cutoff energy, $E_{\rm c}$, the inclination of the slab $i$, the elemental abundance, and the strength of the reflection component relative to that expected from a slab subtending 2$\pi$ solid angle, which we term the reflection fraction $R$ . To this, we add a narrow 6.4 keV emission line with strength as determined using the Monte Carlo calculations of George \& Fabian (1991). In this model, the equivalent width of the iron K$\alpha$ line depends on the photon index of the incident power law, the inclination of the slab, and the abundance of iron relative to the other elements. For a face on slab, $\Gamma=1.9$ and and iron abundance relative to hydrogen of $3.31 \times 10^{-5}$ (Anders \& Ebihara 1982), the George \& Fabian model predicts and equivalent width for the iron K$\alpha$ line of $EW_{\rm 0} = 144$~eV. We parametrerize the photon index-dependence of the equivalent width as:

$$ EW = 9.66 \ EW_{0} \ ( \Gamma^{-2.8} - 0.56 ) \ \ \ \ \ \ \  \  1.1<\Gamma<2.5 $$ 

\noindent
The inclination dependence is well-approximated by a cubic fit:

$$ EW = EW_0 \ \{ 2.20 \cos i - 1.749 (\cos i)^{2} + 0.541 (\cos i)^{3} \}  \\\\\\\\\\    i<85^{\circ} $$ 

\noindent
Finally the abundance dependence is approximately quadratic in log space, over a range of relative abundance $0.1 < A_{\rm Fe}  < 10.0$ :

$$ \log EW = \log EW_0 \ \{ 0.641 \log A_{\rm Fe}  - 0.172 (\log A_{\rm Fe})^{2} \} $$

\noindent
We include a Compton shoulder to the Fe K$\alpha$, approximating this as a Gaussian with energy 6.315 keV and width $\sigma=0.035$ keV.  The strength of the Compton shoulder is given according to the prescription of Matt (2002), which
has an inclination-dependence such that:

$$EW_{\rm cs}= EW_{Fe K\alpha} (0.1 + 0.1 \cos i)$$ 

\noindent
We also include a neutral Fe K$\beta$ line (7.05 keV) with a flux of 11.3~per cent of the iron K$\alpha$ line and Ni K$\alpha$ line (7.47 keV) with 5~per cent of the flux. 

The resulting model, which therefore includes inclination--, index-- and abundance--
dependent Compton reflection, Fe K$\alpha$, Fe K$\beta$ and Ni K$\alpha$ emission and the Fe K$\alpha$ Compton shoulder, was implemented as a local model in \textsc{XSPEC} called {\tt pexmon}. 

In all cases, we adopt an incident continuum which is a power law fixed to the same $\Gamma$ as the observed power law continuum, with high energy cutoff $E_{\rm c} = 1$~MeV. Given the upper bound of 10~keV for the data considered here, the precise choice of cutoff energy has a negligible effect on the fits. We generally adopt solar abundances. When modeling distant, neutral reflection we assume an inclination angle of $60^{\circ}$. This should be a reasonable approximation to the torus in the case of Seyfert 1 galaxies, as long as the opening angle of the torus is of this same order. The only free parameter for the distant reflector is therefore the reflection fraction $R$. 

\subsubsection{Relativistic blurring}

If the emission comes from the inermost regions, e.g. reflection from an accretion disk, it will be blurred by the effects of Doppler motions and gravitation due to the central black hole. We account for this using a blurring  model (Fabian et al. 2002; Crummy et al. 2006) which convolves a given spectrum with the kernel from the relativistic accretion disk line model of Laor (1991). We use this in combination with \textsc{pexmon} thus blurring the emission lines and reflection continuum simultaneously. The particular version of the model we employ here (\textsc{kdblur2}) has 6 parameters: the inner and outer disk radii, $r_{\rm in}$ and $r_{\rm out}$, the inclination, $i$, two power law slopes,  $q_{\rm 1}$ and $q{\rm 2}$, which characterise the emissivity of the disk as $r^{-q}$, and a break radius \rbr\ where the emissivity changes from $q_{\rm 1}$ to $q_{\rm 2}$.  We adopt $r_{out}=400 r_{\rm g}$ throughout, that being the maximum radius in the Laor (1991) computations. $r_{in}$ is fixed at either $6 r_{\rm g}$ when considering a Schwarschild geometry  or $1.24 r_{\rm g}$ for a Kerr geometry. Strictly speaking, the Laor (1991) model is appropriate only for a maximally rotating Kerr geometry, but the differences with the Schwarschild case for $R>6 r_{\rm g}$ are negligible and ignored here. The inclination is left as a free parameter. For a point source at a height $h$ illuminating a slab in a Newtonian geometry, the emissivity is given by $q\propto h/(R^{2}+h^{2})^{3/2}$ (e.g. Vaughan et al. 2004). The emissivity can therefore be approximated by  $q_{\rm 1}=0$ and $q_{\rm 2}=3$.  We adopt these values, leaving $r_{\rm br}$ to be a free parameter. In the ``lampost'' type of geometry $r_{\rm br}$ depends on the height of the source above the slab in the sense that $r_{\rm br} \sim h$.  Regardless of the true geometry, $r_{\rm br}$ represents where the peak of the disk line flux occurs. 

\subsubsection{Ionized absorption}

It is known that many AGN show substantial absorption due to ionized gas in the line-of-sight (Halpern 1984; Crenshaw, Kraemer \& George 2003 and references therein). While this has the strongest effect on the spectrum at soft X-ray energies (e.g. George et al. 1998b), if there is a component of the absorber with a very high column density ($\gg 10^{22}$~cm$^{-2}$) it may have a significant effect even on the spectrum above 2.5~keV (e.g. Nandra \& Pounds 1994). Indeed, the such a ``warm absorber", if sufficiently highly ionized, can introduce curvature into the spectrum mimicking the red wing of a relativistic emission line, and significanty affecting the line parameters derived (Reeves et al. 2004; Turner et al. 2005). 

To model such components in our spectra, we used two \textsc{XSPEC}-compatible  model spectra generated using  the photoionzation code \textsc{XSTAR}\footnote{http://heasarc.gsfc.nasa.gov/docs/software/xstar/xstar.html} (e.g. Kallman et al.  2004). These were created assuming  a powerlaw continuum, with a photon index of 2.  The hydrogen nucleus density was held constant at $10^{6}$~cm$^{-3}$, and the abundances were fixed at their solar values\footnote{The abundance of Ni was set to zero, owing to uncertainties in its atomic rates. This has a negligible effect given the range of parameters considered here.}.  The turbulent velocity was assumed to be $b=100$~km ~s$^{-1}$. This assumption will be discussed further below.

The first table ({\tt cwa18}) is a fine grid covering a wide range of parameter space, and was created 
using version 21kn of \textsc{XSTAR}. The range covered in ionization parameter, $\xi$, is $-4 < \log \xi <4$ and in column density $N_{\rm H}$, $20 < \log N_{\rm H} <24$, both in steps of 0.1. Owing to its wide range in $\xi$ and $N_{\rm H}$, this absorber model  is well suited to modelling the soft X-ray curvature in the spectra of a large variety of objects. The second table ({\tt grid25}) was created using \textsc{XSTAR} version 21ln. This version incorporates  updated information on absorption species around the iron K-complex, particularly for intermediate ionization species between Fe {\sc i} and Fe {\sc xxv}. This grid covers the range $1 < \log \xi  <5$ and $20 < \log N_{\rm H} <24$ in steps of 0.5. Although this grid is coarser compared to  {\tt cwa18}, it is more accurate in modelling the high ionization regime. Therefore, we use this grid to test whether high ionization absorbers can mimic the broad line.

\section{Basic parameterization of the spectra}

\subsection{Continuum modeling}

\begin{figure*}
{
\includegraphics[angle=270,width=58mm]{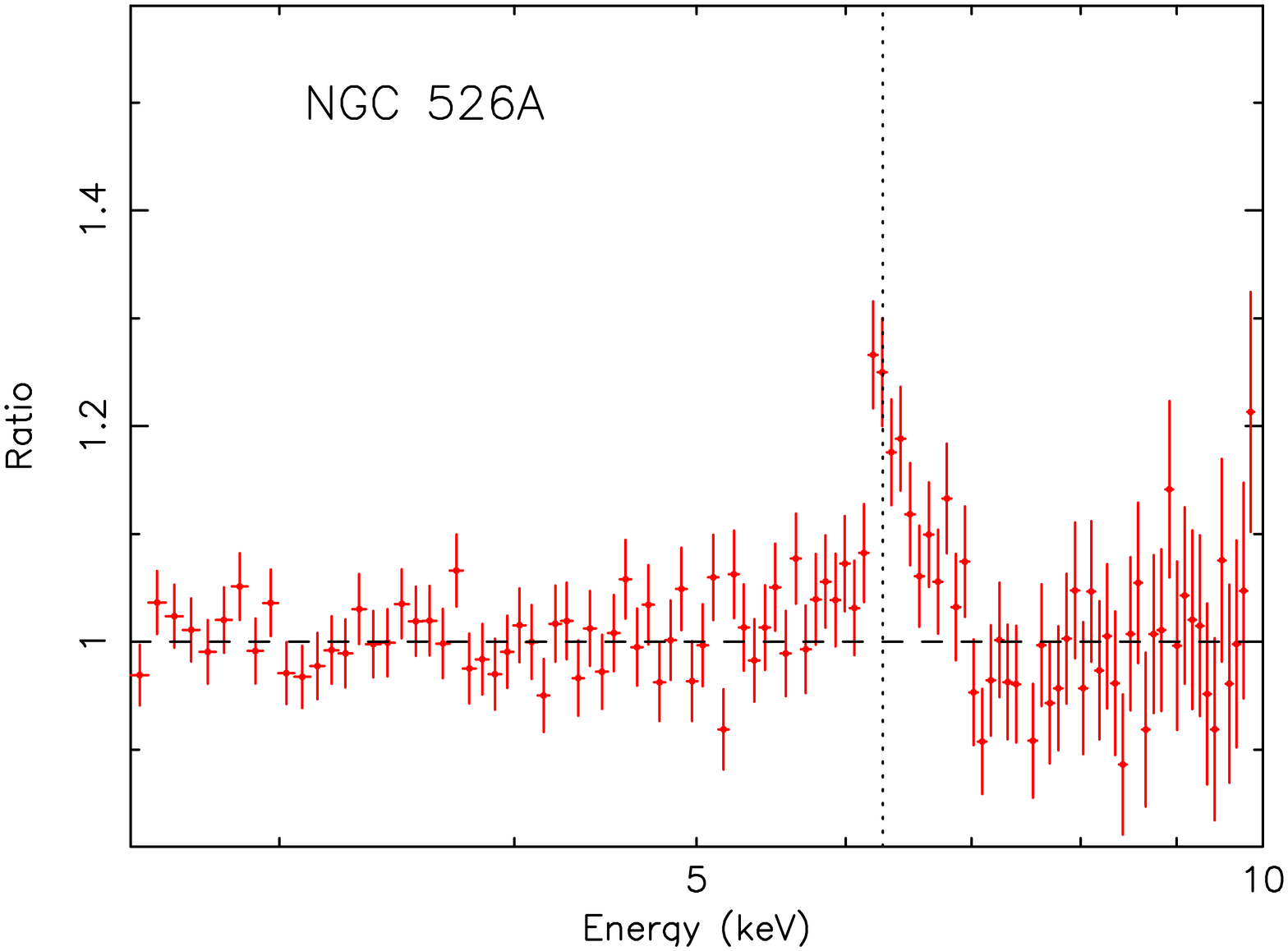}
\includegraphics[angle=270,width=58mm]{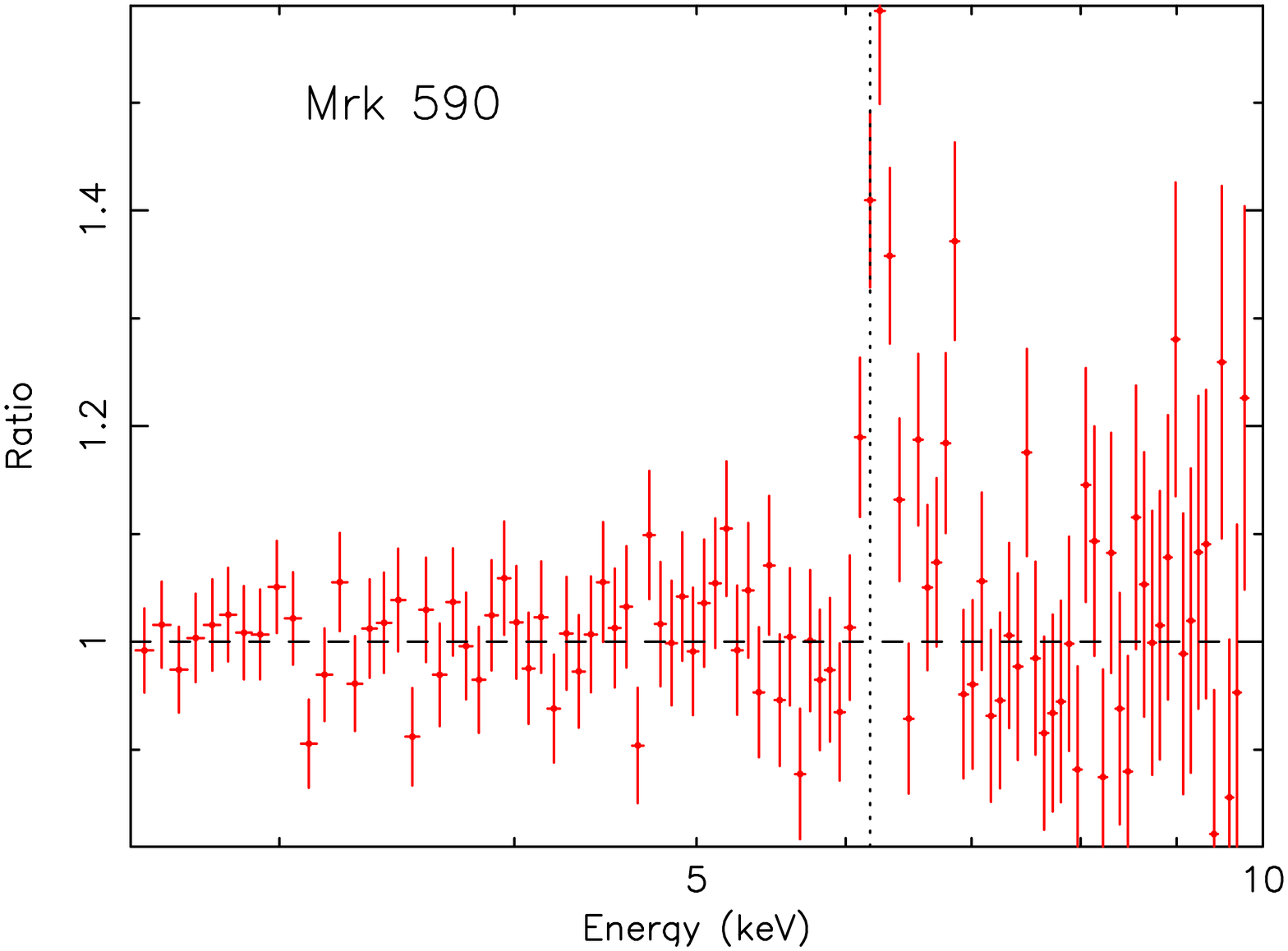}
\includegraphics[angle=270,width=58mm]{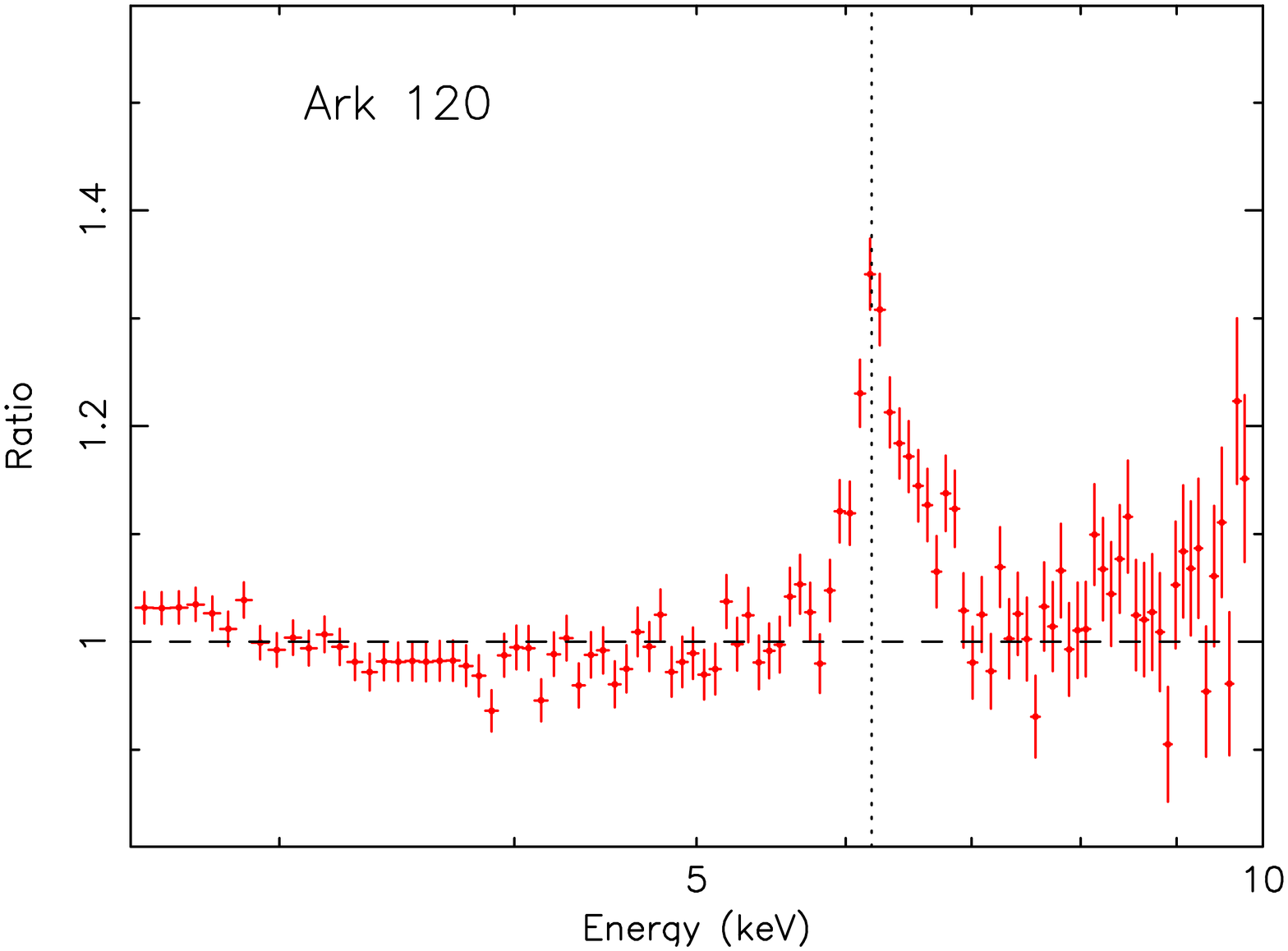}
\includegraphics[angle=270,width=58mm]{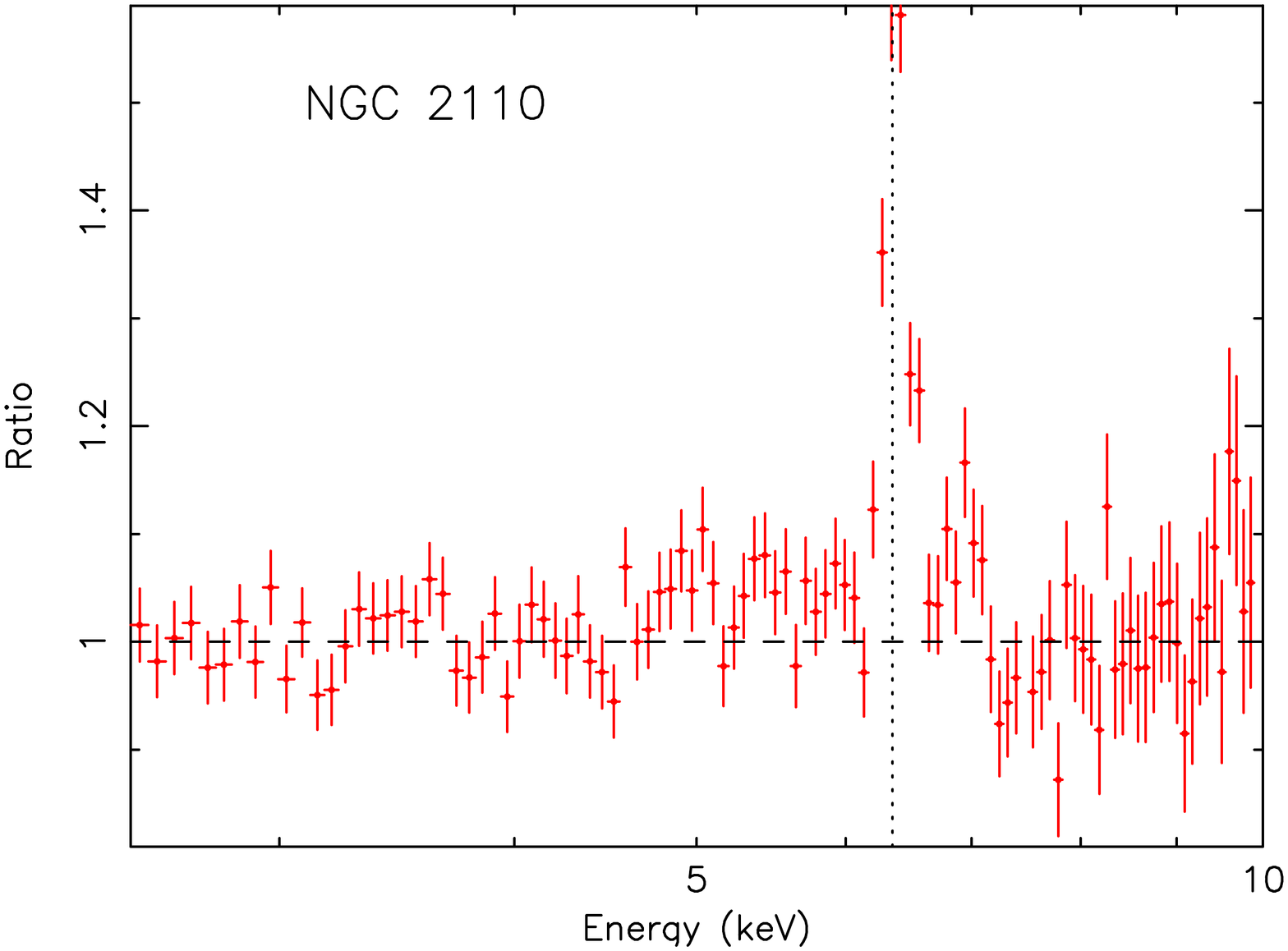}
\includegraphics[angle=270,width=58mm]{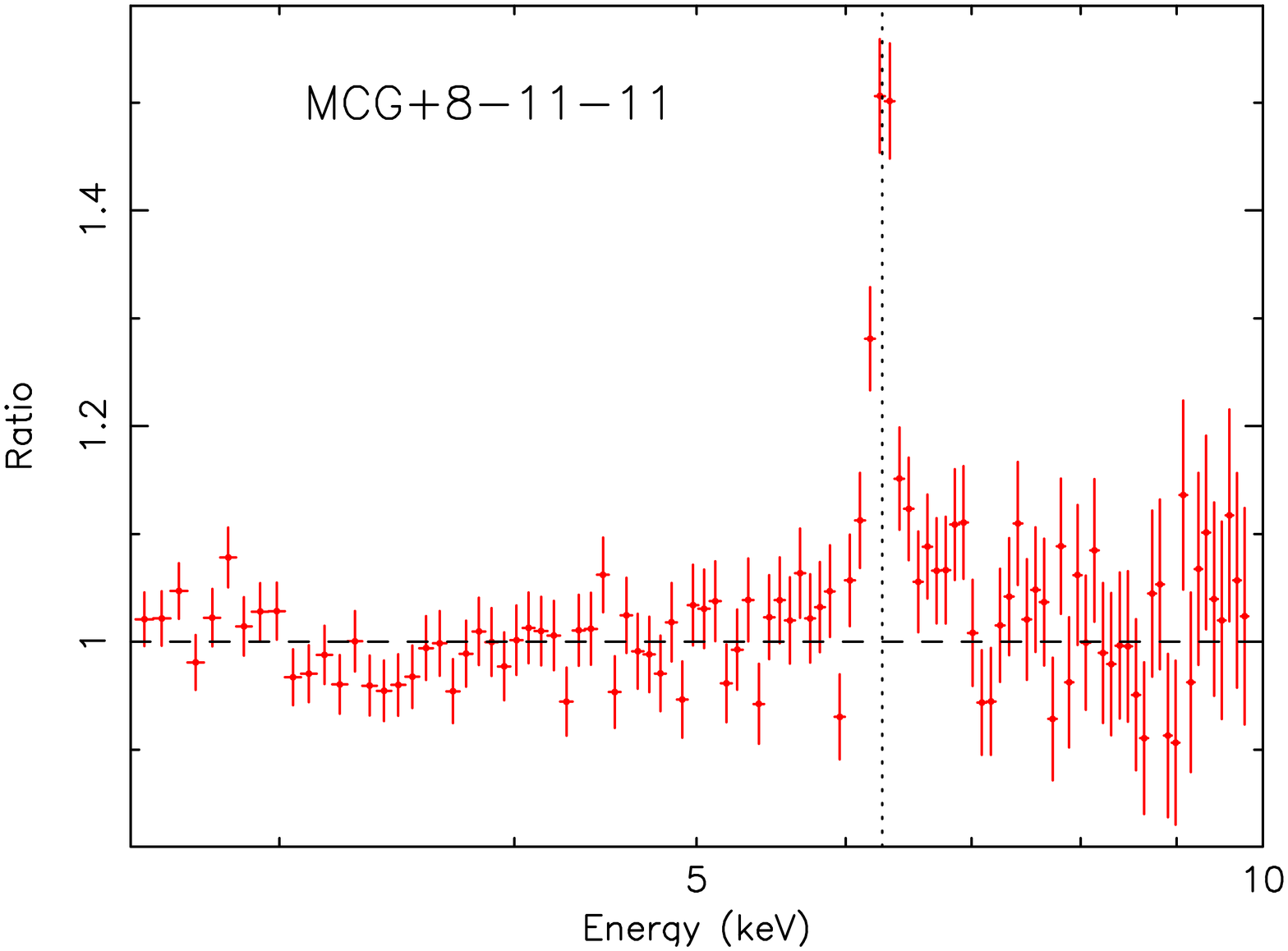}
\includegraphics[angle=270,width=58mm]{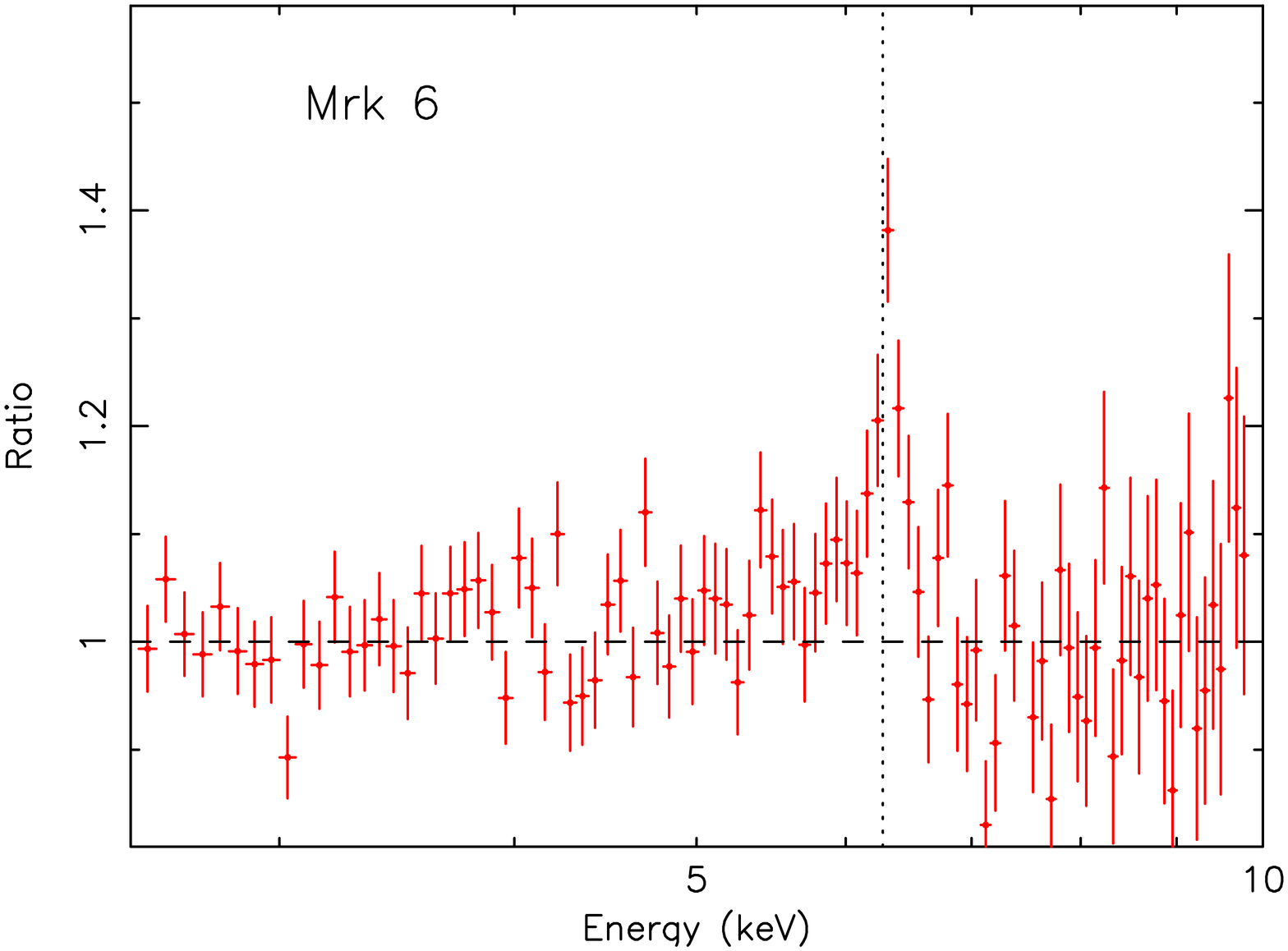}
\includegraphics[angle=270,width=58mm]{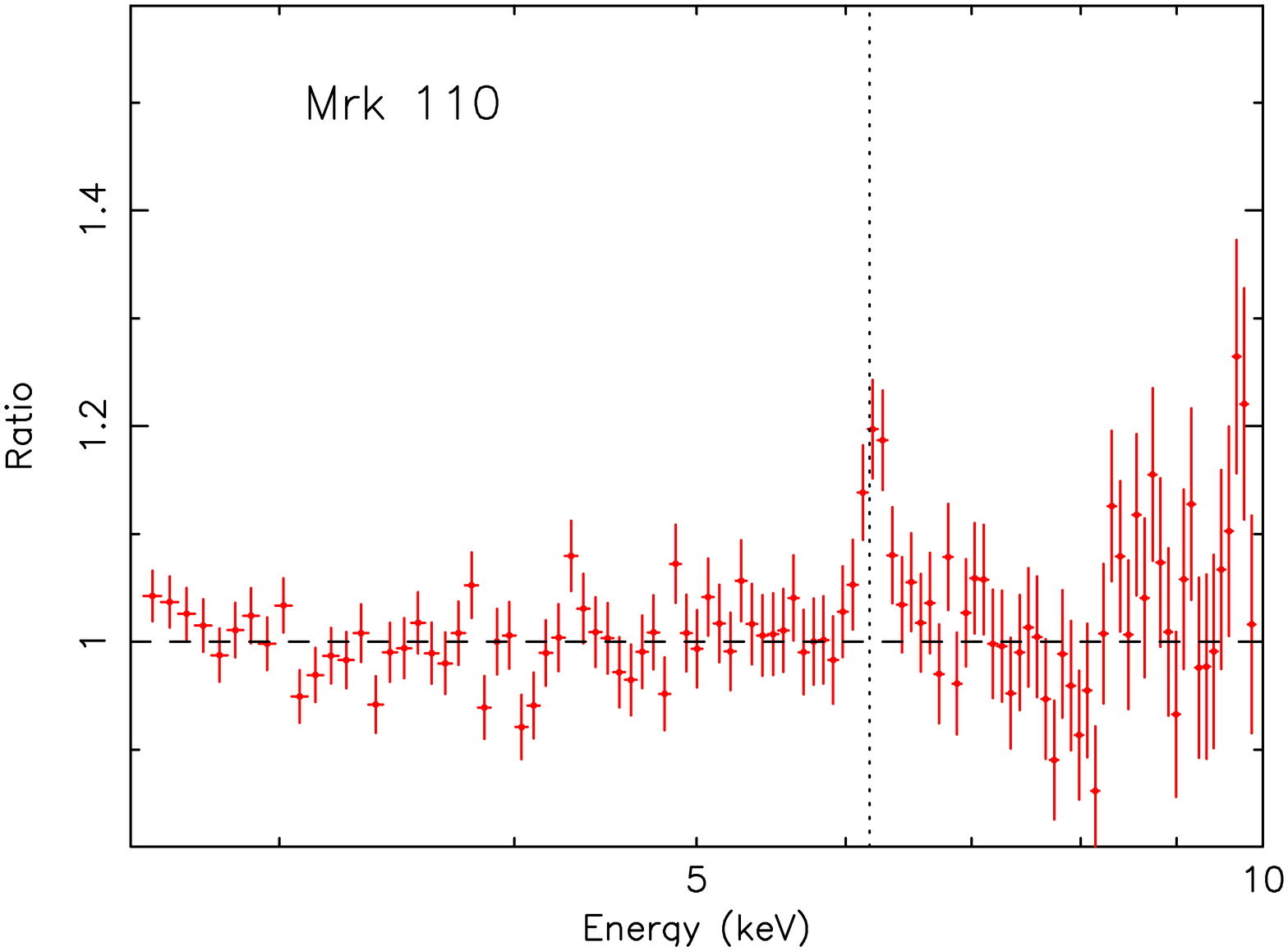}
\includegraphics[angle=270,width=58mm]{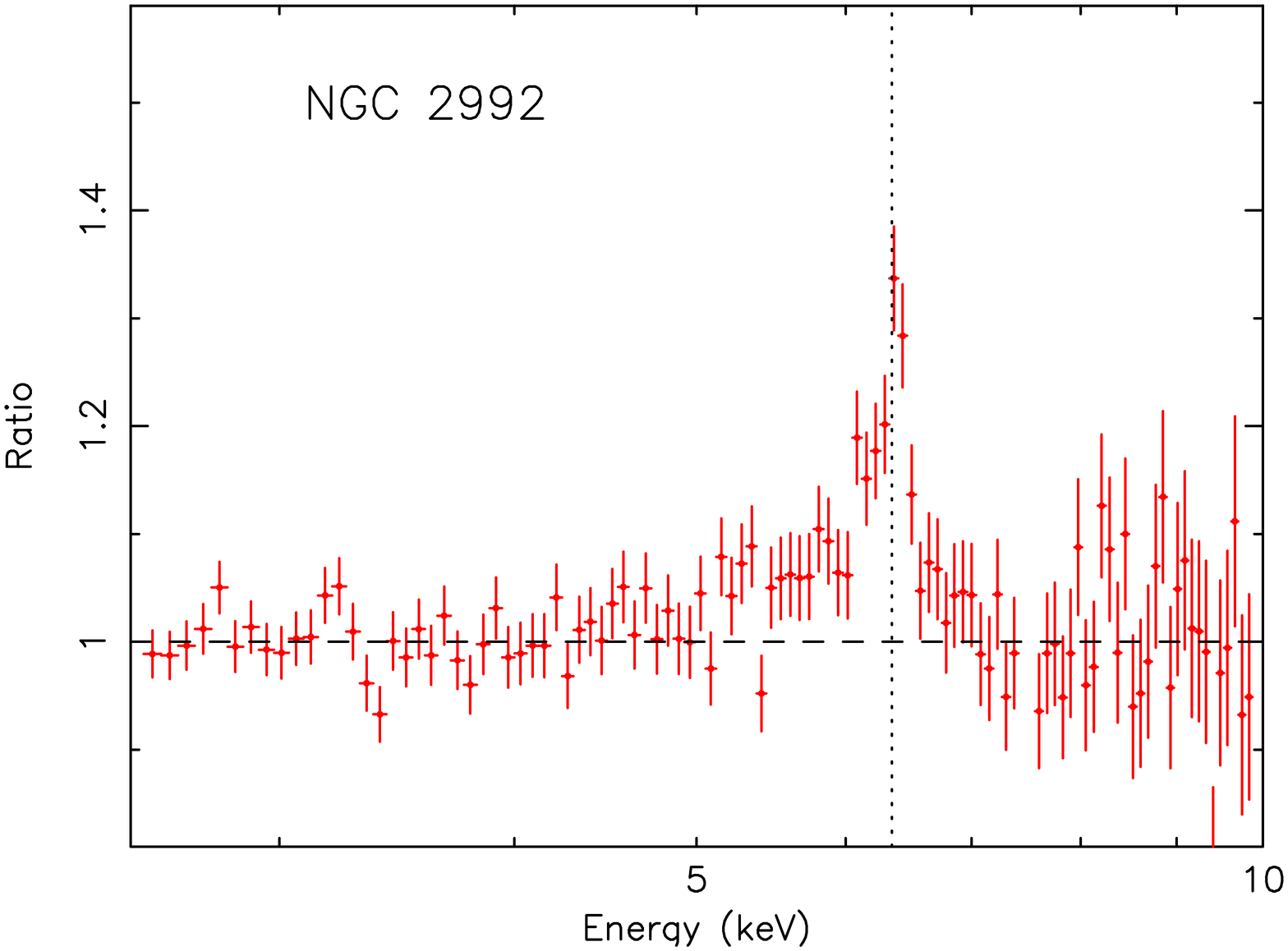}
\includegraphics[angle=270,width=58mm]{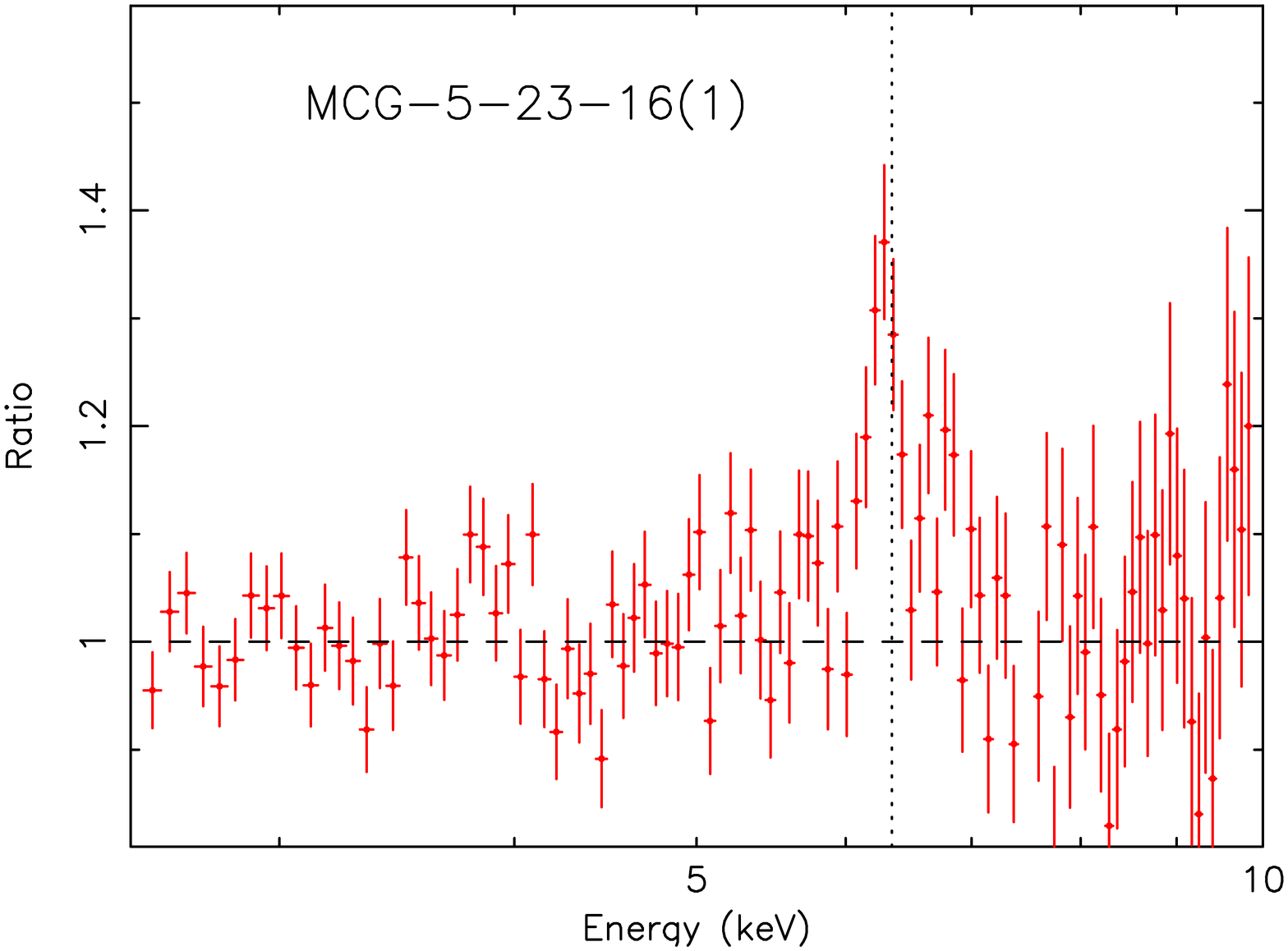}
\includegraphics[angle=270,width=58mm]{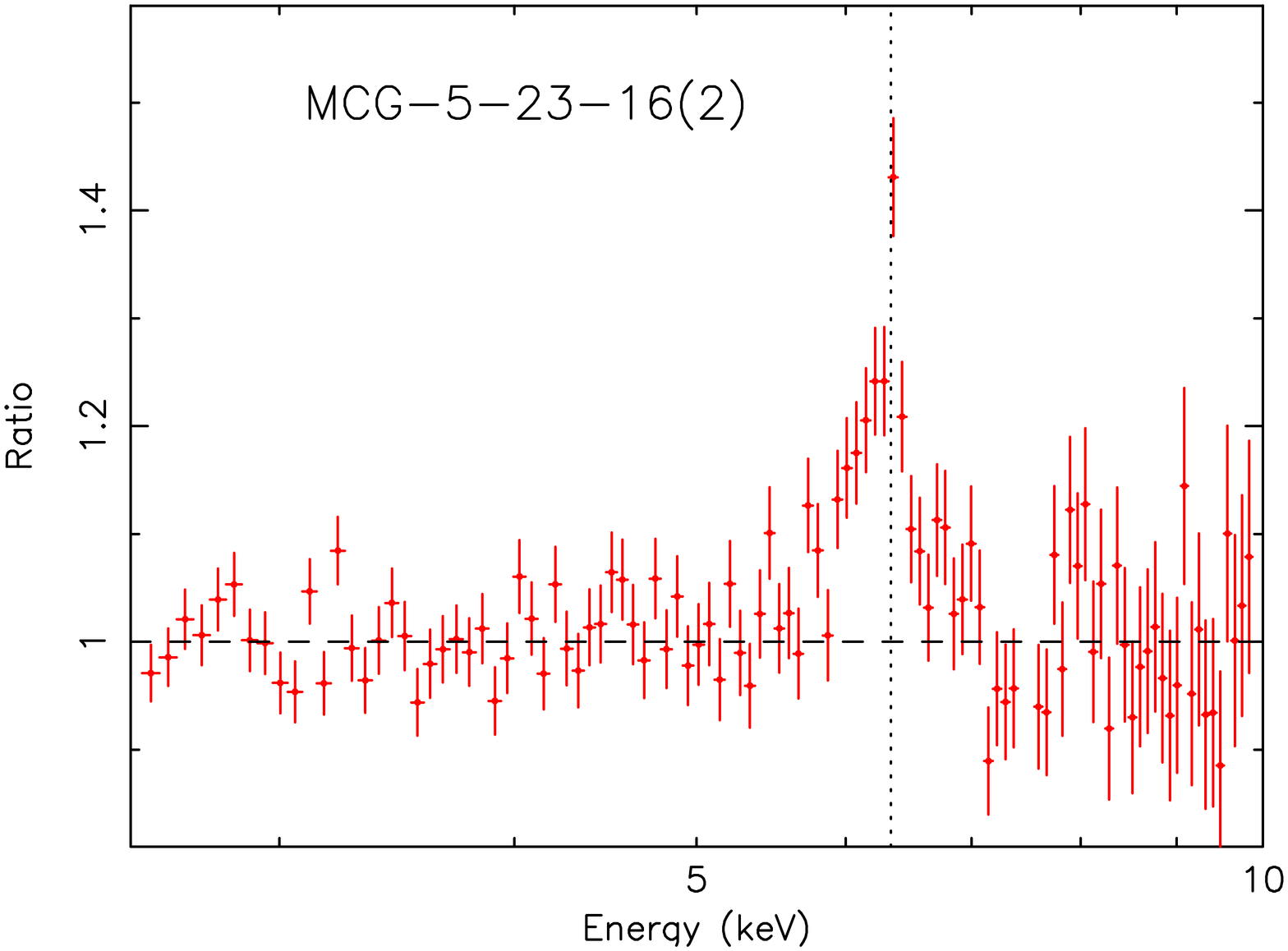}
\includegraphics[angle=270,width=58mm]{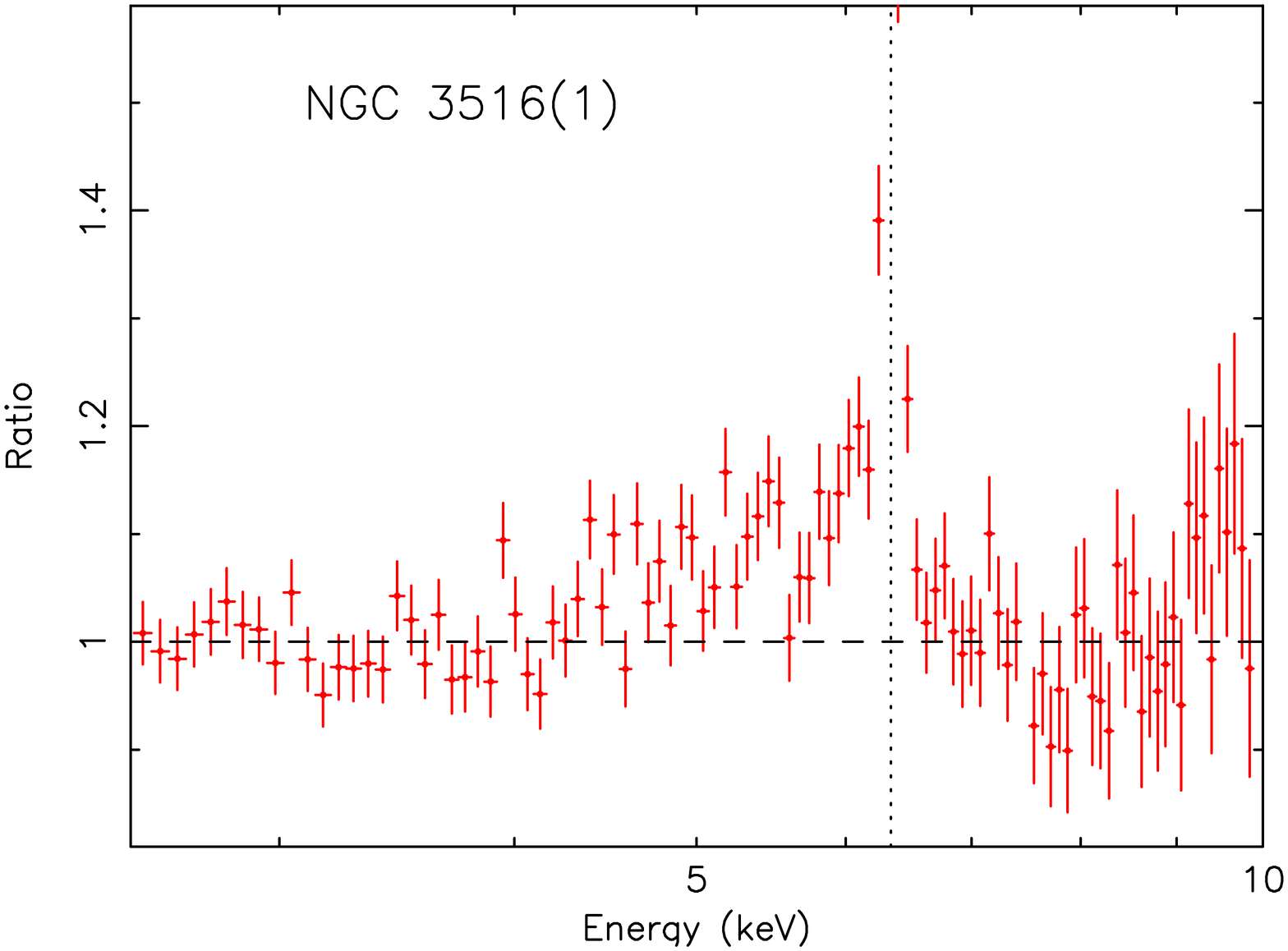}
\includegraphics[angle=270,width=58mm]{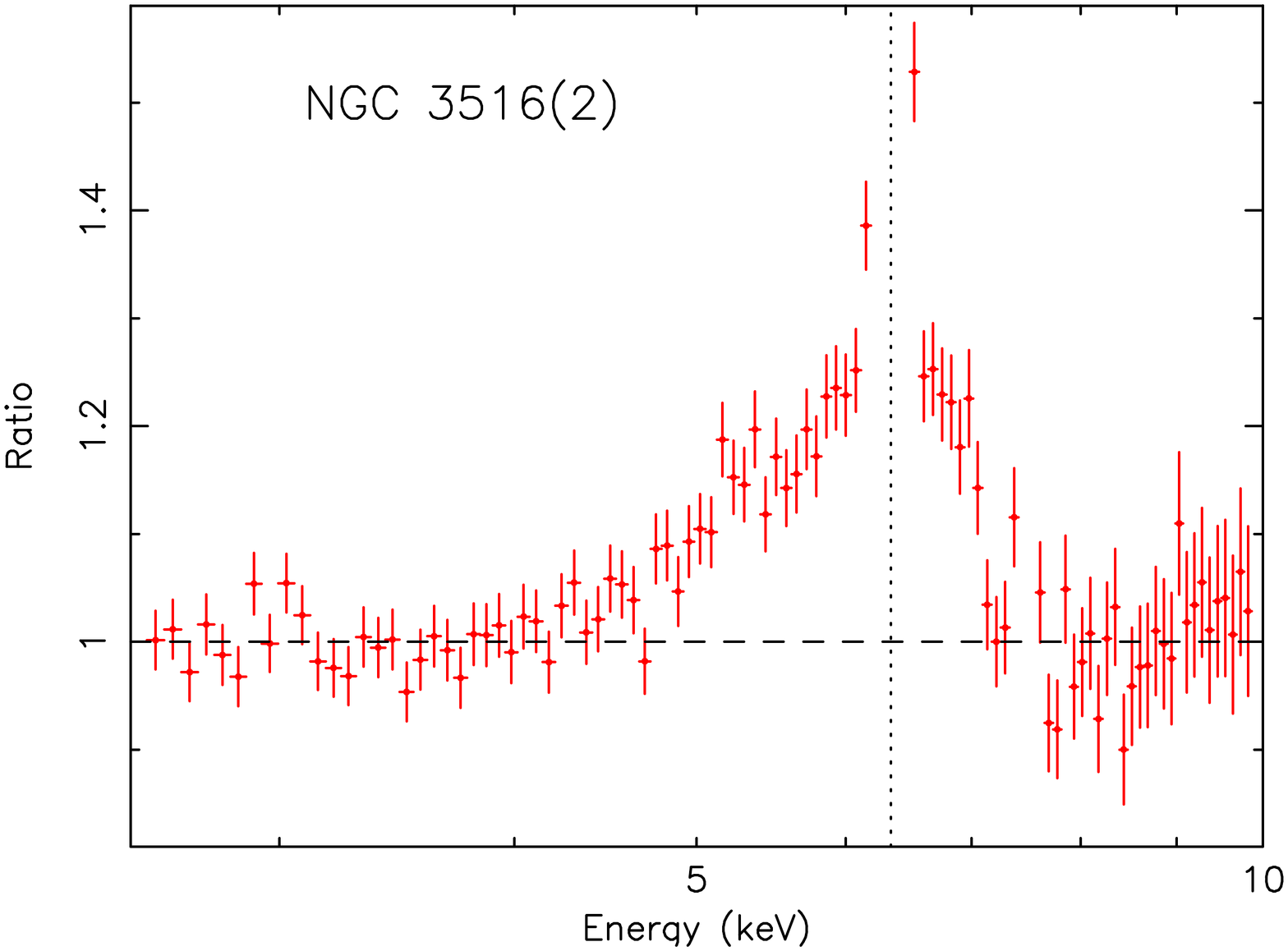}
\includegraphics[angle=270,width=58mm]{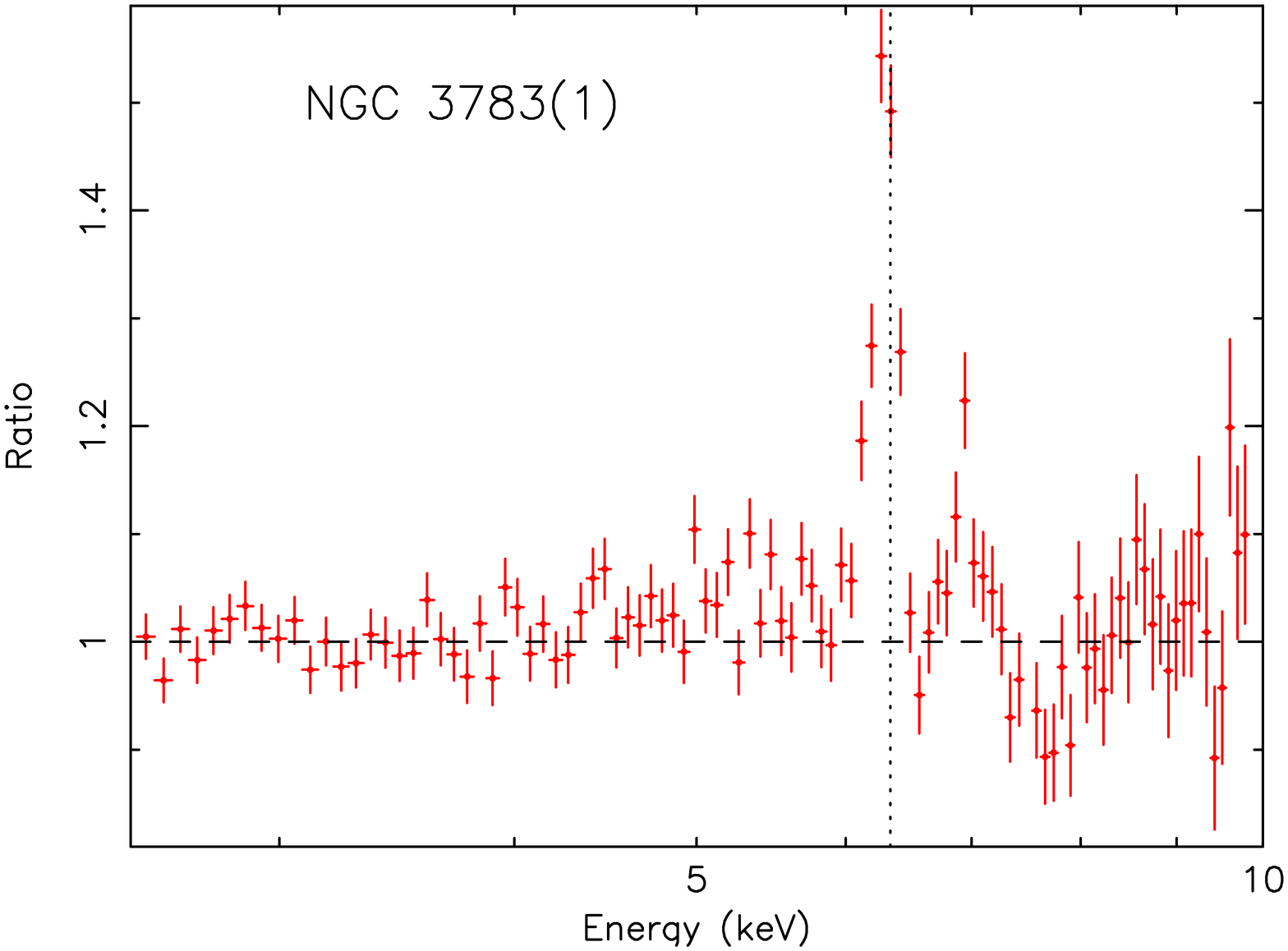}
\includegraphics[angle=270,width=58mm]{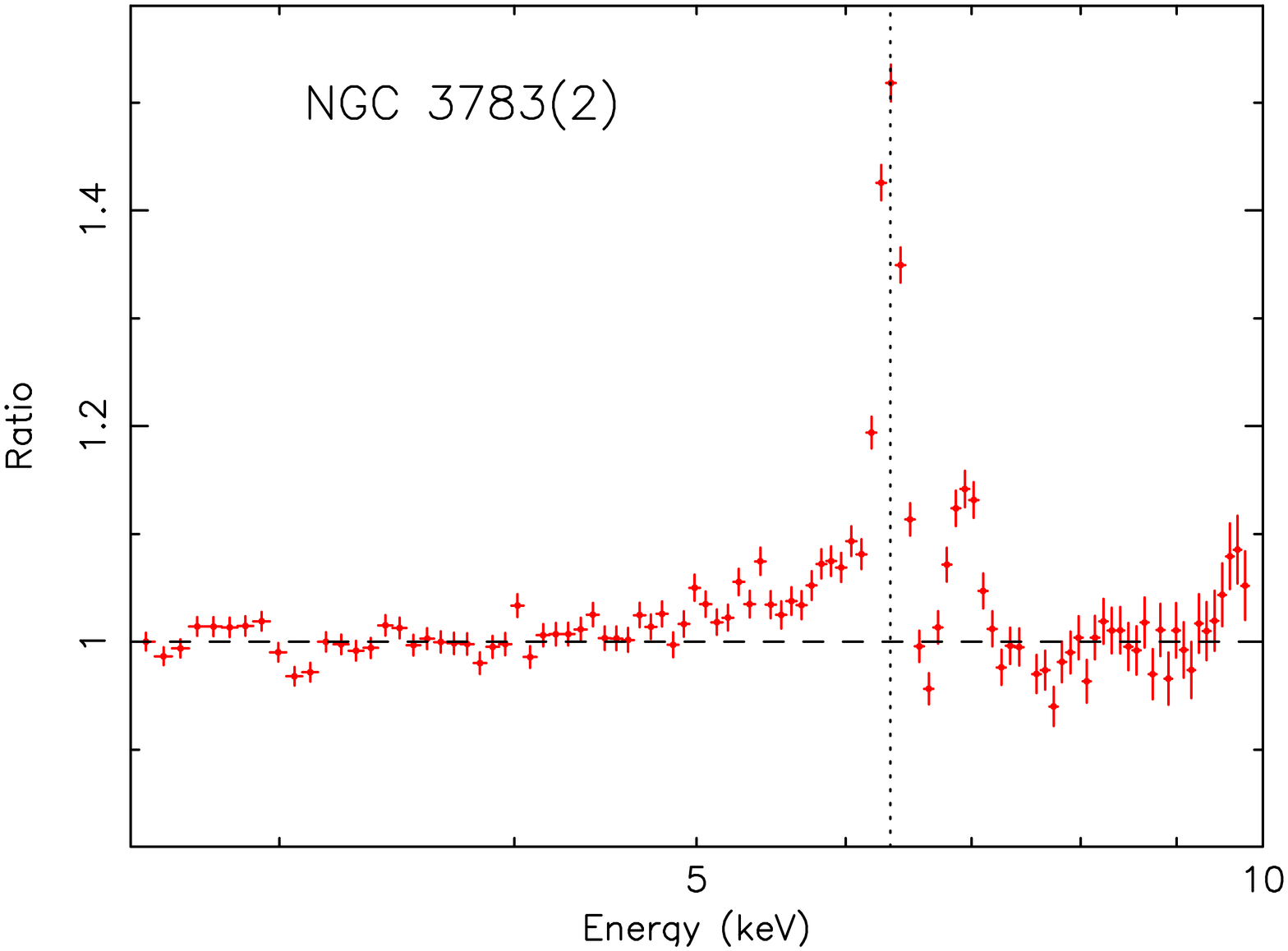}
}
\caption{Data/model ratios, derived excluding excluding the 4.5-7.5 keV region.  The fitted continuum is either a simple power law with Galactic absorption or, in cases where it significantly improves the fit, additional absorption by photoionized gas. The vertical dotted lines show a rest-frame energy of 6.4 keV, which is expected for fluorescence from neutral iron. 
 \label{fig:all_profile}}
\end{figure*}

\setcounter{figure}{2}

\begin{figure*}
{
\includegraphics[angle=270,width=58mm]{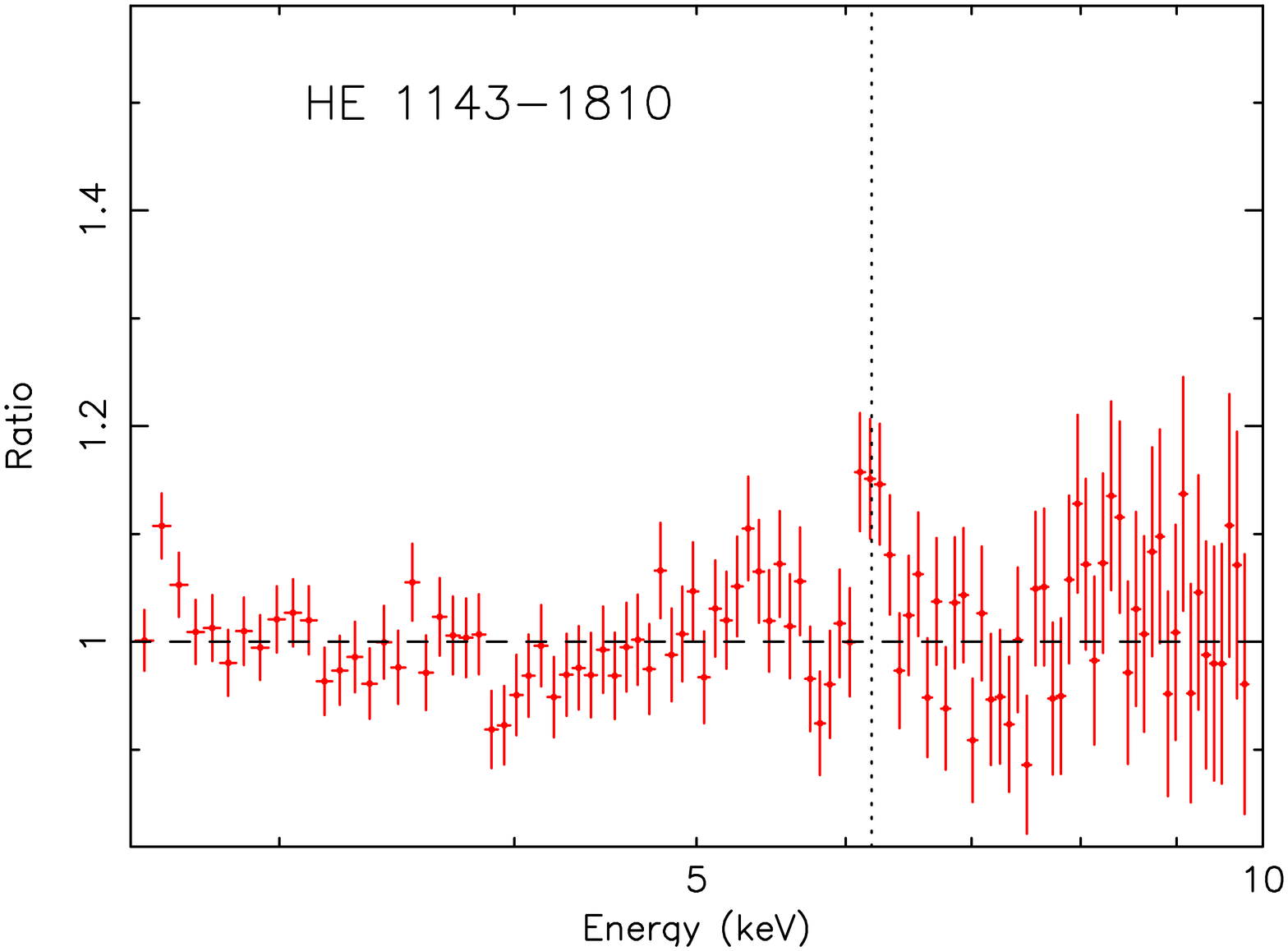}
\includegraphics[angle=270,width=58mm]{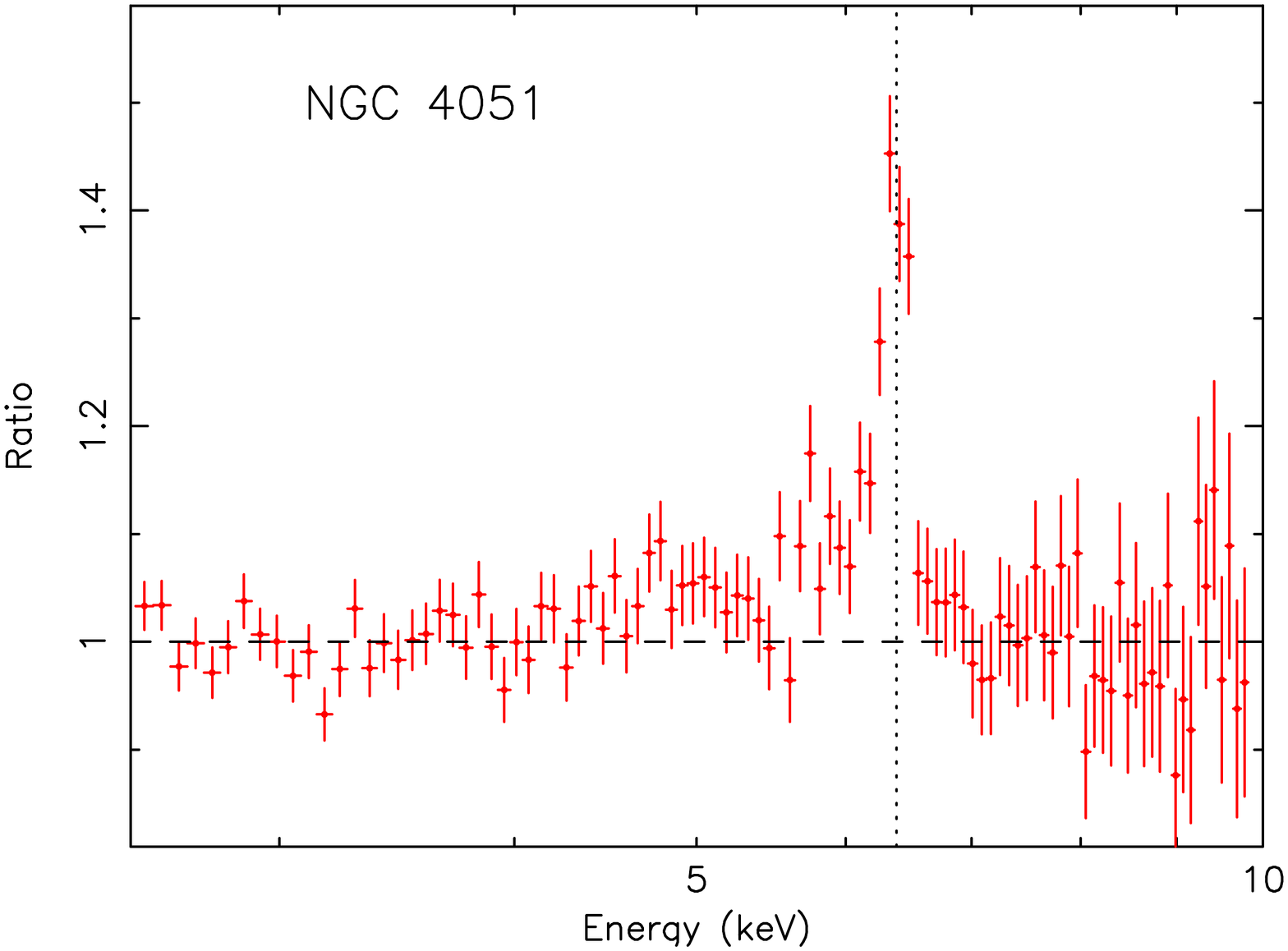}
\includegraphics[angle=270,width=58mm]{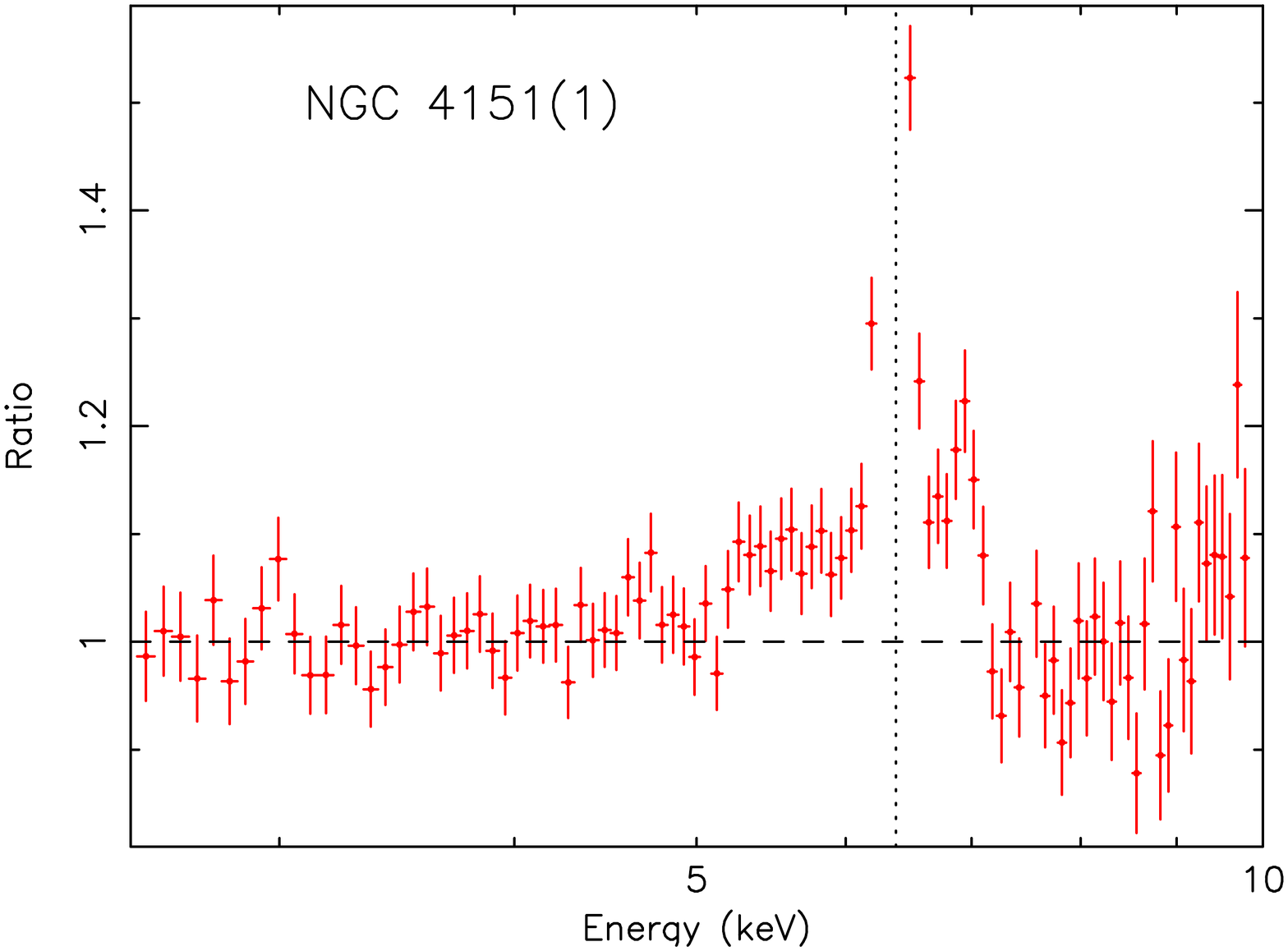}
\includegraphics[angle=270,width=58mm]{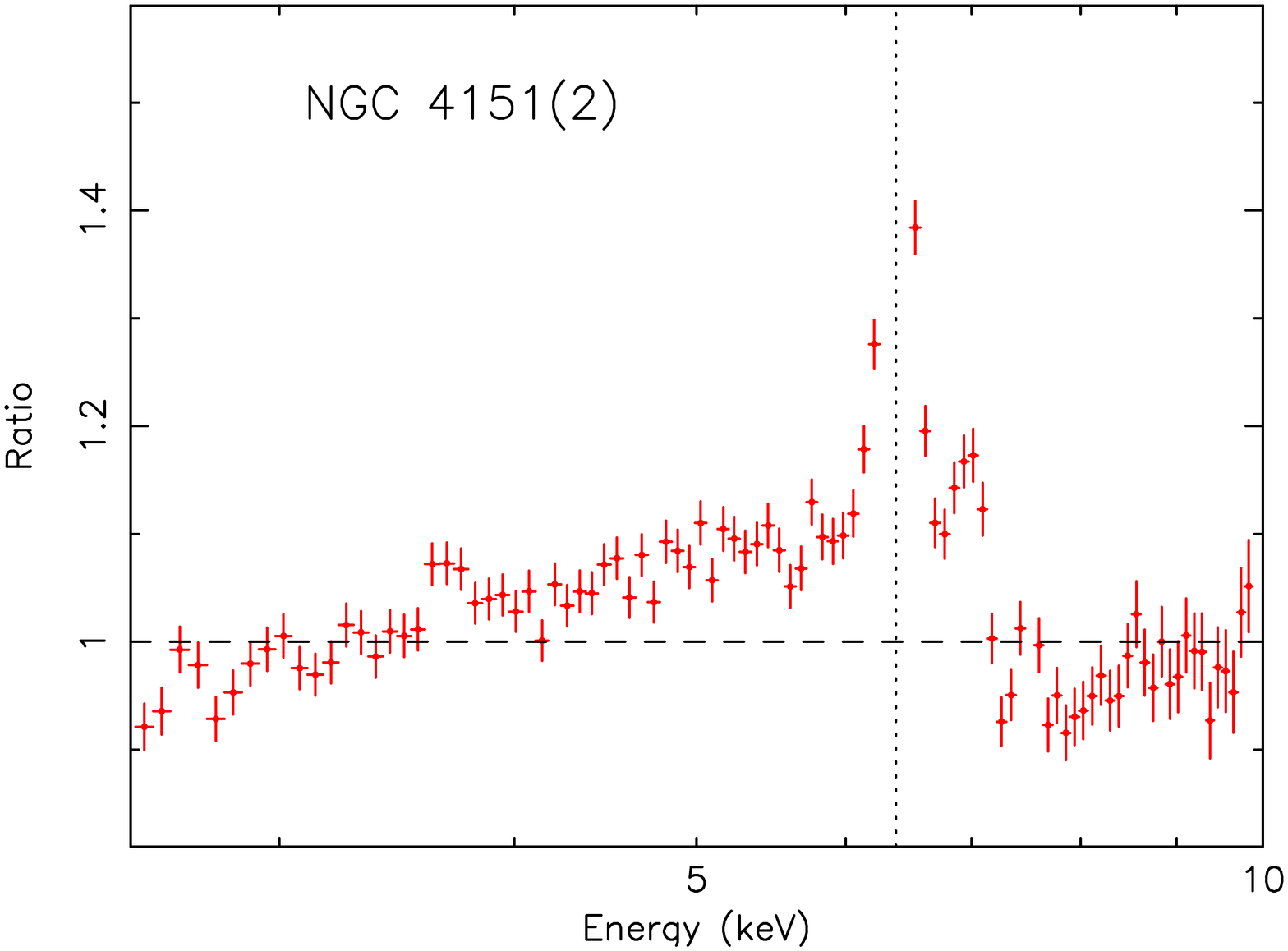}
\includegraphics[angle=270,width=58mm]{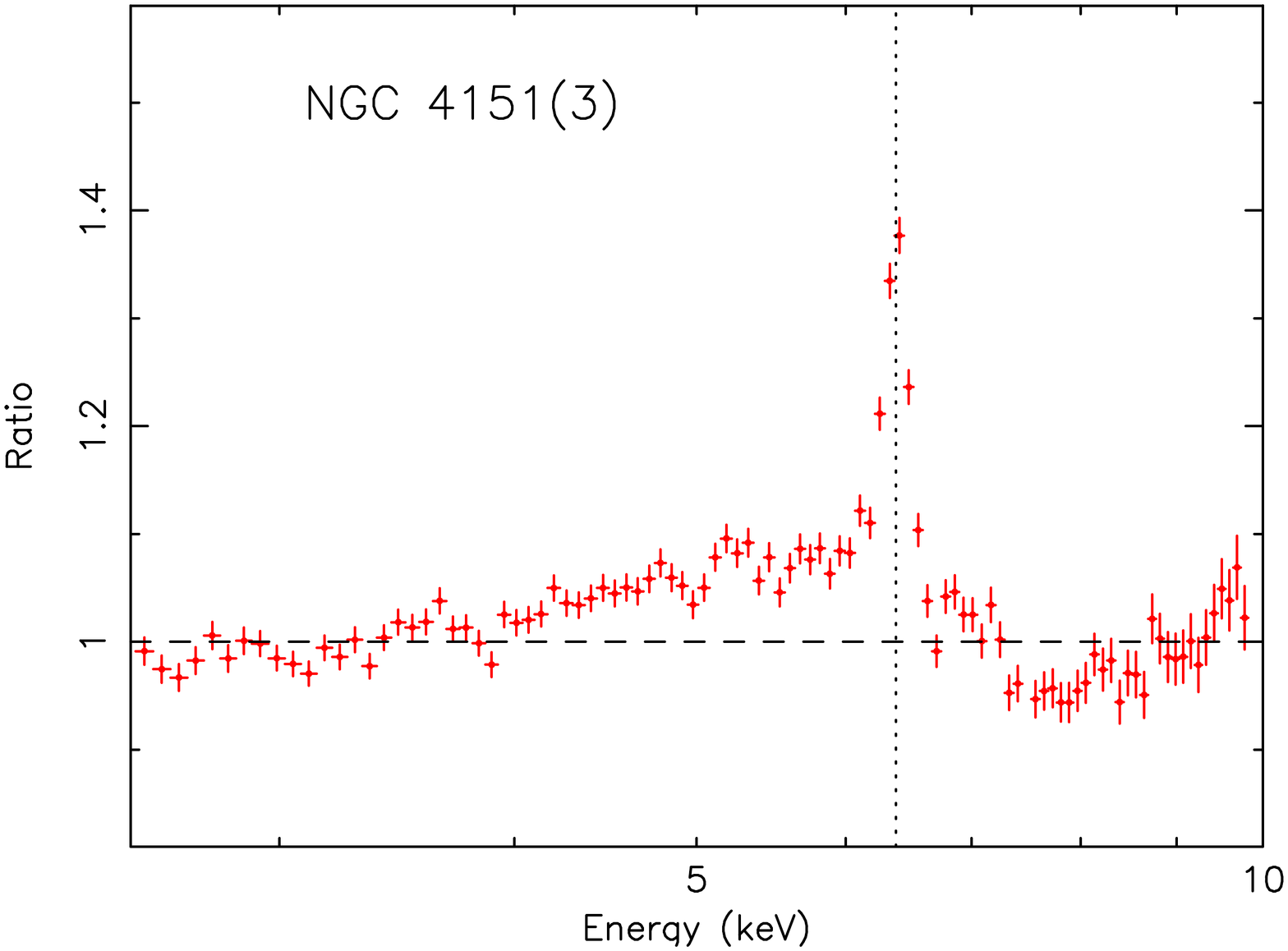}
\includegraphics[angle=270,width=58mm]{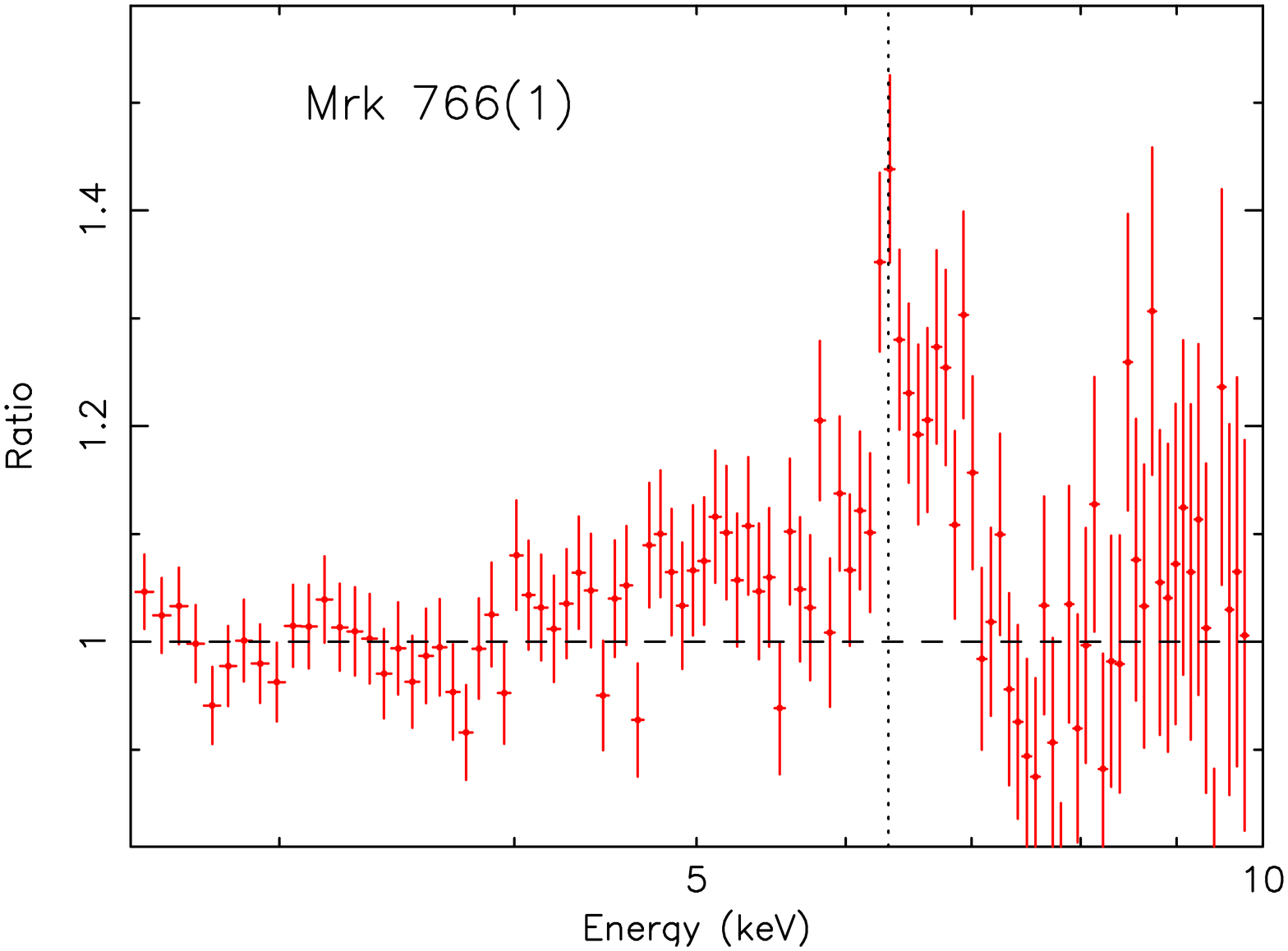}
\includegraphics[angle=270,width=58mm]{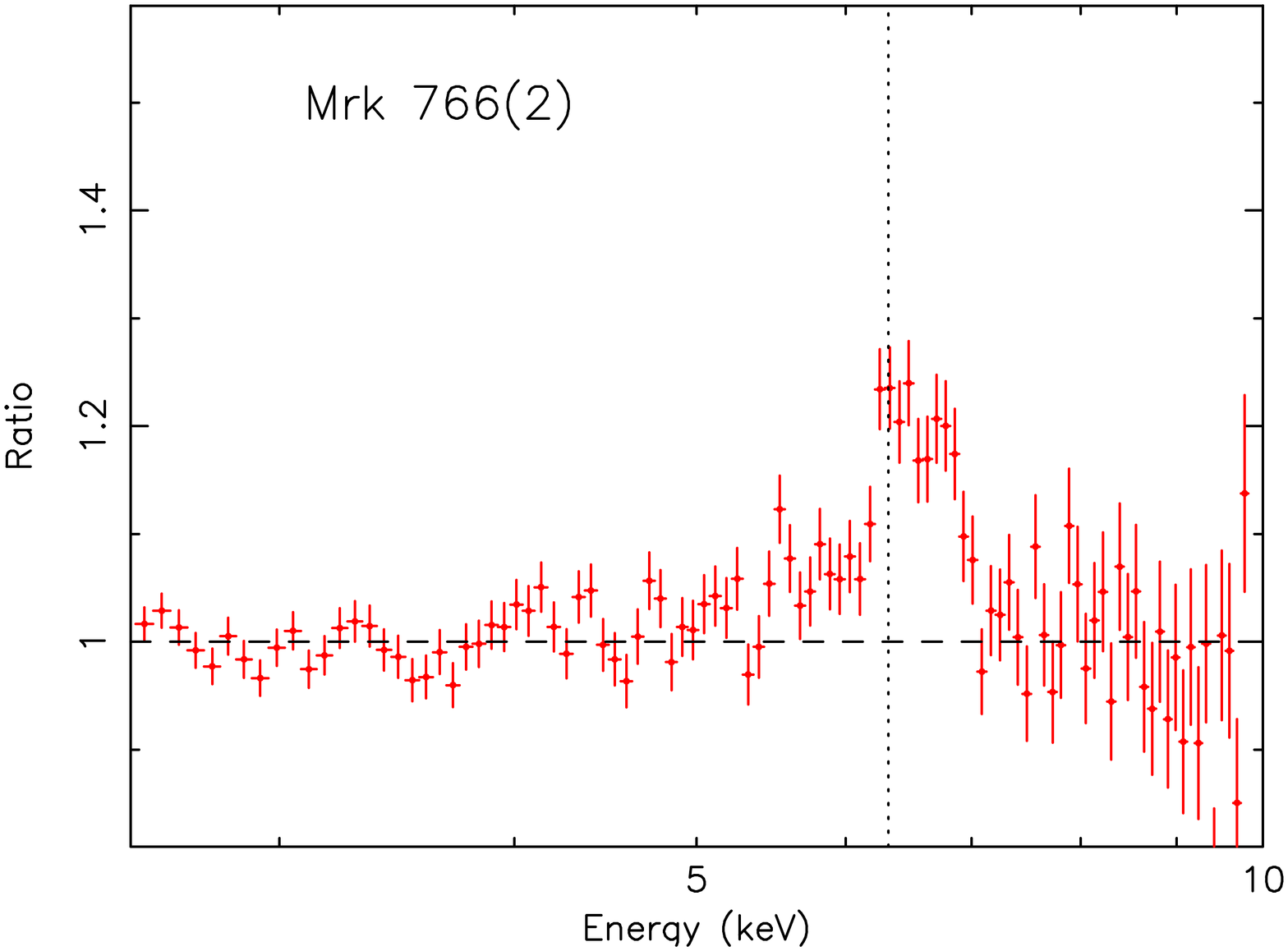}
\includegraphics[angle=270,width=58mm]{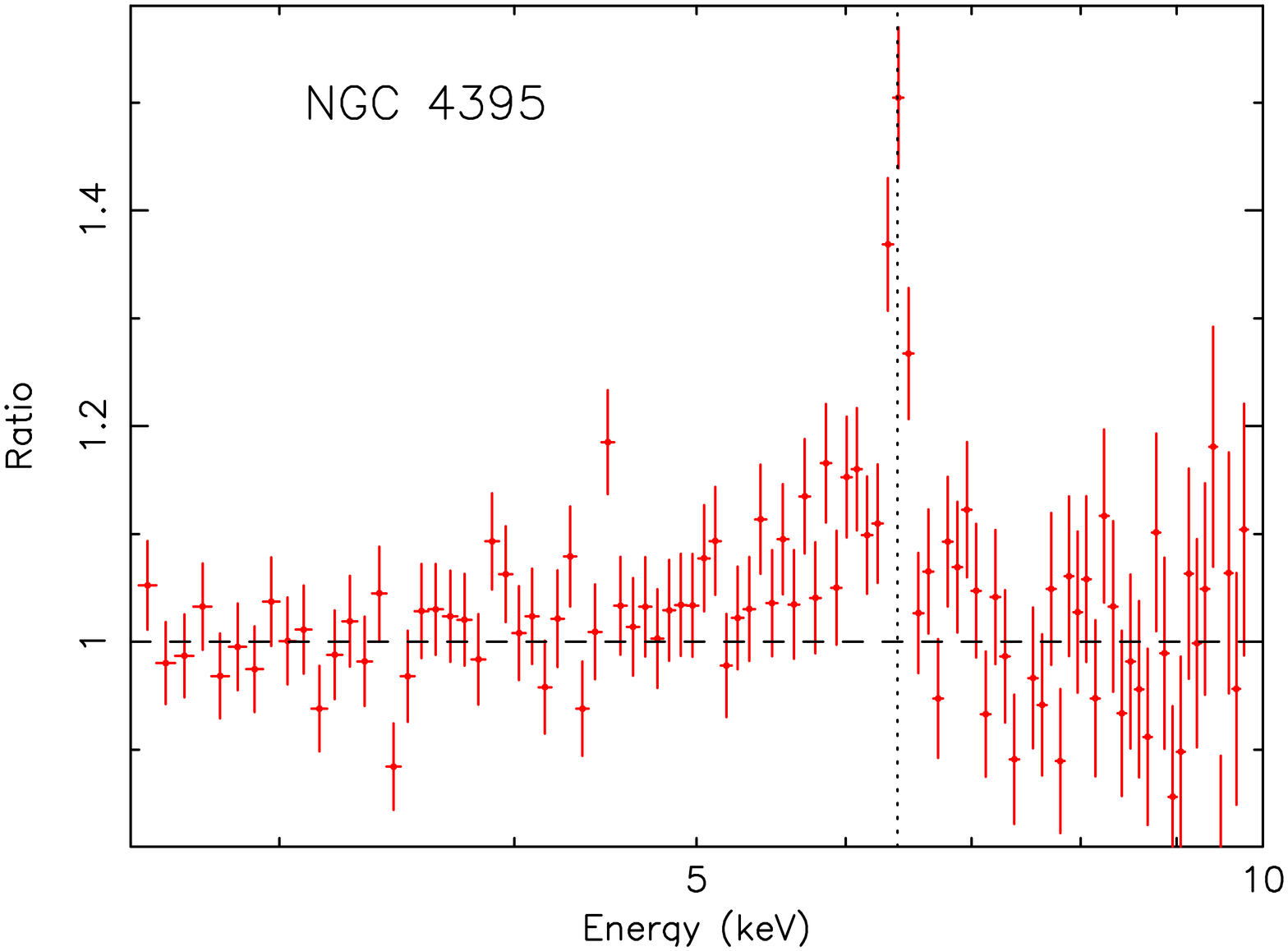}
\includegraphics[angle=270,width=58mm]{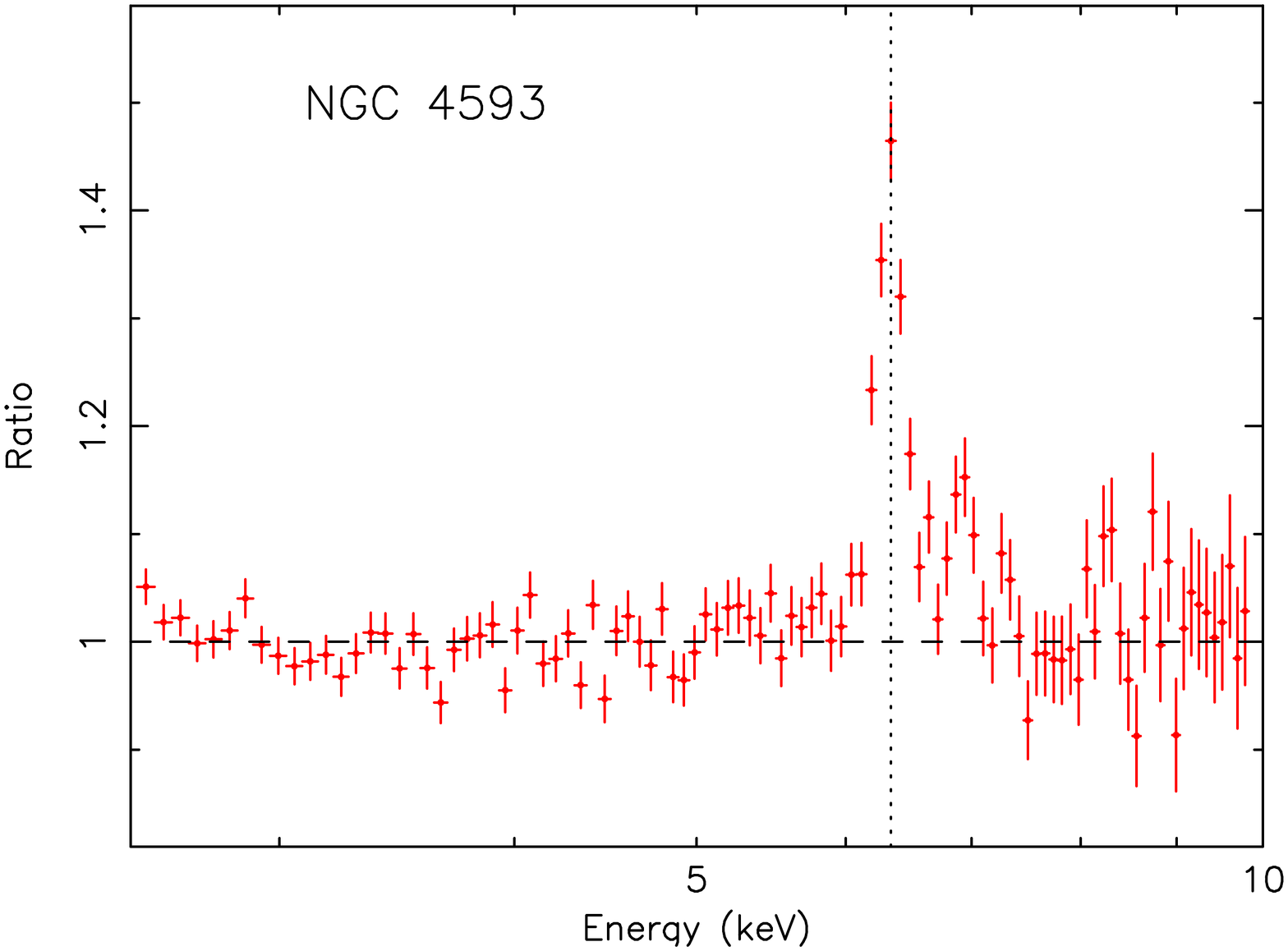}
\includegraphics[angle=270,width=58mm]{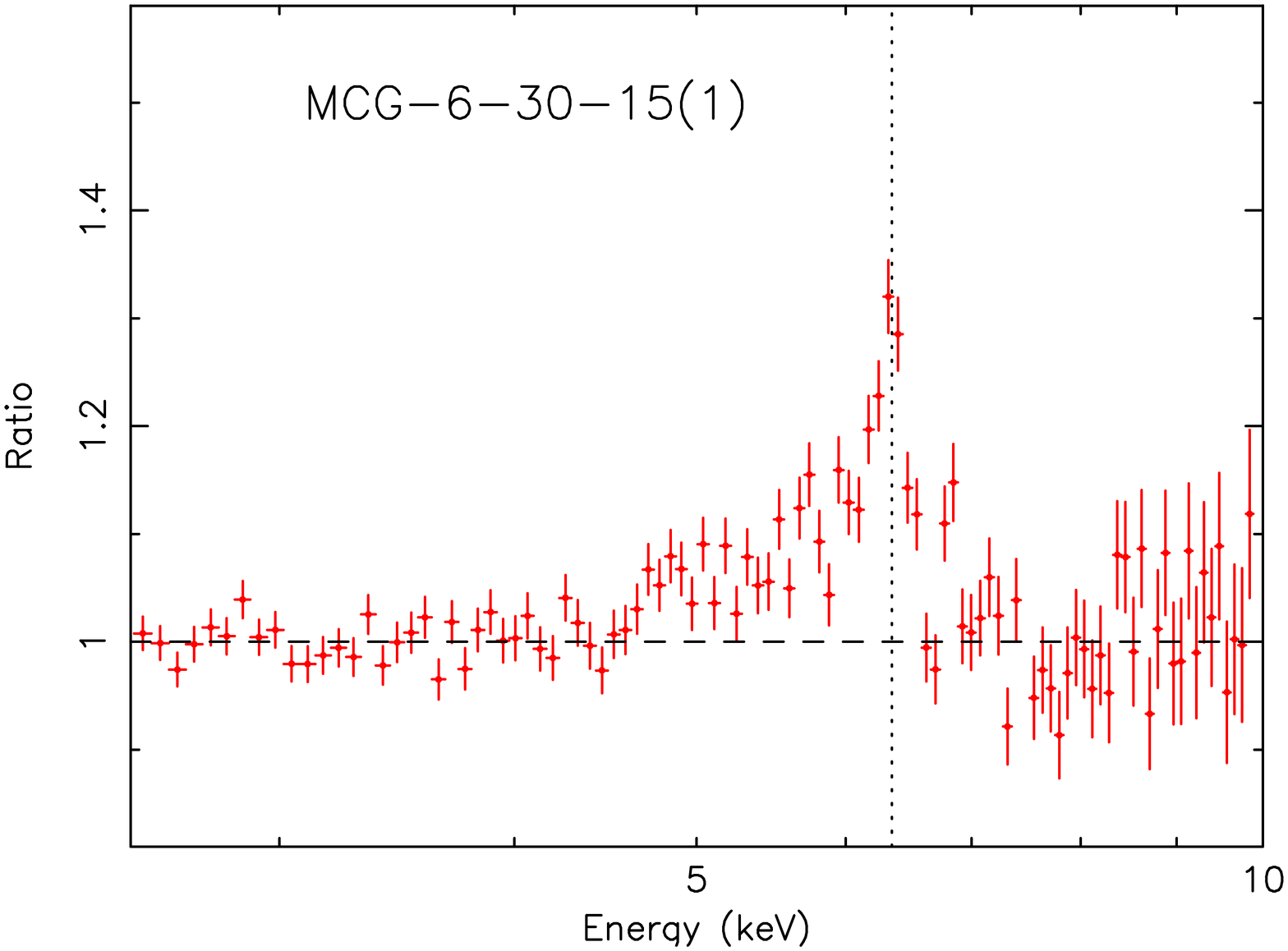}
\includegraphics[angle=270,width=58mm]{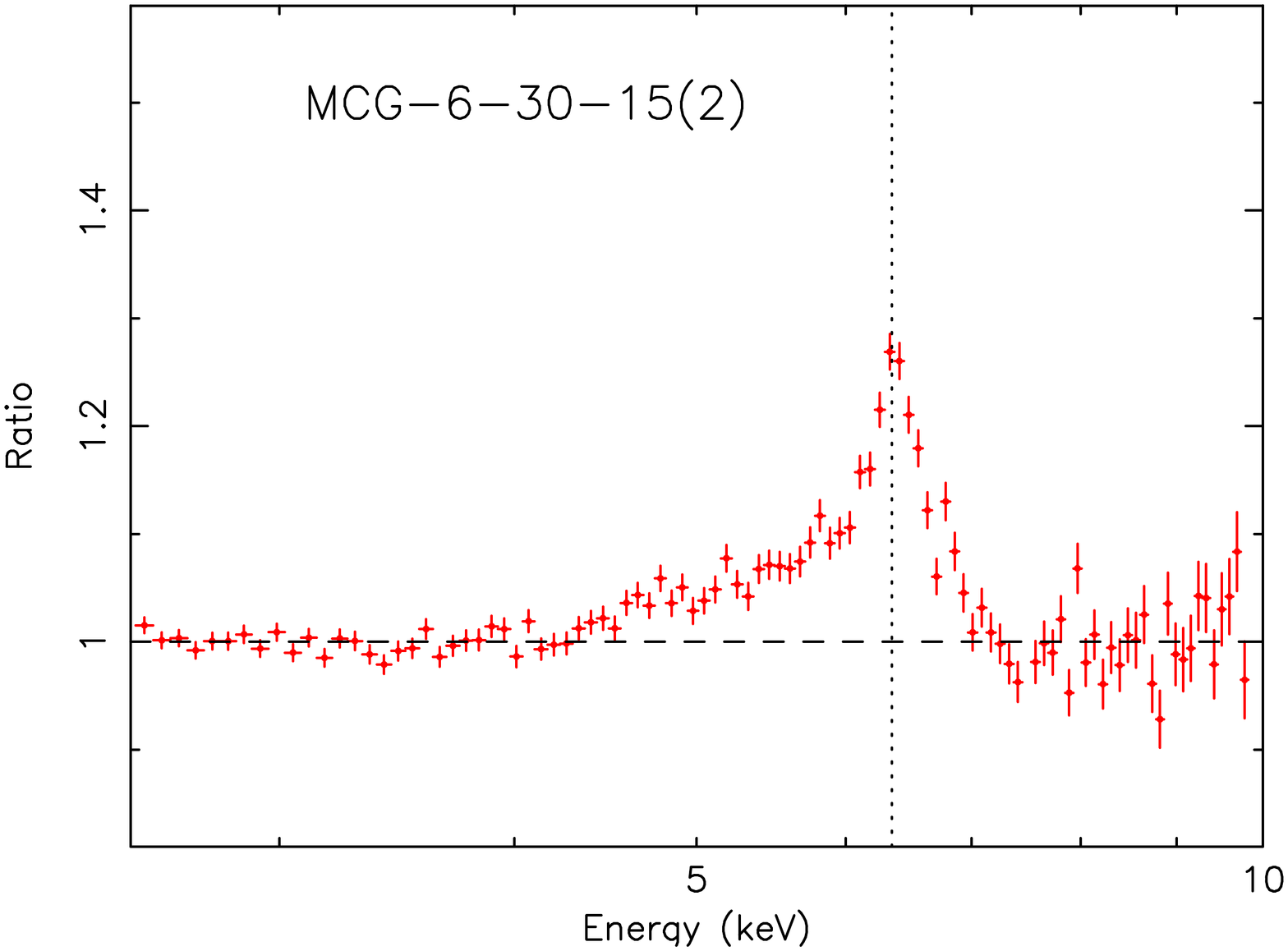}
\includegraphics[angle=270,width=58mm]{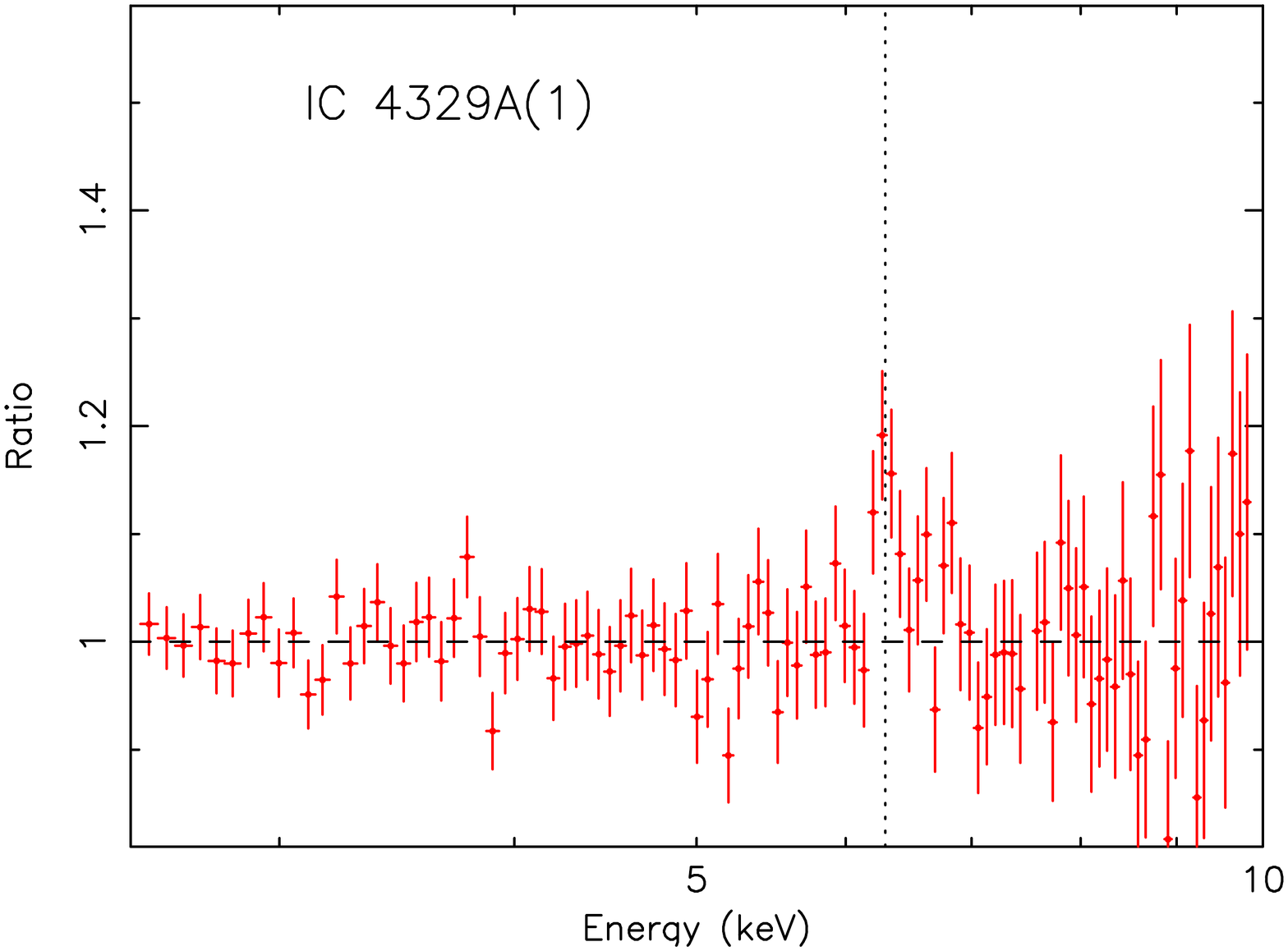}
\includegraphics[angle=270,width=58mm]{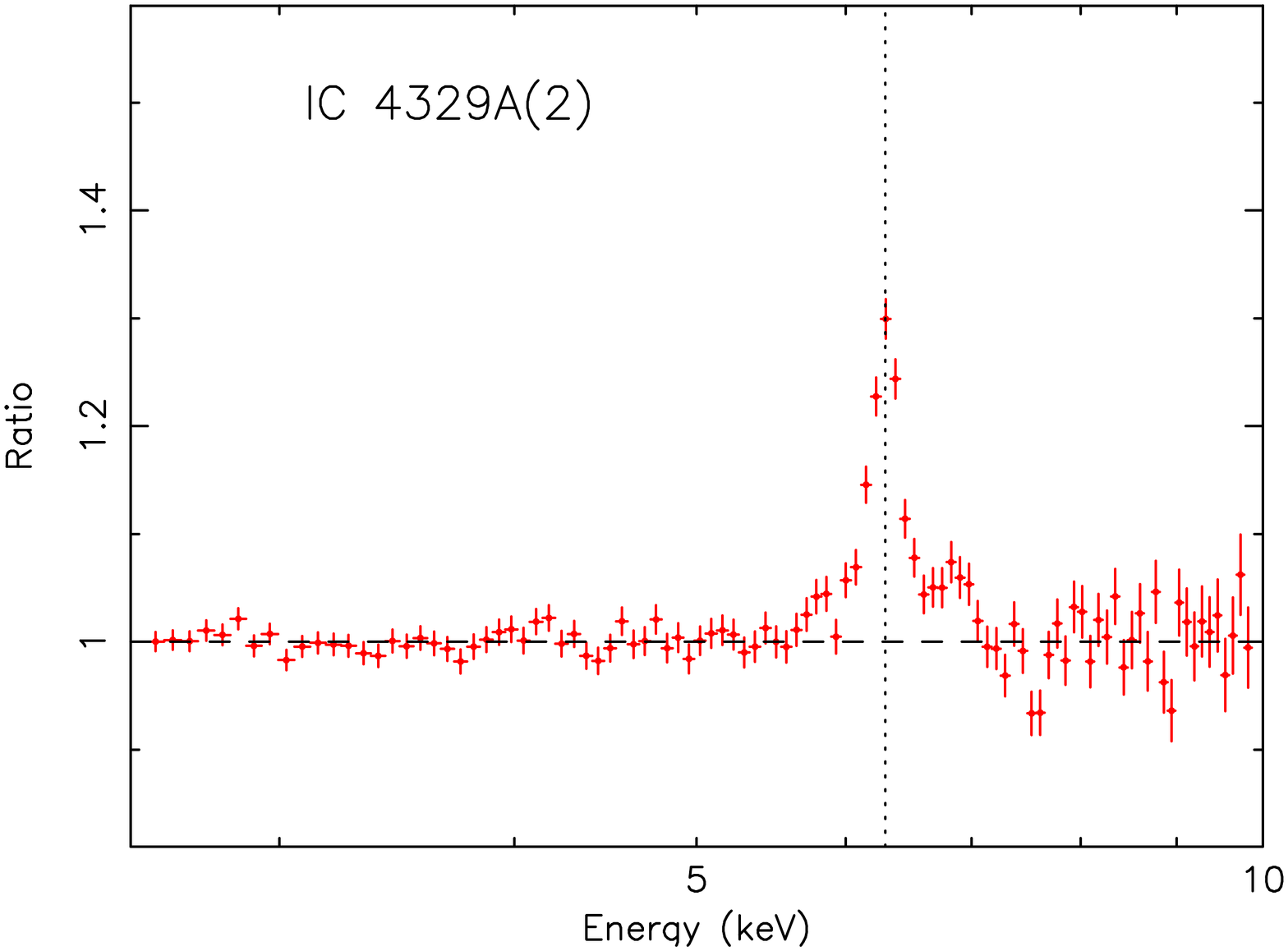}
\includegraphics[angle=270,width=58mm]{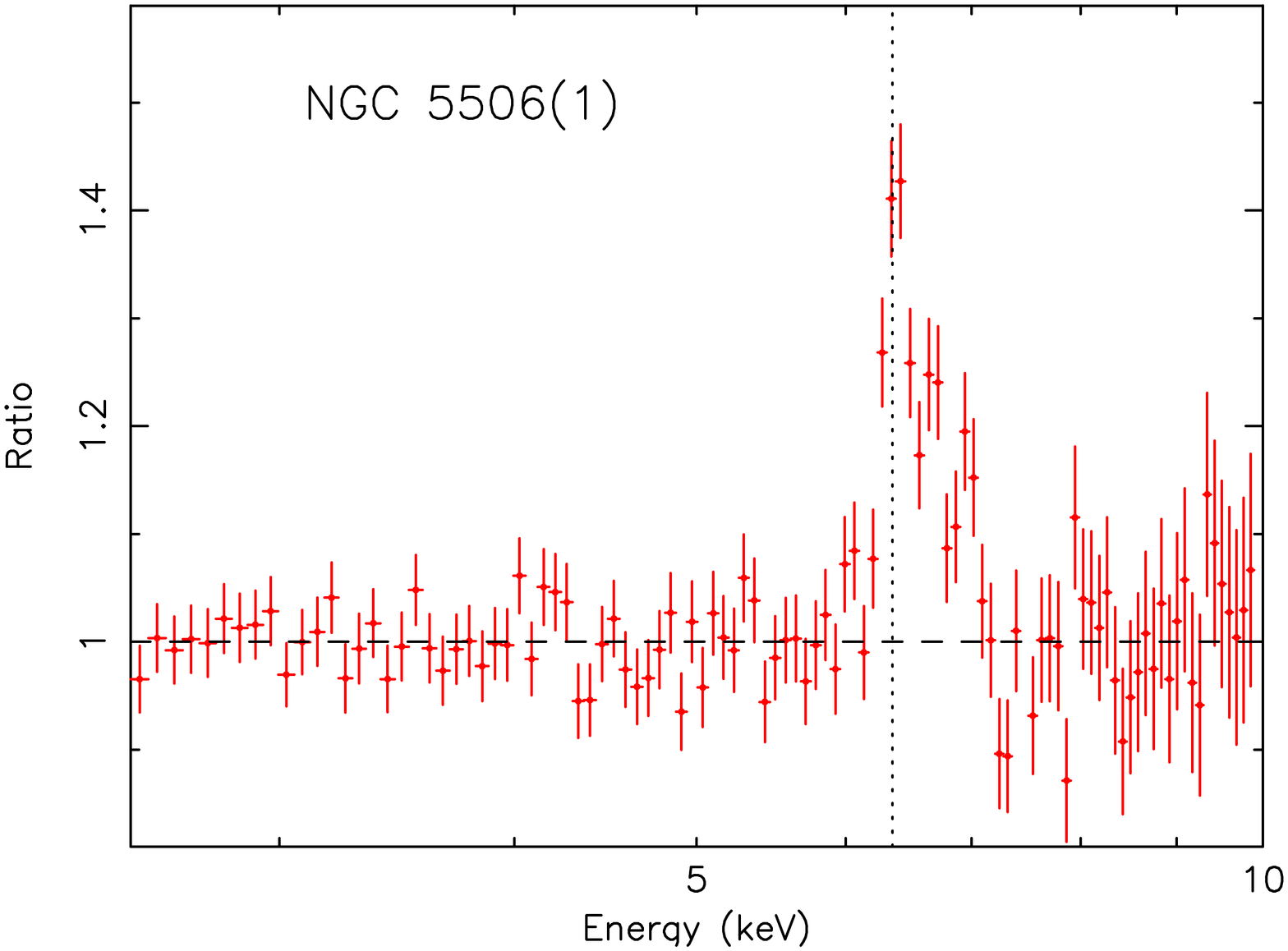}
}
\caption{(continued)}
\end{figure*}

\setcounter{figure}{2}

\begin{figure*}
{
\includegraphics[angle=270,width=58mm]{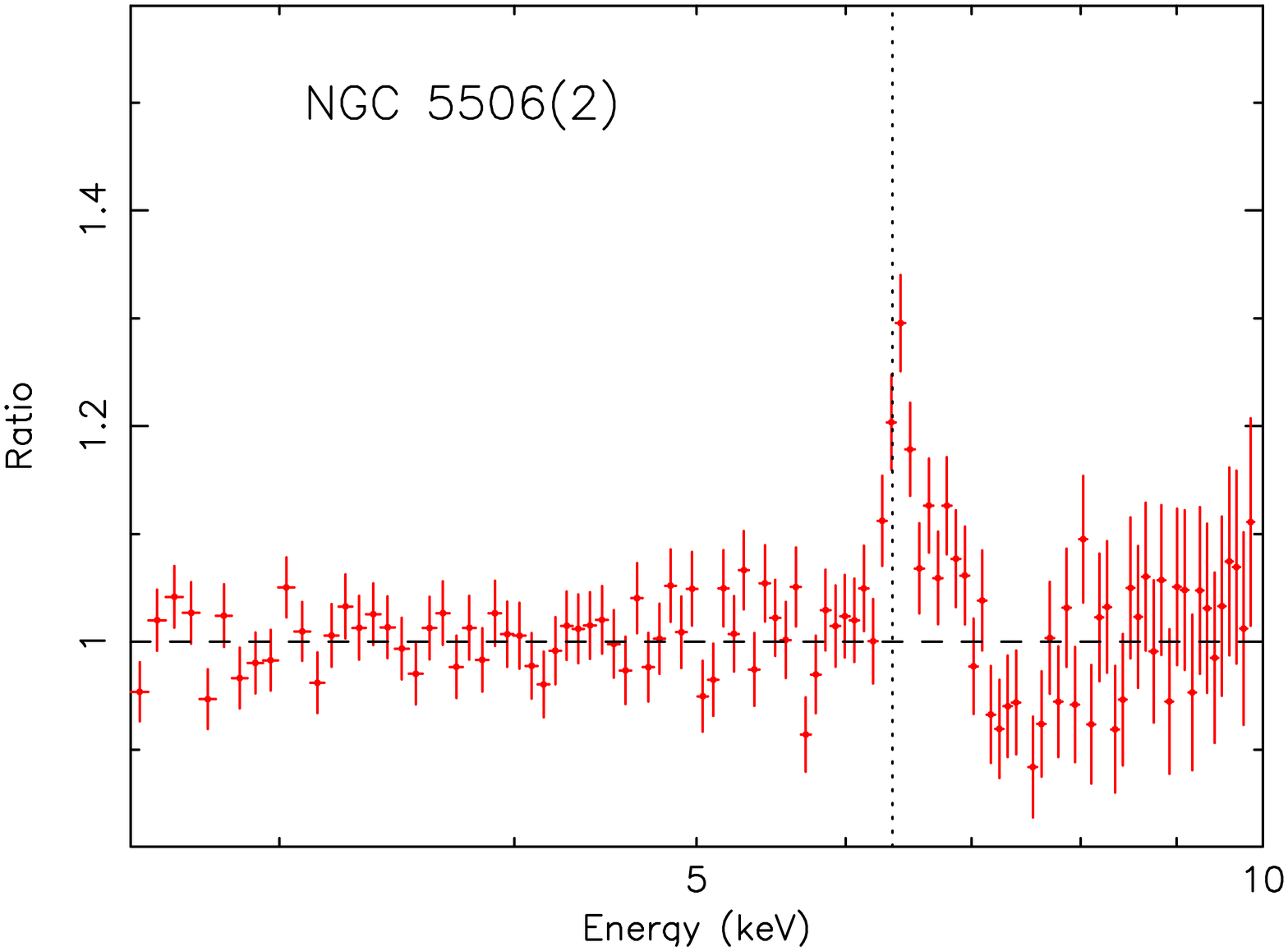}
\includegraphics[angle=270,width=58mm]{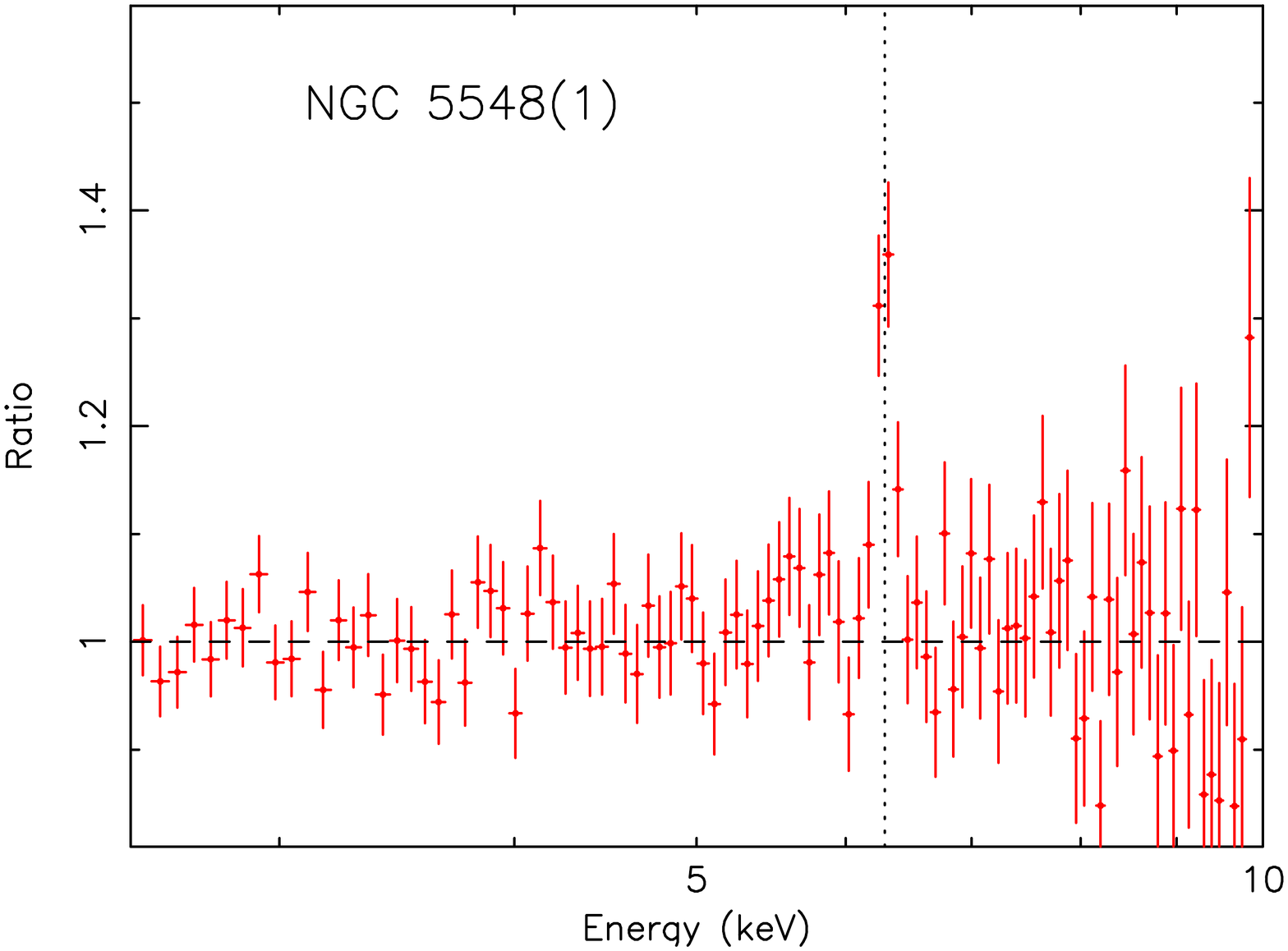}
\includegraphics[angle=270,width=58mm]{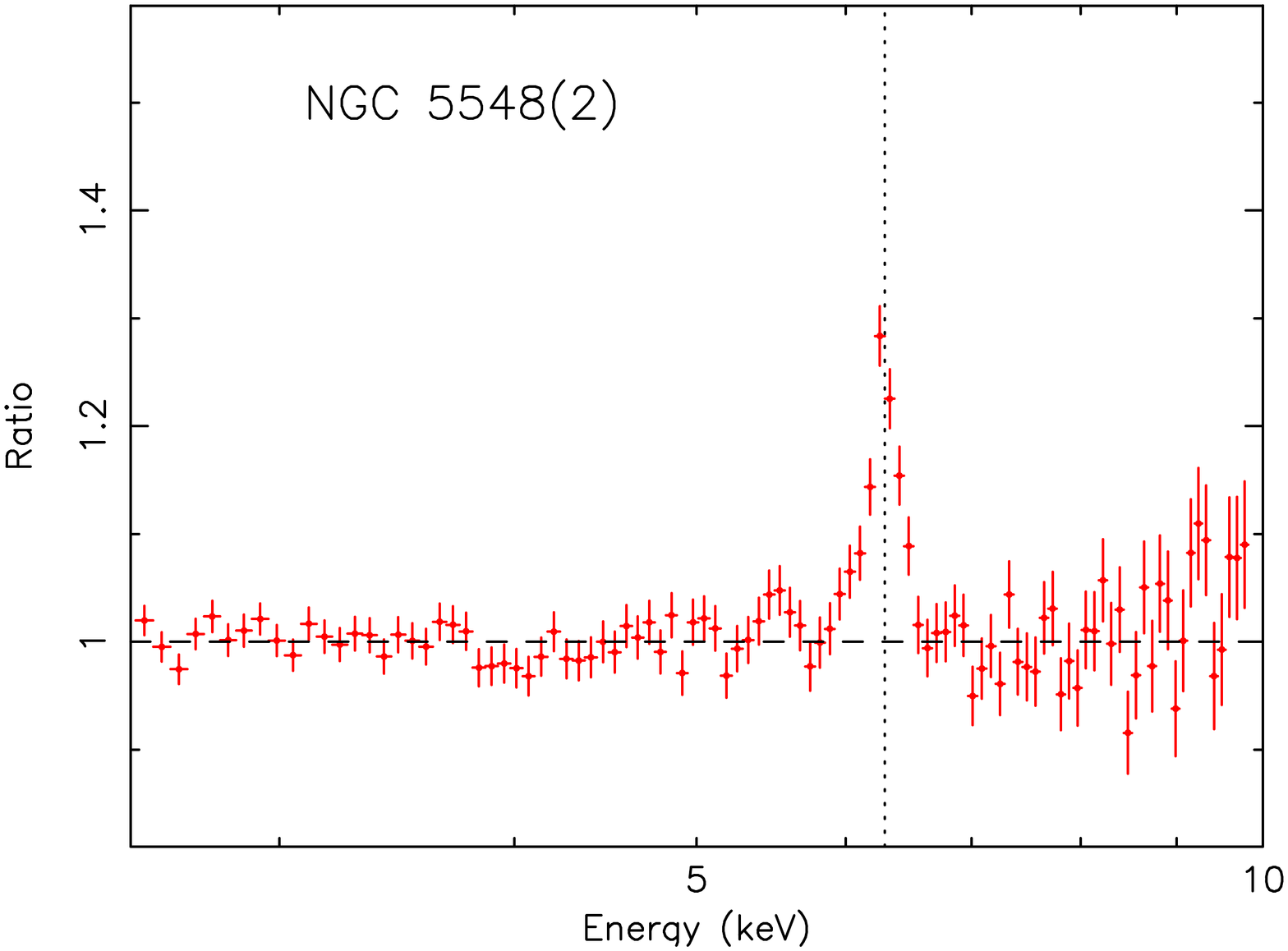}
\includegraphics[angle=270,width=58mm]{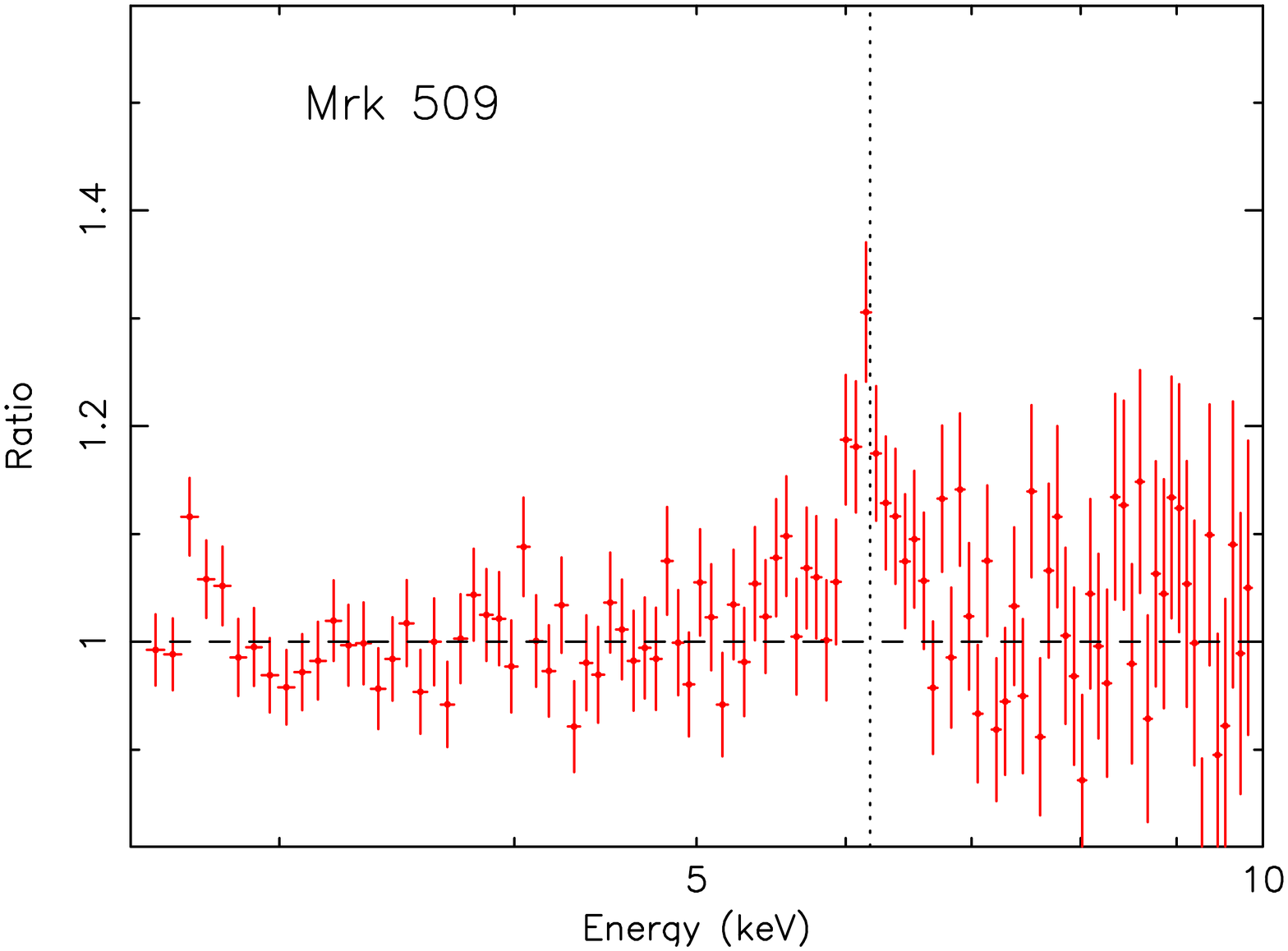}
\includegraphics[angle=270,width=58mm]{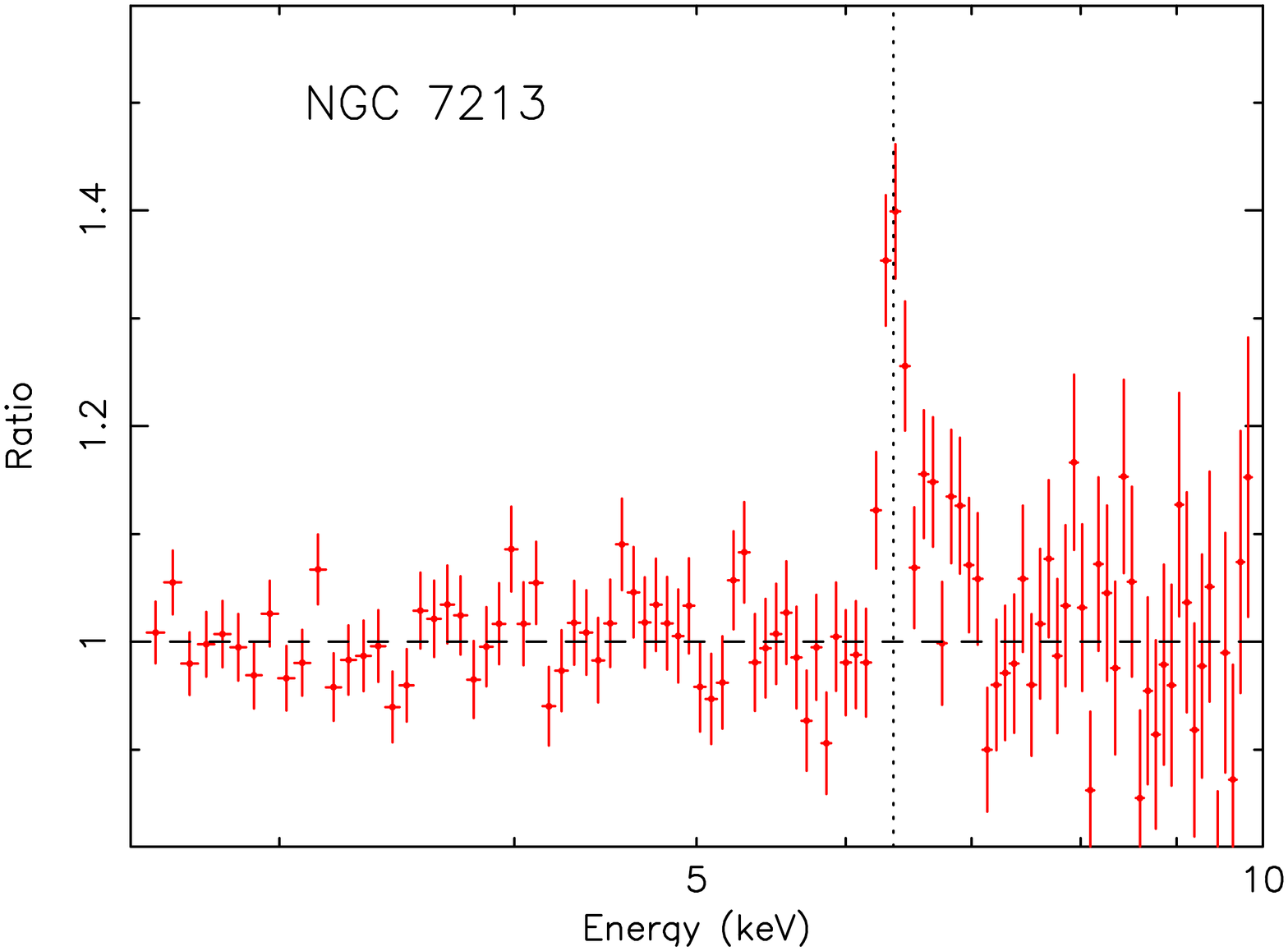}
\includegraphics[angle=270,width=58mm]{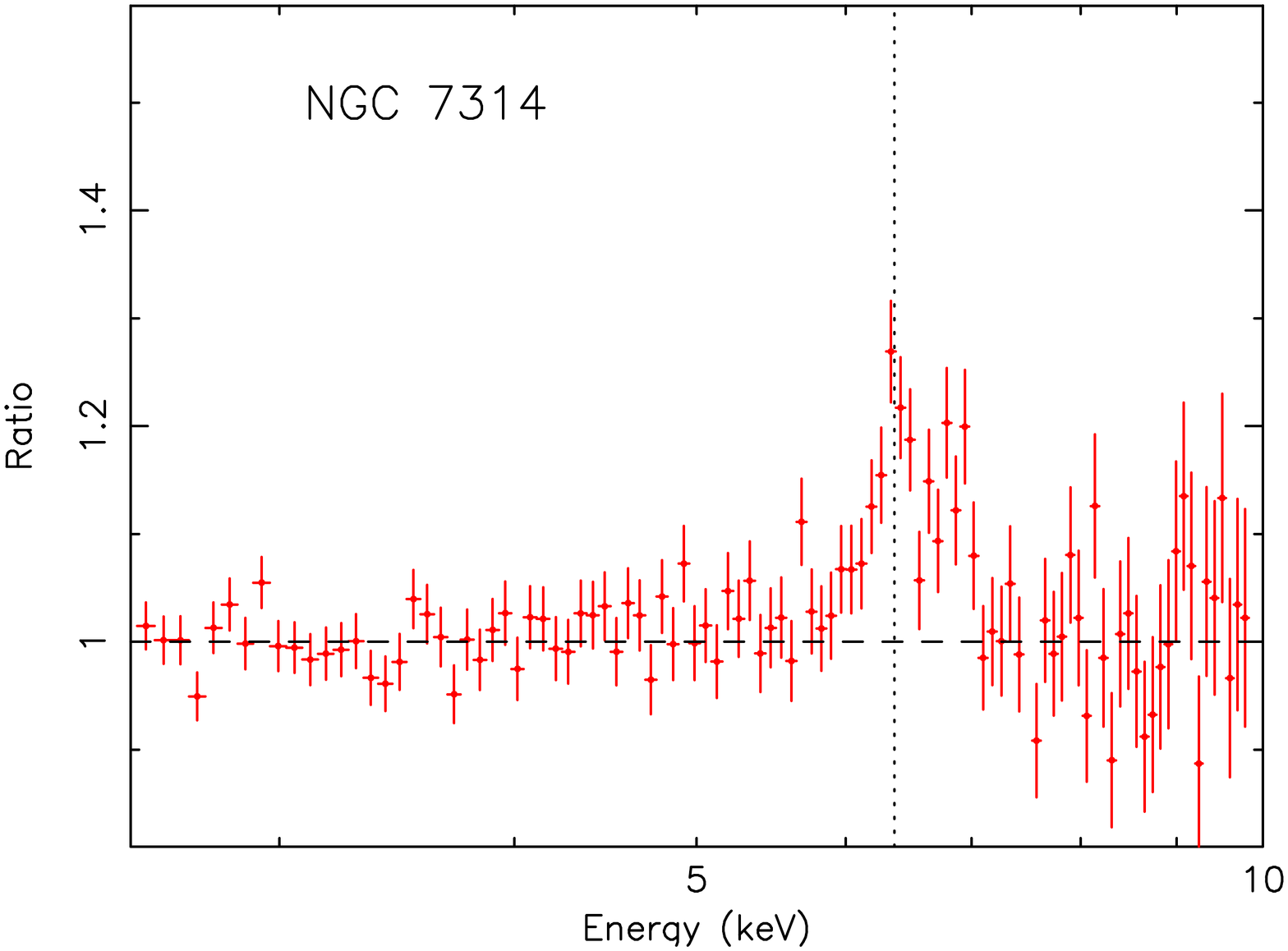}
\includegraphics[angle=270,width=58mm]{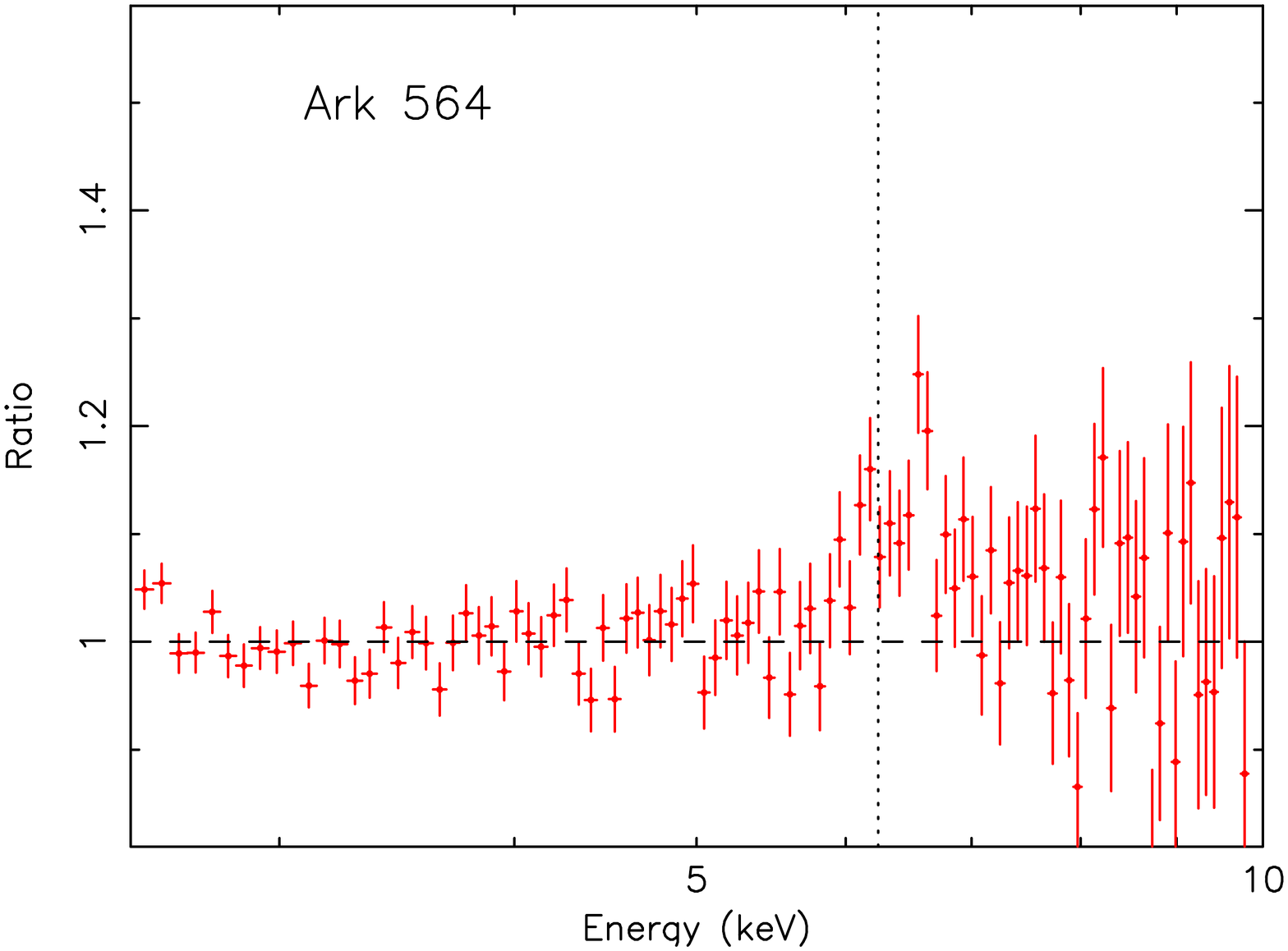}
\includegraphics[angle=270,width=58mm]{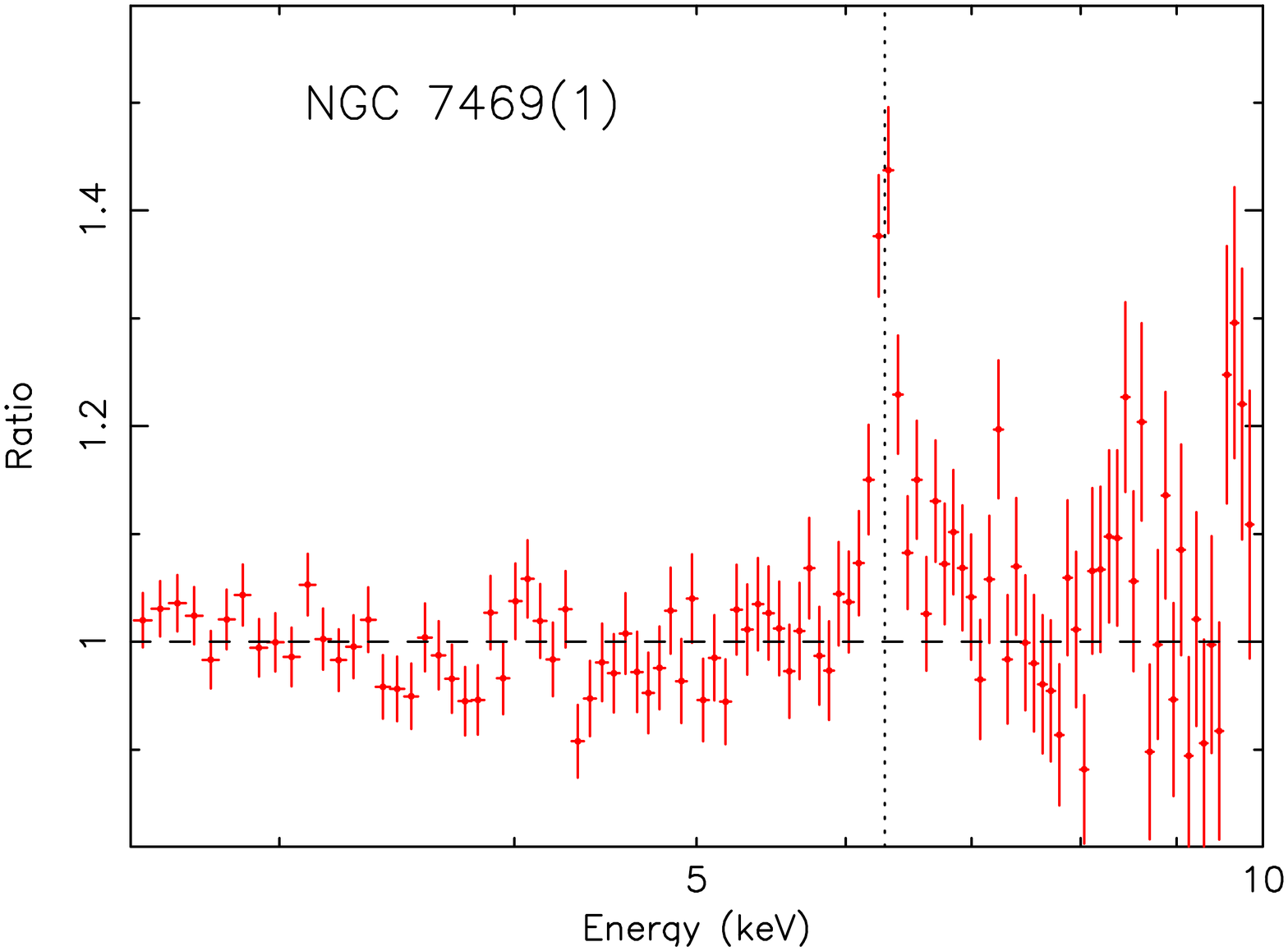}
\includegraphics[angle=270,width=58mm]{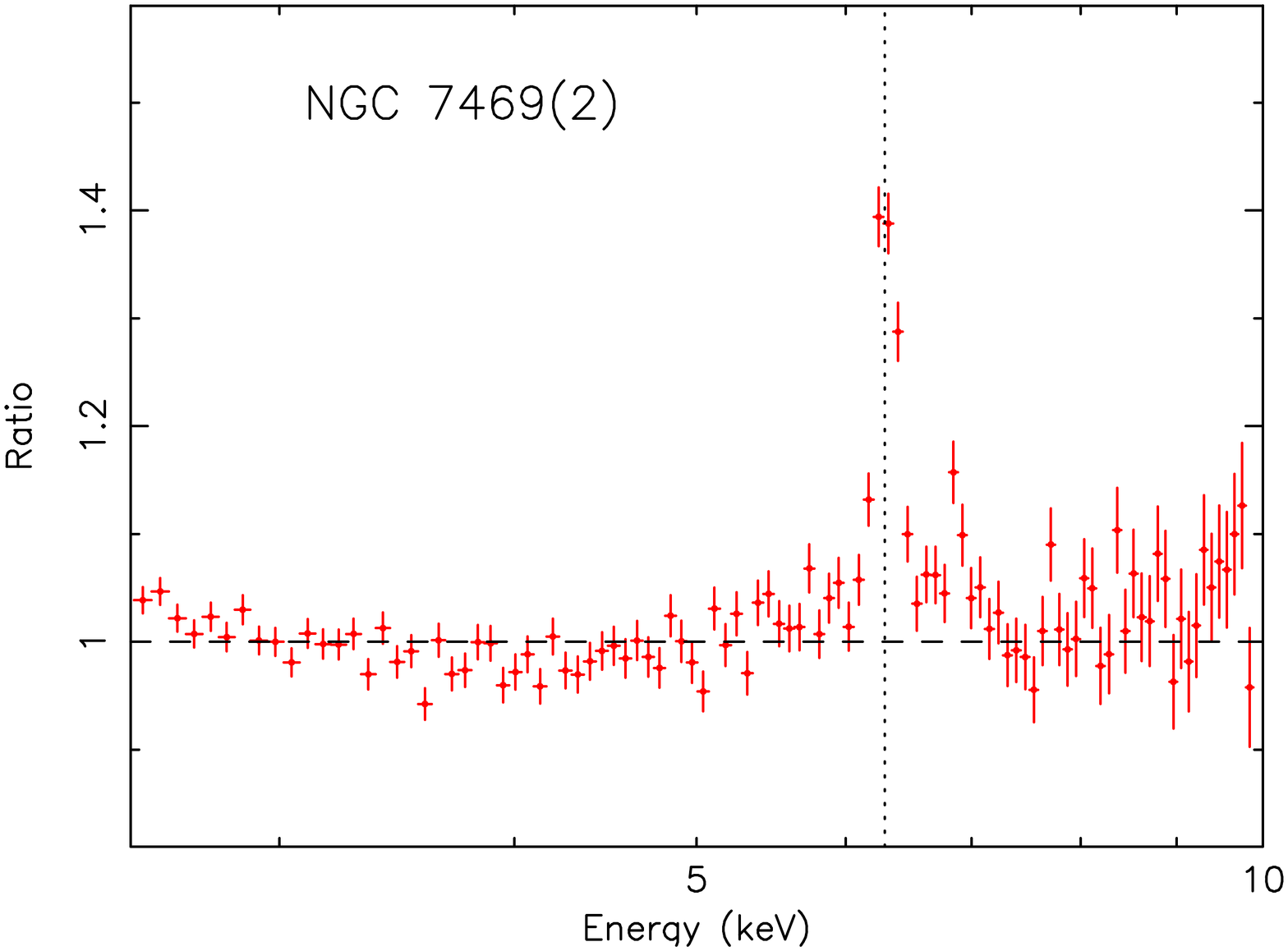}
}
\caption{(continued)}
\end{figure*}

We tested for the presence of absorption in the 2.5--10~keV spectra using  
both a simple, neutral absorber and an ionized absorber. Our intention in adding these components is to provide an adequate model of the continuum, above which any line emission should be evident. We consider more explicitly whether {\it additional} absorbers can mimic, e.g. the broad accretion disk line emission (e.g. Reeves et al. 2004), below. In most cases, we have {\it a priori} information as to which of the three absorption component is likely to be present and/or important in a given source, based on previous observations. In the current analysis, however, we adopt a primarily empirical approach by adding each of the absorption components to the spectra, and seeing if, and by how much, the fits improve. In each case we ignore the 4.5-7.5 keV data, to exclude the iron K$\alpha$ line, and assess the presence or otherwise of the absorption components using the change in $\chi^{2}$. We consider there to be significant absorption in the spectrum above 2.5 keV if any of the following three criteria are satisfied: 1) the addition of a neutral absorber improves the power law fit by $\Delta\chi^{2}>4.0$, 2) an ionized absorber improves the neutral absorber fit by $\Delta\chi^{2}>4.0$, or 3) an ionized absorber improves the fit compared to the power law by $\Delta\chi^{2}>6.2$ (note that 2 additional parameters are added in this last case). These $\Delta\chi^{2}$ thresholds correspond roughly to the 95\% confidence level when the improvement in $\chi^{2}$ is quantified using an F-test. While it is known that the F-test does not always provide an accurate assessment of the chance probability (Protassov et al. 2002) the precise significance of these components is relatively unimportant in our analysis, as we are simply using a continuum parameterization, and not attaching any interpretation to the parameters of the absorber. 

In all cases where, by the above criteria, we conclude that there is significant absorption in the spectrum, we adopt an {\it ionized} absorber to model the continuum. Our ionized absorber model (see \S~\ref{sec:models}) covers a sufficiently wide range of ionization parameters that it adequately models the neutral or near-neutral case where that is appropriate. In others, even though a warm absorber model may not provide a clear improvement in the fit compared to a neutral one, the gas may nonetheless be ionized. As we are restricting the energy range of the fit, many of the key features needed to establish the ionization state of the absorber are excluded. Using a warm absorber in all cases is therefore conservative. About half the observations require significant absorption: NGC 526A, NGC 2110, Mrk 6, NGC 2992, MCG-5-23-16(1,2), NGC 3516 (1,2), NGC 3783(1,2), NGC4151(1,2,3), NGC 4395, MCG-6-30-15(1,2), IC4329A(1), NGC 5506(1,2), and NGC 7314. In most of these cases, the 2.5-4.5,7.5-10 keV fit is acceptable. Exceptions are: MCG-6-30-15(2), NGC 3783(2) and NGC4151 (2,3), all of which have unacceptable $\chi^{2}$ at $>99$~per cent confidence. In these cases the absorption may be more complex than a simple ionized absorber, or otherwise there must be additional spectral complexity. We consider such complexities in \S 6.

There are also objects in which absorption does not improve the fit, but nonetheless a power law (absorbed or otherwise) is not an adequate parameterization. Those with unnacceptable $\chi^{2}$ at worse than $99$~per cent confidence are Ark120, Mrk 110, NGC 4593, NGC 7469(1,2). The majority of these cases exhibit ``concave'' spectra, which may be indicative of continuum Compton reflection (see below).
 
Finally, we stress that in the above analysis we are not attempting to find a model fully consistent with the broad-band spectrum, nor do we consider the absorption lines in the high resolution spectra. We therefore do not necessarily expect our modeling to yield any physical insight into the nature of the absorption. They often do  provide a highly significant improvement in many of the fits, so it is necessary to include the effect of absorption to provide an adequate description of the continuum. However, we make no inferences from the parameter values and we do not consider them ``interesting'' e.g. for calculation of the error bars. This latter comment also applies to the derived photon indices. 

The ratio of the data to the model are shown for the full 2.5-10~keV band in Fig.~\ref{fig:all_profile}. Strong iron K-band emission is evident in all 37 observations. However there are clear differences between observations whereby some appear to show narrow emission lines, some appear to show broad emission lines, and some appear to be skewed, or otherwise complex. 

\subsection{Distant reflection: the narrow K$\alpha$ core (Model~A)}
\label{Sec:Model-A}

\begin{table*}
\centering
\caption{Comparison of models  \label{tab:mcompv2}
(1) Name and Observation number, or (for the lower part of the table) the sample statistic: $\chi^{2}_{\rm tot}$ and $\chi^{2}_{\rm med}$
are the total and median $\chi^{2}$ values; $N_{\rm par}$ is the total number of additional parameters  compared to Model~A;  $N_{\rm good}$ is the number of observations where the fit is acceptable at 99~per cent confidence \
(out of 37 in total); and $N_{\rm bad}$ is the number of catastrophic fits ($\chi^{2}_{\nu}>2.0$). 
(2) Summary for Model~A which consists of an underlying power law continuum, Galactic absorption, a distant reflector (using {\tt pexmon}) and where necessary a soft X-ray warm absorber (using  {\tt cwa18} - see \S~\ref{Sec:Model-A} for details). Acceptable fits are indicated by a tick, catastrophic fits (with $\chi^{2}_{\nu}>2.0$) by a large cross, and other fits by  small cross.
(3) Model B: as for Model~A, but also including a broad Gaussian emission line (see  \S~\ref{Sec:Model-B}).
(4) Model C: as for Model~A, but also including an additional high ionization warm absorber (using {\tt grid25}, if necessary) 
plus, also if necessary, two narrow Gaussians with fixed energies of 6.7 and 6.97~keV (see  \S~\ref{Sec:Model-C}).
(5) Model D: as for Model~C, but also allowing a third narrow Gaussian with free energy in the range 6.4-6.7 keV (see  
\S~\ref{Sec:Model-D}).
(6) Model E: as for Model~A, but now including a  {\tt kdblur} is the relativistically blurred \textsc{pexmon} component
(see  \S~\ref{Sec:Model-E}).
(7) Model F: as for Model~E, but including includes multi-zone warm absorbers and additional narrow line components as needed. 
(see  \S~\ref{Sec:Model-E}).
}
\begin{center}
\begin{tabular}{l rrrrrr }
\hline
Name & \multicolumn{6}{c}{Model} \\
           & A & B & C & D & E & F \\
(1) & (2) & (3) & (4) & (5) & (6) & (7) \\
\hline
\hline
NGC 526A & $\surd$ & $\surd$ & $\surd$ & $\surd$ & $\surd$ & $\surd$ \\
Mrk 590 & $\surd$ & $\surd$ & $\surd$ & $\surd$ & $\surd$ & $\surd$ \\
Ark 120 & X & x & X & x & x & $\surd$ \\
NGC 2110 & $\surd$ & $\surd$ & $\surd$ & $\surd$ & $\surd$ & $\surd$ \\
MCG+8-11-11 & $\surd$ & $\surd$ & $\surd$ & $\surd$ & $\surd$ & $\surd$ \\
Mrk 6 & $\surd$ & $\surd$ & $\surd$ & $\surd$ & $\surd$ & $\surd$ \\
Mrk 110 & $\surd$ & $\surd$ & $\surd$ & $\surd$ & $\surd$ & $\surd$ \\
NGC 2992 & $\surd$ & $\surd$ & $\surd$ & $\surd$ & $\surd$ & $\surd$ \\
MCG-5-23-16(1) & x & $\surd$ & x & x & $\surd$ & $\surd$ \\
MCG-5-23-16(2) & x & $\surd$ & x & x & $\surd$ & $\surd$ \\
NGC 3516(1) & x & $\surd$ & $\surd$ & $\surd$ & $\surd$ & $\surd$ \\
NGC 3516(2) & X & x & X & X & $\surd$ & $\surd$ \\
NGC 3783(1) & X & x & x & x & x & $\surd$ \\
NGC 3783(2) & X & X & X & x & X & $\surd$ \\
HE 1143-1810 & $\surd$ & $\surd$ & $\surd$ & $\surd$ & $\surd$ & $\surd$ \\
NGC 4051 & x & $\surd$ & $\surd$ & $\surd$ & $\surd$ & $\surd$ \\
NGC 4151(1) & x & $\surd$ & $\surd$ & $\surd$ & $\surd$ & $\surd$ \\
NGC 4151(2) & X & x & X & $\surd$ & x & $\surd$ \\
NGC 4151(3) & X & X & X & X & X & $\surd$ \\
Mrk 766(1) & $\surd$ & $\surd$ & $\surd$ & $\surd$ & $\surd$ & $\surd$ \\
Mrk 766(2) & X & x & x & x & $\surd$ & $\surd$ \\
NGC 4395 & $\surd$ & $\surd$ & $\surd$ & $\surd$ & $\surd$ & $\surd$ \\
NGC 4593 & x & x & x & x & x & $\surd$ \\
MCG-6-30-15(1) & X & x & x & x & x & x \\
MCG-6-30-15(2) & X & X & X & X & x & x \\
IC 4329A(1) & $\surd$ & $\surd$ & $\surd$ & $\surd$ & $\surd$ & $\surd$ \\
IC 4329A(2) & x & $\surd$ & x & x & $\surd$ & $\surd$ \\
NGC 5506(1) & x & $\surd$ & $\surd$ & $\surd$ & $\surd$ & $\surd$ \\
NGC 5506(2) & x & $\surd$ & $\surd$ & $\surd$ & $\surd$ & $\surd$ \\
NGC 5548(1) & $\surd$ & $\surd$ & $\surd$ & $\surd$ & $\surd$ & $\surd$ \\
NGC 5548(2) & $\surd$ & $\surd$ & $\surd$ & $\surd$ & $\surd$ & $\surd$ \\
Mrk 509 & $\surd$ & $\surd$ & $\surd$ & $\surd$ & $\surd$ & $\surd$ \\
NGC 7213 & $\surd$ & $\surd$ & $\surd$ & $\surd$ & $\surd$ & $\surd$ \\
NGC 7314 & $\surd$ & $\surd$ & $\surd$ & $\surd$ & $\surd$ & $\surd$ \\
Ark 564 & x & $\surd$ & $\surd$ & $\surd$ & $\surd$ & $\surd$ \\
NGC 7469(1) & $\surd$ & $\surd$ & $\surd$ & $\surd$ & $\surd$ & $\surd$ \\
NGC 7469(2) & X & x & X & x & x & x \\
\hline
\multicolumn{7}{l}{Summary Statistics for Whole Sample} \\
$\chi^{2}_{\rm tot}$      & 6472.2 & 4583.0 & 5605.2 &   4792.3 & 4364.2 & 3838.7 \\
$\chi^{2}_{\rm med}$   & 139.9   &  108.6  &   120.5  &   112.3  &   112.2  & 100.5 \\  
$N_{\rm par}$              & 0          &  111     &   148    &    222     &   111    & 137      \\
$N_{\rm good}$           & 17        &   26      &   23       &     24     &     28     &   34 \\
$N_{\rm bad}$             & 10        &   3        &    7        &      3      &     2       &    0 \\
\hline
\end{tabular}
\end{center}
\end{table*}

We first test to see if the iron line emission evident in Fig.~\ref{fig:all_profile} can be accounted for solely by a  distant, neutral reflector. We do this using the \textsc{pexmon} model applied to the full 2.5-10~keV spectra, and hence self-consistently account for the line emission, Compton reflection and Compton shoulder.
A summary of the results of this model (hereafter Model~A) are given in Table~\ref{tab:mcompv2}.
There is a strong requirement for such a component in almost all the spectra. However such a model does not completely account for all the complexity seen in many of the observations. 
Indeed this model provides an acceptable fit to only 17 of the 37 observations (denoted by "$\surd$" in 
in Table~\ref{tab:mcompv2}). The remaining 20 observations are unacceptable at 99~per cent confidence
(10 of them with $\chi^{2}_{\nu}>2.0$; denoted by "X"  in Table~\ref{tab:mcompv2}).
For these observations there are clearly systematic and correlated residuals to these fits particularly around the iron K-shell region.

\begin{table*}
\centering
\caption{Broad gaussian fits. 
Col.(1): Name and Observation number.
Col.(2): Base continuum model, being either a power law (PL) or warm absorber (WA). Note that in either case the {\tt pexmon} model of distant reflection is always included;
Col.(3): 2-10 keV observed-frame model flux in units of $10^{-11}$~erg cm$^{-2}$ s$^{-1}$. 
Col.(4): log of 2-10 keV rest frame luminosity in erg s$^{-1}$; 
Col.(5): Central energy of the broad Gaussian, with 68\% upper and lower bounds for 3 interesting parameters ($\Delta\chi^{2}=3.5$). Note that when the reduced $\chi^{2}$ is $>2$, which is the case for NGC 3783, MCG-6-30-15(2) and NGC 4151(3), the confidence ranges is not quoted; 
Col.(6): 1$\sigma$ width of the broad Gaussian;
Col.(7): Equivalent width of the broad Gaussian;
Col.(8): Reflection fraction of the {\tt pexmon} component, where $R_{\rm tor}=1$ corresponds to a semi-infinite slab seen at an inclination of 60$^{\circ}$ and subtending 2$\pi$ solid angle at the X-ray source; 
Col.(9): F-statistic for addition of the Gaussian. For 3 additional parameters $F>4.0$ represents an improvement at $>99$~per cent confidence based on this test. 
Col. (10): $\chi^{2}$ and degrees of freedom for this model. Col.(10): Null hypothesis probability that this $\chi^{2}$ is acceptable by chance. 
\label{tab:gauss}}
\begin{center}
\begin{tabular}{lcccccrcrrc}
\hline
Name & Base & Flux & $\log L_{\rm X}$ & $E_{\rm K\alpha}$ & $\sigma_{\rm K\alpha}$ & 
 \multicolumn{1}{c}{$EW_{\rm K \alpha}$} & $R_{\rm tor}$ & \multicolumn{1}{c}{$F$} & $\chi^{2}$/dof & prob\\
 & Model & (2-10 keV) & (2-10 keV) &   (keV) & (keV) & (eV) & & & & \\
(1) & (2) & (3) & (4) & (5) & (6) & \multicolumn{1}{c}{(7)} & (8) & \multicolumn{1}{c}{(9)} & \multicolumn{1}{c}{(10)} & (11) \\
\hline
\hline
NGC 526A         & WA & 2.23 & 43.25 & $6.49 ^{+0.18 }_{-0.17 }$ & $0.31 ^{+0.19 }_{-0.13 }$ & $80^{+43}_{-40}$ & $0.20 ^{+0.15 }_{-0.12 }$ & 7.16 & 79.4/98 &  0.92 \\
Mrk 590          & PL & 0.69 & 43.27 & $6.50 ^{+0.03 }_{-0.04 }$ & $0.00 ^{+0.07 }_{-0.00 }$ & $64^{+47}_{-23}$ & $0.38 ^{+0.12 }_{-0.35 }$ & 9.15 & 95.6/96 &  0.49 \\
Ark 120          & PL & 3.76 & 43.97 & $6.56 ^{+0.08 }_{-0.07 }$ & $0.32 ^{+0.08 }_{-0.06 }$ & $119^{+29}_{-28}$ & $0.34 ^{+0.11 }_{-0.10 }$ & 19.83 & 142.1/96 &  0.00 \\
NGC 2110         & WA & 2.50 & 42.42 & $6.54 ^{+0.05 }_{-0.08 }$ & $0.04 ^{+0.08 }_{-0.04 }$ & $31^{+32}_{-12}$ & $0.73 ^{+0.10 }_{-0.26 }$ & 8.35 & 98.6/98 &  0.46 \\
MCG+8-11-11      & PL & 4.47 & 43.59 & $6.63 ^{+0.63 }_{-0.27 }$ & $0.18 ^{+0.82 }_{-0.18 }$ & $27^{+79}_{-23}$ & $0.71 ^{+0.15 }_{-0.19 }$ & 2.27 & 88.3/96 &  0.70 \\
Mrk 6    & WA & 1.43 & 43.06 & $6.50 ^{+0.11 }_{-0.14 }$ & $0.02 ^{+0.56 }_{-0.02 }$ & $30^{+60}_{-22}$ & $0.29 ^{+0.20 }_{-0.29 }$ & 2.34 & 90.4/89 &  0.44 \\
Mrk 110          & PL & 2.85 & 43.93 & $6.53 ^{+0.14 }_{-0.11 }$ & $0.00 ^{+0.23 }_{-0.00 }$ & $17^{+41}_{-12}$ & $0.22 ^{+0.11 }_{-0.22 }$ & 2.17 & 108.7/97 &  0.20 \\
NGC 2992         & WA & 8.03 & 43.05 & $6.19 ^{+0.15 }_{-0.25 }$ & $0.40 ^{+0.25 }_{-0.15 }$ & $100^{+50}_{-37}$ & $0.25 ^{+0.13 }_{-0.13 }$ & 10.47 & 101.6/100 &  0.43 \\
MCG-5-23-16(1)   & WA & 8.08 & 43.05 & $6.25 ^{+0.25 }_{-0.05 }$ & $0.04 ^{+0.54 }_{-0.04 }$ & $37^{+67}_{-9}$ & $0.34 ^{+0.21 }_{-0.19 }$ & 4.17 & 133.0/100 &  0.02 \\
MCG-5-23-16(2)   & WA & 7.11 & 43.10 & $6.25 ^{+0.09 }_{-0.12 }$ & $0.32 ^{+0.12 }_{-0.10 }$ & $120^{+40}_{-39}$ & $0.25 ^{+0.14 }_{-0.13 }$ & 9.41 & 116.4/100 &  0.13 \\
NGC 3516(1)      & WA & 2.27 & 42.60 & $5.23 ^{+0.16 }_{-0.19 }$ & $0.91 ^{+0.09 }_{-0.19 }$ & $277^{+115}_{-89}$ & $0.89 ^{+0.12 }_{-0.23 }$ & 8.85 & 120.2/95 &  0.04 \\
NGC 3516(2)      & WA & 1.51 & 42.40 & $6.05 ^{+0.10 }_{-0.12 }$ & $0.72 ^{+0.11 }_{-0.09 }$ & $394^{+72}_{-63}$ & $1.57 ^{+0.62 }_{-0.18 }$ & 54.96 & 139.8/94 &  0.00 \\
NGC 3783(1)      & WA & 5.35 & 43.07 & $5.28 ^{+0.44 }_{-0.47 }$ & $0.91 ^{+0.09 }_{-0.22 }$ & $104^{+69}_{-44}$ & $0.74 ^{+0.18 }_{-0.13 }$ & 2.61 & 179.4/94 &  0.00 \\
NGC 3783(2)      & WA & 4.92 & 43.03 & $6.06 ^{+0.08 }_{-0.35 }$ & $0.22 ^{+0.47 }_{-0.08 }$ & $29^{+45}_{-9}$ & $0.73 ^{+0.07 }_{-0.07 }$ & 7.58 & 264.1/94 &  0.00 \\
HE 1143-1810     & PL & 2.83 & 43.85 & $5.53 ^{+0.14 }_{-0.22 }$ & $0.14 ^{+0.24 }_{-0.14 }$ & $30^{+28}_{-21}$ & $0.28 ^{+0.12 }_{-0.11 }$ & 3.71 & 84.2/96 &  0.80 \\
NGC 4051         & PL & 2.32 & 41.30 & $5.41 ^{+0.35 }_{-0.36 }$ & $0.99 ^{+0.01 }_{-0.30 }$ & $193^{+61}_{-80}$ & $0.64 ^{+0.15 }_{-0.13 }$ & 10.75 & 106.0/96 &  0.23 \\
NGC 4151(1)      & WA & 4.47 & 41.91 & $6.37 ^{+0.09 }_{-0.31 }$ & $0.28 ^{+0.34 }_{-0.15 }$ & $81^{+48}_{-27}$ & $1.45 ^{+0.34 }_{-0.37 }$ & 9.81 & 101.6/94 &  0.28 \\
NGC 4151(2)      & WA & 4.54 & 41.92 & $6.54 ^{+0.03 }_{-0.03 }$ & $0.01 ^{+0.07 }_{-0.01 }$ & $24^{+10}_{-5}$ & $1.41 ^{+0.13 }_{-0.15 }$ & 12.96 & 157.8/90 &  0.00 \\
NGC 4151(3)      & WA & 22.84 & 42.64 & $5.53 ^{+0.16 }_{-0.22 }$ & $0.79 ^{+0.14 }_{-0.11 }$ & $154^{+59}_{-36}$ & $0.45 ^{+0.05 }_{-0.02 }$ & 31.22 & 236.2/94 &  0.00 \\
Mrk 766(1)       & PL & 1.55 & 42.68 & $6.76 ^{+0.11 }_{-0.35 }$ & $0.18 ^{+0.34 }_{-0.11 }$ & $87^{+112}_{-43}$ & $0.61 ^{+0.26 }_{-0.43 }$ & 6.62 & 91.5/96 &  0.61 \\
Mrk 766(2)       & PL & 2.41 & 42.88 & $6.54 ^{+0.10 }_{-0.12 }$ & $0.38 ^{+0.19 }_{-0.11 }$ & $162^{+53}_{-48}$ & $0.11 ^{+0.11 }_{-0.11 }$ & 18.98 & 142.6/96 &  0.00 \\
NGC 4395         & WA & 0.56 & 40.08 & $5.88 ^{+0.25 }_{-0.64 }$ & $0.30 ^{+0.70 }_{-0.23 }$ & $65^{+148}_{-40}$ & $0.55 ^{+0.11 }_{-0.20 }$ & 4.06 & 97.6/89 &  0.25 \\
NGC 4593         & PL & 4.09 & 42.86 & $6.52 ^{+0.17 }_{-0.16 }$ & $0.43 ^{+0.26 }_{-0.15 }$ & $75^{+36}_{-30}$ & $0.56 ^{+0.10 }_{-0.10 }$ & 6.29 & 136.4/96 &  0.00 \\
MCG-6-30-15(1)   & WA & 3.21 & 42.65 & $5.77 ^{+0.20 }_{-0.21 }$ & $0.63 ^{+0.20 }_{-0.16 }$ & $139^{+58}_{-42}$ & $0.39 ^{+0.13 }_{-0.10 }$ & 15.16 & 148.3/95 &  0.00 \\
MCG-6-30-15(2)   & WA & 4.22 & 42.77 & $5.99 ^{+0.11 }_{-0.11 }$ & $0.71 ^{+0.10 }_{-0.11 }$ & $198^{+33}_{-30}$ & $0.28 ^{+0.05 }_{-0.05 }$ & 50.12 & 244.9/94 &  0.00 \\
IC 4329A(1)      & WA & 10.63 & 43.78 & $6.68 ^{+0.82 }_{-2.18 }$ & $0.00 ^{+1.00 }_{-0.00 }$ & $9^{+65}_{-9}$ & $0.27 ^{+0.12 }_{-0.10 }$ & 0.89 & 74.1/100 &  0.97 \\
IC 4329A(2)      & PL & 9.51 & 43.73 & $6.42 ^{+0.07 }_{-0.07 }$ & $0.32 ^{+0.09 }_{-0.07 }$ & $64^{+15}_{-14}$ & $0.30 ^{+0.05 }_{-0.05 }$ & 22.77 & 107.9/96 &  0.19 \\
NGC 5506(1)      & WA & 5.85 & 42.80 & $6.66 ^{+0.05 }_{-0.08 }$ & $0.16 ^{+0.10 }_{-0.07 }$ & $84^{+20}_{-17}$ & $0.45 ^{+0.12 }_{-0.22 }$ & 19.84 & 92.2/98 &  0.65 \\
NGC 5506(2)      & WA & 9.96 & 43.03 & $6.61 ^{+0.11 }_{-0.10 }$ & $0.19 ^{+0.10 }_{-0.08 }$ & $60^{+38}_{-24}$ & $0.22 ^{+0.11 }_{-0.13 }$ & 10.06 & 107.0/98 &  0.25 \\
NGC 5548(1)      & PL & 3.21 & 43.31 & $5.68 ^{+0.24 }_{-1.18 }$ & $0.06 ^{+0.94 }_{-0.06 }$ & $16^{+106}_{-15}$ & $0.45 ^{+0.14 }_{-0.13 }$ & 1.64 & 80.1/97 &  0.89 \\
NGC 5548(2)      & PL & 4.26 & 43.44 & $6.37 ^{+0.05 }_{-0.08 }$ & $0.16 ^{+0.08 }_{-0.06 }$ & $49^{+23}_{-19}$ & $0.20 ^{+0.11 }_{-0.16 }$ & 6.70 & 103.5/96 &  0.28 \\
Mrk 509          & PL & 3.01 & 43.92 & $6.30 ^{+0.25 }_{-0.37 }$ & $0.42 ^{+0.46 }_{-0.29 }$ & $87^{+67}_{-54}$ & $0.23 ^{+0.17 }_{-0.22 }$ & 3.54 & 90.9/96 &  0.63 \\
NGC 7213         & PL & 2.17 & 42.23 & $6.69 ^{+0.16 }_{-0.19 }$ & $0.17 ^{+0.15 }_{-0.17 }$ & $40^{+41}_{-29}$ & $0.53 ^{+0.17 }_{-0.20 }$ & 3.18 & 101.6/96 &  0.33 \\
NGC 7314         & WA & 4.01 & 42.34 & $6.54 ^{+0.11 }_{-0.16 }$ & $0.38 ^{+0.17 }_{-0.08 }$ & $112^{+50}_{-46}$ & $0.20 ^{+0.37 }_{-0.12 }$ & 8.23 & 93.3/94 &  0.50 \\
Ark 564          & PL & 1.66 & 43.37 & $6.73 ^{+0.04 }_{-0.19 }$ & $0.01 ^{+0.12 }_{-0.01 }$ & $37^{+15}_{-13}$ & $0.37 ^{+0.16 }_{-0.16 }$ & 7.62 & 118.5/96 &  0.06 \\
NGC 7469(1)      & PL & 2.67 & 43.23 & $6.65 ^{+0.85 }_{-0.30 }$ & $0.38 ^{+0.62 }_{-0.38 }$ & $57^{+102}_{-50}$ & $0.57 ^{+0.18 }_{-0.18 }$ & 2.06 & 114.7/97 &  0.11 \\
NGC 7469(2)      & PL & 2.92 & 43.27 & $6.78 ^{+0.49 }_{-0.47 }$ & $1.00 ^{+0.00 }_{-0.74 }$ & $112^{+47}_{-81}$ & $0.62 ^{+0.07 }_{-0.13 }$ & 3.54 & 184.7/97 &  0.00 \\
\hline
\end{tabular}
\end{center}
\end{table*}

\subsection{A simple parameterization of the iron K$\alpha$ complexity (Model~B)}
\label{Sec:Model-B}

As a first systematic test for additional complexity beyond narrow emission from distant Compton thick material, we have introduced a Gaussian component to the fit, with free energy ($E_{\rm K\alpha}$), width ($\sigma_{\rm K\alpha}$) and strength (parameterized by its equivalent width $EW_{\rm K \alpha}$). The energy $E_{\rm K \alpha}$ is restricted to be in the range 4.5-7.5 keV.  At this stage we do not identify this Gaussian with any particular physical component, but use it as a crude parameterization of any excess emission in the line region.  

These fits are summarized in Table~\ref{tab:mcompv2} (as Model~B), and the best-fitting parameters listed 
in Table~\ref{tab:gauss}. For each object Table~\ref{tab:gauss} also lists the observed flux in the 2-10~keV band, the 
luminosity of the underlying continuum in the 2--10~keV rest frame, and the  strength of the 
distant reflection component (parameterized in \textsc{pexmon} by $R_{\rm tor}$, where $R_{\rm tor}=1$ corresponds to a semi-infinite slab seen at an inclination of 60$^{\circ}$ and subtending 2$\pi$ solid angle at the X-ray source.
As listed in Table~\ref{tab:mcompv2}, Model~B provides total and median $\chi^2$ values 
($\chi^{2}_{\rm tot}$ and $\chi^{2}_{\rm med}$) for superior to those obtained for Model~A.
Based on the F-test and given the addition of 3 free parameters, the fit is improved at $>99$~per cent confidence in 26/37 spectra. Indeed the number of observations for which Model~B provides an acceptable fit ($N_{\rm good}$)
is 26, while only three observations have $\chi^{2}_{\nu}>2.0$ (i.e. $N_{\rm bad}$=3)  and hence are 
``catastrophic'' fits (compared to $N_{\rm good}$=17 and  $N_{\rm bad}$=10 for Model~A).

\begin{figure*}
{
\includegraphics[angle=0,width=58mm,height=50mm]{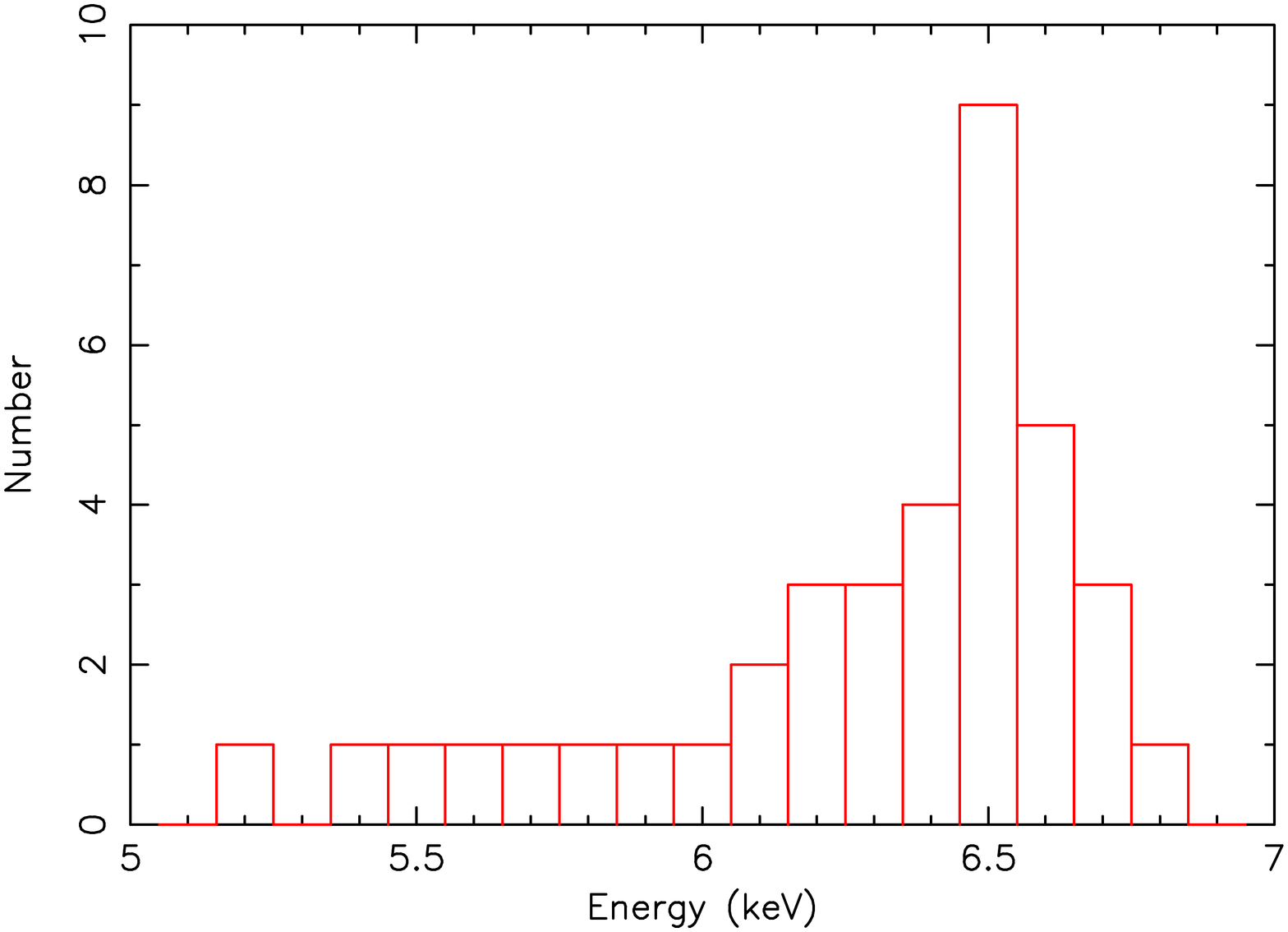}
\includegraphics[angle=0,width=58mm,height=50mm]{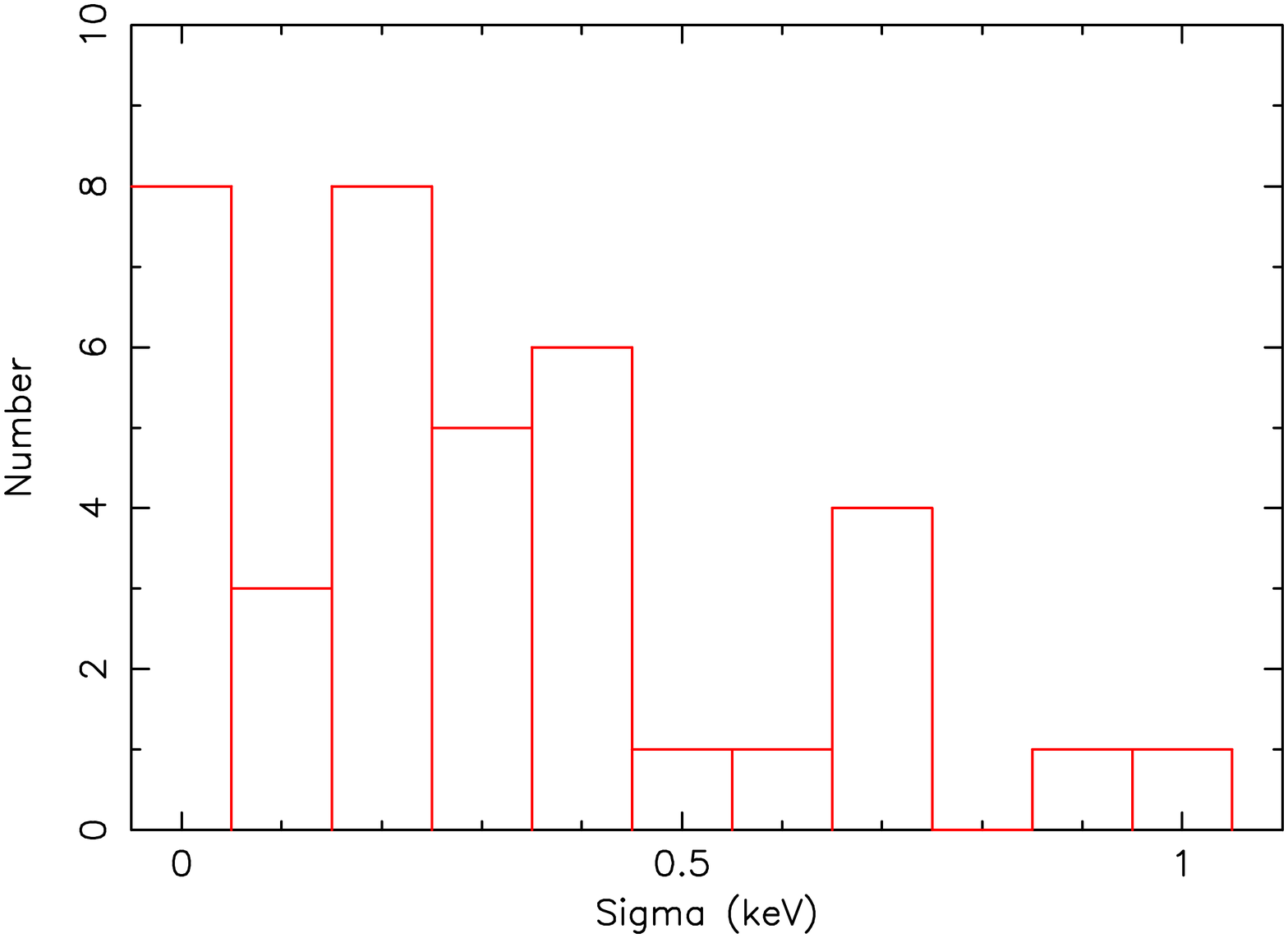}
\includegraphics[angle=0,width=58mm,height=50mm]{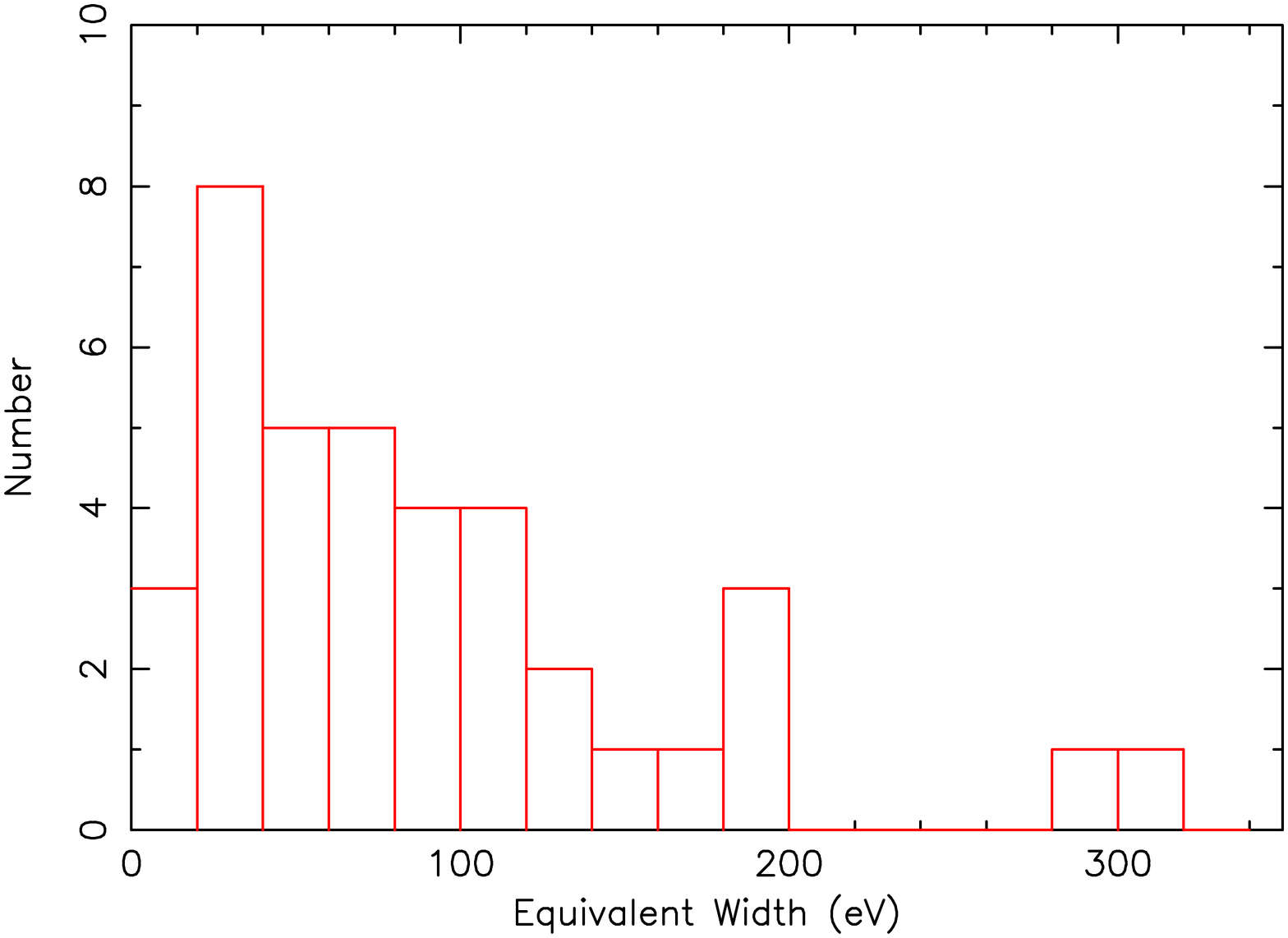}
}
\caption{Histograms of the rest-frame energy (top panel), gaussian $1\sigma$ width (middle panel) and equivalent width (bottom panel). The base model is a power law, with warm absorber as needed, and the \textsc{pexmon} component accounting for neutral reflection from distant material, incluing any narrow component of the emission line at 6.4 keV (see text).
\label{fig:gauss}}
\end{figure*}

Complex emission at iron-K evidently extremely common in Seyfert galaxies, being observed in 77 per cent (20/26) of the objects. Model~B provides a crude, model-independent characterization of the properties of the broad and/or complex emission lines. For 65 per cent of the sample (17/26 objects, 23/37 of the observations), there is evidence that the line is broad (i.e. the width of the additional Gaussian is constrained to be $\sigma>0$). Histograms of the relevant parameters are give in Fig.~\ref{fig:gauss}. The line energies range from $\sim 5-7$~keV, with a peak at 6.5 keV. Following Nandra \& Pounds (1994), we have used the maximum-likelihood method of Maccacaro et al. (1988) to determine the mean and intrinsic spread of the distribution of the line parameters, after accounting for the measurement errors. The results are shown in Table~\ref{tab:meanpars}, which also shows the unweighted mean and standard error, and the mean determined from a constant fit. The results are broadly consistent but we place most emphasis on the maximum-likelihood values as they both account for the measurement errors and allow an estimate of  any intrinsic spread in the distribution.

\begin{table*}
\centering
\caption{Average parameters for the sample derived assuming different models.
The upper values are for the broad Gaussian fits (Section~3.3) and the lower the neutral disk line (Section 4.1) \label{tab:meanpars}
(1) Fit parameter;
(2) Unit;
(3) Unweighted mean and standard error;
(4) Expectation allowing for measurement errors (see text);
(5) Intrinsic dispersion of distribution (see text)
}
\begin{center}
\begin{tabular}{lllll}
\hline
Parameter & Unit & \multicolumn{1}{c}{$\mu$} & \multicolumn{1}{c}{$<\mu>$} & \multicolumn{1}{c}{$\sigma_{\rm i}$} \\
(1) & (2) & \multicolumn{1}{c}{(3)} & \multicolumn{1}{c}{(4)} & \multicolumn{1}{c}{(5)} \\
\hline
\hline
\multicolumn{5}{l}{\bf Parameters of the Gaussian emission line from Model B (\S~\ref{Sec:Model-B})} \\
\hline
$E_{K\alpha}$      & keV & $6.27   \pm 0.07  $ & $6.30  ^{+0.10  }_{-0.11  }$ & $0.34  ^{+0.10  }_{-0.07  }$ \\
$\sigma_{K\alpha}$ & keV & $0.34   \pm 0.05  $ & $0.34  ^{+0.08  }_{-0.08  }$ & $0.26  ^{+0.07  }_{-0.05  }$ \\
$EW_{K\alpha}$     & eV  & $91.3  \pm 12.8 $ & $76.6^{+16.2 }_{-14.2 }$ & $41.0^{+17.3}_{-13.8}$ \\
$R_{tor}$ & $\Omega/2\pi$  & $0.51   \pm 0.06  $ & $0.47^{+0.07  }_{-0.07  }$ & $0.23^{+0.07  }_{-0.05  }$  \\
\hline
\multicolumn{5}{l}{\bf Fe line \& Compton Reflection parameters from Model E (\S~\ref{Sec:Model-E})} \\
\hline
$R_{disk}$     & $\Omega/2\pi$ & $0.95   \pm 0.16  $ & $0.54  ^{+0.11  }_{-0.10  }$  & $0.24^{+0.11  }_{-0.08  }$ \\
$R_{tor}$       & $\Omega/2\pi$  & $0.50   \pm 0.06  $ & $0.45  ^{+0.08  }_{-0.08  }$ & $0.27^{+0.07  }_{-0.06  }$  \\
$\log R_{br}$ & $r_{\rm g}$ & $1.20   \pm 0.14  $      & $1.20  ^{+0.24  }_{-0.26  }$ & $0.66 ^{+0.23  }_{-0.18  }$  \\
$\cos i$         & deg & $0.73   \pm 0.05  $                   & $0.81  ^{+0.06  }_{-0.07  }$  & $0.20  ^{+0.06  }_{-0.04  }$ \\
\hline
\multicolumn{5}{l}{\bf Fe line \& Compton Reflection parameters from Models E \& F (\S~\ref{Sec:Model-F})} \\
\hline
$R_{disk}$ & $\Omega/2\pi$ & $1.00   \pm 0.18  $    & $0.48  ^{+0.12  }_{-0.10  }$ & $0.21  ^{+0.12  }_{-0.09  }$ \\
$R_{tor}$ & $\Omega/2\pi$  & $0.46   \pm 0.06  $    & $0.38  ^{+0.07  }_{-0.06  }$ & $0.17 ^{+0.07  }_{-0.06  }$ \\
$\log R_{br}$ & $r_{\rm g}$ & $1.20   \pm 0.15  $   & $1.25  ^{+0.26  }_{-0.27  }$ & $0.72  ^{+0.23  }_{-0.20  }$ \\
$\cos i$ & deg & $0.73   \pm 0.05  $                       & $0.79  ^{+0.06}_{-0.07  }$ & $0.20  ^{+0.06  }_{-0.04  }$ \\
$\Gamma$ & -- & $1.86\pm 0.04$ & $1.86^{+0.06}_{-0.06}$ & $0.22^{+0.05}_{-0.04}$ \\
\hline
\end{tabular}
\end{center}
\end{table*}

In the case of the line energy we find $<E_{\rm K \alpha}> = 6.30\pm 0.11$~keV. The \xmm\ data therefore clearly show that the residual broad emission is associated with iron. A low ionization state is favoured by the peak energy and indeed there is even a hint of redshift. 
Unlike the \asca\ data of N97, the \xmm\ spectra show a significant spread to the distribution of 
$\sigma_{E_{\rm K\alpha}} =0.31$, in other words there are many observations in which 
the peak line emission is found significantly above and below both the mean and the neutral value. The distribution (Fig.~\ref{fig:gauss})  is clearly skewed towards redshifted lines, showing the possible presence of the characteristic ``red wing'' profiles of a relativistic accretion disk (Fabian et al. 1989; Tanaka et al. 1995). In cases where a peak energy higher than 6.4 keV is seen, this could be due to ionized iron, or Doppler blueshifting. We investigate these scenarios explicitly below. 
The average line width is $<\sigma_{\rm K\alpha}>=0.34\pm 0.08$~keV, corresponding to a FWHM velocity of $\sim 37,000$~km s$^{-1}$. The typical line width is therefore mildly relativistic ($\sim 0.12$c). As with the energy, a significant spread in the width distribution is required. This was not detected by \asca, a fact that we can attribute to the \xmm\ line widths being considerably better determined. 
The typical equivalent width is $EW \sim 77 \pm 16$~eV, yet again with a significant spread. 
Table~\ref{tab:gauss} also shows the mean value of the reflection fraction $R_{\rm tor}$ of the distant, neutral reflector. We find $R=0.46\pm 0.06$ which, assuming solar abundances and an intrinsic spectrum $\Gamma=1.9$, implies a narrow line equivalent width of  EW$_{\rm tor}$=$49\pm6$~eV. 
The equivalent width of the additional gaussian component implies that the material resposible for its production subtends approximately half the sky at the X-ray source. Specifically, assuming a slab geometry with an inclination of $60^{\circ}$, expected in the angle--averaged case, we deduce a solid angle of $1.6\pm 0.3\pi$~sr for the broad line reflector

\begin{figure}
{
\includegraphics[angle=0,width=90mm]{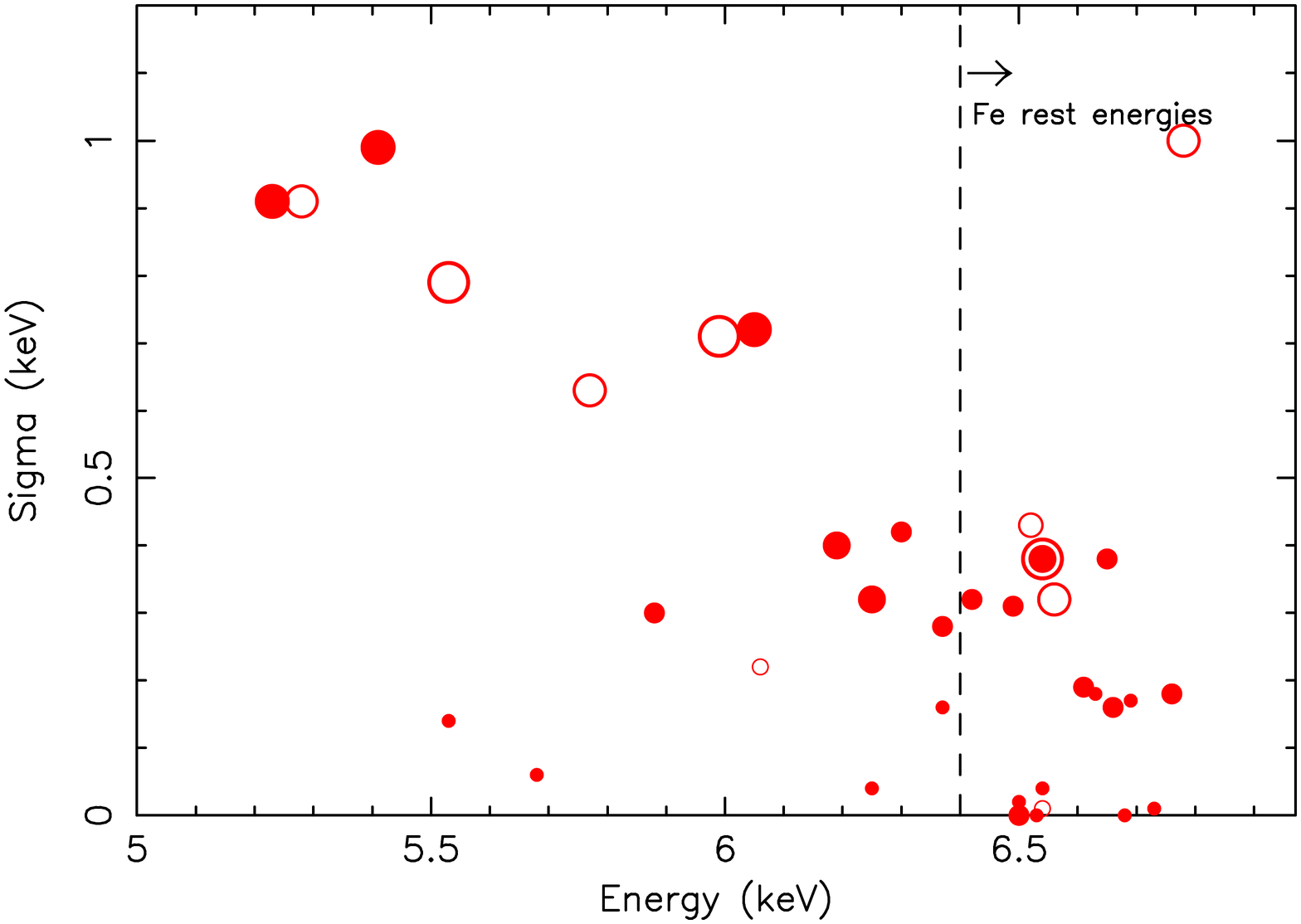}
}
\caption{Width of the additional Gaussian line component versus the line energy. Symbol sizes reflect the strength of the lines (0-50 eV, 50-100 eV, 100-150 eV, $>150$~eV), with larger symbols for higher equivalent width. The vertical dashed line is at 6.4 keV, at or above which the physical energy for unshifted iron K$\alpha$ lines are expected, up to the limit of the plot (6.97 keV, the expected energy for Fe {\sc xxvi}). Open symbols show fits which are formally poor, and hence in which the determination of the energy and $\sigma$ may be unreliable. 
\label{fig:en_sigma}}
\end{figure}

Further insight into the nature of the line complexity can be can be gained by an examination of Fig.~\ref{fig:en_sigma}. This shows the width of the additional Gaussian component of the line versus the line energy. The majority of objects are clustered around the mean values mentioned above. The spread in values appears primarily due to two effects. Firstly, there are four observations (Mrk 110, NGC 2110, Mrk590 and NGC 5548(1)) which show relatively weak, narrow lines at energies consistent with the rest values for iron K$\alpha$ species. Rather than necessarily showing broad iron lines, these object may simply have an additional component from ionized material far from the nucleus, without any requirement for an accretion disk component. Additionally, there are 7 observations which exhibit strong lines which are both significantly broader and significantly more redshifted than the typical values. 
These are NGC 3516(1,2),  NGC 3783(1), NGC 4051, NGC 4151(3), and MCG-6-30-15(1,2). These objects have unusually strong ``red wings" to the iron lines. As shown by, e.g., Yaqoob et al. (2002), if one tries to model an asymmetric profile such as a disk line with a Gaussian, one may either end up modeling the core of the line or the red wing, and this will give a misleading picture of the line widths overall. It is therefore necessary to proceed to more detailed modeling of the iron lines.

\section{Non--relativistic models (Models~C \& D)}
\label{Sec:Model-C}
\label{Sec:Model-D}

The complexities in the iron band need not necessarily imply relativistic broadening. Blends of several narrow emission lines may present, and it has also been proposed that complex absorption effects may introduce spectral curvature in this energy range. The chief suggestion, made by Reeves et al. (2004) and Turner et al. (2005),  is that high ionization components of the warm absorber can modify the spectral shape in the 4-6 keV region by mimicing the ``red wing" characteristic of low inclination disk line models. Obviously such a scenario can significantly affect the derived parameters of an accretion disk line, and in the extreme may account entirely for the complexity without recourse to relativistic effects at all. Indeed  Reeves et al. found that the inclusion of an absorption line apparently from He-like iron at 6.67~keV dramatically changed the parameters derived in the case of NGC~3783.
Since ionized absorption is present in this and many other objects in our sample, the idea that there would be a thick, highly ionized component is not implausibe. Since absorption by iron is critical in modeling these types of absorber we model this component with the \textsc{grid25} model from \textsc{XSTAR}, which includes updated atomic data for iron. 

It is clear from an examination of Fig.~1, that an absorption component {\it alone} would have no chance of modeling the spectra of most of our objects. However, it may be possible to combine this model with a series of unshifted, narrow emission lines and model the whole of the spectral excess. To test this we combine the high ionization warm absorber with a blend of narrow lines. The first, already included in the prior fits, is the 6.4 keV component from distant, optically thick material.  It is furthermore plausible that more highly ionized gas exists at large distances, and we account for this possibility initially by adding two narrow emission lines at the rest energies of helium-like iron (6.70 keV) and hydrogen-like iron (6.97 keV). 

In a large number of observations, this model (which we designate Model C)  provides a very significant improvement in the fit compared to the simple, distant reflection model (Model A).  However by all the statistical measures 
(see Table~\ref{tab:mcompv2}) the fit to the whole sample is dramatically worse for this model compared to the broad Gaussian model (Model B). 
For 
15 observations (11 objects) $\chi^{2}$ indicates that model C is inconsistent with the data at $>99$~per cent confidence, with 7 of these catastrophically poor ($\chi_{\nu}^{2}>2$). Additionally, in no individual case does the model significantly improve the fit compared to Model B. Thus we conclude that Model  C is worse description of the data 
than our simple broad-Gaussian parameterization.

A possible reason for the bad fits is the existence of excess emission in the 6.4--6.7 keV range  (e.g. see 
Fig~\ref{fig:all_profile}),  between the neutral and helium-like line energies. To account for this, we have introduced an additional line with free energy, but constrained to be in the range 6.4-6.7 keV. This model (Model D) improves the fits markedly in a large number of cases compared to Model C, and reduces the number of catastrophic failures ($N_{\rm bad}$) from 7 to 3. The same 15 fits remain statistically unacceptable at $>99$ per cent confidence, however, and the comparison with the Model B is still highly unfavourable. Once again, all the statistical measures in Table~\ref{tab:mcompv2} favour the broad Gaussian model. This comparison is all the more striking when one considers that Model B has three fewer free parameters than the absorber/blend Model D. 

A clear conclusion of our work is therefore that a combination of narrow line blends and complex ionized absorption cannot alone explain the spectra of bright Seyfert 1 galaxies, and overall provides a worse description than a simple Gaussian. The main problem with the Gaussian parameterization is the lack of any physical motivation. We therefore now proceed to test models of relativistic disk lines against the data, to see if they can account for the iron K$\alpha$ complexity. 

\section{Relativistic broad-line models (Models~E)}
\label{Sec:Model-E}

\begin{figure}
\includegraphics[angle=0,width=90mm]{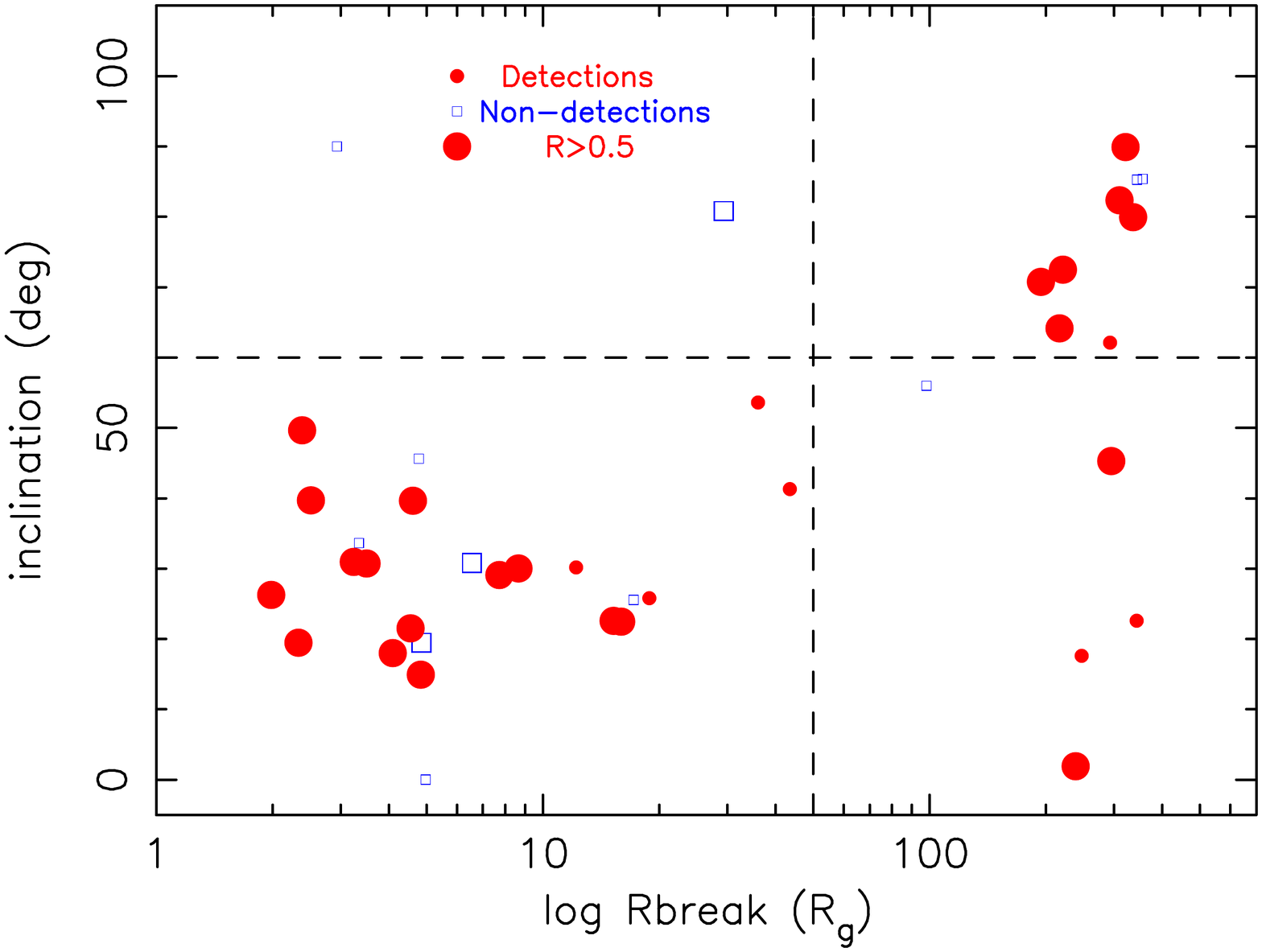}
\caption{Characteristic emission radius for the relativistic iron K$\alpha$ lines versus disk inclination. This plot is divided into four quadrants, defined at $50 r_{\rm g}$, notionally dividing the ``inner disk" from more distant material and $60^{\circ}$, distinguishing between face-on and edge-on disks. The filled circles are for observations in which the F-test indicates a significant improvement of the fit, whereas the open squares are ones in which no significant improvement in $\chi^{2}$ is found. Larger symbols show cases where strong reflection $R_{\rm disk}>0.5$ is observed. Most objects populate the bottom left part of the diagram, where one would expect to see broad lines with significant gravitational redshift.  A number of objects show no evidence for such effects, however, particularly those in the bottom right of the diagram. See text for further discussion.  
}
\label{fig:rbreak}
\end{figure}

\begin{table*}
\centering
\caption{Disk line parameters derived using Model E. 
Col.(1): Name and Observation number;
Col.(2): Reflection fraction of the blurred {\tt pexmon} component, where $R=1$ corresponds to a semi-infinite slab seen at an inclination $i$ subtending 2$\pi$ solid angle at the X-ray source; 
Col.(3): Inclination of the slab representing the blurred reflector (i.e. the accretion disk); 
Col.(4): Break radius for the emissivity law. Note that the inner radius has been fixed to $6 r_{\rm g}$, appropriate to the Schwarzschild cases, so while allowed in the model values of $r_{\rm br} < 6 r_{\rm g}$ are statistically indistinguishable from those with  $r_{\rm 
br} = 6 r_{\rm g}$ 
Col.(5): Reflection fraction of the unblurred {\tt pexmon} component, where $R=1$ corresponds to a semi-infinite slab seen at an inclination of 60$^{\circ}$ and subtending 2$\pi$ solid angle at the X-ray source; 
Col. (6): F-statistic for the addition of the blurred reflection component (3 additional d.o.f.). $F>4.0$ is significant at $>99$~per cent confidence. 
Col. (7): $\chi^{2}$ and degrees of freedom including all components. 
Col. (8): Probability of obtaining $\chi^{2}$ by chance
\label{tab:kdpex}}
\begin{center}
\begin{tabular}{llcllcccc}
\hline
Name &  $R_{\rm disk}$ & $i$ & $r_{\rm br}$ &  $R_{\rm tor}$ & F & $\chi^{2}$/dof & prob \\
 & ($\Omega/2\pi$) & (deg) & ($r_{\rm g}$) & ($\Omega/2\pi$) & & & & \\
(1) & (2) & (3) & (4) & (5) & (6) & (7) & (8) \\
\hline
\hline
NGC 526A & $0.39 ^{+2.22 }_{-0.23 }$ & $43 ^{+42 }_{-20 }$ & $44.4 ^{+355.6 }_{-43.2 }$ & $0.28 ^{+0.15 }_{-0.13 }$ & 6.65 & 80.4/98 & 0.90 \\
Mrk 590 & $0.44 ^{+0.63 }_{-0.44 }$ & $46 ^{+39 }_{-46 }$ & $4.8 ^{+395.2 }_{-3.5 }$ & $0.63 ^{+0.24 }_{-0.22 }$ & 1.10 & 118.9/96 & 0.06 \\
Ark 120 & $2.28 ^{+1.45 }_{-1.67 }$ & $82 ^{+3 }_{-30 }$ & $216.9 ^{+183.1 }_{-52.9 }$ & $0.45 ^{+0.11 }_{-0.11 }$ & 20.67 & 139.9/96 & 0.00 \\
NGC 2110 & $0.41 ^{+0.30 }_{-0.25 }$ & $31 ^{+11 }_{-6 }$ & $12.2 ^{+48.6 }_{-11.0 }$ & $0.74 ^{+0.13 }_{-0.16 }$ & 3.93 & 110.5/98 & 0.18 \\
MCG+8-11-11 & $3.18 ^{+4.04 }_{-2.84 }$ & $85 ^{+0 }_{-60 }$ & $9.3 ^{+390.7 }_{-8.1 }$ & $0.87 ^{+0.31 }_{-0.23 }$ & 2.78 & 87.0/96 & 0.73 \\
Mrk 6 & $0.36 ^{+0.45 }_{-0.30 }$ & $26 ^{+59 }_{-26 }$ & $21.7 ^{+378.3 }_{-20.5 }$ & $0.28 ^{+0.23 }_{-0.10 }$ & 2.29 & 90.6/89 & 0.43 \\
Mrk 110 & $0.80 ^{+1.02 }_{-0.80 }$ & $85 ^{+0 }_{-85 }$ & $295.7 ^{+104.3 }_{-294.5}$ & $0.23 ^{+0.12 }_{-0.12 }$ & 1.08 & 112.3/97 & 0.14 \\
NGC 2992 & $0.62 ^{+0.40 }_{-0.28 }$ & $24 ^{+7 }_{-7 }$ & $15.7 ^{+56.7 }_{-14.5 }$ & $0.20 ^{+0.17 }_{-0.18 }$ & 12.38 & 97.4/100 & 0.56 \\
MCG-5-23-16(1) & $0.50 ^{+0.17 }_{-0.26 }$ & $0 ^{+19 }_{0 }$ & $114.5 ^{+285.5 }_{-77.9 }$ & $0.02 ^{+0.17 }_{-0.02 }$ & 5.44 & 128.7/100 & 0.03 \\
MCG-5-23-16(2) & $0.63 ^{+0.12 }_{-0.21 }$ & $17 ^{+5 }_{-17 }$ & $32.9 ^{+55.3 }_{-16.5 }$ & $0.19 ^{+0.25 }_{-0.18 }$ & 9.13 & 117.1/100 & 0.12 \\
NGC 3516(1) & $0.91 ^{+0.42 }_{-0.37 }$ & $19 ^{+7 }_{-19 }$ & $3.6 ^{+6.4 }_{-2.3 }$ & $0.81 ^{+0.24 }_{-0.23 }$ & 7.40 & 124.7/95 & 0.02 \\
NGC 3516(2) & $2.42 ^{+0.75 }_{-0.65 }$ & $31 ^{+2 }_{-4 }$ & $8.5 ^{+6.0 }_{-7.2 }$ & $1.78 ^{+0.45 }_{-0.39 }$ & 101.59 & 90.8/94 & 0.57 \\
NGC 3783(1) & $0.44 ^{+0.30 }_{-0.21 }$ & $0 ^{+19 }_{0 }$ & $176.3 ^{+223.7 }_{-118.7}$ & $0.52 ^{+0.33 }_{-0.34 }$ & 12.63 & 138.5/94 & 0.00 \\
NGC 3783(2) & $0.29 ^{+0.12 }_{-0.07 }$ & $15 ^{+7 }_{-15 }$ & $10.3 ^{+7.0 }_{-9.1 }$ & $0.71 ^{+0.09 }_{-0.08 }$ & 11.28 & 241.2/94 & 0.00 \\
HE 1143-1810 & $0.19 ^{+0.16 }_{-0.19 }$ & $0 ^{+85 }_{-0 }$ & $3.5 ^{+396.5 }_{-2.3 }$ & $0.24 ^{+0.13 }_{-0.08 }$ & 1.04 & 91.0/96 & 0.62 \\
NGC 4051 & $0.85 ^{+0.44 }_{-0.35 }$ & $22 ^{+6 }_{-15 }$ & $1.2 ^{+14.2 }_{-0.0 }$ & $0.52 ^{+0.18 }_{-0.17 }$ & 12.11 & 102.8/96 & 0.30 \\
NGC 4151(1) & $0.65 ^{+0.28 }_{-0.24 }$ & $17 ^{+12 }_{-17 }$ & $209.6 ^{+190.4 }_{-113.4}$ & $1.18 ^{+0.40 }_{-0.34 }$ & 13.29 & 93.7/94 & 0.49 \\
NGC 4151(2) & $0.63 ^{+0.20 }_{-0.18 }$ & $31 ^{+4 }_{-3 }$ & $9.7 ^{+6.8 }_{-8.5 }$ & $1.45 ^{+0.16 }_{-0.07 }$ & 10.76 & 166.3/90 & 0.00 \\
NGC 4151(3) & $0.63 ^{+0.10 }_{-0.10 }$ & $23 ^{+2 }_{-3 }$ & $1.2 ^{+7.2 }_{-0.0 }$ & $0.38 ^{+0.05 }_{-0.06 }$ & 40.55 & 205.6/94 & 0.00 \\
Mrk 766(1) & $1.50 ^{+1.36 }_{-0.79 }$ & $40 ^{+5 }_{-6 }$ & $1.2 ^{+25.8 }_{-0.0 }$ & $0.57 ^{+0.39 }_{-0.32 }$ & 10.27 & 83.6/96 & 0.81 \\
Mrk 766(2) & $1.32 ^{+0.49 }_{-0.39 }$ & $40 ^{+2 }_{-2 }$ & $3.6 ^{+16.6 }_{-2.3 }$ & $0.27 ^{+0.14 }_{-0.14 }$ & 24.88 & 127.8/96 & 0.02 \\
NGC 4395 & $0.50 ^{+0.38 }_{-0.30 }$ & $15 ^{+12 }_{-15 }$ & $1.2 ^{+22.8 }_{-0.0 }$ & $0.52 ^{+0.26 }_{-0.20 }$ & 4.28 & 97.0/89 & 0.26 \\
NGC 4593 & $0.35 ^{+0.18 }_{-0.16 }$ & $24 ^{+61 }_{-17 }$ & $247.6 ^{+152.4 }_{-246.3}$ & $0.42 ^{+0.31 }_{-0.22 }$ & 6.68 & 135.0/96 & 0.01 \\
MCG-6-30-15(1) & $0.92 ^{+0.35 }_{-0.28 }$ & $20 ^{+4 }_{-20 }$ & $2.3 ^{+10.6 }_{-1.1 }$ & $0.34 ^{+0.19 }_{-0.13 }$ & 19.69 & 135.2/95 & 0.00 \\
MCG-6-30-15(2) & $1.32 ^{+0.23 }_{-0.12 }$ & $30 ^{+2 }_{-2 }$ & $7.6 ^{+3.2 }_{-6.3 }$ & $0.18 ^{+0.07 }_{-0.07 }$ & 96.52 & 156.0/94 & 0.00 \\
IC 4329A(1) & $0.16 ^{+0.98 }_{-0.16 }$ & $60 ^{+25 }_{-60 }$ & $107.1 ^{+292.9 }_{-105.8}$ & $0.24 ^{+0.16 }_{-0.14 }$ & 0.67 & 74.6/100 & 0.97 \\
IC 4329A(2) & $0.27 ^{+0.08 }_{-0.07 }$ & $31 ^{+20 }_{-10 }$ & $225.1 ^{+174.9 }_{-181.9}$ & $0.27 ^{+0.09 }_{-0.10 }$ & 21.03 & 111.4/96 & 0.13 \\
NGC 5506(1) & $0.71 ^{+3.48 }_{-0.28 }$ & $58 ^{+27 }_{-19 }$ & $135.3 ^{+264.7 }_{-134.1}$ & $0.54 ^{+0.10 }_{-0.14 }$ & 10.48 & 112.2/98 & 0.15 \\
NGC 5506(2) & $0.51 ^{+0.33 }_{-0.19 }$ & $38 ^{+5 }_{-5 }$ & $1.2 ^{+34.7 }_{-0.0 }$ & $0.25 ^{+0.12 }_{-0.12 }$ & 7.09 & 115.0/98 & 0.12 \\
NGC 5548(1) & $0.06 ^{+0.33 }_{-0.06 }$ & $15 ^{+70 }_{-15 }$ & $1.2 ^{+398.8 }_{-0.0 }$ & $0.43 ^{+0.16 }_{-0.15 }$ & 0.09 & 83.9/97 & 0.83 \\
NGC 5548(2) & $0.22 ^{+0.11 }_{-0.10 }$ & $23 ^{+19 }_{-12 }$ & $130.5 ^{+269.5 }_{-129.3}$ & $0.24 ^{+0.11 }_{-0.16 }$ & 6.33 & 104.5/96 & 0.26 \\
Mrk 509 & $0.40 ^{+0.62 }_{-0.28 }$ & $29 ^{+56 }_{-29 }$ & $28.2 ^{+371.8 }_{-27.0 }$ & $0.23 ^{+0.19 }_{-0.23 }$ & 3.65 & 90.6/96 & 0.64 \\
NGC 7213 & $0.16 ^{+1.49 }_{-0.16 }$ & $68 ^{+17 }_{-68 }$ & $356.1 ^{+44.0 }_{-354.8}$ & $0.55 ^{+0.17 }_{-0.22 }$ & 0.32 & 110.6/96 & 0.15 \\
NGC 7314 & $0.85 ^{+0.20 }_{-0.38 }$ & $42 ^{+3 }_{-4 }$ & $3.9 ^{+69.3 }_{-2.7 }$ & $0.32 ^{+0.16 }_{-0.14 }$ & 11.04 & 87.1/94 & 0.68 \\
Ark 564 & $1.63 ^{+2.63 }_{-1.10 }$ & $77 ^{+8 }_{-32 }$ & $310.1 ^{+89.9 }_{-291.8}$ & $0.07 ^{+0.22 }_{-0.07 }$ & 6.57 & 121.7/96 & 0.04 \\
NGC 7469(1) & $3.69 ^{+5.00 }_{-2.63 }$ & $85 ^{+0 }_{-53 }$ & $1.2 ^{+398.8 }_{-0.0 }$ & $0.77 ^{+0.36 }_{-0.23 }$ & 2.34 & 113.8/97 & 0.12 \\
NGC 7469(2) & $3.91 ^{+2.15 }_{-1.84 }$ & $85 ^{+0 }_{-2 }$ & $1.2 ^{+71.3 }_{-0.0 }$ & $0.76 ^{+0.15 }_{-0.12 }$ & 7.38 & 166.8/97 & 0.00 \\
\hline
\end{tabular}
\end{center}
\end{table*}

The prevailing paradigm established by \asca\ is that these broad emission features arise from a relativistic acccretion disk (Tanaka et al. 1995; Nandra et al. 1997). To test this, we have replaced the broad Gaussian line with a disk line model. We reiterate that in the fits described below we account for soft X-ray ionized absorption, where needed, and add a distant neutral reflector {\tt pexmon} in all cases. 

We begin with the assumption that the accretion disk is in a low state of ionization, modeling this case with the {\tt pexmon} model with relativistic blurring with the Laor (1991) kernel (Fabian et al. 2002), but fixing the inner radius to $r_{\rm in} = 6 r_{\rm g}$, appropriate for the Schwarzschild case. As described above the parameters are the strength of the reflection $R_{\rm disk}$, the inclination, and the break radius $r_{\rm br}$.  Formally we allow the break radius to vary below the inner radius of the disk, but in practice the profile so derived will not differ from one with $r_{\rm br} = 6 r_{\rm g}$. We designate this Model E and give the results in Table~\ref{tab:kdpex}. 
As illustrated in Table~\ref{tab:mcompv2}, we find this model provides an excellent description of the spectra, with $N_{\rm good}$=28, and only two catastrophic failures (NGC 3783 (2) and NGC 4151(3)). The other six fits unacceptable at $99$~per cent confidence are to Ark 120, MCG-6-30-15(1,2), NGC 4151(2), NGC 4593 and NGC 7469(2). In these objects (and perhaps others) there could also be additional unmodelled components and we investigate this in detail below. In terms of the statistical comparison to our earlier models, Model E provides a better description of the data, with the single exception that the median $\chi^{2}$ for Model B (the broad Gaussian) is very slightly lower. At the ensemble level, then, the relativistic disk model is clearly preferred over the alternatives considered above. Further insight can be obtained by comparing the goodness-of-fit of the models on an observation-by-observations basis. The statistical improvement over simple, distant reflection (Model A) is once again very clear, even when considering that the disk line model has 3 additional parameters,
with the statistic improving by $\Delta\chi^{2}=58.1$ per observation. Comparing to Model B (the broad gaussian) the goodness-of-fit is often similar.  However the physically-motivated disk line model provides a substantial improvement in many cases. $\Delta\chi^{2}>10$ is found in NGC 3516(2), NGC 3783 (2), NGC 4151(1,3), Mrk 766(2), MCG-6-30-15(1,2) and NGC 7469(2), with the inference being that an asymmetric line profile is preferred in those objects. A substantially {\it worse} fit compared to the Gaussian is seen for Mrk 590 and NGC 5506 (1,2). 

Comparing Model E to those which combine complex absorption and line blends, a slightly more complex picture emerges. We have already noted that overall the disk line model fares better and this is confirmed in a large number of individual cases. When an intermediate (6.4-6.7 keV) line is allowed, however, there are a number of observations which fit better to the absorber/blend model than the disk line model. The most striking example is NGC 3783 (Reeves et al. 2004) in which Model D is very clearly preferred 
by $\Delta\chi^{2} \sim 90$. It should be noted, however, that even Model D fails to provide an {\it acceptable} fit to this spectrum. Several other spectra are significantly better fit with Model D than Model E (at $>99$~per cent confidence based on the F-test), these being Mrk 590, NGC 2110, NGC 4151(2), NGC 5506(1), with NGC 7213 just failing to meet the 99~per cent threshold. There is considerable overlap between this list of objects and those in which previous work  has suggested complex absorption or line blends as an alternative to the disk lines. 

Our overall assessment is therefore that, while a relativistic disk provide the best {\it single} model for the iron line profiles seen in the sample, the alternative descriptions in terms of complex absorption and line blends are surely important in some of the objects. This motivates our investigation of additional complexities in the next  section. In the meantime, because the disk provides a decent description of the data with very few catastrophically poor fits, it is legitimate to investigate the typical parameters inferred from the disk line fits. We can then see how and if these conclusions change when we do consider further complications in spectra. 

Some of the more salient features of the parameters of the relativistic line model E are shown in Fig.~\ref{fig:rbreak}. This figure plots the derived disk inclination against the break radius $r_{\rm br}$. For clarity we have split the plot into four quadrants with dividing points at $r_{\rm br}  = 50 \ r_{\rm g}$ and $i=60$. It can be seen that the majority of the objects lie in the bottom left quadrant with low inclinations and small break radii. It this regime where we expect ``classic'' disk lines, of the type seen in MCG-6-30-15, to occur. The emission is concentrated in the innermost regions, so relativistic effects are important and the line will be broad and skewed, and the inclination is relatively low, so the gravitational redshifted is not swamped by Doppler blueshifting. 

The upper left portion of the diagram is where we expect weak and very broad lines from highly inclined disks with the emission concentrated in the inner regions. It is sparsely populated, which is expected as such lines are difficult to detect.  We return to this point in the discussion.

The upper right portion shows several strong disk lines with apparently high inclinations but at relatively large radii. This indicates that the lines are broad but that the asymmetric broad emission is predominantly due to a strong ``blue wing''.  These are likely candidates for a highly ionized disk, which is in reality at lower inclination than inferred in fits which assume the disk is neutral. We investigate this possibility in the following section. They may also be examples of objects in which there is a blend of ionized lines. 

Finally, at the bottom right of the diagram we see a few objects which show emission apparently at low inclination and large radius. In these objects the lines will be relatively narrow and not strongly shifted. For these, there is no requirement for the line to arise in the inner accretion disk and no clear evidence for relativistic effects at all. There are a number of possible explanations for this which we explore in the Discussion. 

The average parameters for this model are shown in Table~\ref{tab:meanpars}.  We find that on average the material responsible for the the blurred reflector covers a solid angle  ($2 \pi \times <R_{disk}>$) of $1.1\pm0.2 \pi$ at the X-ray source. This is lower than $2 \pi$ solid angle expected for  a canonical flat (semi-infinite) accretion disk. However we also find that on average   the distant neutral reflector covers a similar (and additional) solid angle at the source ($2\pi \times <R_{tor}>$ = $0.9\pm0.2 \pi$).  However we note that both these values show a significant dispersion, with  $\sigma_i \sim 0.4\pi$ and 0.6$\pi$ (respectively). The average inclination derived for the blurred reflection component is $<i> =36^{\circ}\pm6$. We note this value  is consistent to that found for the \asca\ sample by N97. Finally, turning to the characteristic radius at which the broad line is emitted, we find a mean value of $r_{\rm br}=18\pm10$ R$_{\rm g}$.  At this radius in a Schwarzschild geometry, the gravitational redshift is of order $\sim 6$~per cent. This can be compared to the instrumental FWHM of the EPIC-pn camera of $<2$~per cent, so the strong gravitational effects of the black hole are typically observable in the profile. 
We note that both $i$ and $r_{\rm br}$ exhibits a wide dispersion of values for the sample (see also Table~\ref{tab:kdpex}).
The possible implications for these values and distributions, and the implications for the geometry of the central regions,
will be discussed below.

\section{Additional complexities (Model F)}
\label{Sec:Model-F}

In \S~\ref{Sec:Model-C} we found that in general the combination of complex absorption and line blends does not provide a good fit to all the spectra. These components may nonetheless be present in addition to the disk lines.  For example, a highly ionized, high column density absorber has been inferred in NGC 3783 (Reeves et al. 2004) and we found a better fit to this object with Model D, which incorporates such a component.  
The emission line complexes in NGC 5506 and NGC 7213 have also been fitted with blends of highly ionized lines (Matt et al. 2001; Bianchi et al. (2003), and again our fitting also indicates a possible preference for this interpretation. Most importantly, poor fits are obtained with the disk line model (Model-E) in 9 of our 37 observations.
We must therefore strive to find a better description of the data in these cases. The important question for our study is then not whether absorbers or blends can generally account for the spectral complexity at iron K$\alpha$, as we have shown explicitly that they can not, but how they affect the derived parameters of the disk lines and in particular the evidence for relativistic effects. The 'customized' models described in this section are referred to as Model F. In all cases below, and as in model E, we fix the inner radius for the blurred reflector to $6 r_{\rm g}$, i.e. the Schwarzschild case. 

\subsection{High ionization warm absorbers}

We start by testing whether an additional screen of ionized absorption improves the disk line fits (Model E and Table~\ref{tab:kdpex}) significantly. We used both the 
{\tt cwa18} warm absorber and the highly-ionized {\tt grid25} absorber in this test. 
Adding an additional {\tt cwa18} component to Model E we found a significant improvement at $>99$ per cent confidence in four observations, NGC 3516(1), NGC 3783 (1,2) and NGC 4151(3). The same four (and only these four) sources also show an improvement over the Model E fits when the {\tt cwa18} component was replaced by the {\tt grid25} component.
In fact the latter are very marginally better fits to the data than the former. For simplicity we therefore adopt the {\tt grid25} model for the high ionization warm absorber. A large improvement is also seen in NGC 4151(2) with \delchi$=12.7$. While this is not formally significant according to the F-test because the fit is poor, we favour the
 inclusion of this component as the reduction is $\chi^{2}$ is large and because there is evidence for it in another of the observations of the same source. 
It is very interesting to note at this point that these three sources are the prime examples in the literature in which complex absorption has been invoked as an alternative to the disk line in the literature (Schurch et al. 2003; Reeves et al. 2004; Turner et al. 2005), but they are the only objects in our sample which require such absorption.

\subsection{Narrow emission and absorption lines}

We have already tested the possibility of emission from He-like or H-like iron from distant material as an alternative to broad lines, and there are sometimes even greater improvements when intermediate species in the physical rest-frame range of 6.4-6.7 keV are allowed. It is perfectly reasonably to expect weak, narrow emission lines from ionized species to arise from gas at large scales, particularly at the He-like and H-like energies of 6.70 and 6.97 keV (e.g. Krolik \& Kallman 1987). Furthermore, such emission has been reported in previous work. 

In addition, {\it absorption} features at these same energies have been reported in a few objects, notably NGC 3783 and MCG-6-30-15 (Reeves et al. 2004; Young et al. 2005). Such features indicate large columns of ionized gas in the line-of-sight. They should therefore be modeled by our warm absorber tables, but it is prudent to test whether any residual features are present in the spectra, which would indicate that  while we may have got the curvature correct, we have not modeled all the details of this gas correctly.  

Assessing the significance of narrow features in the X-ray spectra of AGN is notoriously difficult, and it has been shown specifically that the F-test does not provide an accurate estimate of the chance probability of detecting such features, particularly if multiple energies are permitted (Protassov et al. 2002). Correctly accounting for the number of trials is best done using Monte Carlo simulations, and we take such an approach here. Adding an emission or absorption line to the model will result in some change in $\chi^{2}$, but to correctly determine the significance of that change we need to perform the same test on a large number of synthetic spectra.  

To determine the appropriate significance of our $\Delta\chi^{2}$ values we have therefore simulated power law spectra, adding appropriate noise using the \textsc{XSPEC} {\tt fakeit} command. We then refit the power law to the simulated spectra, and then perform a search for narrow emission lines and absorption lines using the fake spectra. We assumed $\Gamma=1.9$ for the power law, and have tested two cases, one with $\sim 500,000$ counts in the 2.5-10 keV spectrum, appropriate to our observation with the highest signal-to-noise ratio (MCG-6-30-15(2)) and one with 50,000 counts, which represents a typical-to-low signal-to-noise ratio for our sample. We found this choice made only a small difference and we adopt the results from the high quality spectrum here. 

\subsubsection{He and H-like lines}

Since we later consider the possibility of other emission and absorption features, rather than test for the significance only of the H-like and He-like lines, we performed a grid search using a narrow ($\sigma = 10$ eV) Gaussian  stepping through energies between 2.8 and 9.4 keV, in steps of 30 eV. The step is small enough to sample the full instrumental resolution, and it and the limits are chosen to sample precise energies of 6.70 and 6.97 keV (as well as 6.40 keV, the neutral line). Ten thousand  such simulations were performed. The significance of any particular feature depends on the number of trial energies. If we restrict the search only to the physical He-like and H-like energies, which we discuss first, 99~per cent of our simulations gave $\Delta\chi^{2} < 7.4$.  Thus if the real data exhibit a change in $\chi^{2}$ greater than this the feature can be considered significant at this confidence level. 

We also performed precisely the same grid search on the real spectra, noting the $\chi^{2}$ deviation at 6.70 and 6.97 keV for each spectrum. The base model was was Model E (Table~\ref{tab:kdpex}), to which we add the {\tt grid25} warm absorber in the four observations noted in the previous subsection (i.e. NGC 3516(1), NGC 3783 (1,2), NGC 4151(2,3)). 

Significant emission from H-like iron was found in NGC 3783(1,2) and NGC 4593, and emission from He-like iron was found in NGC 5506(1).  Significant {\it absorption} features at 6.7 keV were seen in two cases NGC 3516(2) and MCG-6-30-15(2). The He-like absorption feature at 6.7 keV is seen in MCG-6-30-15(1) just below our 99~per cent significance level ($\Delta \chi^{2} = 6.3$), 
which we nonetheless include because it is also present in the other spectrum of the same source which has higher signal-to-noise ratio. All the above features have been noted in the same objects in previous work (Matt et al. 2001, Reeves et al. 2004,  Reynolds et al. 2004a, Turner et al. 2005, and  Young et al. 2005). We include them in all subsequent fits, modeling them as narrow Gaussians. This is probably a reasonable assumption in the cases of the emission lines, but more caution is required when considering the absorption lines as in reality they indicate an unmodelled or mismodelled screen of very highly ionized gas in the line-of-sight. We have already attempted explicitly to include such a component in the spectral fit using the {\tt grid25} warm absorber model above, but this failed to improve the fit in the observations in question (NGC 3516 and MCG-6-30-15). We note that Reeves et al. (2004) have reported a highly significant He-like absorption feature in NGC 3783(2), this is also present in our spectrum, but in this case our {\tt grid25} model accounts for it successfully. 

\begin{table}
\centering
\caption{Additional narrow line components
Col.(1): Name and Observation number;
Col.(2): Energy of line; 
Col.(3): Equivalent width of line;
Col.(4): Change in $\chi^{2}$ when the component is added to the model; 
Col. (5): $\chi^{2}$ and degrees of freedom with the line added;  
\label{tab:addlines}}
\begin{center}
\begin{tabular}{lclcc}
\hline
Name & $E_{line}$ & EW & $\Delta\chi^{2}$ & $\chi^{2}$/dof \\
(1) & (2) & (3) & (4) & (5)\\
\hline
\hline
Mrk 590               & $6.49$ & $+73^{+33}_{-32}$ & 26.1 & 0.98/95 \\
Ark 120                & $6.01$ & $-21^{+10}_{-10}$ & 21.3 &  1.25/95  \\ 
NGC 2110           & $6.55$ & $+27^{+15}_{-13}$ & 13.5 & 01.00/97 \\
NGC 3516(2)      & $6.70$ & $-39^{+20}_{-20}$ & 14.2 & 0.82/93 \\
NGC 3783 (1)     &  $6.97$ & $+21^{+16}_{-16}$ & 7.3 & 1.26/91  \\
NGC 3783 (2)     &  $6.97$ & $+20^{+11}_{-9}$ & 32.7 & 1.47/91  \\
                               & $6.52$ & $+19^{+13}_{-12}$ & 24.9 & 1.21/90 \\
NGC 4151(2)       & $6.52$ & $+23^{+9}_{-12}$ & 28.5 & 1.35/87 \\
		             & $7.33$ & $-15^{+8}_{-8}$ &  16.8 & 1.17/86 \\
                               & $3.70$ &  $+10^{+6}_{-5}$ & 15.6 & 1.00/85 \\
NGC 4151(3)       & $7.45$ & $-16^{+7}_{-6}$ & 28.0 & 1.47/91 \\
                               & $5.23$ & $+8^{+4}_{-4}$ &18.9 & 1.27/90 \\
NGC 4593             &  $6.97$ &$+18^{+11}_{-12}$ & 11.6 & 1.30/95 \\
MCG-6-30-15(1)  & $6.70$ & $-16^{+}_{-}$ & 6.9 & 1.37/94 \\
MCG-6-30-15(2)  & $6.70$ & $-25^{+10}_{-15}$ &  11.1 & 1.52/93 \\
IC 4329A(2)          & $7.69$ &  $-15^{+7}_{-7}$ & 21.0 & 0.95/95 \\
NGC 5506(1)        & $6.70$ & $+32^{+17}_{-21}$ & 15.0 & 1.00/97 \\
                                & $6.49$ & $+26^{+20}_{-16}$ & 12.0 &  0.88/96 \\
NGC 5506(2)        & $6.46$ & $+38^{+22}_{-20}$ & 14.2 & 1.04/97 \\
NGC 7469(2)        & $6.52$ & $+15^{+9}_{-9}$ & 13.7 & 1.60/96 \\ 
\hline
\end{tabular}
\end{center}
\end{table}

\subsubsection{Intermediate line features}

In the comparison of models described in previous sections (and e.g. Table~\ref{tab:mcompv2}) we considered the possibility of line emission in between the neutral and He-like energies of 6.4 and 6.7 keV and Fig.~1 shows that in some cases these appear to be present. The results of the grid search above can be used to test this quantitatively. If we widen the energy range of the simulations to 6.4--6.97 keV, the range spanned by iron, then we find that 1~per cent of the simulations exceed a $\Delta\chi^{2}$ of 10.6. 

Adopting this as our significance criterion, we find line emission from intermediate ionization species of iron ($\sim 6.5$ keV) in Mrk 590, NGC 2110, NGC 3783(2),  NGC 4151(2) and NGC 5506(1). 
The precise energies are given in Table~\ref{tab:addlines}. As discussed above it is currently unclear whether these features are signatures of gas with intermediate ionization, or represent some other mismodeling of the line complex. For the moment, we assume that they are real, discrete, narrow emission features. In assessing the constraints on any broader relativistic line emission, this approach is conservative. 

\begin{table*}
\centering
\caption{Disk line parameters derived using Model F for the 17 observations which require an additional multi-zone warm absorber and/or  narrow
line components.
Col.(1): Name and Observation number;
Col.(2): Reflection fraction of the blurred {\tt pexmon} component, where $R=1$ corresponds to a semi-infinite slab seen at an inclination $i$ subtending 2$\pi$ solid angle at the X-ray source; 
Col.(3): Inclination of the slab representing the blurred reflector (i.e. the accretion disk); 
Col.(4): Break radius for the emissivity law
Col.(5): Reflection fraction of the unblurred {\tt pexmon} component, where $R=1$ corresponds to a semi-infinite slab seen at an inclination of 60$^{\circ}$ and subtending 2$\pi$ solid angle at the X-ray source; 
Col. (6): F-statistic for the addition of the blurred reflection component (3 additional d.o.f.). $F>4.0$ is significant at $>99$~per cent confidence. 
Col. (7): $\chi^{2}$ and degrees of freedom including all components. 
Col. (8): Probability of obtaining $\chi^{2}$ by chance
\label{tab:kdbest}}
\begin{center}
\begin{tabular}{llcllcccc}
\hline
Name &  $R_{\rm disk}$ & $i$ & $r_{\rm br}$ &  $R_{\rm tor}$ & F & $\chi^{2}$/dof & prob \\
 & ($\Omega/2\pi$) & (deg) & ($r_{\rm g}$) & ($\Omega/2\pi$) & & & & \\
(1) & (2) & (3) & (4) & (5) & (6) & (7) & (8) \\
\hline
\hline
Mrk 590 & $0.41 ^{+0.58 }_{-0.41 }$ & $46 ^{+39 }_{-46 }$ & $4.8 ^{+395.2 }_{-3.5 }$ & $0.28 ^{+0.25 }_{-0.24 }$ & 0.90 & 92.8/95 & 0.55 \\
Ark 120 & $3.74 ^{+0.84 }_{-2.86 }$ & $84 ^{+1 }_{-23 }$ & $216.9 ^{+183.1 }_{-96.8 }$ & $0.40 ^{+0.12 }_{-0.12 }$ & 29.53 & 118.6/95 & 0.05 \\
NGC 2110 & $0.14 ^{+0.34 }_{-0.14 }$ & $0 ^{+85 }_{-0 }$ & $6.4 ^{+393.6 }_{-5.1 }$ & $0.76 ^{+0.17 }_{-0.19 }$ & 0.35 & 96.5/97 & 0.50 \\
NGC 3516(1) & $1.07 ^{+0.59 }_{-0.52 }$ & $33 ^{+3 }_{-9 }$ & $3.6 ^{+7.8 }_{-2.3 }$ & $0.70 ^{+0.32 }_{-0.16 }$ & 8.64 & 100.5/93 & 0.28 \\
NGC 3516(2) & $3.23 ^{+1.03 }_{-0.96 }$ & $34 ^{+2 }_{-2 }$ & $7.7 ^{+7.1 }_{-6.5 }$ & $1.87 ^{+0.45 }_{-0.39 }$ & 90.71 & 76.5/93 & 0.89 \\
NGC 3783(1) & $0.56 ^{+0.23 }_{-0.22 }$ & $3 ^{+18 }_{-3 }$ & $263.9 ^{+136.1 }_{-194.9}$ & $0.22 ^{+0.29 }_{-0.22 }$ & 13.29 & 114.4/91 & 0.05 \\
NGC 3783(2) & $0.49 ^{+0.11 }_{-0.13 }$ & $29 ^{+4 }_{-4 }$ & $11.4 ^{+10.5 }_{-10.2 }$ & $0.63 ^{+0.06 }_{-0.06 }$ & 36.29 & 119.7/91 & 0.02 \\
NGC 4151(2) & $0.11 ^{+0.14 }_{-0.10 }$ & $21 ^{+69 }_{-21 }$ & $118.2 ^{+281.8 }_{-117.0}$ & $1.15 ^{+0.27 }_{-0.27 }$ & 6.77 & 81.9/85 & 0.57 \\
NGC 4151(3) & $0.75 ^{+0.09 }_{-0.15 }$ & $33 ^{+1 }_{-3 }$ & $8.5 ^{+4.1 }_{-7.3 }$ & $0.33 ^{+0.06 }_{-0.03 }$ & 44.46 & 113.9/90 & 0.05 \\
Mrk 766(2) & $1.19 ^{+0.43 }_{-0.38 }$ & $40 ^{+2 }_{-2 }$ & $3.6 ^{+16.9 }_{-2.3 }$ & $0.26 ^{+0.13 }_{-0.13 }$ & 23.18 & 112.3/95 & 0.11 \\
NGC 4593 & $0.36 ^{+0.17 }_{-0.16 }$ & $24 ^{+28 }_{-15 }$ & $247.6 ^{+152.4 }_{-160.3}$ & $0.41 ^{+0.22 }_{-0.22 }$ & 9.60 & 123.4/95 & 0.03 \\
MCG-6-30-15(1) & $0.99 ^{+1.12 }_{-0.31 }$ & $24 ^{+14 }_{-8 }$ & $2.3 ^{+13.5 }_{-1.1 }$ & $0.31 ^{+0.17 }_{-0.15 }$ & 21.43 & 128.9/94 & 0.01 \\
MCG-6-30-15(2) & $1.72 ^{+0.38 }_{-0.33 }$ & $34 ^{+1 }_{-2 }$ & $1.8 ^{+7.8 }_{-0.6 }$ & $0.16 ^{+0.08 }_{-0.07 }$ & 99.97 & 141.6/93 & 0.00 \\
IC 4329A(2) & $0.26 ^{+0.08 }_{-0.07 }$ & $30 ^{+18 }_{-9 }$ & $225.1 ^{+174.9 }_{-167.5}$ & $0.26 ^{+0.09 }_{-0.10 }$ & 25.88 & 90.4/95 & 0.61 \\
NGC 5506(1) & $0.37 ^{+0.36 }_{-0.22 }$ & $45 ^{+4 }_{-5 }$ & $3.7 ^{+137.6 }_{-2.4 }$ & $0.52 ^{+0.23 }_{-0.20 }$ & 3.68 & 84.3/96 & 0.80 \\
NGC 5506(2) & $0.45 ^{+0.24 }_{-0.18 }$ & $40 ^{+5 }_{-5 }$ & $1.2 ^{+42.3 }_{-0.0 }$ & $0.07 ^{+0.16 }_{-0.07 }$ & 8.30 & 100.7/97 & 0.38 \\
NGC 7469(2) & $3.65 ^{+2.06 }_{-1.82 }$ & $85 ^{+0 }_{-2 }$ & $1.2 ^{+16.1 }_{-0.0 }$ & $0.69 ^{+0.15 }_{-0.12 }$ & 9.88 & 153.1/96 & 0.00 \\
\hline
\end{tabular}
\end{center}
\end{table*}

\subsubsection{Shifted line features}

Narrow emission and absorption lines in AGN have been claimed over a very wide range of X-ray energies. Some of these are associated with unshifted transitions of iron, as tested above, but in others velocity shifts are invoked and the features can be either blueshifted or redshifted (e.g. Nandra et al. 1999; Turner et al. 2002).  Again the grid search described above will reveal such features if present. The simulations shows that surprisingly large reductions in $\chi^{2}$ can be obtained by chance if they are allowed at any energy, which under normal circumstances one may have been tempted to interpret as real emission or absorption features. Performing 10,000 simulation, $\sim 1$~per cent of our simulations shows a $\Delta\chi^{2} > 14.3$, and we adopt this as our significance criterion.

To ensure we do not simply rediscover features found in the previous analysis, we repeated the grid search over the whole 2.8-9.4 keV for the objects in which a significant feature had already been found above, including that feature, or, in the case of NGC 5506(1), both features. 

Significant absorption features outside the nominal range appropriate for iron K$\alpha$ were found in Ark 120 (6.01 keV), NGC 4151(2) (7.33 keV), NGC4151(3) (7.42 keV) and  IC4329A (7.69 keV). Significant {\it emission} features were also found in NGC4151(2) at 3.70 keV, and in NGC4151(3) at 5.23 keV.  The features in Ark 120 and IC4329A have been noted previously by Vaughan et al. 2004 and Markowitz et al. 2006 respectively. The other features have not previously been reported. 

The interpretation of these features is potentially very exciting, and is discussed further below, but we caution that there significance depends substantially on whether we have modeled the overall spectra correctly. Once again, however, we include them in all subsequent fits to be conservative.

\subsection{Iron K-edges}

We performed a search very similar to the narrow line search for bound-free absorption edges, associated with iron-K, stepping from 7.10 to 9.30 keV in steps of 50 eV.  The baseline model is the same as in the previous section, but with the narrow emission lines include where needed. Monte Carlo simulations similar to those described above show that, for a random energy in the 7.1-9.3 keV, we require $\Delta\chi^{2}>9.4$ for a detection at 99~per cent confidence. 
 
The edge test reveals significant additional or mismodeling of the absorption in NGC 4151(1), which shows a significant feature at 7.10 keV.  This additional edge feature has been noted previously by Schurch et al. (2003), who interpreted it as an iron overabundance in a neutral (or near-neutral) absorber. As described in the previous subsection, the other two observations also show an improvement in the fit when additional absorption is allowed around 7.3-7.4 keV, but this has been modeled as an absorption line. In reality these could also be edges. We discuss the possible origin of these features in greater detail later, but for the time being the apparent edge in NGC 4151(1) is ignored.

Our test also reveal an additional absorption edge at the He-like energy in Mrk 766(2). This edge indicates that there is a column of very highly ionized gas in this object which is again not well modeled with the {\tt grid25} warm absorber, which fails to improve the spectral fit in this observation. Evidently, while there appears to be a very high ionization (and high column density) absorber in this source it introduces very little spectral curvature below the edge. For the time being, therefore, we simply model this as a single absorption edge.

\subsection{Hard tails}

Earlier, we highlighted the issue that there are a number of objects in our sample for which a simple power law is not an adequate description of the data outside the iron band. In some of these the spectra are ``convex" and the complexity is modeled as absorption. In others the continuum was found to be ``concave" with the existence of a hard tail. If our modeling is correct, this should be accounted for by the Compton reflection continuum associated with both the narrow and broad part of the iron K$\alpha$ line.   Of the objects highlighted in Section 3.1 which have such concave spectra and hence are poorly modeled with a power law outside the iron band, we find that the combination of the distant and blurred  {\tt pexmon} components successfully accounts for the curvature in Ark 120, Mrk 110, NGC 4593 and NGC 7469(1). The spectrum of NGC 7469(2) is still poorly modeled even with the additional complexities described above, and there may be other objects with residual concave spectra, in which have unmodeled reflection present.

We have tested this by adding an additional {\tt pexrav} component to the spectra, with free normalization and solar abundance, but without any associated line components. This produced a large improvement in the fit in 2 spectra: NGC 4151(1)  ($\Delta \chi^{2}$=14.0) and NGC 7469(2) ($\Delta \chi^{2}$=17.7). It is interesting to note that the additional edge in NGC 4151(1) found in the edge search may therefore associated with additional Compton reflection, rather than absorption. As with the additional edges, this additional Compton reflection is not self-consistent because with neutral, solar abundances reflection it should always be accompanied by an iron K$\alpha$ emission line. Rather the hard tails seen in the three quoted objects are indicative of some other mismodelling. Possible examples include a sub-solar abundance of iron, extreme relativistic blurring, or very high ionization in the reflector. All of these would tend to produce a hard reflection component, without any apparent emission line. 

\begin{figure}
\includegraphics[angle=0,width=90mm]{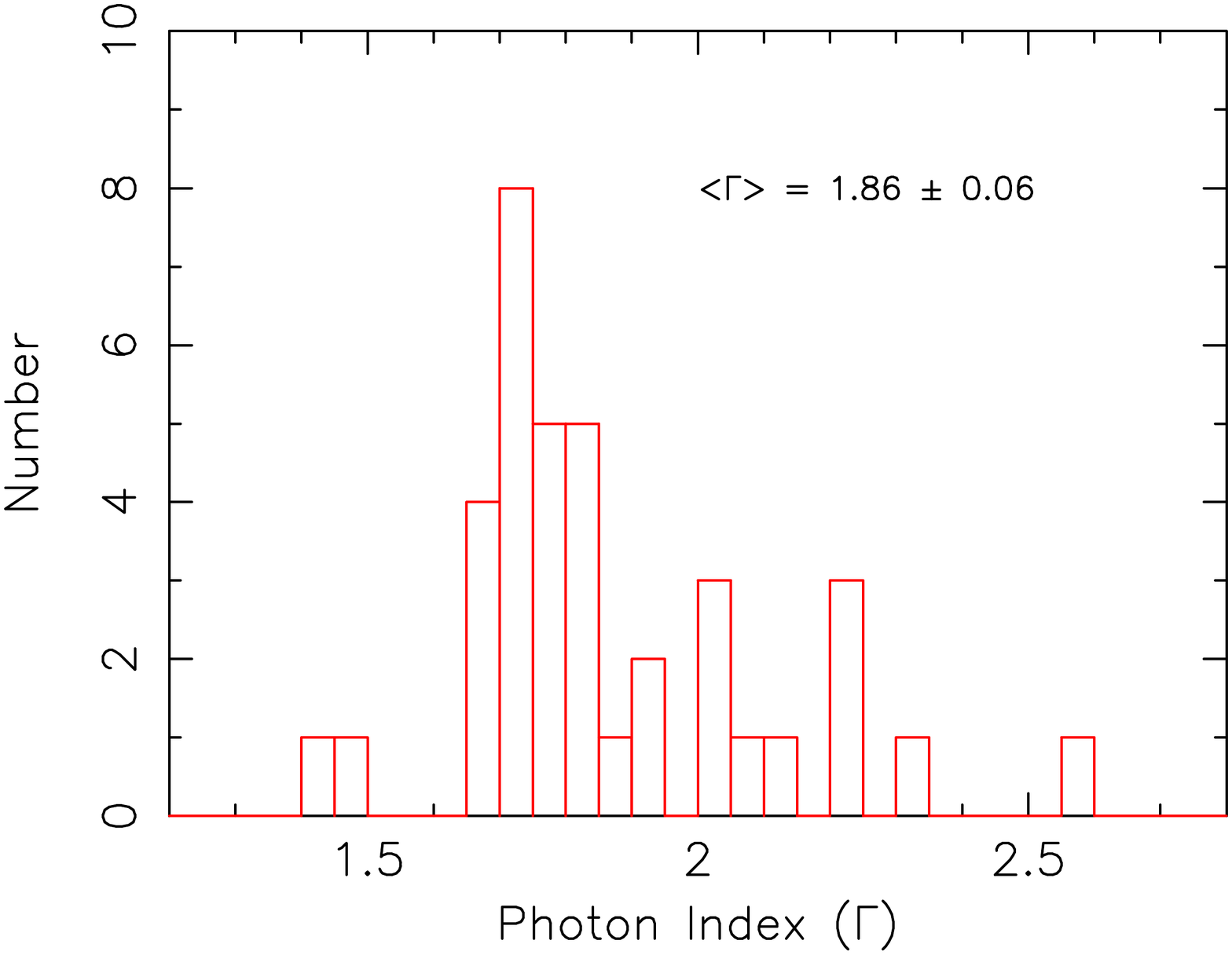}
\caption{Histogram of spectral indices obtained from our best-fit models (E or F). The mean value is similar to that found in previous work, and there is a significant dispersion of $\sigma_{\Gamma}=0.22$}
\label{fig:ghist}
\end{figure}

\section{Blurred-Reflection properties of the sample}

In \S~\ref{Sec:Model-E} we found a preference for a model which includes a blurred reflection component (Model~E). However we found that such a model was unacceptable for several of the observations. In \S~\ref{Sec:Model-F} we investigated a number of causes of the this additional complexity by  adding (a limited number of plausible) additional spectral components, and found that indeed improved  fits were obtained. Once the above complexities are accounted for we can be much more confident in the constraints we obtain on the parameters associated with the blurred reflection component. The revised (Model F) values for the 17 observations affected are reported in Table~\ref{tab:kdbest}. There remain 3 observations for which our fit is unacceptable at 99 per cent confidence, but the worst $\chi^2_{\nu}$ is now only 1.6. These values listed in Table~\ref{tab:kdbest} have been combined with those for the other observations from Table~\ref{tab:kdpex} to derive the  statistics for the sample listed in  Table~\ref{tab:mcompv2}. This, our final model, clearly outperforms all previous ones in terms of statistical quality. 

\subsection{Average Parameters}

Having accounted for the further complexities we can also recompute the average parameter values for the blurred reflection model, and these are listed in Table~\ref{tab:meanpars}. We find that these change very little compared to the simpler model of Table~\ref{tab:kdpex}, showing that the complex absorbers and narrow components have relatively little impact on modeling of the disk lines overall. We find again that on average the material responsible for the blurred reflector apparently covers a smaller solid angle than expected from a canonical flat disk, with the average being $1.0\pm0.2 \pi$ at the X-ray source.  The most obvious conclusion to draw from this is that the true geometry differs from that of a point source above a semi-infinite slab, as we have assumed so far. Many alternative geometries would be expected to give a relatively smaller reflected flux for a given covering fraction (e.g. Nandra \& George 1994; Zdziarski, Lubinski \& Smith 1999). There are also clear differences from object-to-object in the strength of the reflection, demonstrated by a significant dispersion in this value ($\sim 0.4\pi$).

In our fits, a distant, neutral reflector is found to be almost ubiquitous. The distant reflector subtend an apparent solid angle of $0.8 \pm 0.1 \pi$ at the X-ray source. This can be translated formally into an opening angle for the torus of $\sim 60^{\circ}$. A more accurate statement in this regard could be made by fitting the data with a physical model of the torus (if one existed), rather than a slab. Our estimate also relies on the assumption that the Thomson optical depth of the torus is $>>1$, which is not necessarily the case. Nonetheless the ubiquity of these narrow components supports the idea that the majority of these objects contain an obscuring torus as envisaged in the unification schemes. A significant dispersion in the value of $R_{\rm tor}$ is seen. However, this need not necessarily reflect true differences in the solid angle of the distant reflector. We have assumed a single inclination for the distant reflecting slab and in e.g., a torus geometry, the strength of the observed reflection depends on the viewing angle and the degree of self-covering of the reflecting material. 

The average inclination derived for the slab responsible for the blurred reflection component is $38\pm6$ degrees. This implies that the objects are viewed relatively-face on compared to random inclinations. This is perhaps to be expected given that we have excluded Seyfert 2s from the sample, and hence presumably have selected against edge-on systems. There is a further bias in that for inclined accretion disks, the reflection component becomes both weak and strongly blurred by Doppler effects. The upshot is that the observational constraints on highly inclined systems will typically be much worse, and hence they will be given relatively low weight in determining the average values. Despite this, we find a very significant dispersion in the inclinations derived from our fits. This may be attributed in large part to a few observations in which very high inclinations are inferred. The most extreme example is NGC 7469(2), in which the inclination pegs at the maximum value allowed by the model ($85^{\circ}$) with a very small error bar. The apparently extreme broadening of the reflection component in this object is discussed further below, but it is far from clear that in reality the disk is at such a high inclination. 

The characteristic radius at which the broadened reflection is emitted is again $r_{\rm br}=18\pm10$ R$_{\rm g}$, essentially unchanged compared to the simple reflection Model E. As before this implies that, typically, the broad emission comes from close enough to the black hole that effects of gravitational redshift are measurable. Once again, however, there is a highly significant dispersion in the derived values of $r_{\rm br}$. Taken at face value, this implies that there are some objects in the sample in which the emission is concentrated much closer to the black hole than the average, and others in which the emission is much more distant. 

In Fig.~\ref{fig:ghist} we plot the histogram of spectral indices from our best fit models (E or F). Employing the maximum likelihood method (Table~\ref{tab:meanpars} we find that the mean underlying continuum for the sample has a spectral index $\Gamma=1.86 \pm 0.06$ with a significant dispersion $\sigma_{\Gamma} = 0.22^{+0.05}_{-0.04}$. The underlying continuum is in good agreement with estimates from previous work (e.g. Nandra \& Pounds 1994; N97) and provides an important check on whether our parameterization of the spectra is reasonable.

\begin{figure*}
{
\includegraphics[angle=270,width=52mm]{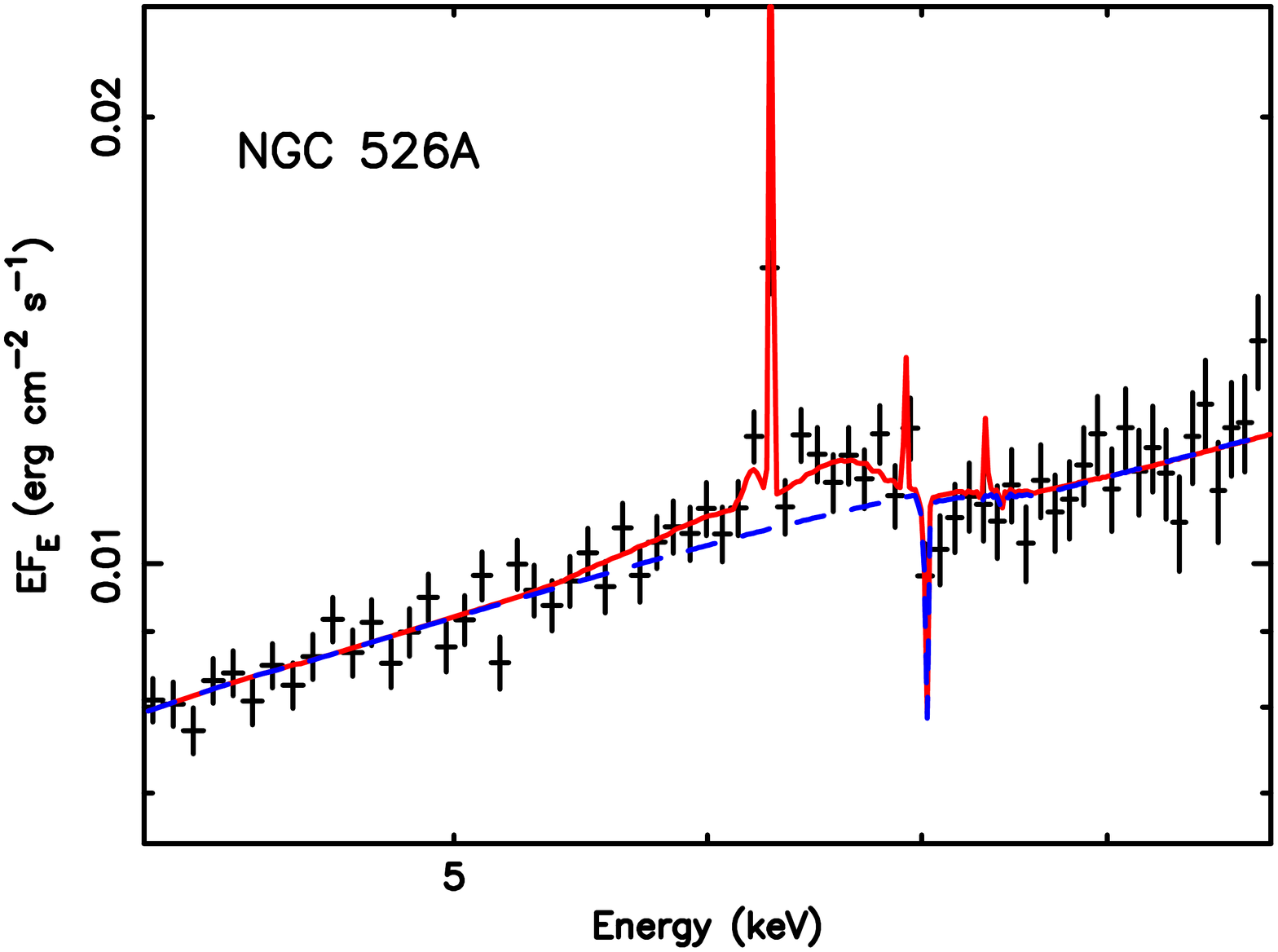}
\includegraphics[angle=270,width=52mm]{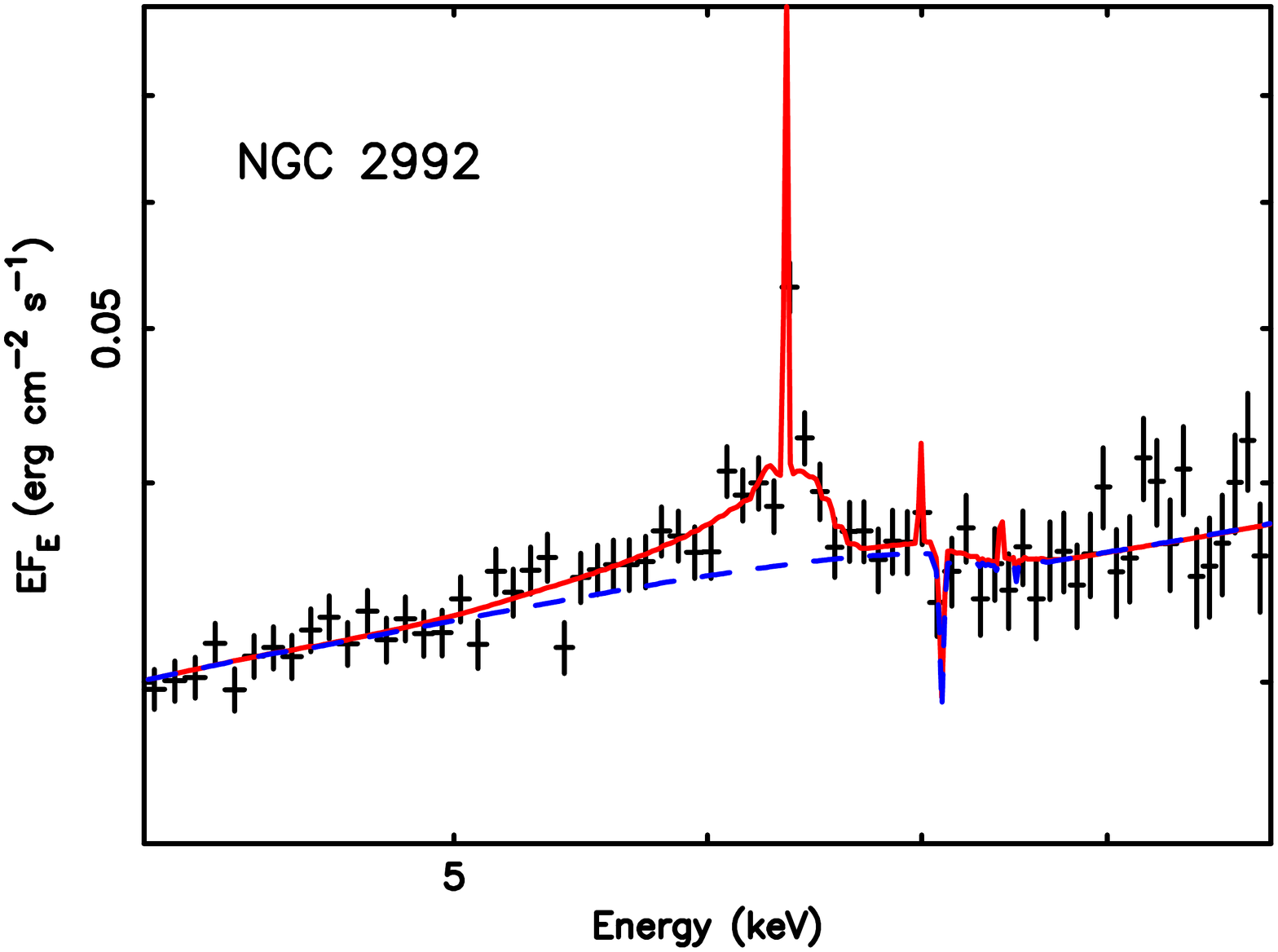}
\includegraphics[angle=270,width=52mm]{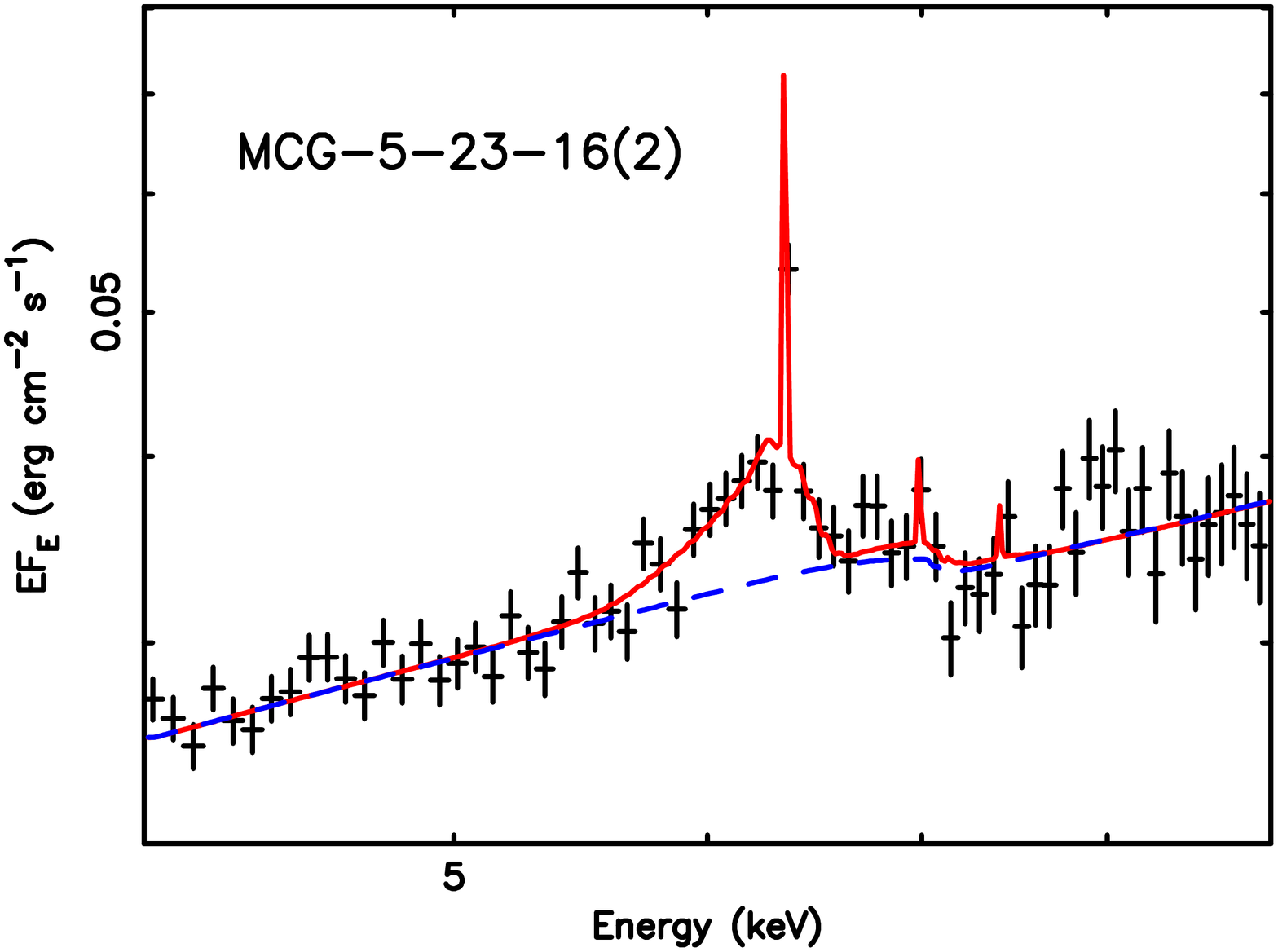}
\includegraphics[angle=270,width=52mm]{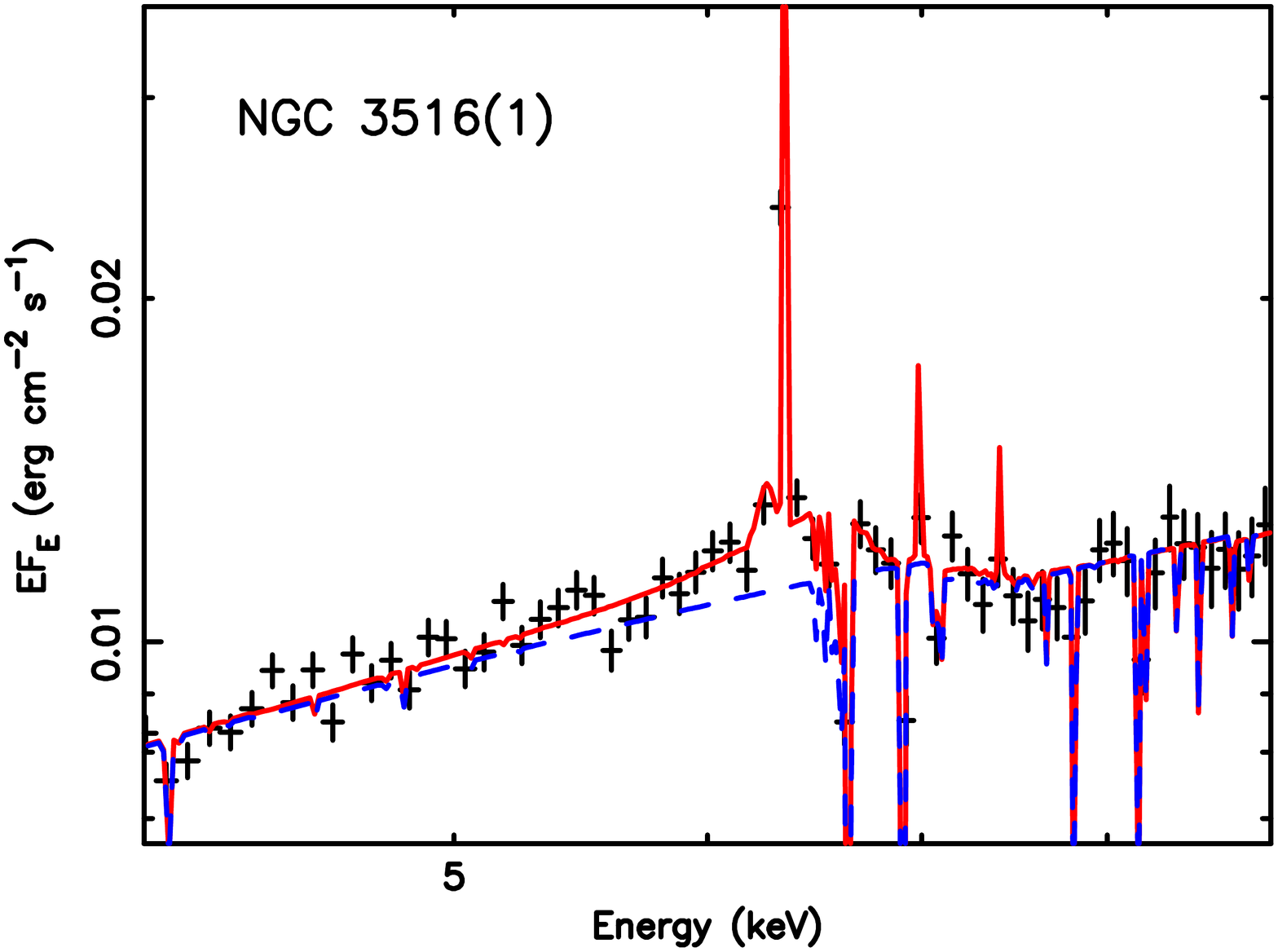}
\includegraphics[angle=270,width=52mm]{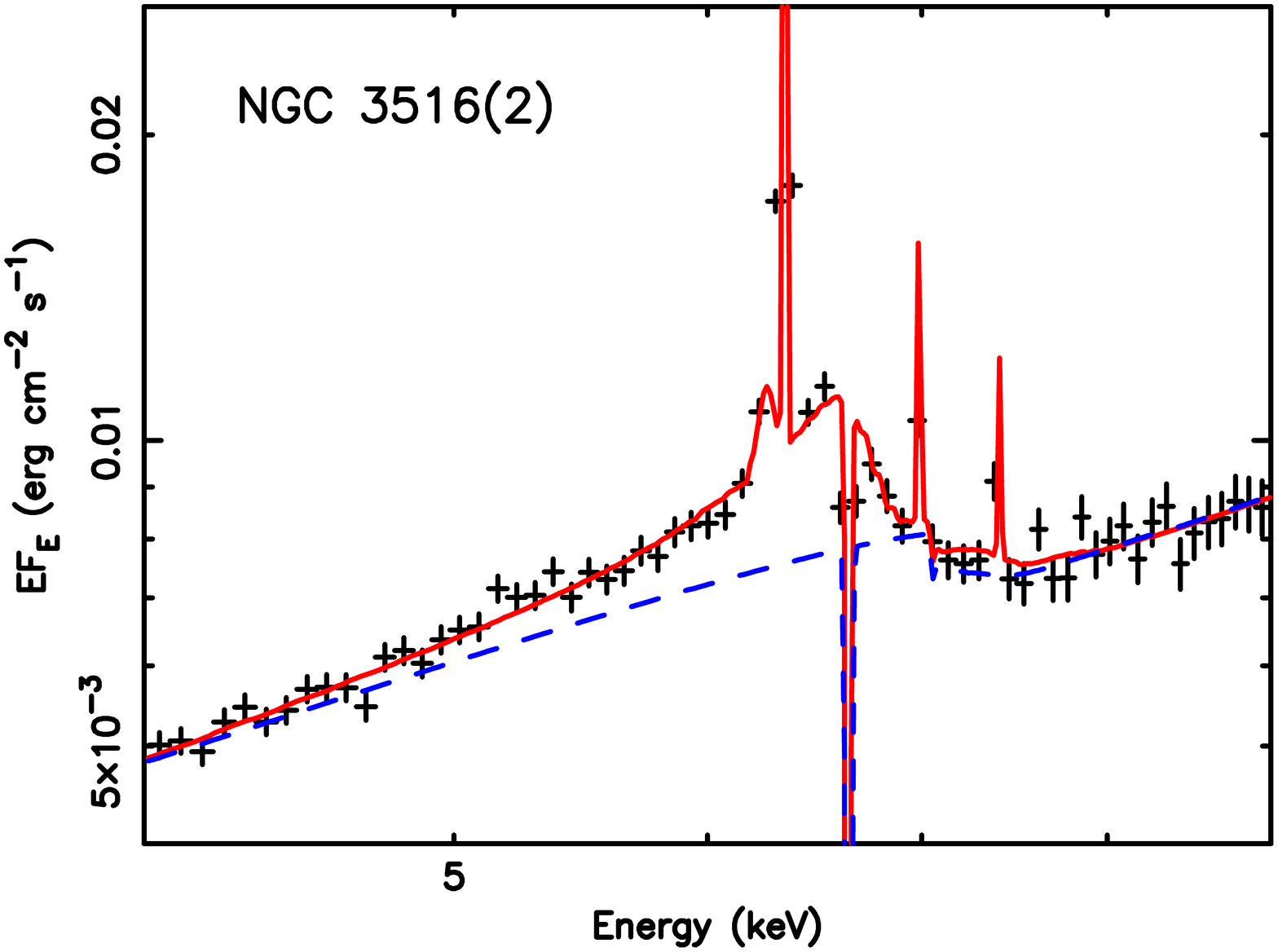}
\includegraphics[angle=270,width=52mm]{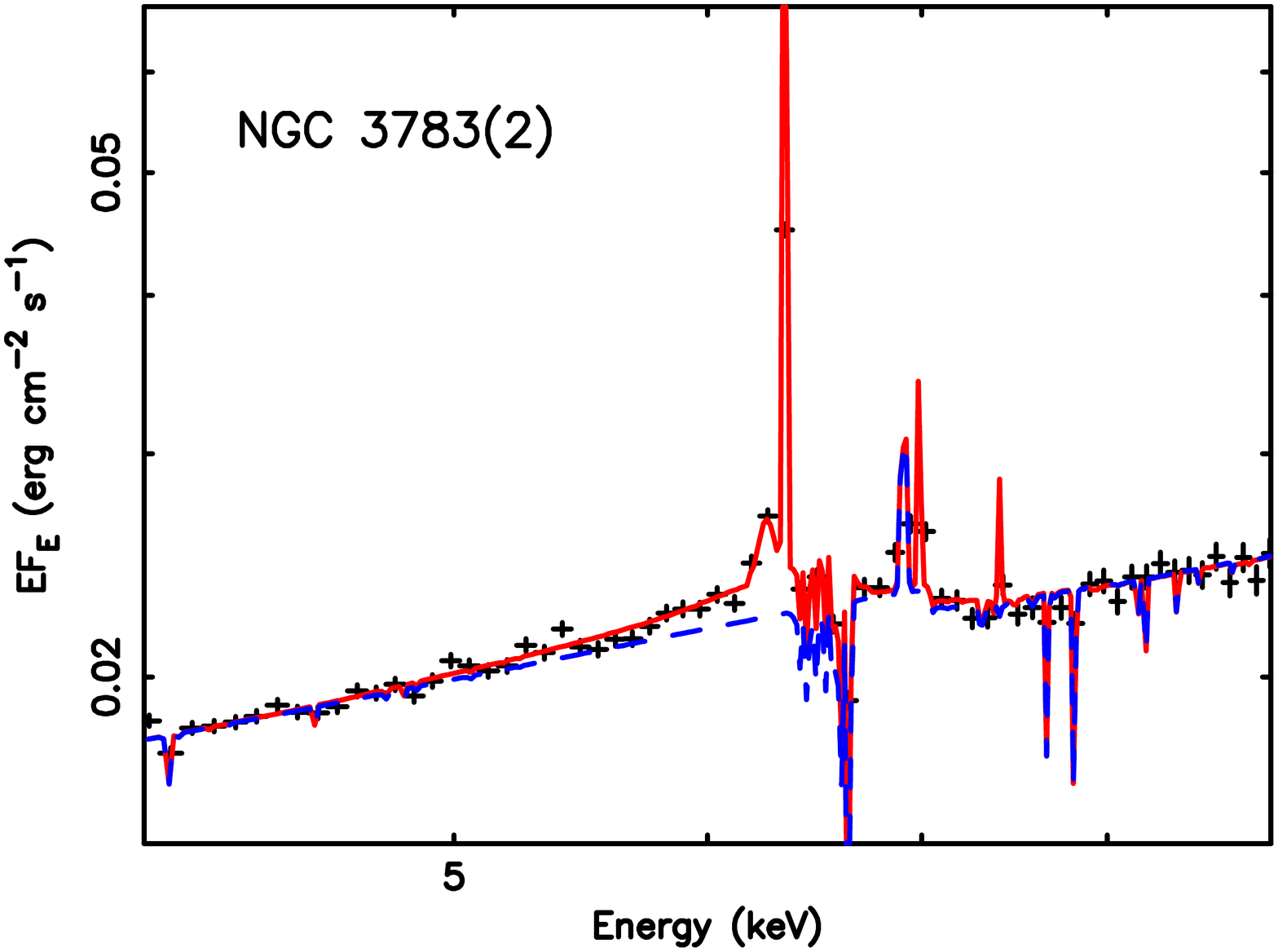}
\includegraphics[angle=270,width=52mm]{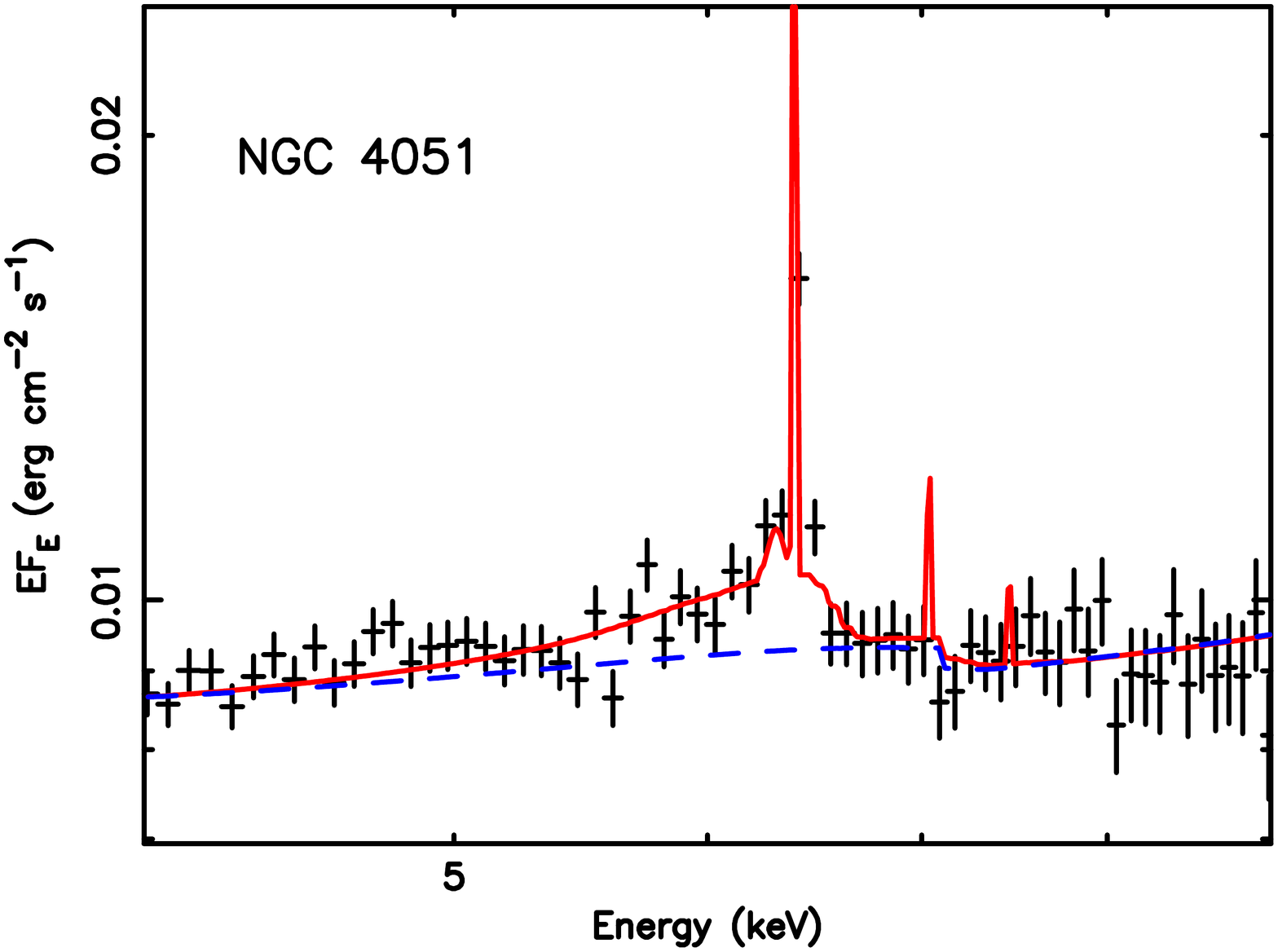}
\includegraphics[angle=270,width=52mm]{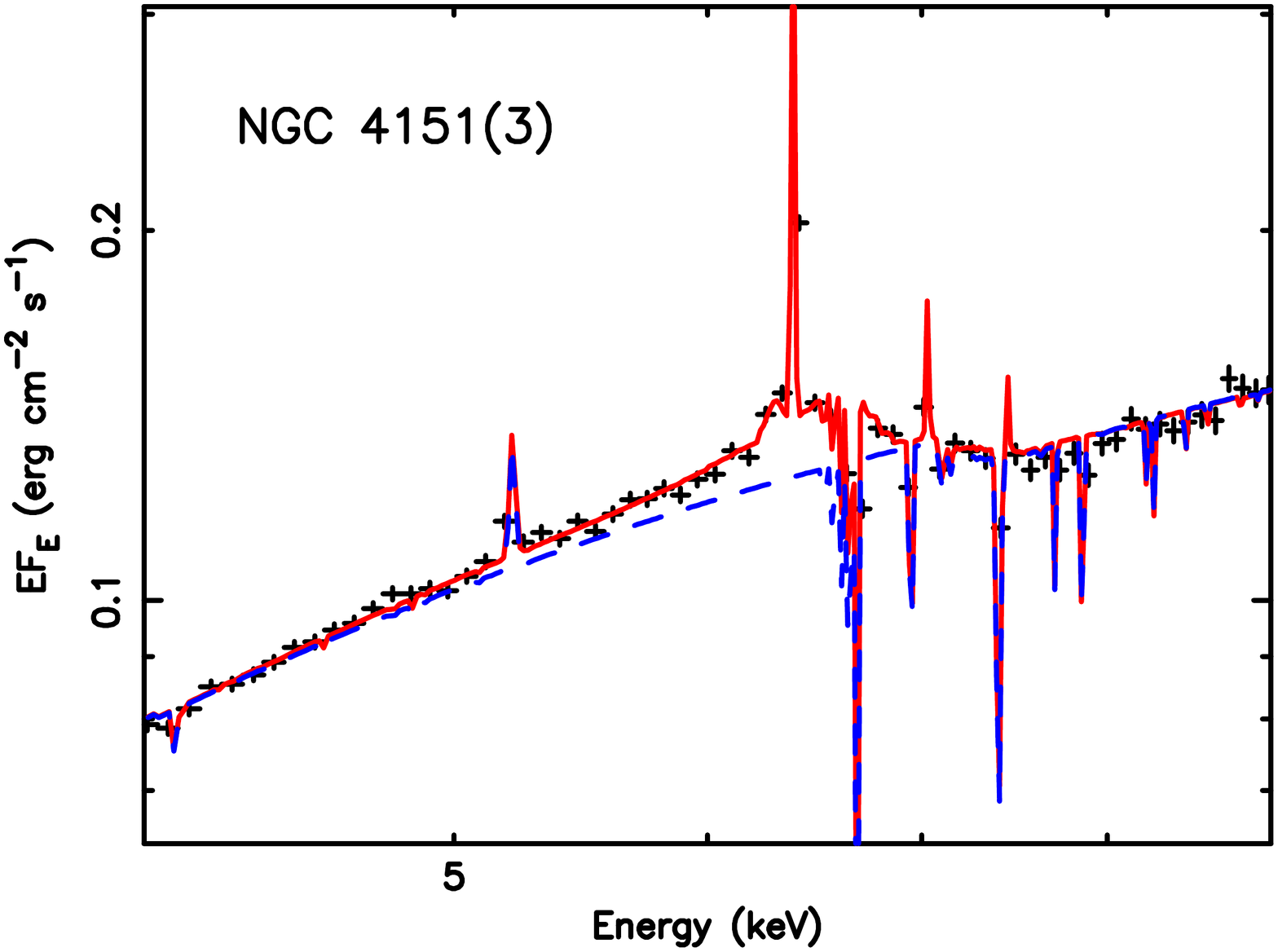}
\includegraphics[angle=270,width=52mm]{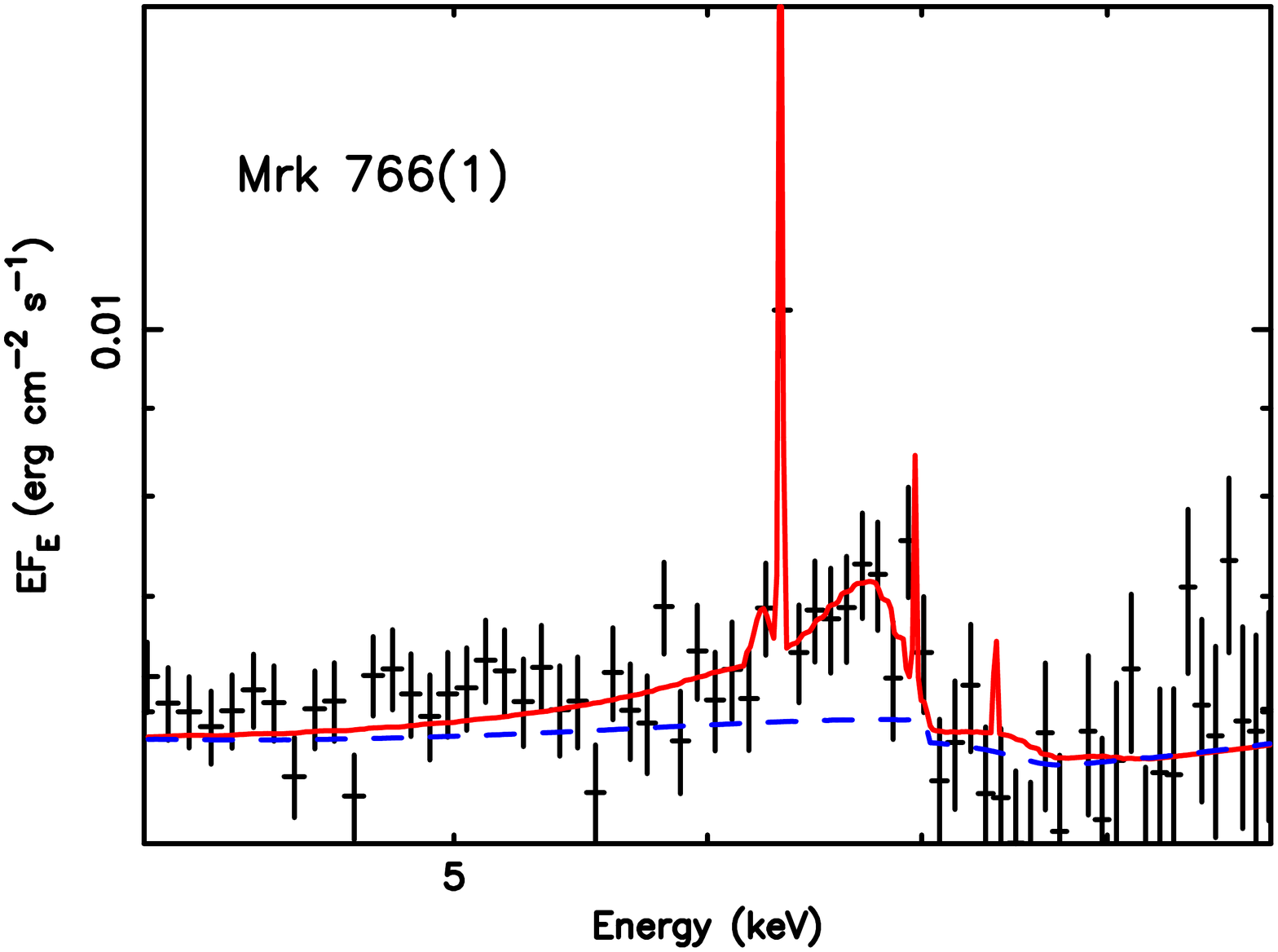}
\includegraphics[angle=270,width=52mm]{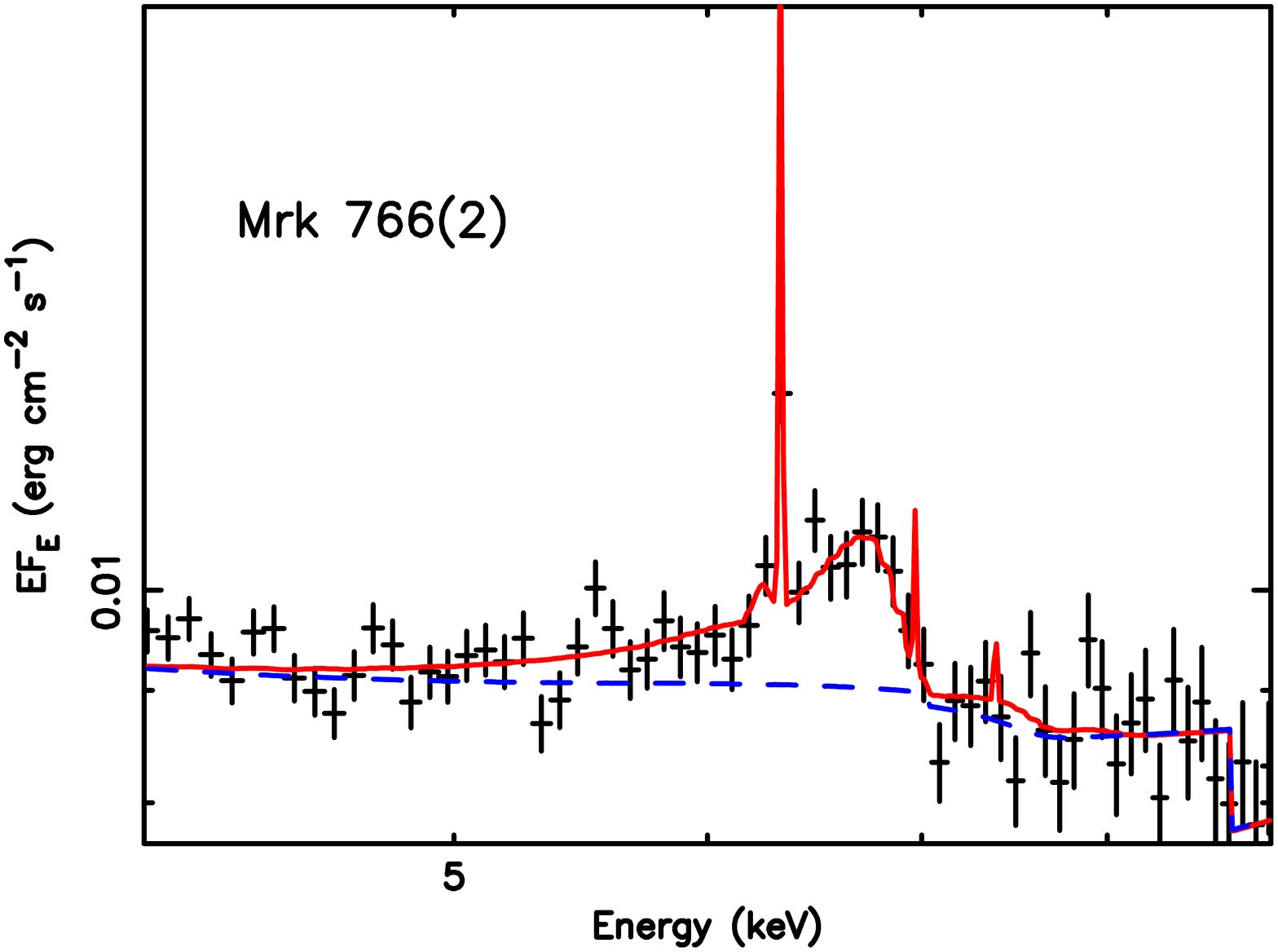}
\includegraphics[angle=270,width=52mm]{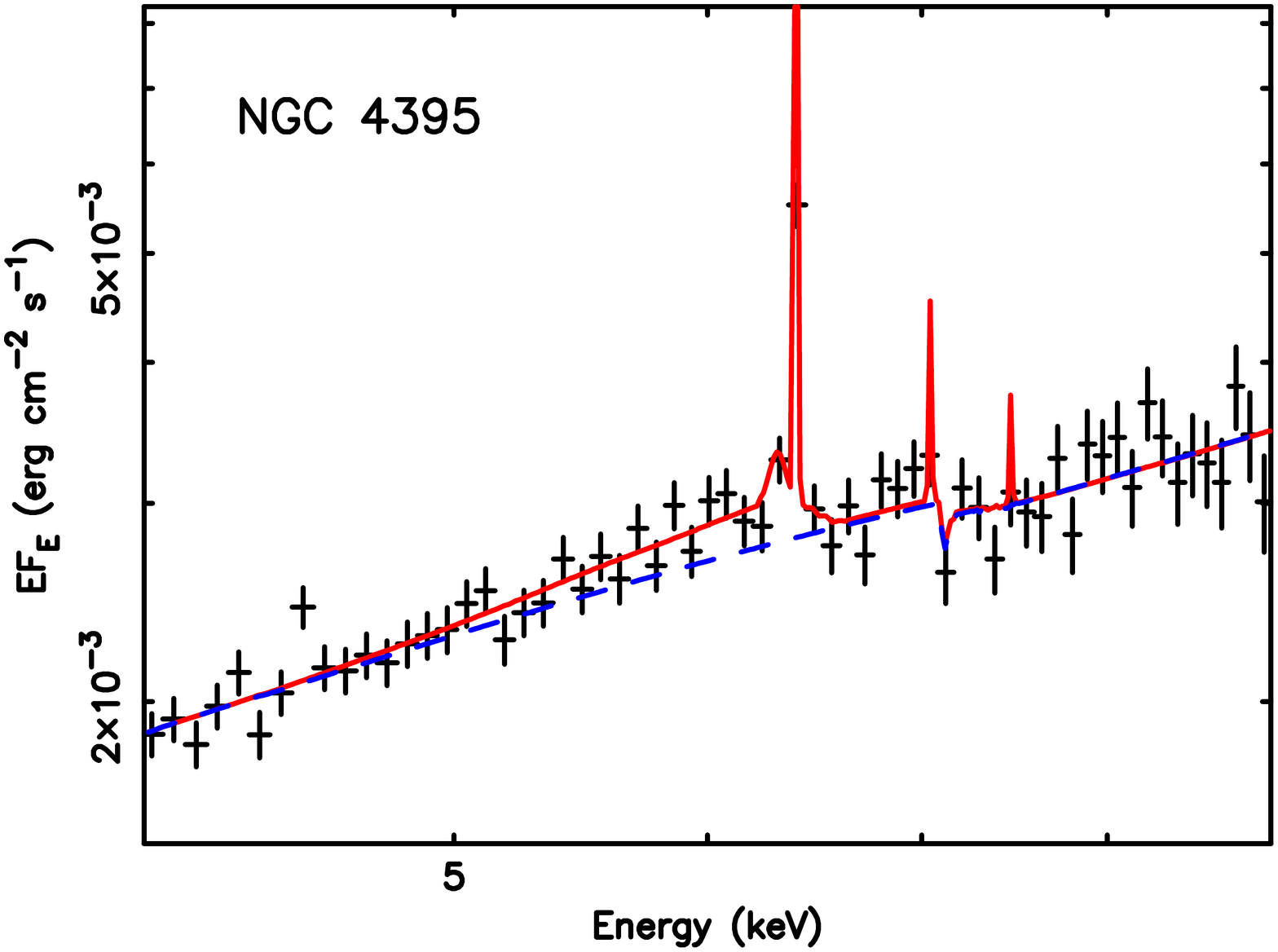}
\includegraphics[angle=270,width=52mm]{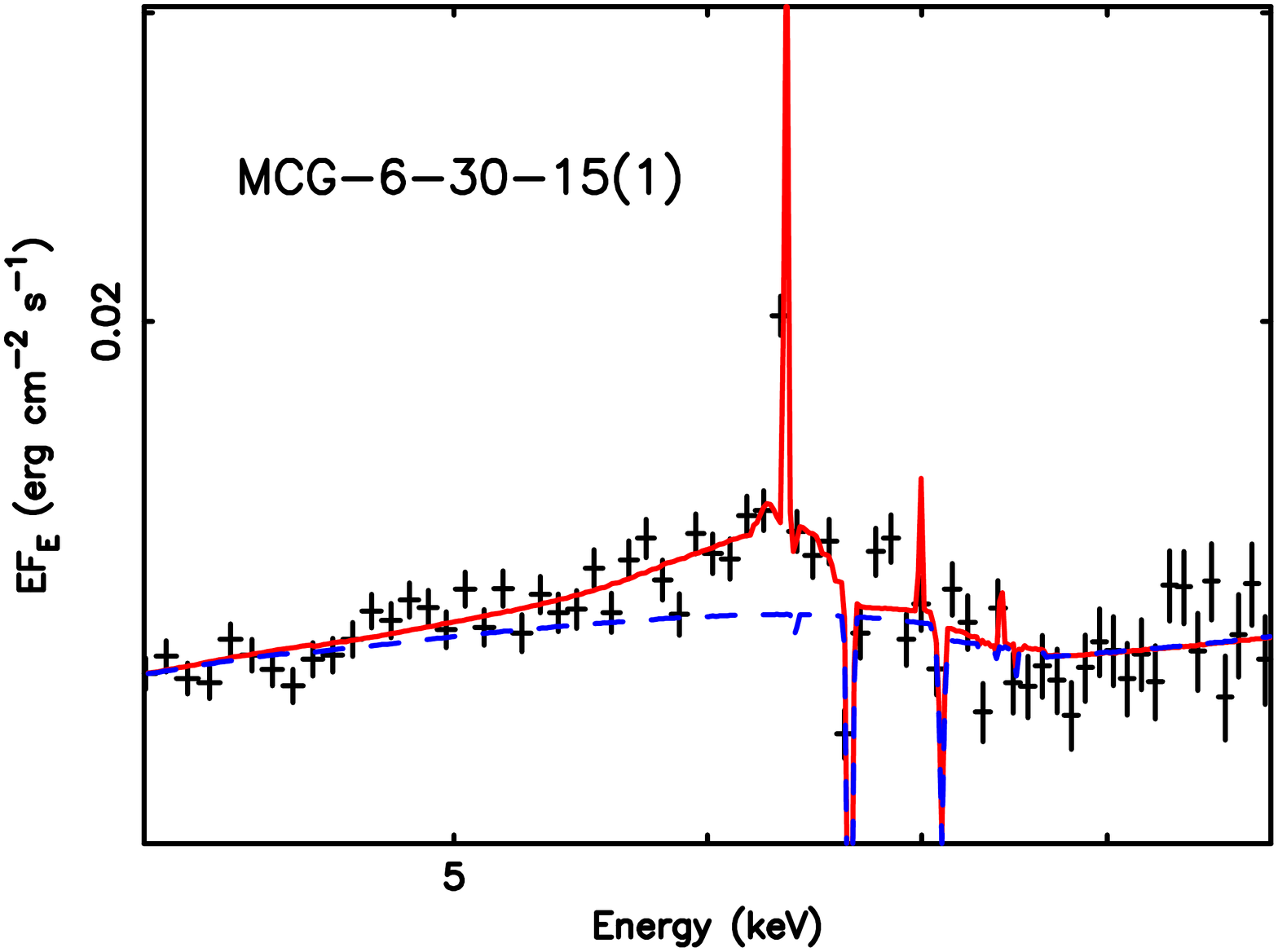}
\includegraphics[angle=270,width=52mm]{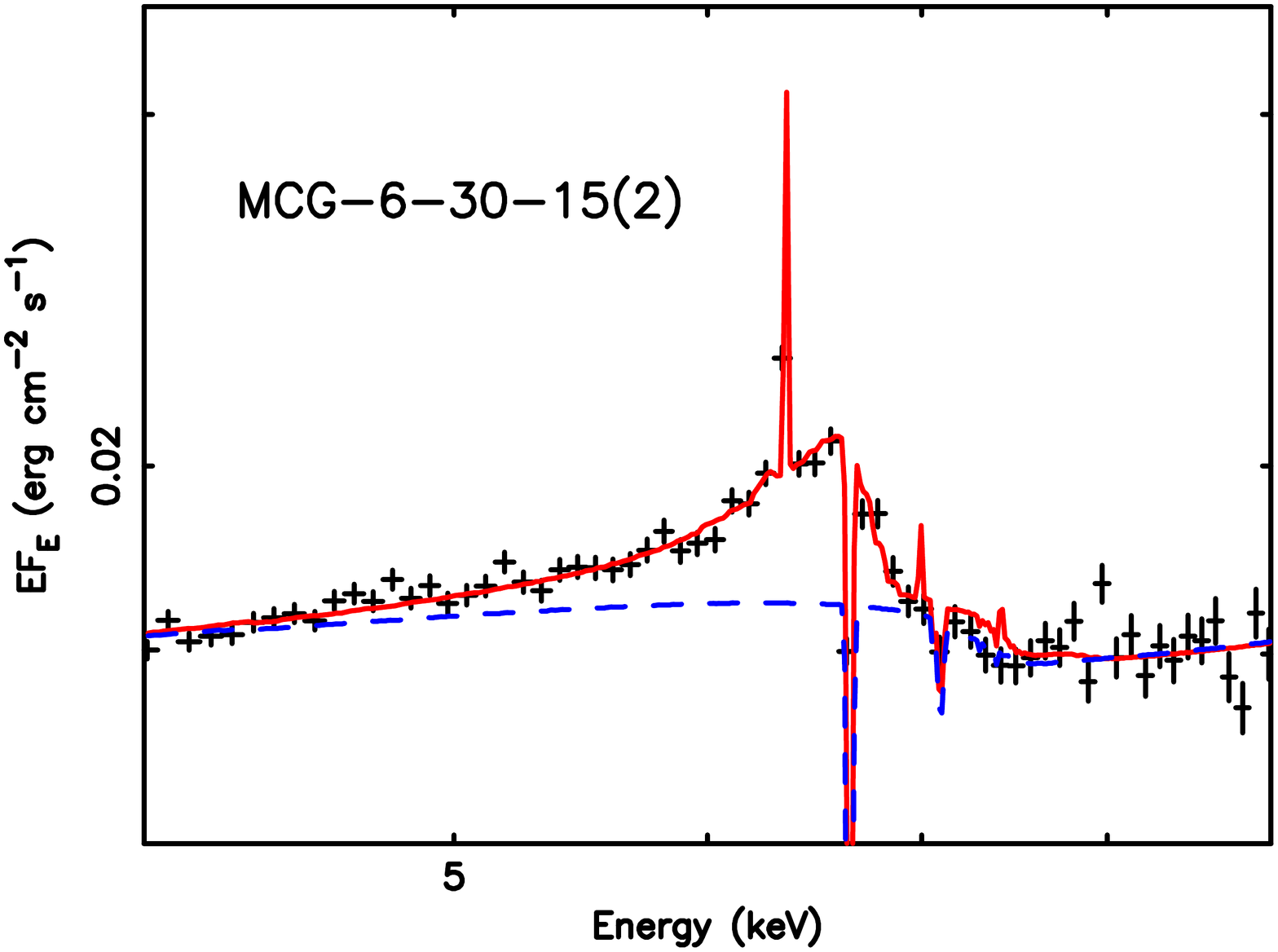}
\includegraphics[angle=270,width=52mm]{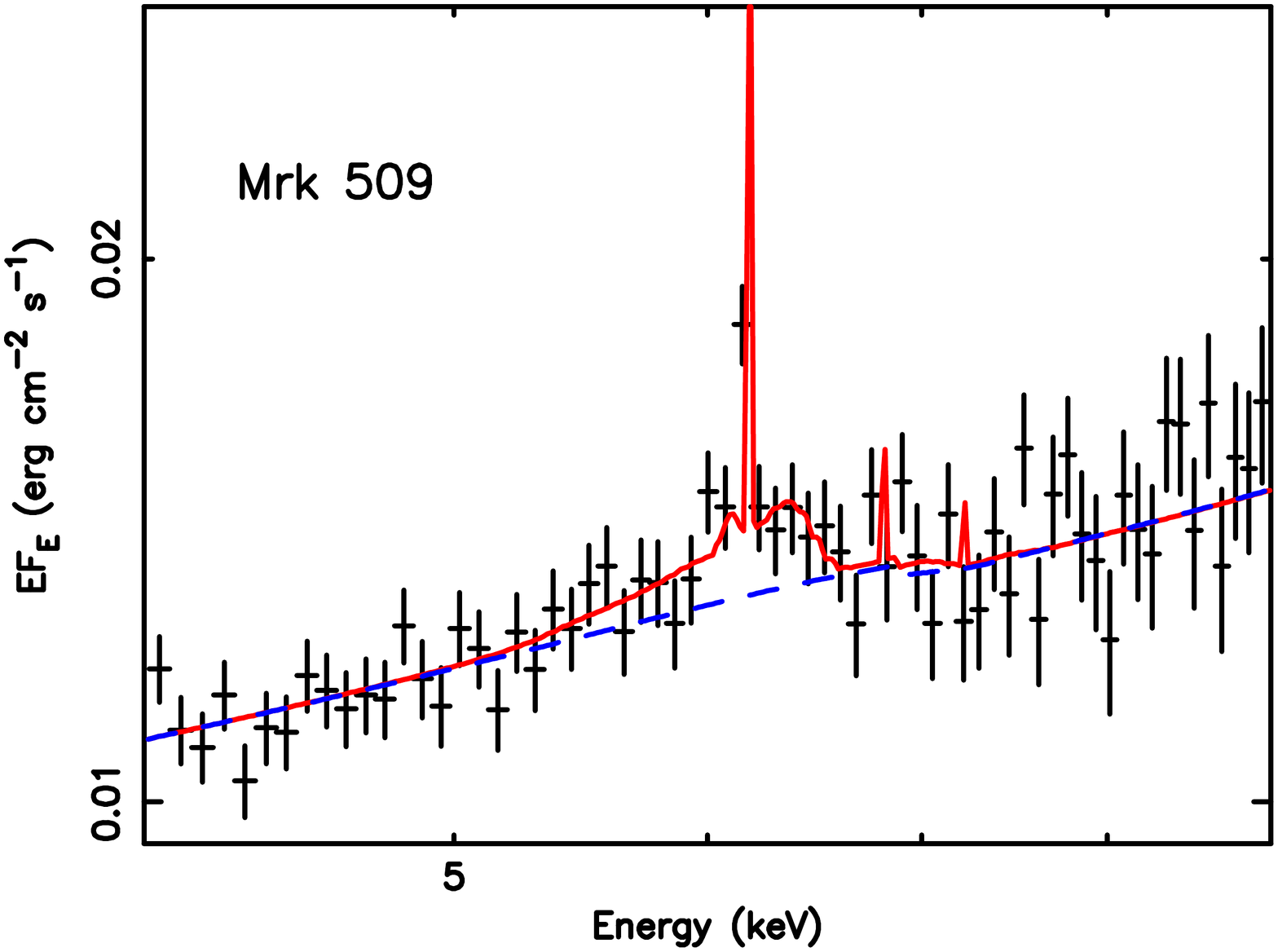}
\includegraphics[angle=270,width=52mm]{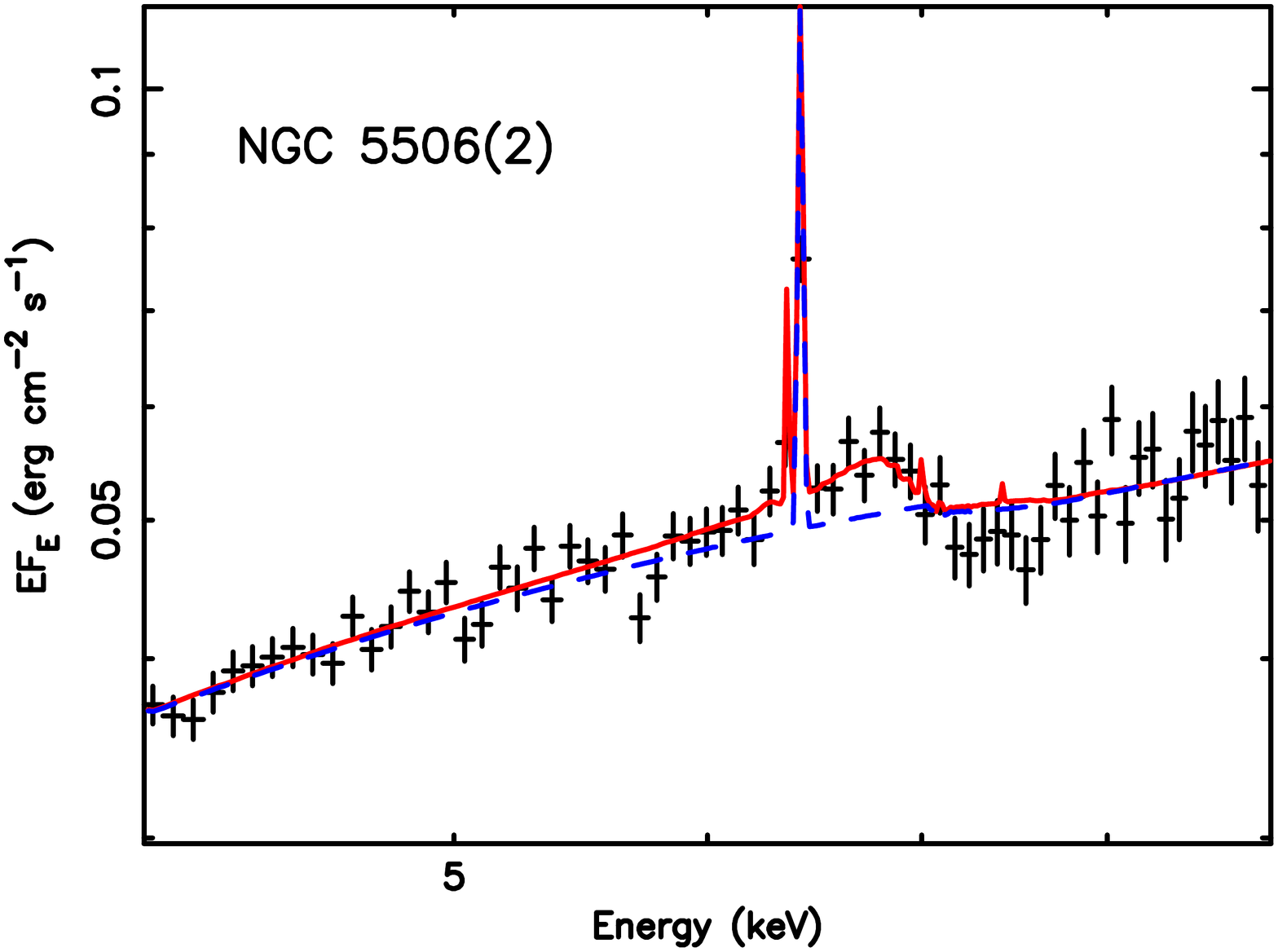}
\includegraphics[angle=270,width=52mm]{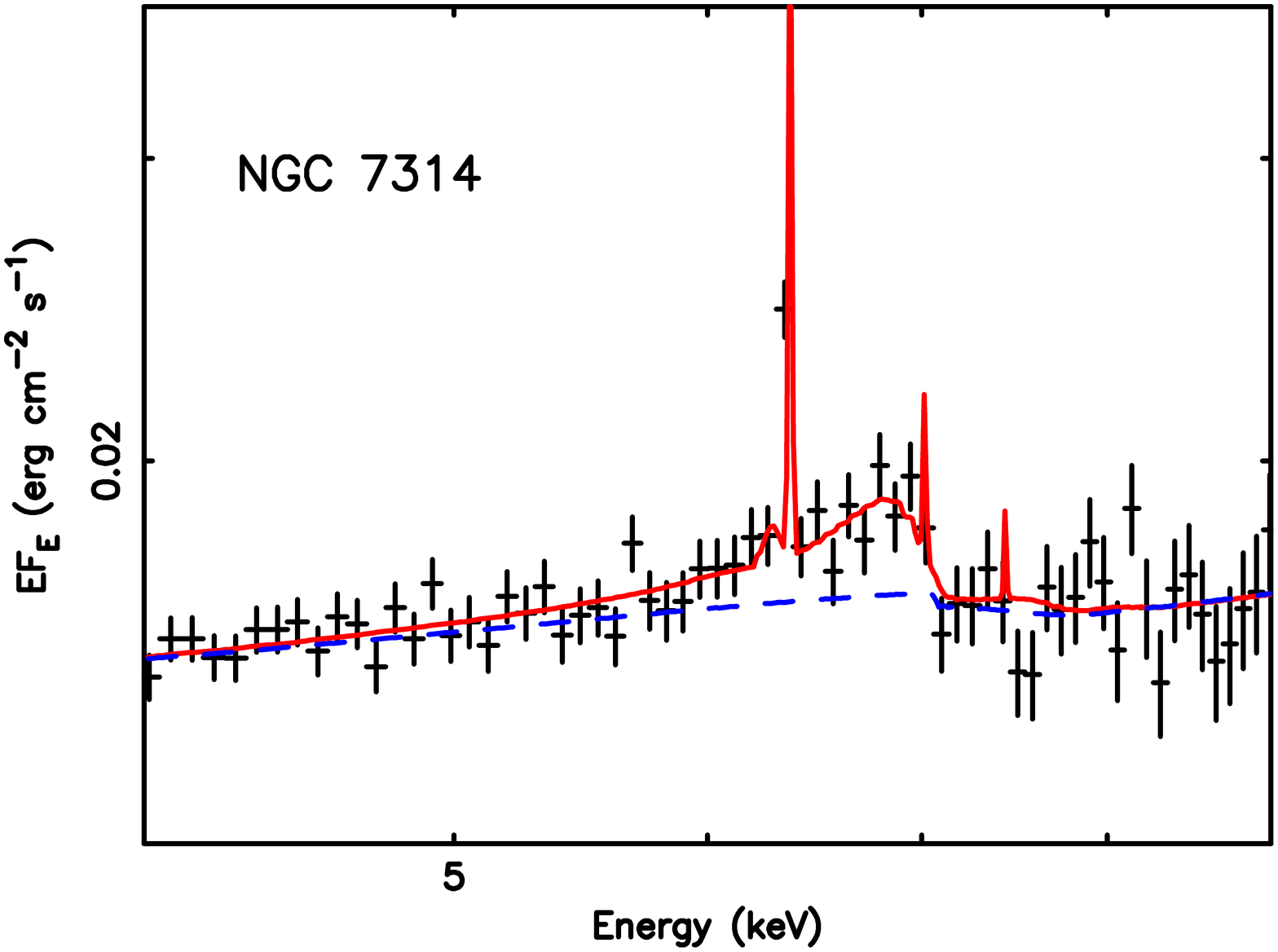}
\includegraphics[angle=270,width=52mm]{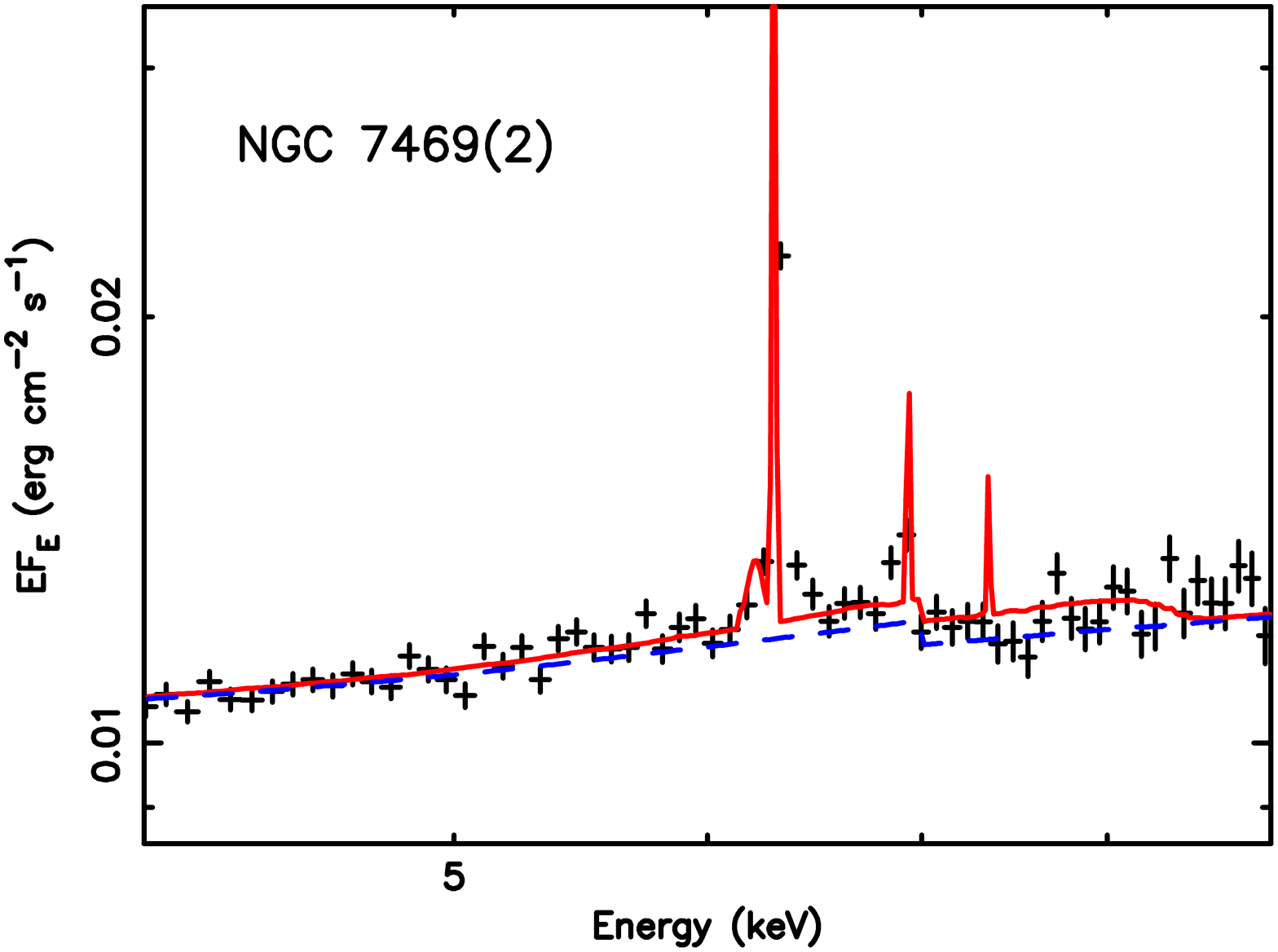}
}
\caption{Relativistic disk lines. The unfolded spectra centered around the iron band are shown for observations in which the additional of the blurred reflection component improves the fit at $>99$~per cent confidence and for which the best-fit characteristic emission radius is $<50~r_{\rm g}$.The solid line shows our best-fit model. The dashed line shows the model excluding the {\tt pexmon} line components (distant and blurred). 
\label{fig:dream}}
\end{figure*}

\begin{figure*}
{
\includegraphics[angle=270,width=52mm]{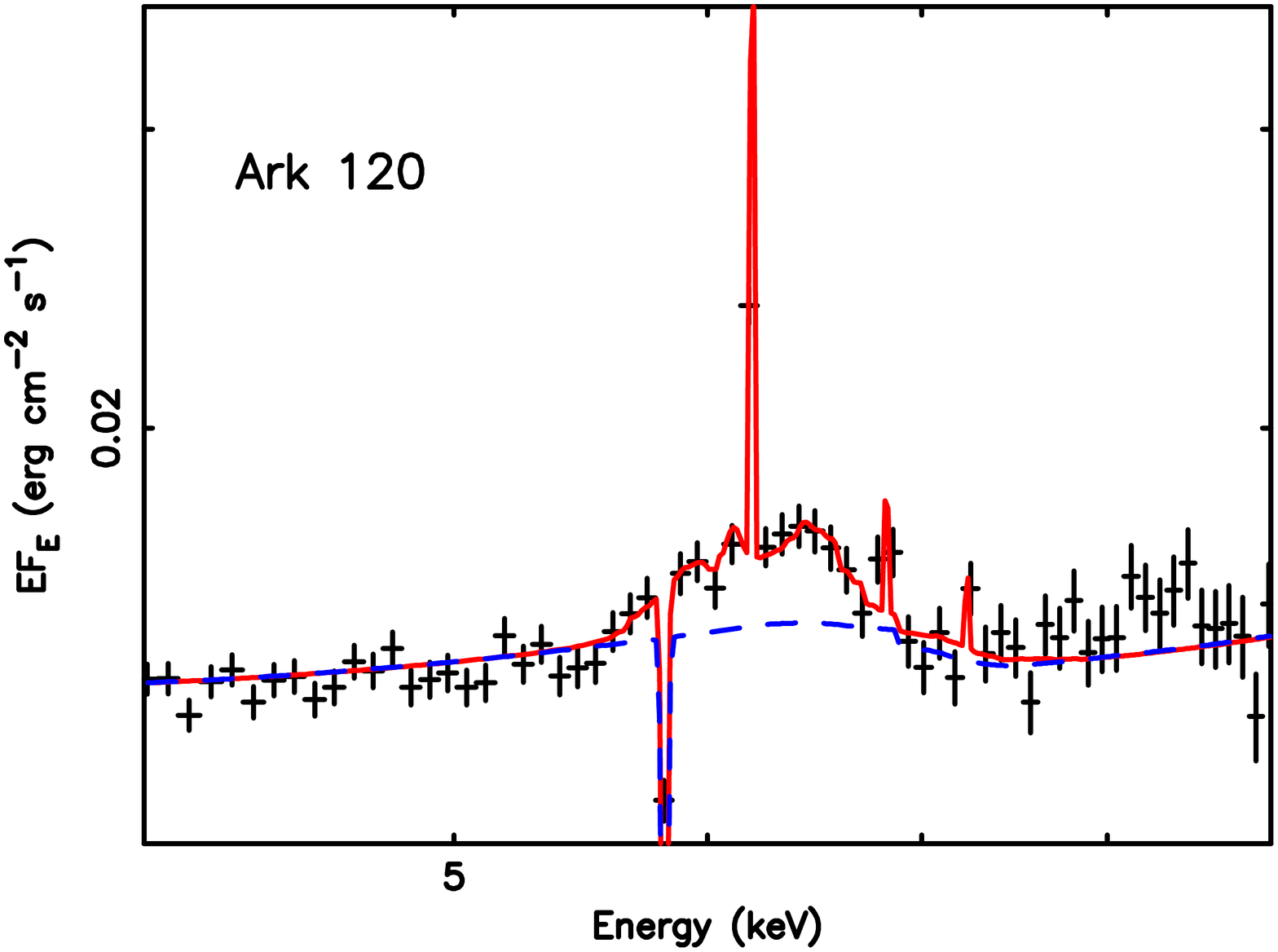}
\includegraphics[angle=270,width=52mm]{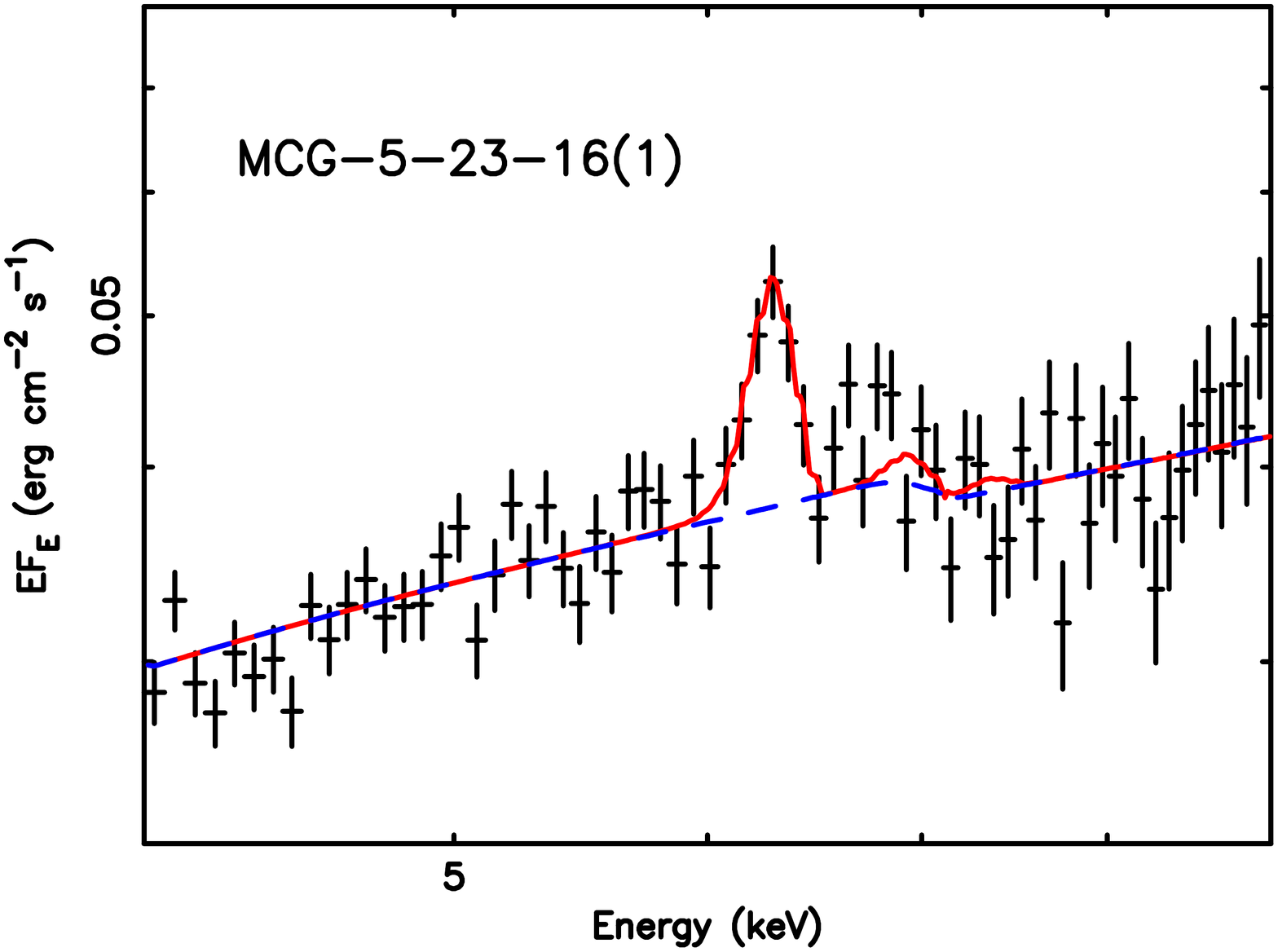}
\includegraphics[angle=270,width=52mm]{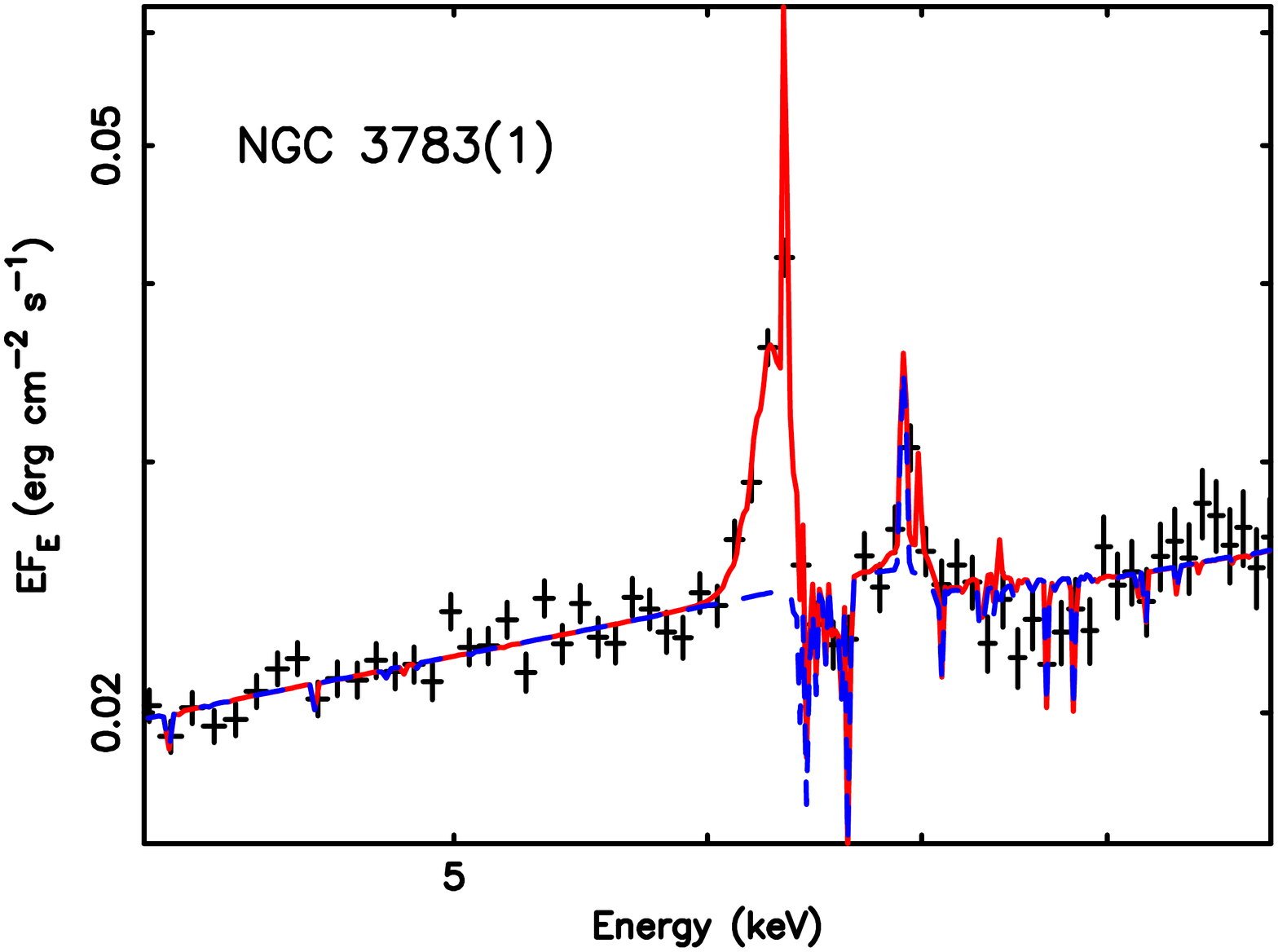}
\includegraphics[angle=270,width=52mm]{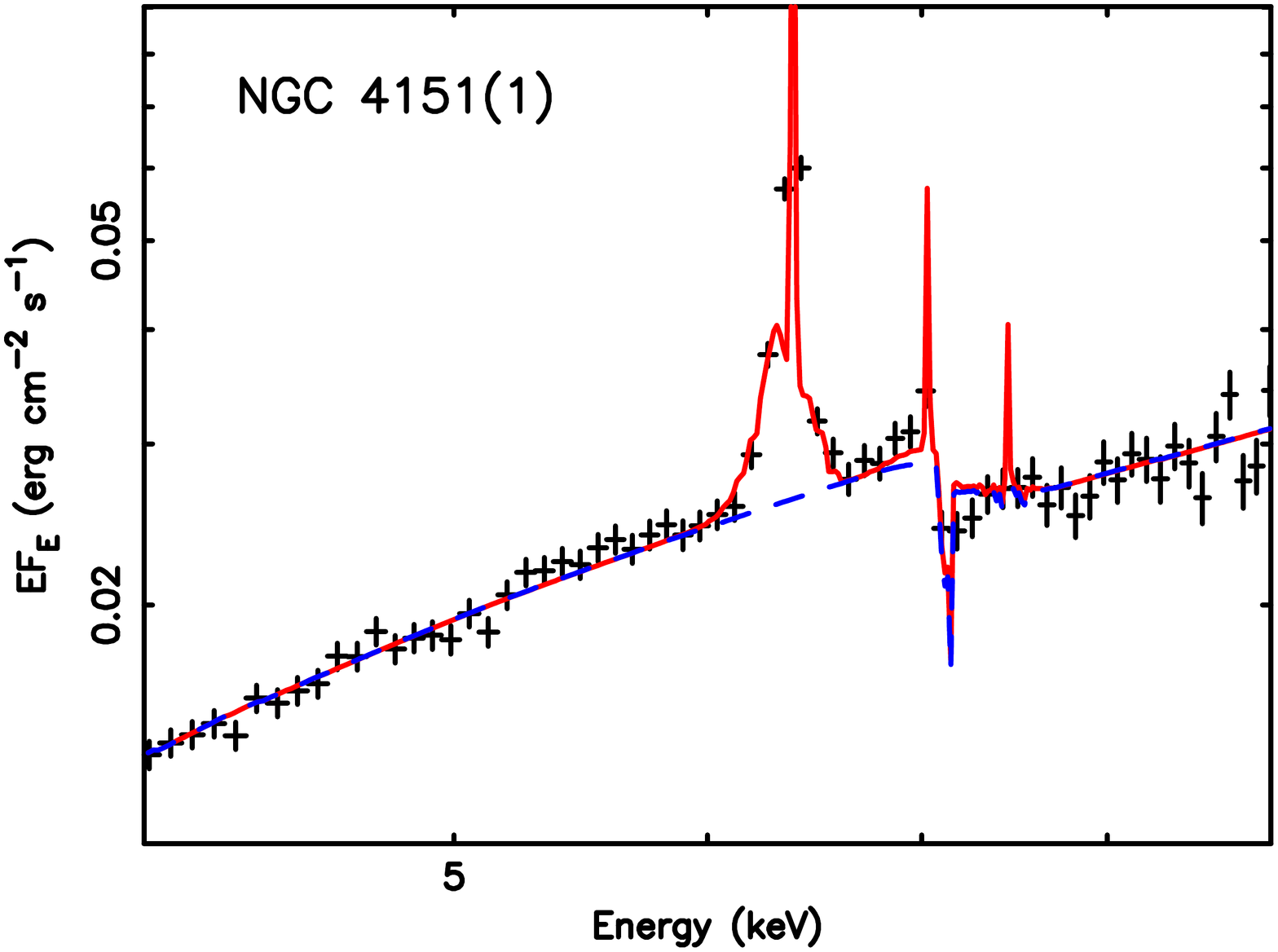}
\includegraphics[angle=270,width=52mm]{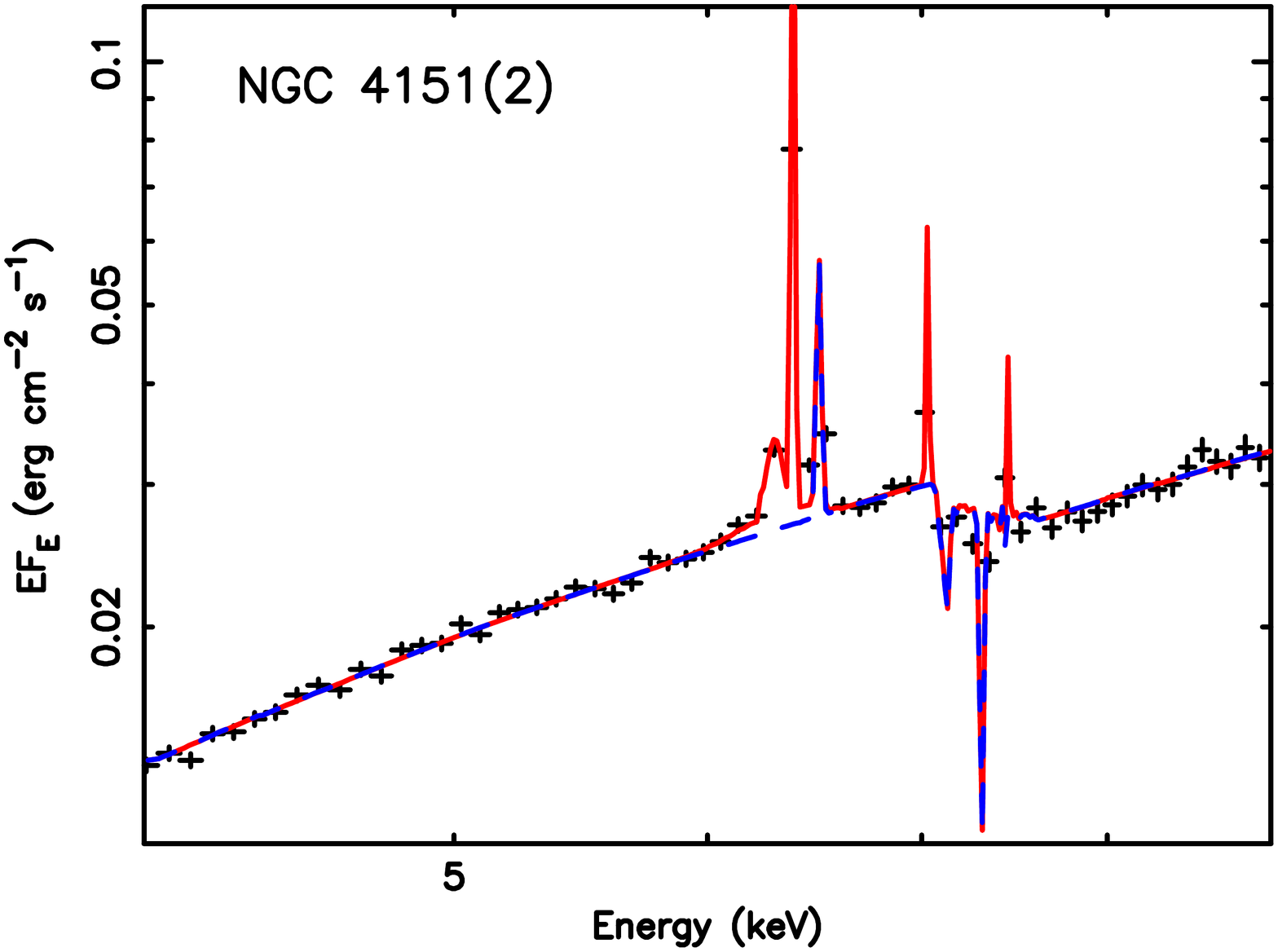}
\includegraphics[angle=270,width=52mm]{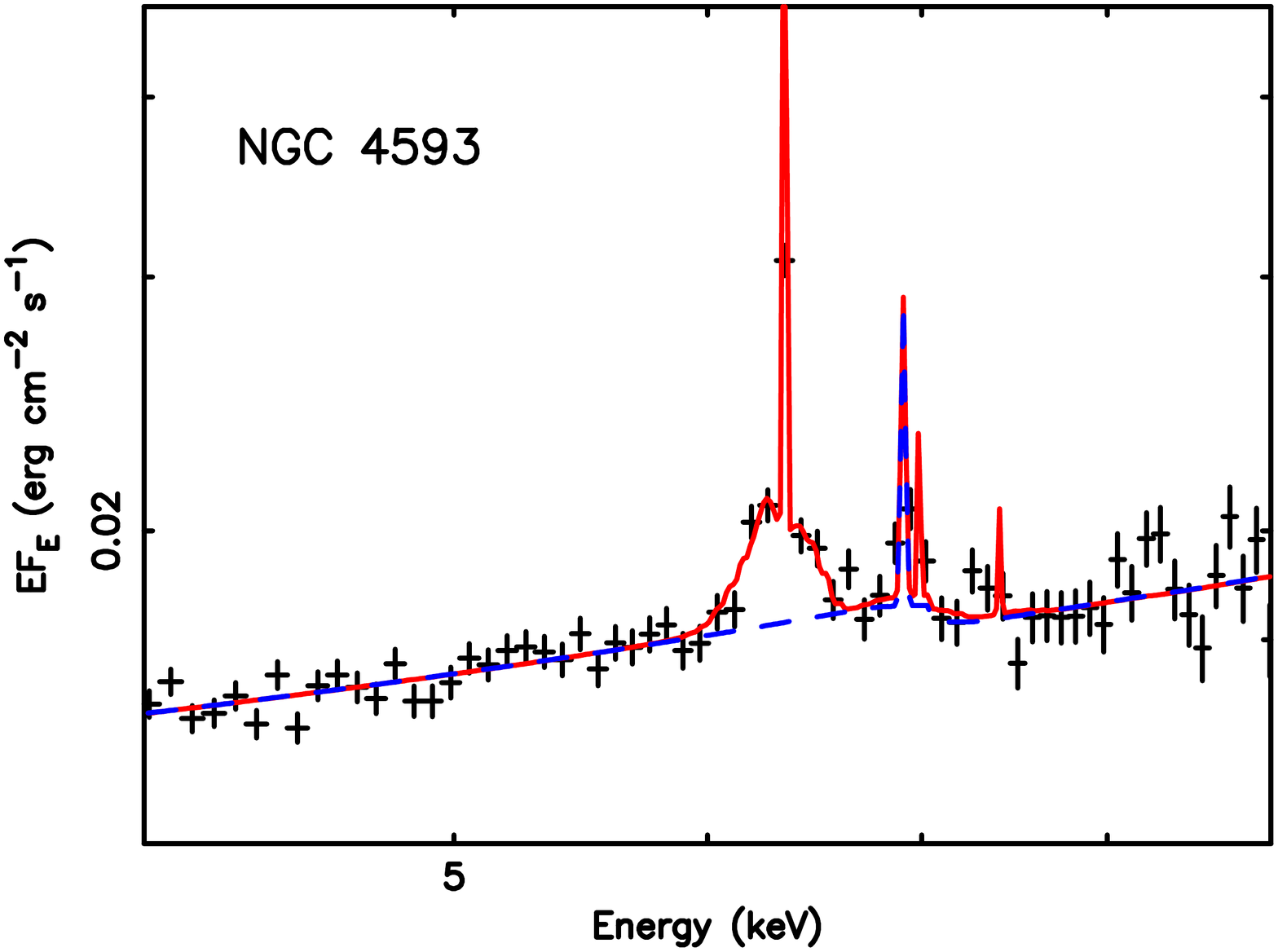}
\includegraphics[angle=270,width=52mm]{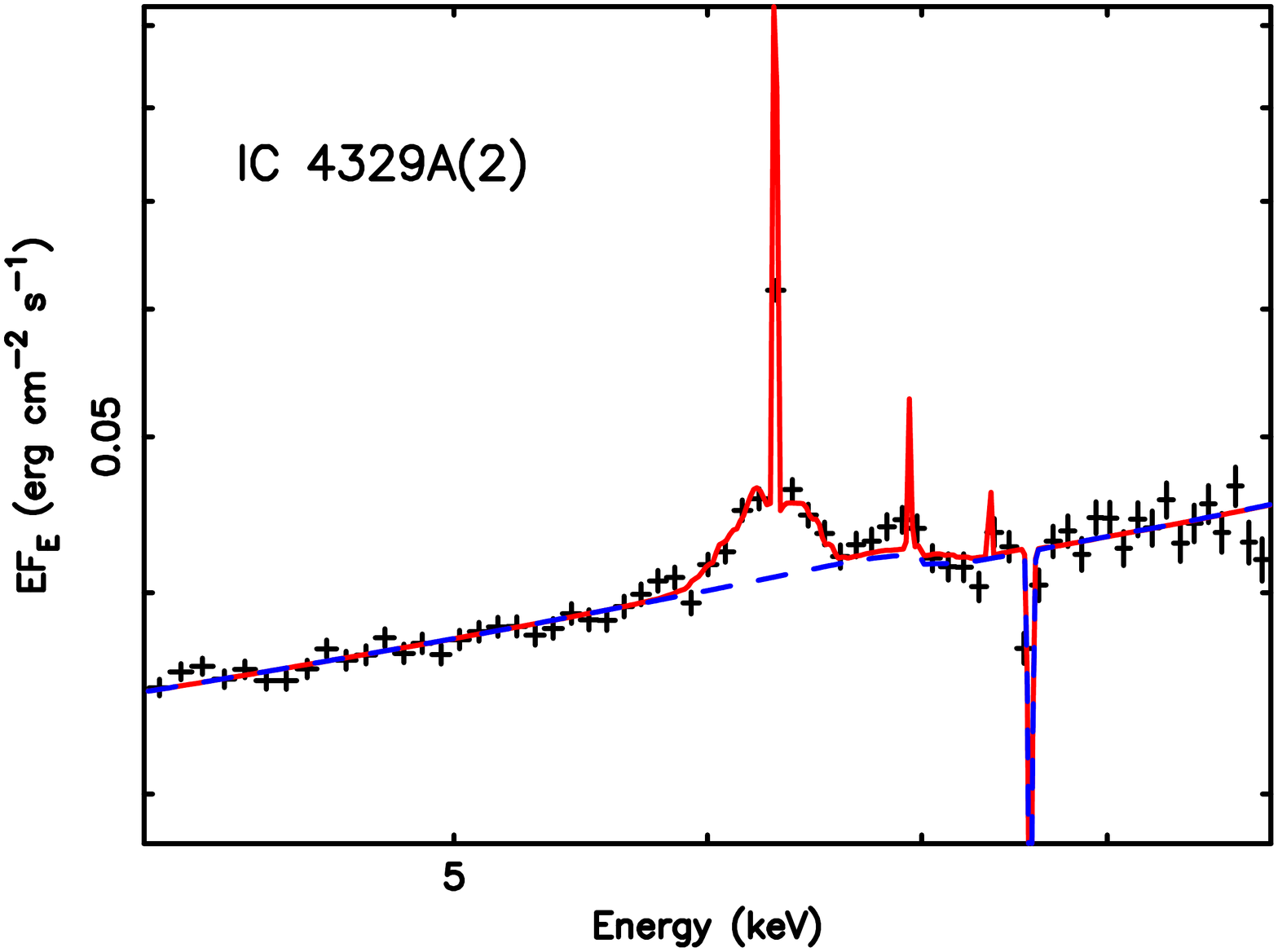}
\includegraphics[angle=270,width=52mm]{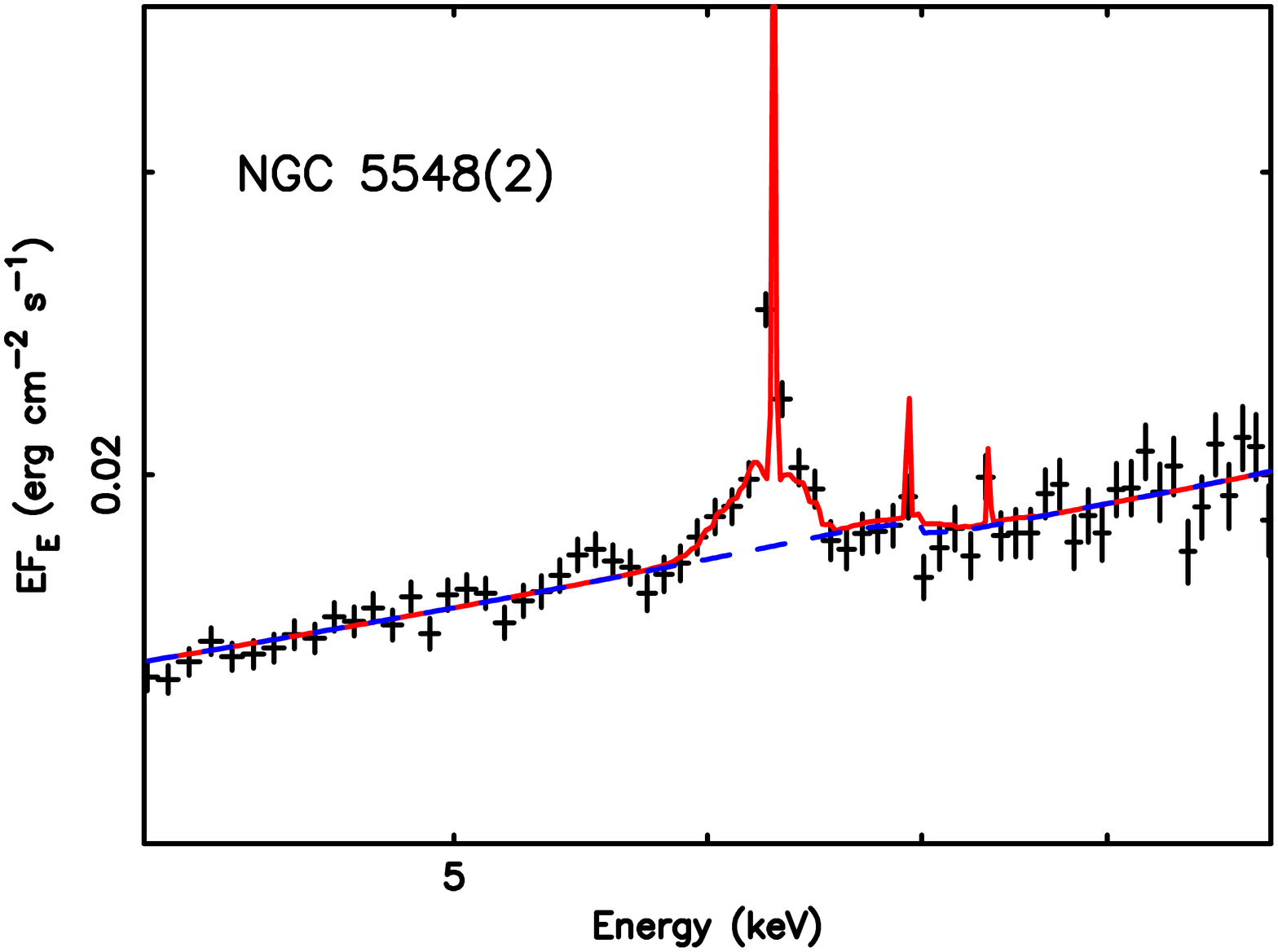}
\includegraphics[angle=270,width=52mm]{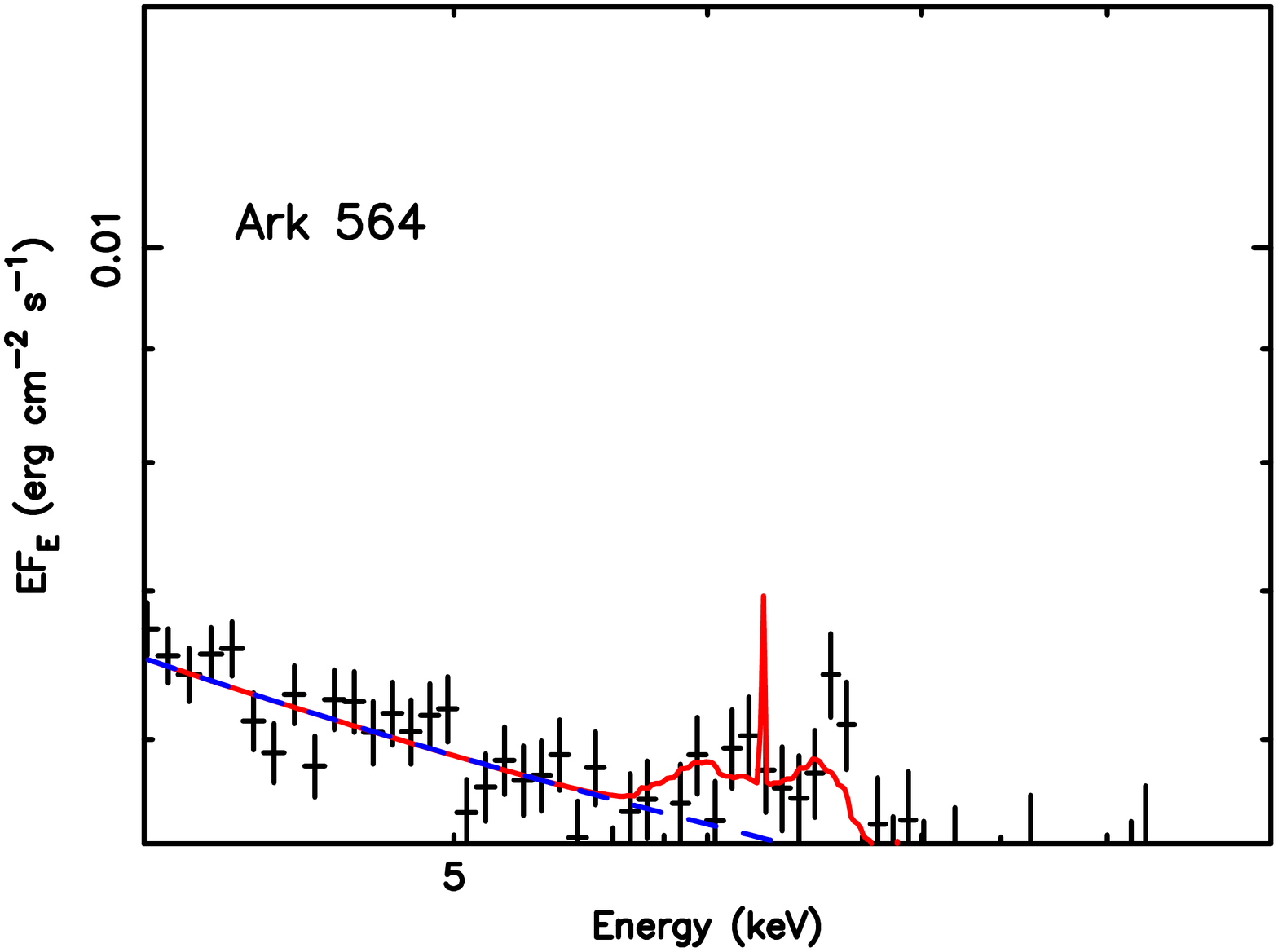}
}
\caption{Broad, but non-relativistic lines. Unfolded spectra and models for observations in which a velosty-broadened neutral refelection component improved the fits significantly, but that had a characteristic emission radius outside $50 r_{\rm g}$. Model lines are as in Fig.~\ref{fig:dream}. 
 \label{fig:yawn}}
\end{figure*}

\begin{figure*}
{
\includegraphics[angle=270,width=52mm]{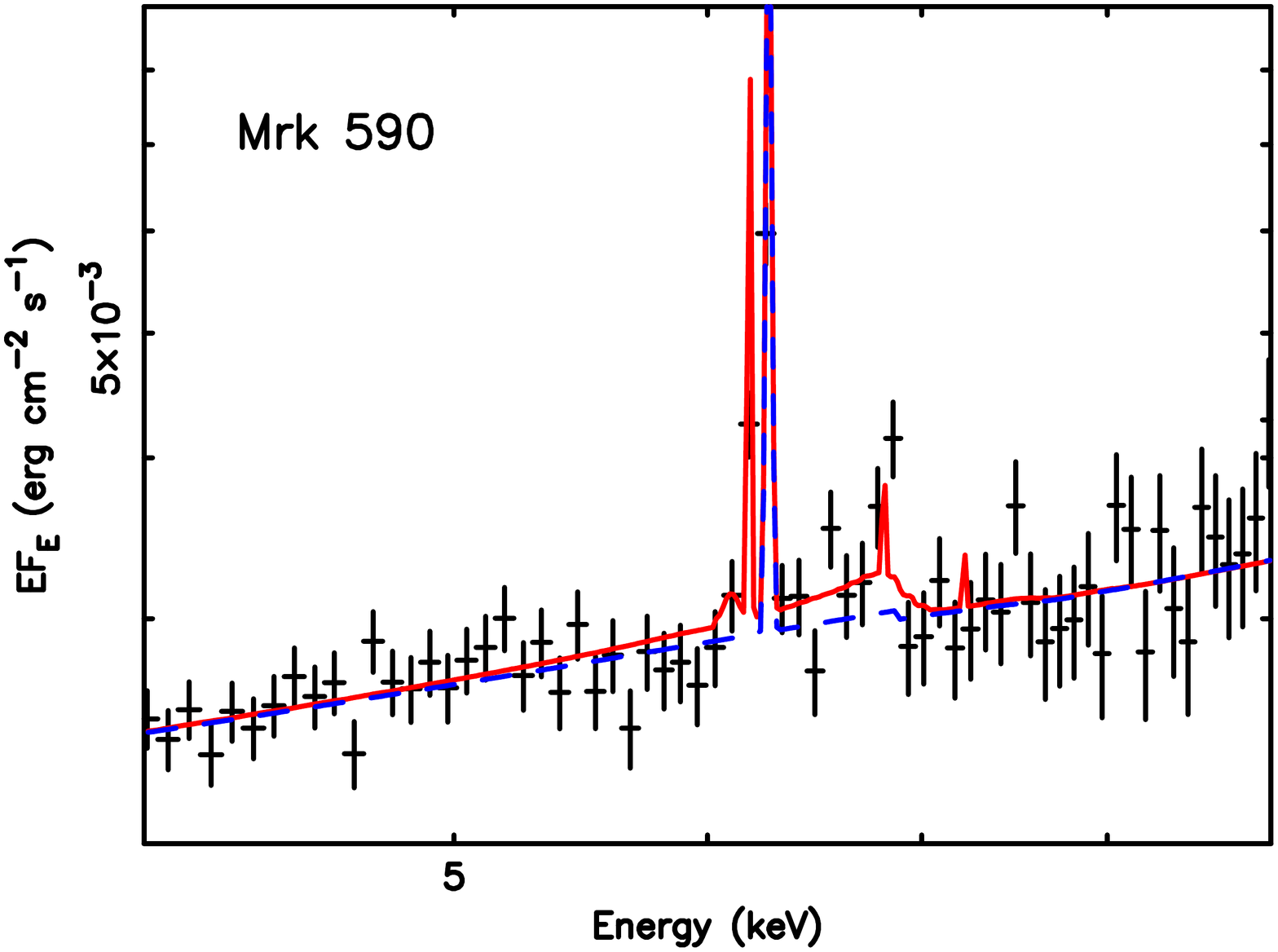}
\includegraphics[angle=270,width=52mm]{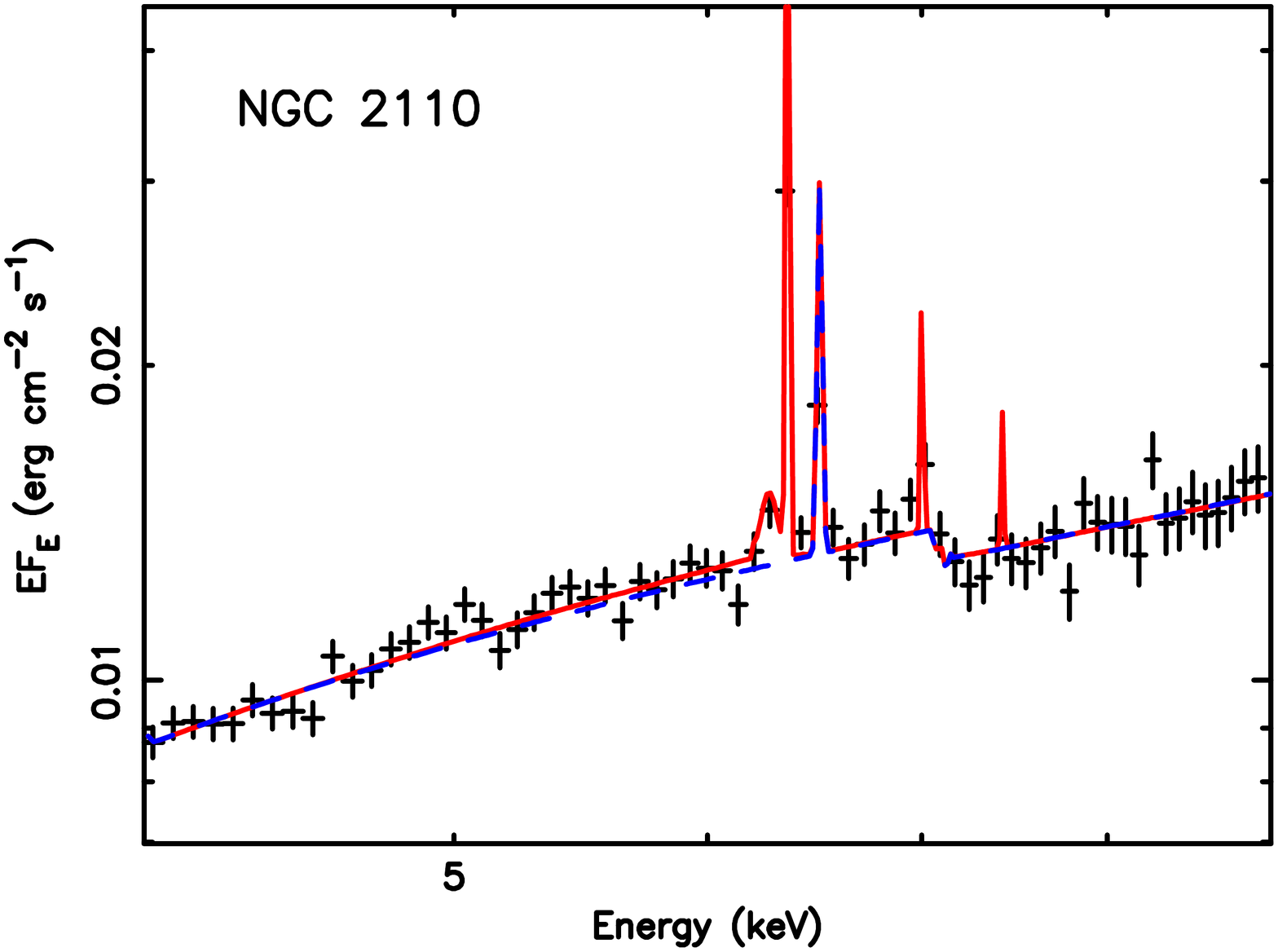}
\includegraphics[angle=270,width=52mm]{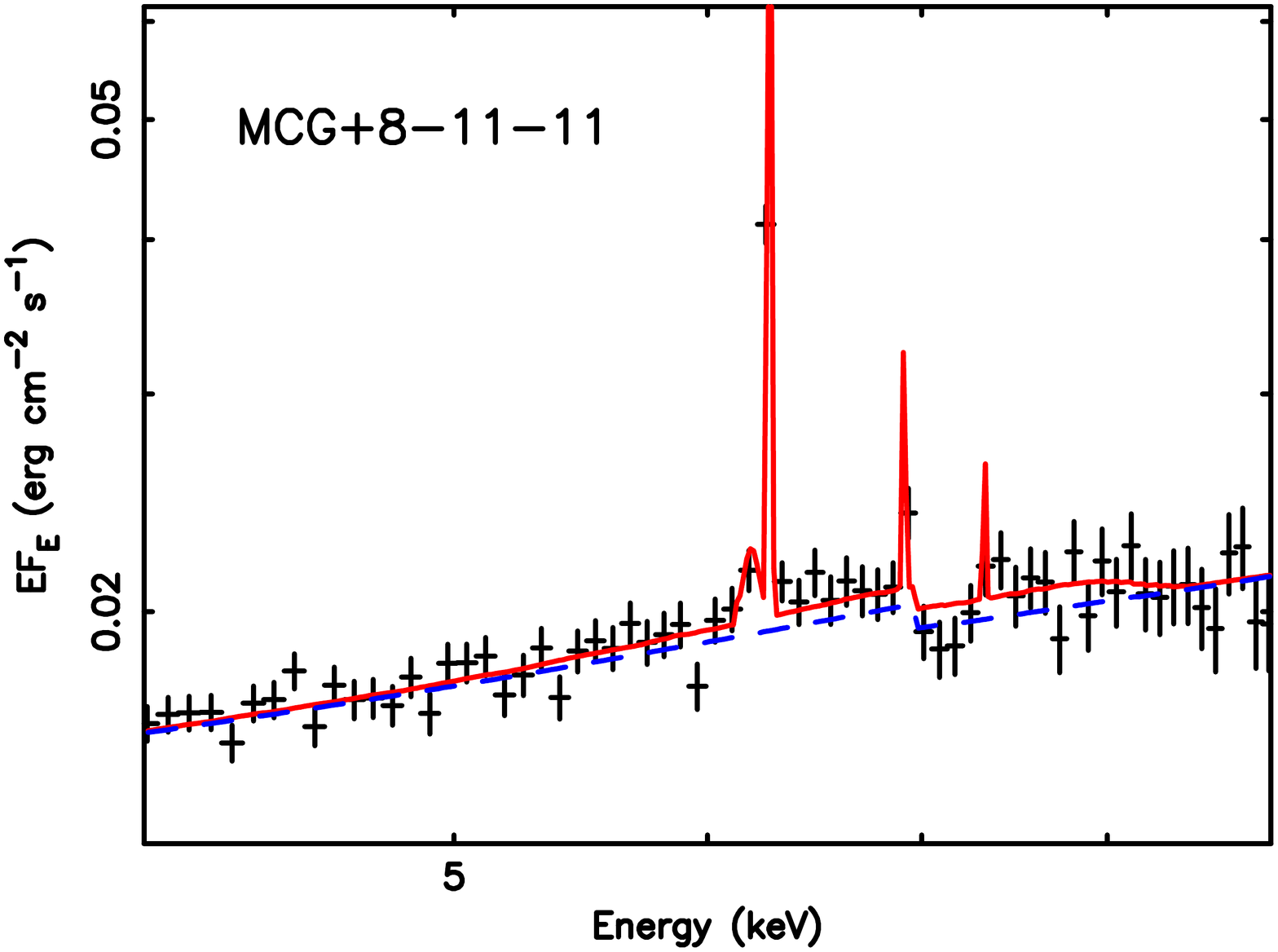}
\includegraphics[angle=270,width=52mm]{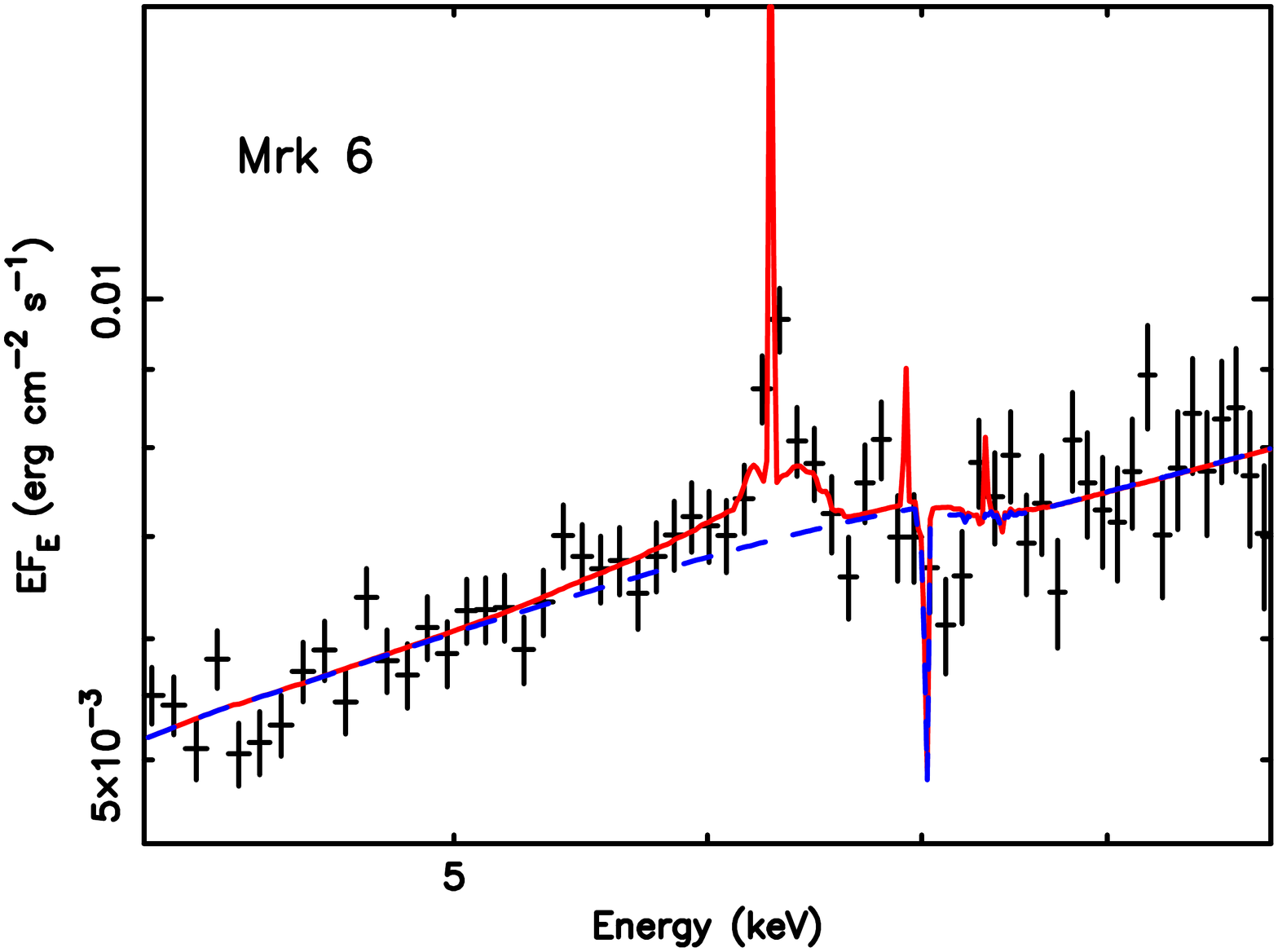}
\includegraphics[angle=270,width=52mm]{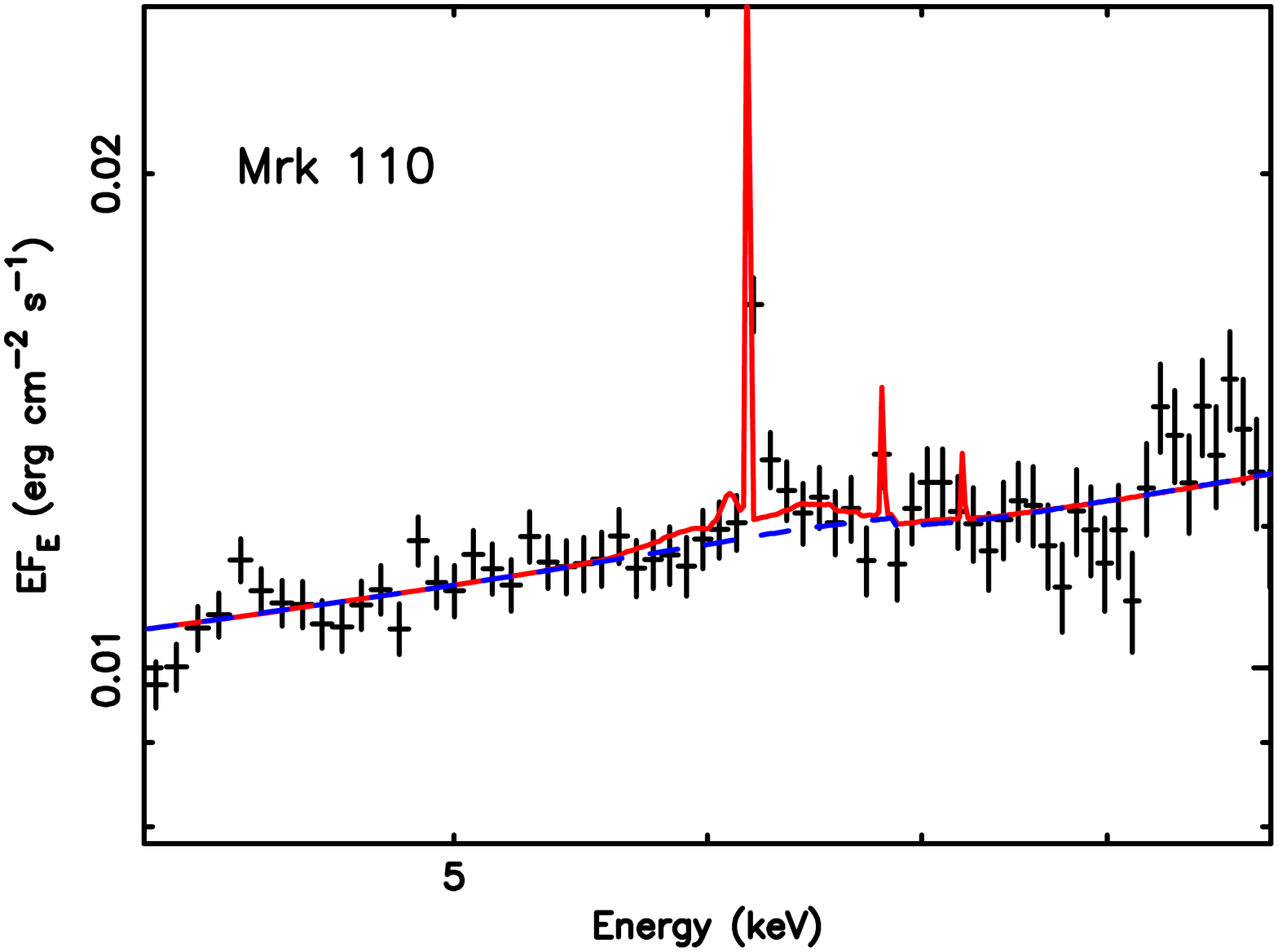}
\includegraphics[angle=270,width=52mm]{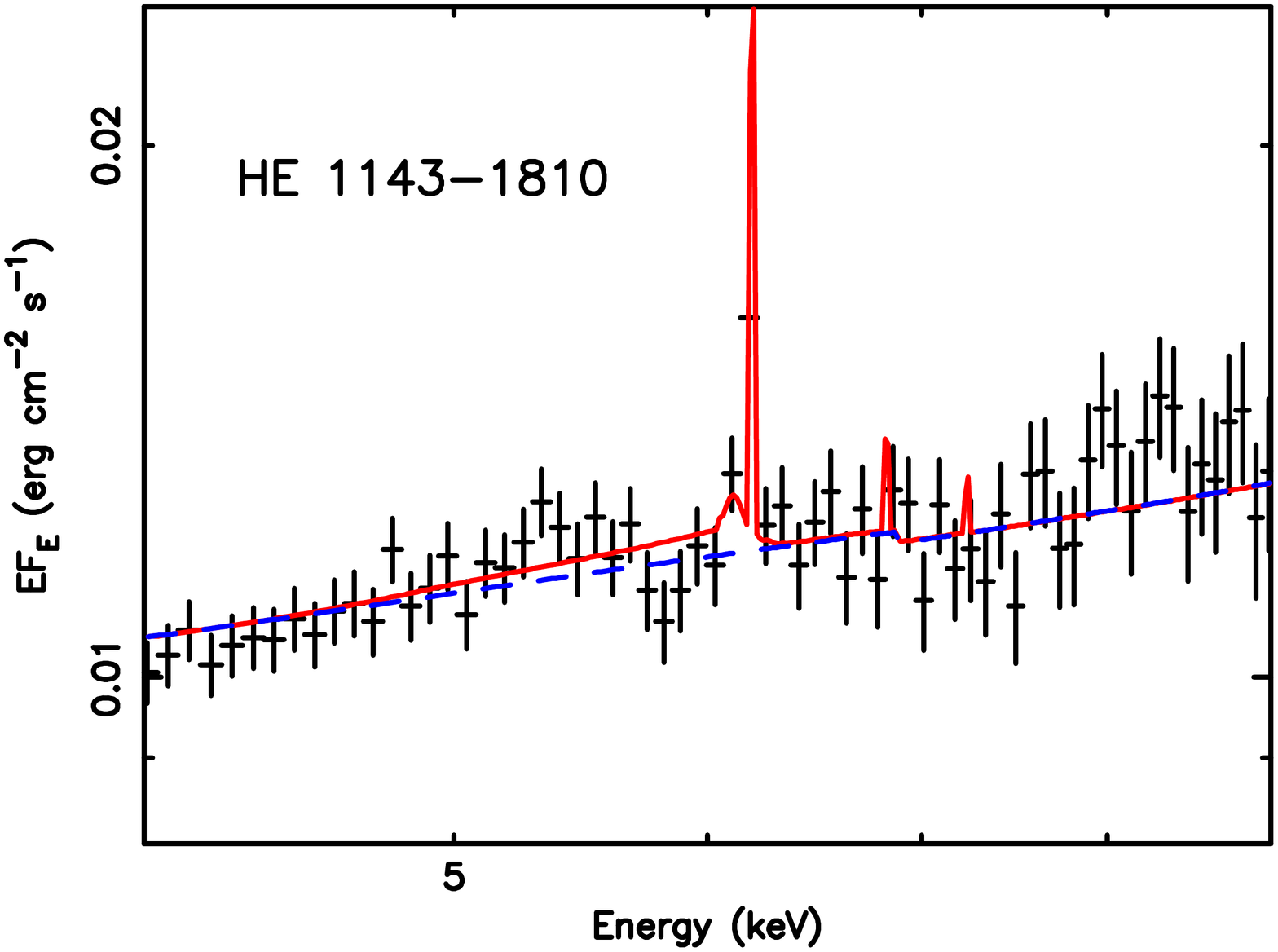}
\includegraphics[angle=270,width=52mm]{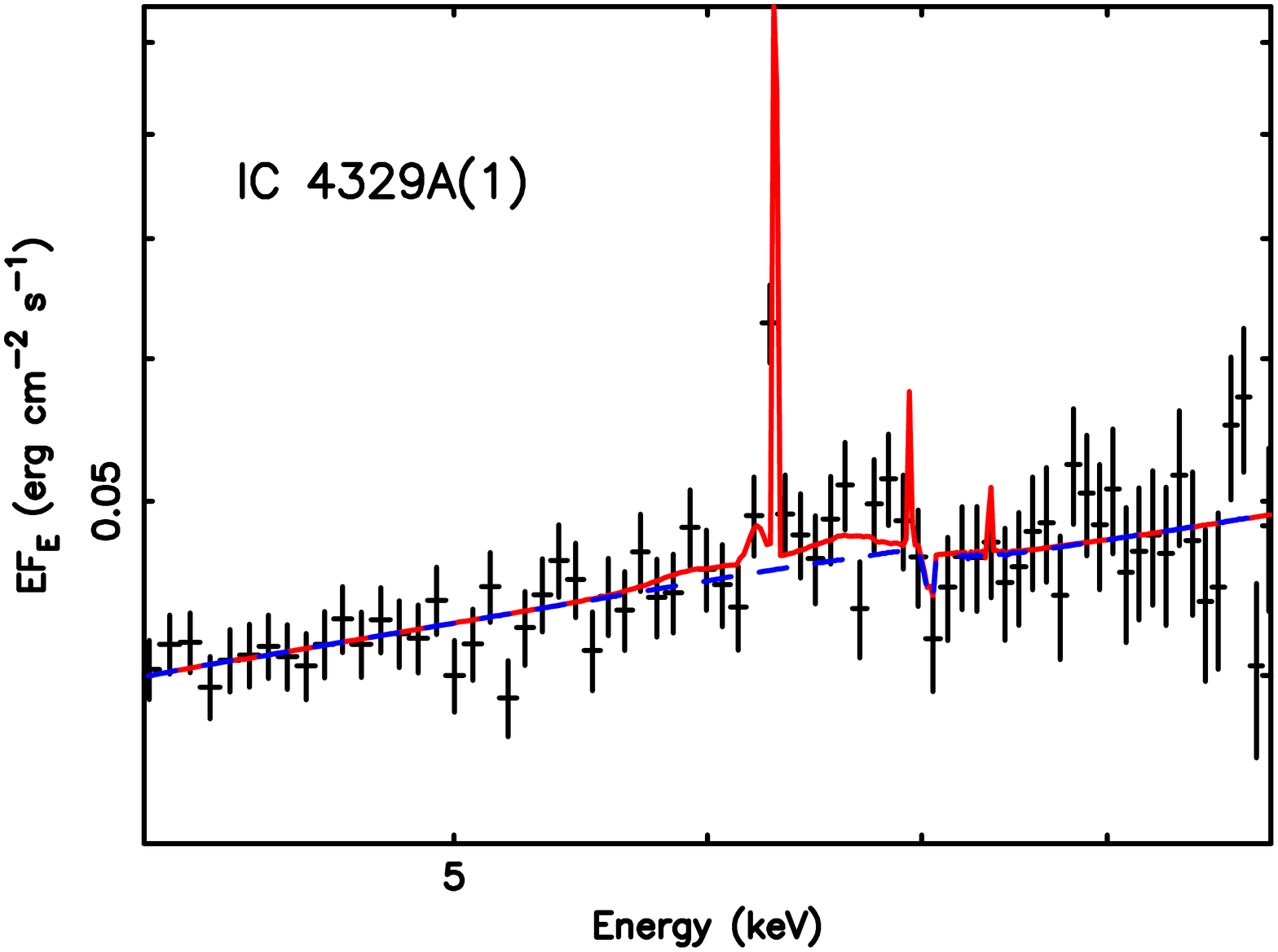}
\includegraphics[angle=270,width=52mm]{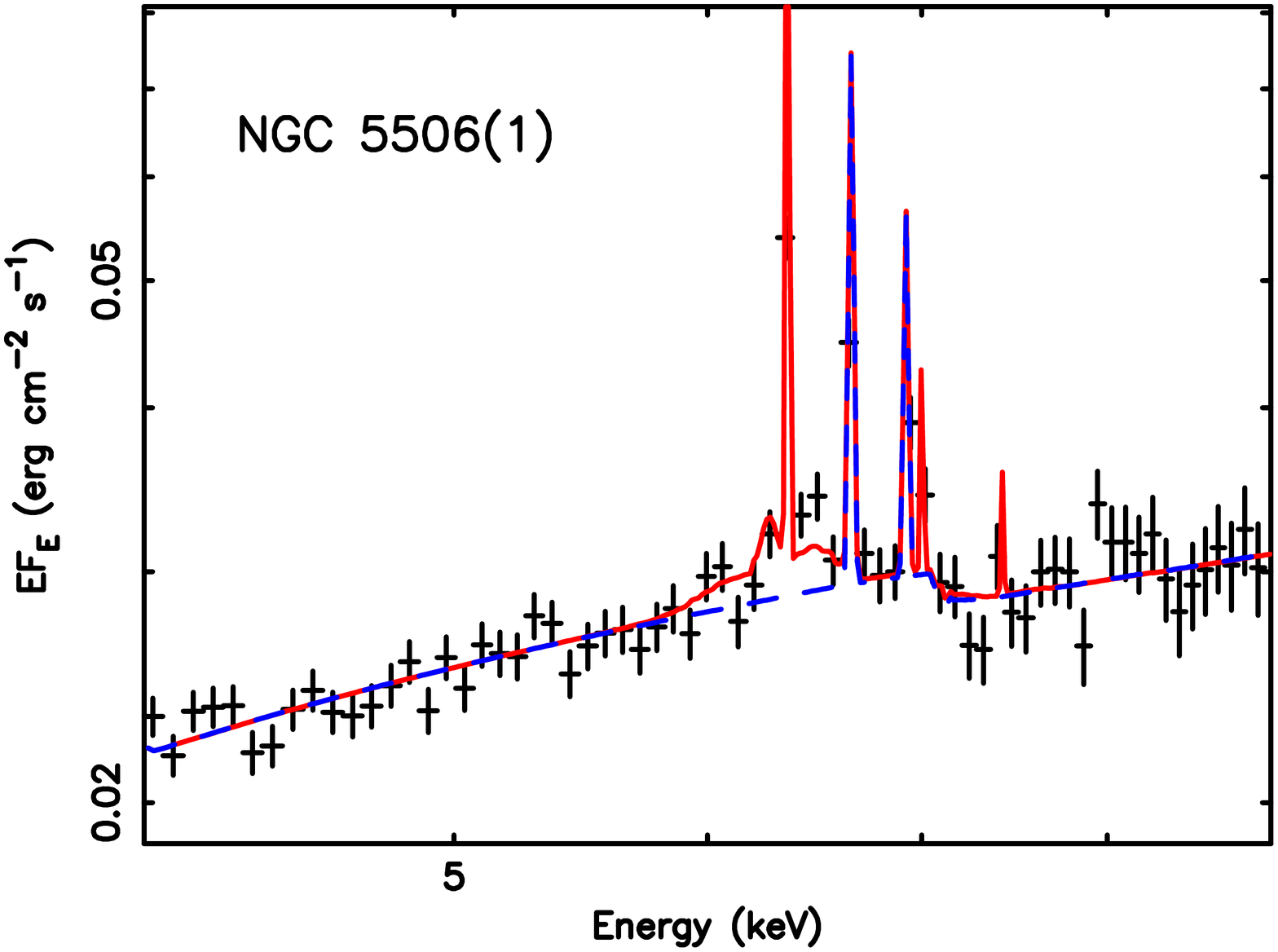}
\includegraphics[angle=270,width=52mm]{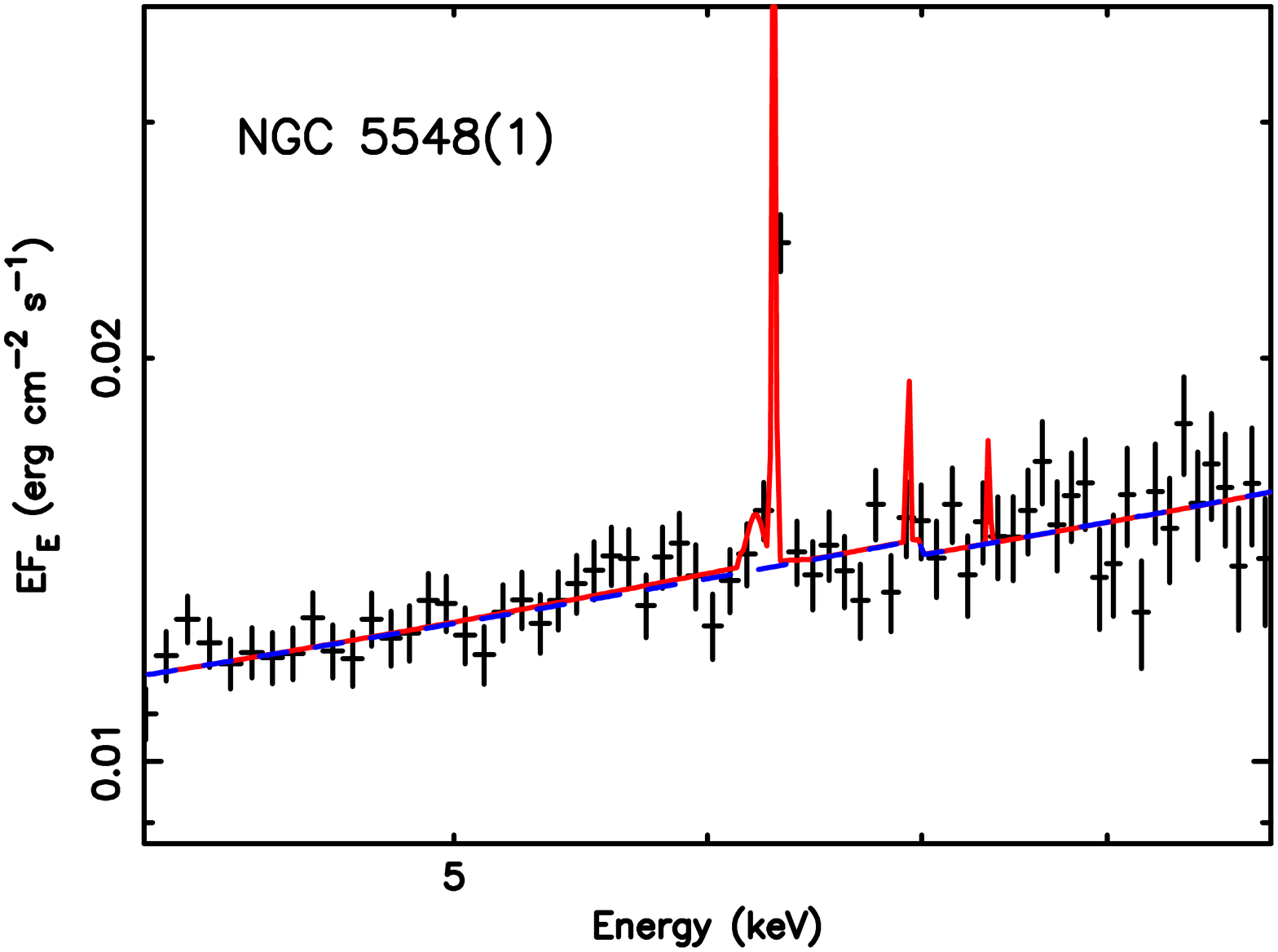}
\includegraphics[angle=270,width=52mm]{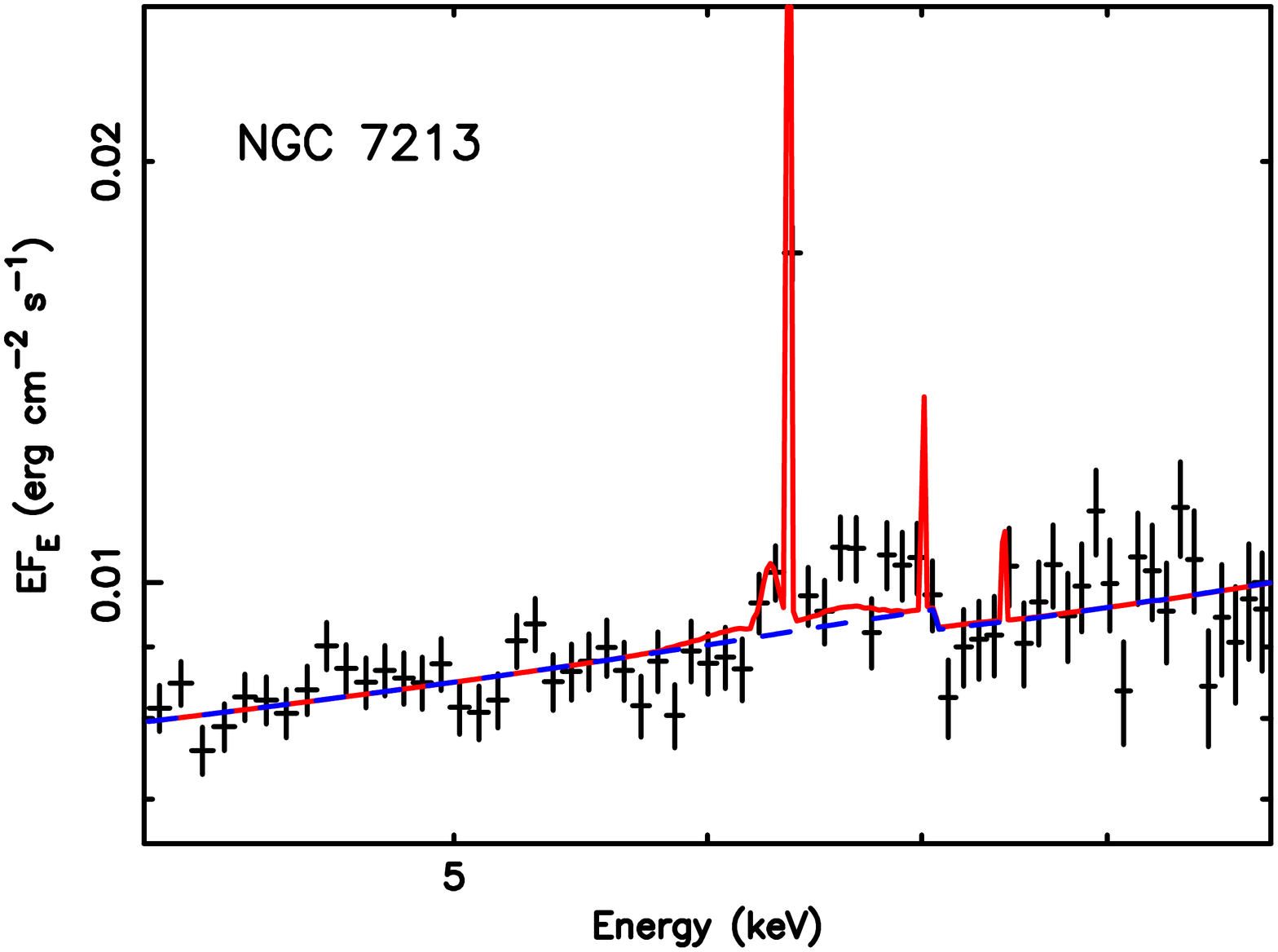}
\includegraphics[angle=270,width=52mm]{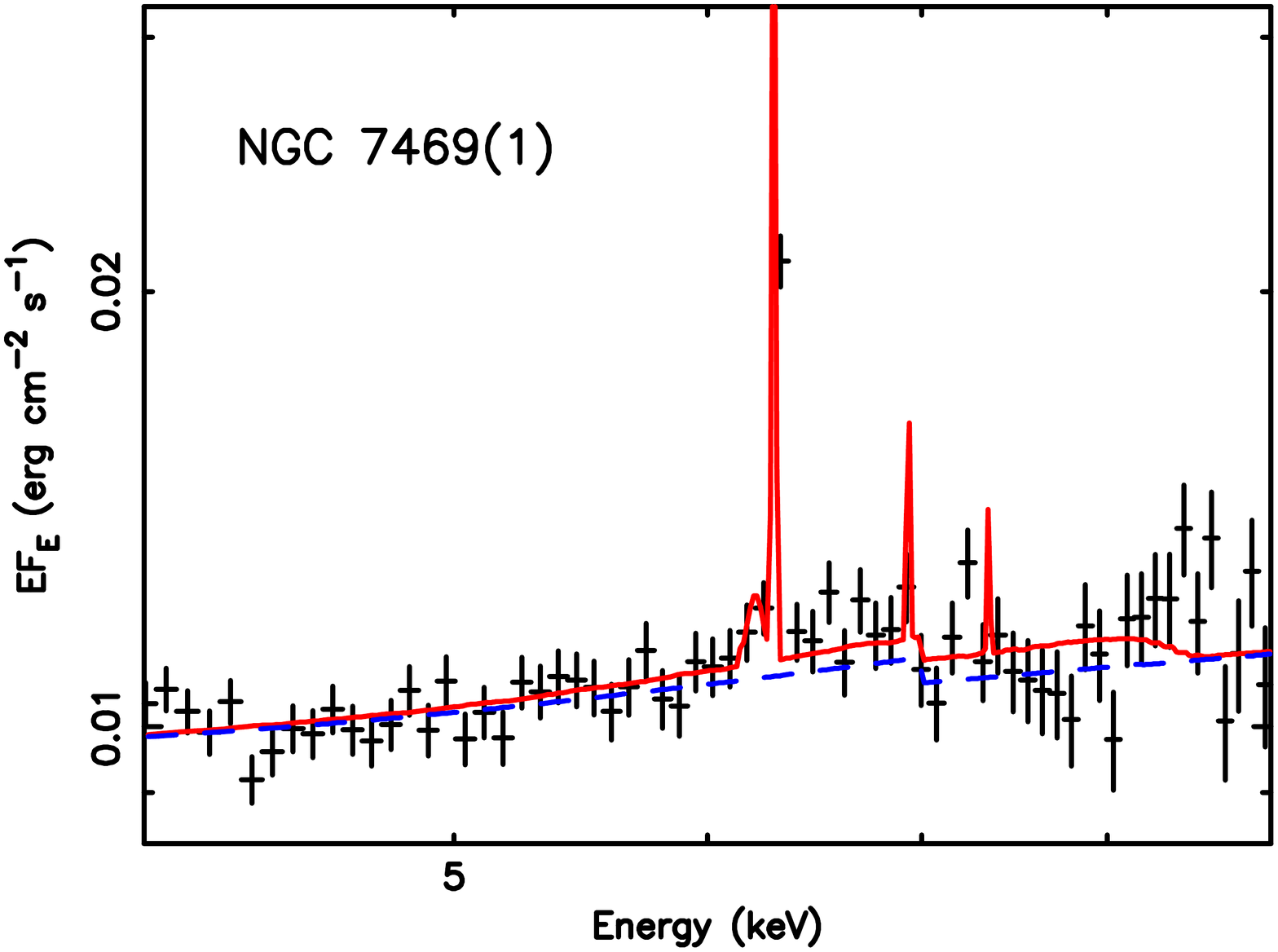}
}
\caption{Narrow lines. Unfolded spectra and models for observations in which a blurred, neutral reflection component did not improve the fit significantly.  Model lines are as in Fig.~\ref{fig:dream}
 \label{fig:nightmare}}
\end{figure*}

\subsection{How many objects have evidence for broad lines?} 

According to the F-test, the blurred-reflection component can be considered significant at 99 per cent confidence if this statistic exceeds $\sim 4.0$, given the addition of the three disk parameters (i.e. $i$, $r_{\rm br}$ and $R_{\rm disk}$).  Simulations show that, unlike the case of the narrow features, the F-test is probably conservative in the case of the disk lines, which indicate $F=3.0$ for 99 per cent confidence. However here we adopt the F-test value as a conservative measure of the significance of blurred reflection. By this criterion, 70~per cent of the sample (18/26 objects and 24/37 spectra) show evidence for broadened emission. 

While there is therefore evidence for broadened Fe emission in a very large fraction of the sample, several caveats must be borne in mind. The first is that there are a number of cases in which the F-test indicates a significant broad line, but that the alternative model consisting of a multi-zone warm absorber and a blend of narrow lines (Model D) gives as good or a better fit to the spectrum. This is true in the cases of NGC 526A, Ark 564 and NGC 5506(1), where the absorber/blend model is barely distinguishable in statistical terms from the blurred reflector. Despite this, as we have tested for, and compared explicity with models with narrow line emission in our fitting, we can conclude robustly that over half the sample show evidence for broadened emission. 

An examination of Tables~\ref{tab:kdpex} \& ~\ref{tab:kdbest} shows that, in a number of cases, the characteristic emission radius $R_{br}$ can often be relatively large. Indeed in many cases the radius approaches the upper limit of the model ($400$~$r_{\rm g}$). While there is therefore very clear evidence for broadening in a large number of cases, whether or not the broadening should be considered relativistic is a more subjective assessment. In this regard, we define two criteria based on the expected gravitational redshift as a function of radius. We define {\it relativistic} lines as those which originate within a characteristic radius $<50 r_{\rm g}$. This is the approximate radius where the gravitational redshift is equal to the FWHM resolution of the EPIC-pn CCDs, and hence within which this redshift can be robustly measured.  We further consider there to be evidence for {\it strong gravity} in cases where $R_{br}<20 r_{\rm g}$, being the radius at which the gravitational shift is approximately 5 per cent.

Based on these criteria, the best-fit parameters indicate {\it relativistic} line emission in 17 observations (of 14 objects). Thus, our best estimate of the proportion of the sample observations (objects) showing relativistic lines is $45\pm 8$~per cent ($54 \pm 10$~per cent), where the errors have been calculated using a multinomial distribution $\sigma_{p} = \sqrt{p(1-p)/N}$, where $p$ is the proportion and $N$ the total number of observations (or objects). 14 observations (11 objects) showing evidence for strong gravity. The 17 spectra showing a significant improvement at 99~per cent confidence when a blurred {\tt pexmon} component is added, and which have a characteristic emission radius $< 50 r_{\rm g}$ are shown in Figure~\ref{fig:dream}. It should be noted that by no means all of these observations can constrain the emission radius sufficiently well to {\it require} relativistic or strong gravity effects. Indeed, an examination of Table~\ref{tab:kdpex} shows that in for two objects, NGC 526A and Mrk 509, in which $r_{\rm br}$  barely constrained by the data, and a further three where $r_{br}$ could formally exceed $50 r_{\rm g}$ considering the error bars. The number of observations for which $r_{\rm br} <50 r_{\rm g}$ is statistically {\it required} based on the errors is 12, and by the same criterion 8 observations require strong gravity. Despite the poor constrains in some cases, the percentage of relativistic lines just quoted should still be representative of the sample. It is as likely for the uncertainties to scatter an object out of the ``relativistic" regime as into it, so with a sufficient number of observations the percentage should converge on the true value. 

Nine observations (of 8 objects), or $24 \pm 7$~per cent of the sample, show evidence for velocity broadening, in the form of a significant improvement when the blurred reflection  model is applied, but in which the characteristic emission radius is $> 50 r_{\rm g}$, and hence there is no requirement for substantial relativistic effects. These are plotted in Fig.~\ref{fig:yawn}. Three objects, MCG-5-23-16, NGC 3783 and NGC 4151 appear both in both Figs.~\ref{fig:dream} and \ref{fig:yawn} i.e. one observation shows evidence for relativistic effects while another does not. 

Finally, in Fig.~\ref{fig:nightmare}, we show the remaining 11 observations, $30 \pm 8$~per cent of the sample, in which no signficant improvement is obtained when the blurred reflection component is added. These observations can all be described satisfactorily with just narrow emission lines. Again, there are observations of objects which appear in this figure that have already appeared in with Fig.~\ref{fig:dream} or Fig.~\ref{fig:yawn},  i.e. one observations shows evidence for broadened emission but another shows none. These are IC4329A, NC 5506, NGC 5548 and NGC 7469. Indeed, for every objects in our sample observed multiple times, at least one observation shows evidence for broadening, and in all four cases cited above it is the observation with higher signal--to--noise ratio which shows evidence for the broad line. This indicates that the non-detection of broad emission may in many cases simply be due to low signal-to--noise ratio.  We now therefore discuss the limits on disk line emission for these objects. 

\subsection{Are there cases blurred-reflection is absent?}
\label{sec:noblur}

There are two categories of observation just mentioned in which a relativistically-blurred reflection component  may be absent. The first are those that, in the fitting above, show no significant improvement in $\chi^{2}$ for the blurred reflector. This applies to 11 of the 37 observations in the sample. One must, however, also consider the possibility that for these sources the signal-to-noise ratio of the spectra is too low to allow for the detection of a blurred reflection component. Indeed, given the nature of the models we are fitting to the data, it is very difficult to exclude formally a disk line in any individual case. This is because the disk may be viewed at a high inclination angle, when the reflection is expected to be weak. 

\begin{figure*}
{
\includegraphics[angle=0,width=58mm]{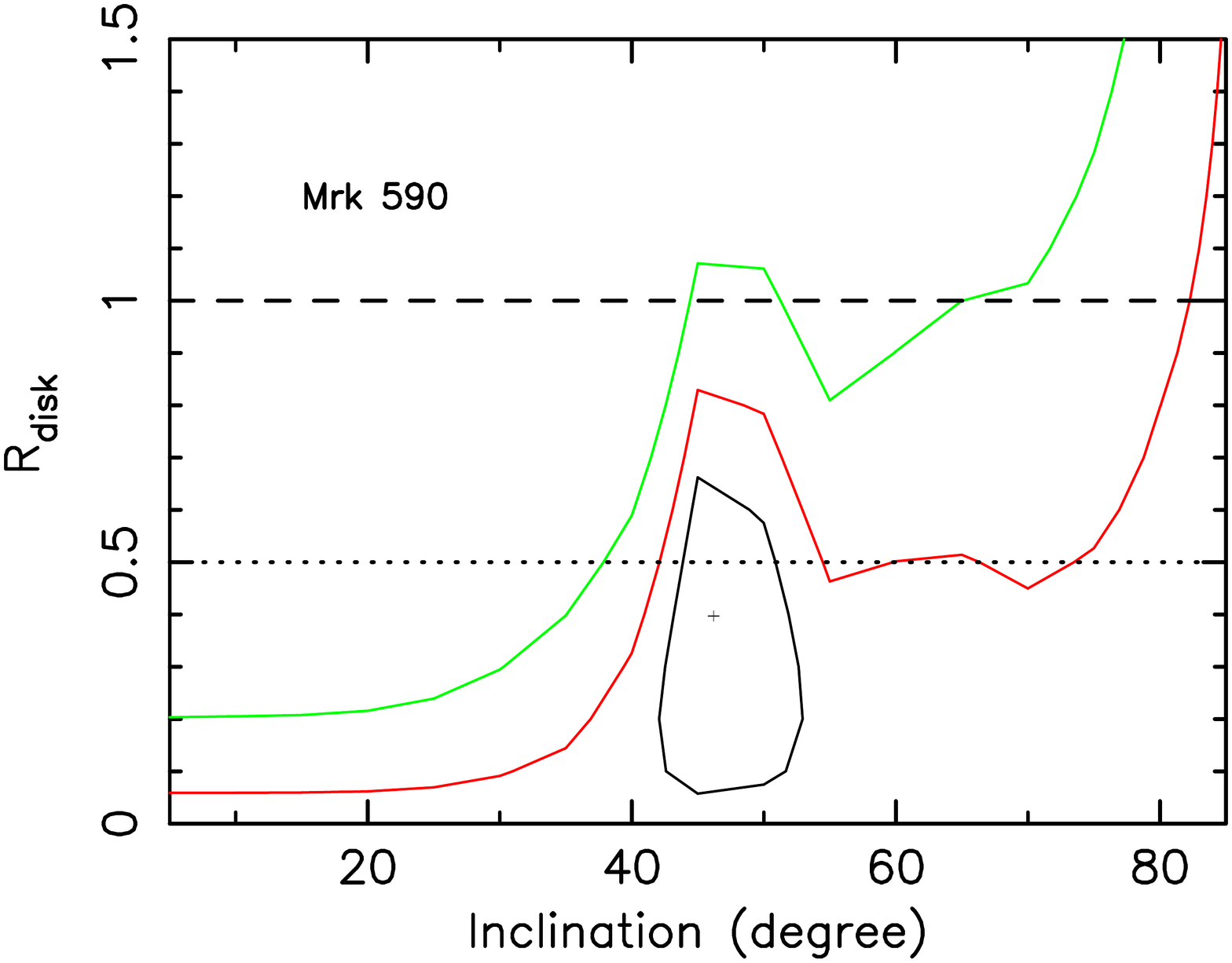}
\includegraphics[angle=0,width=58mm]{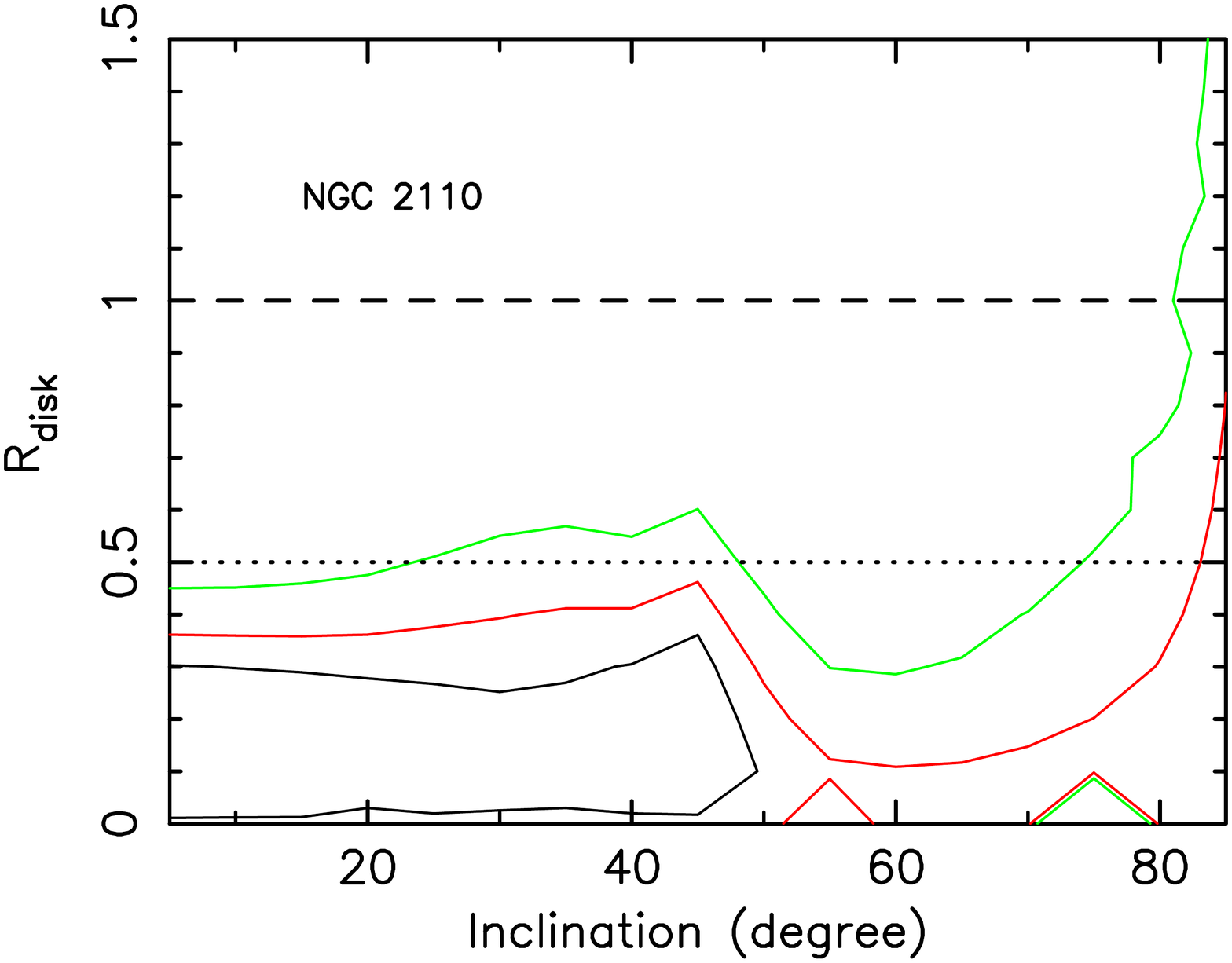}
\includegraphics[angle=0,width=58mm]{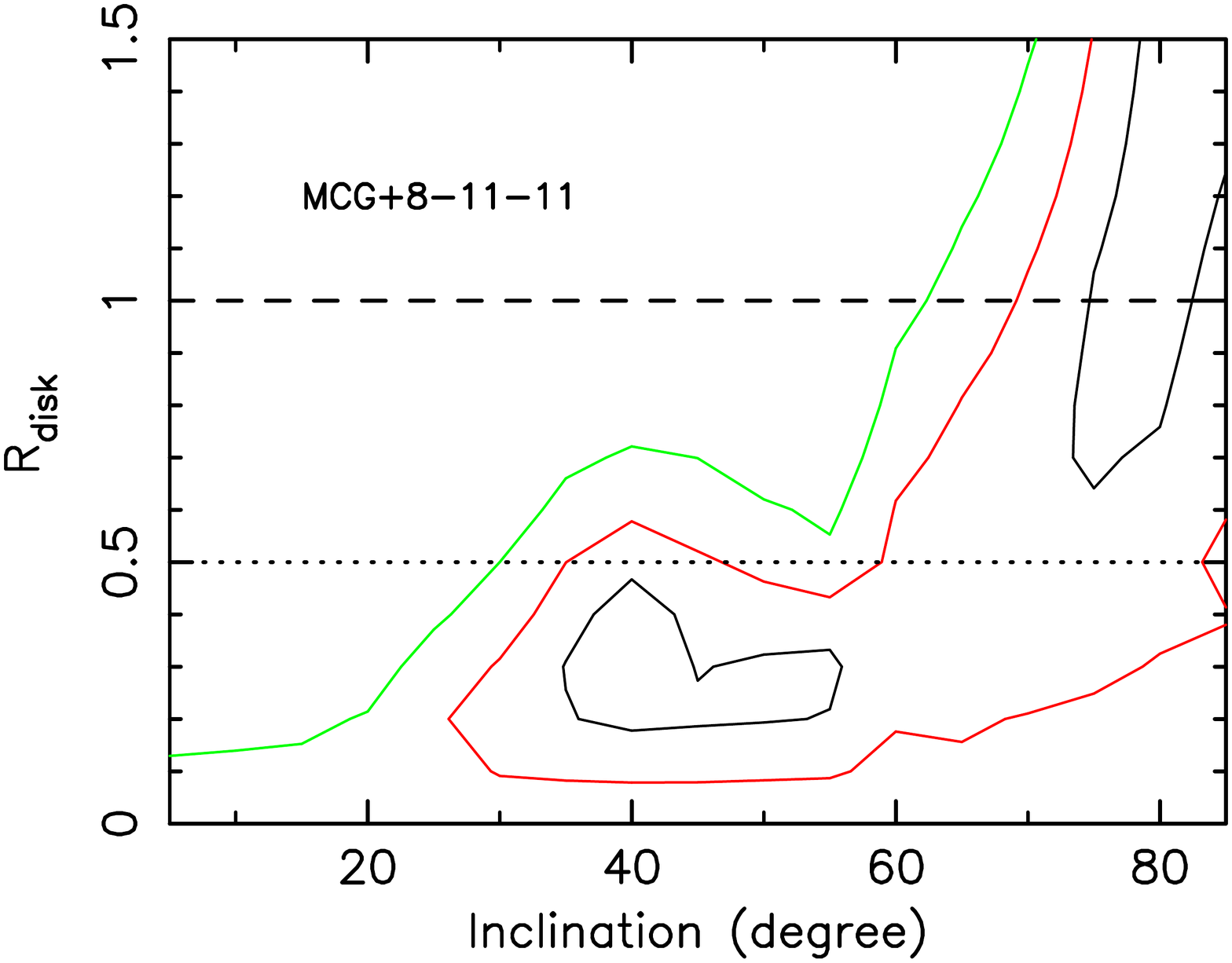}
\includegraphics[angle=0,width=58mm]{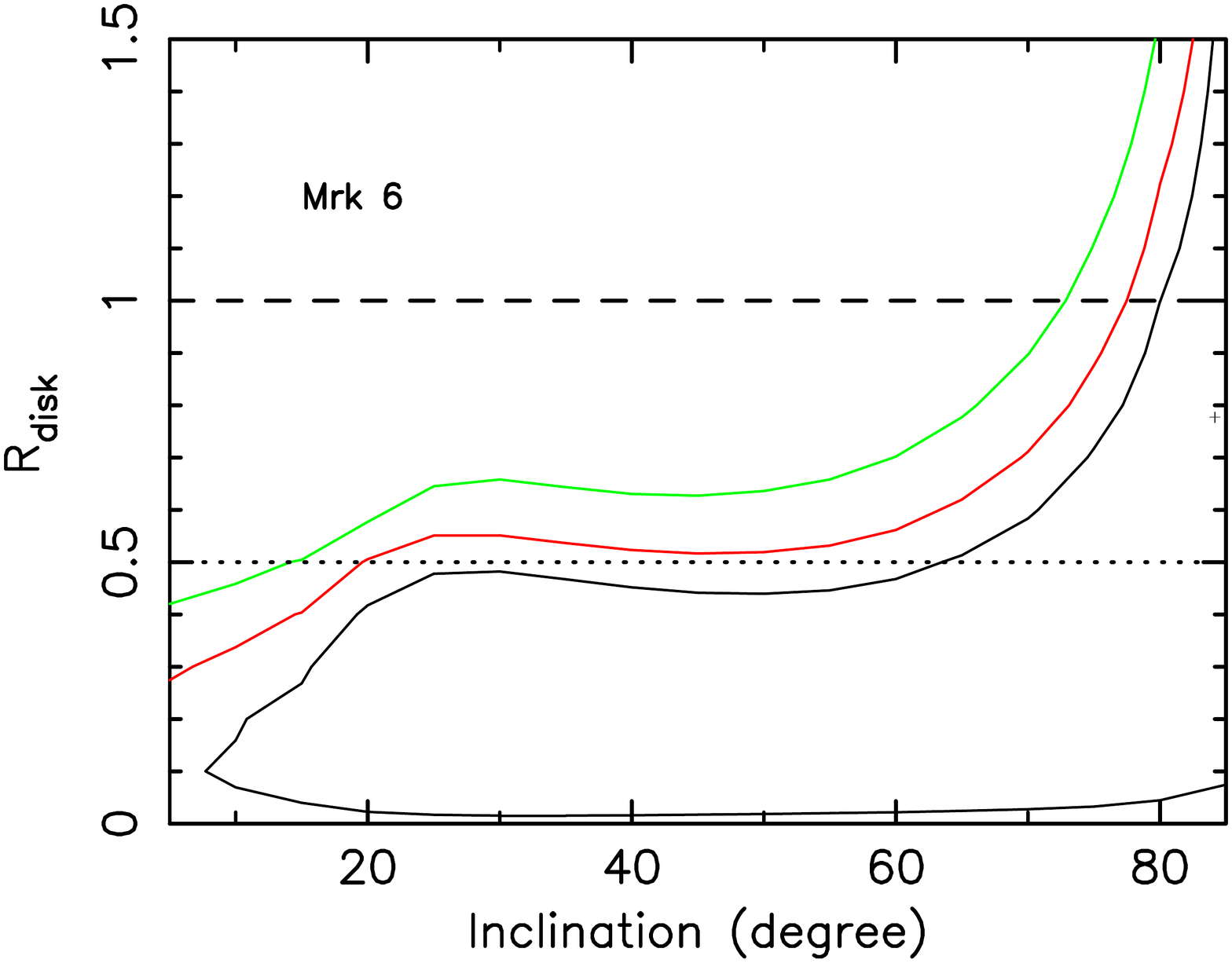}
\includegraphics[angle=0,width=58mm]{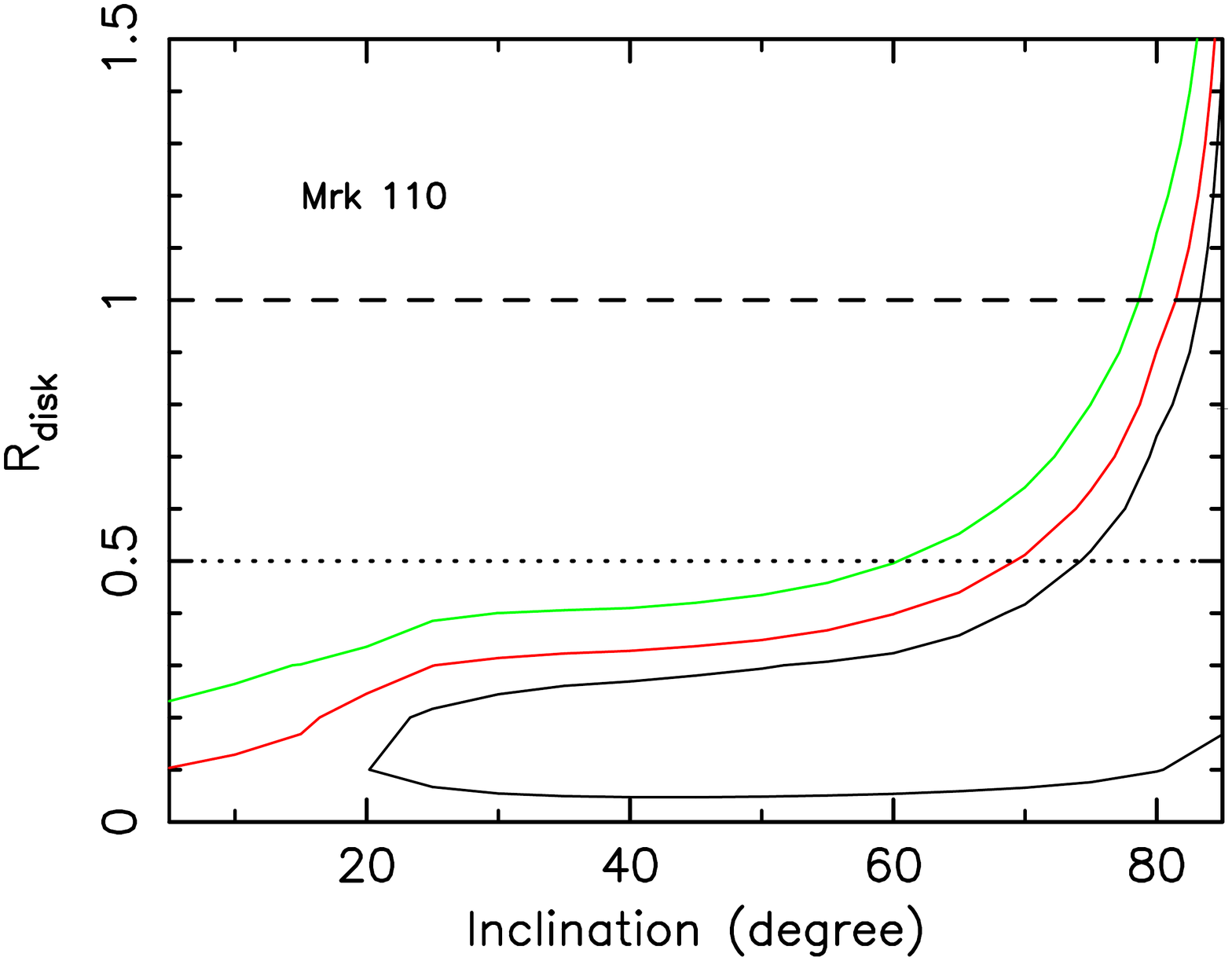}
\includegraphics[angle=0,width=58mm]{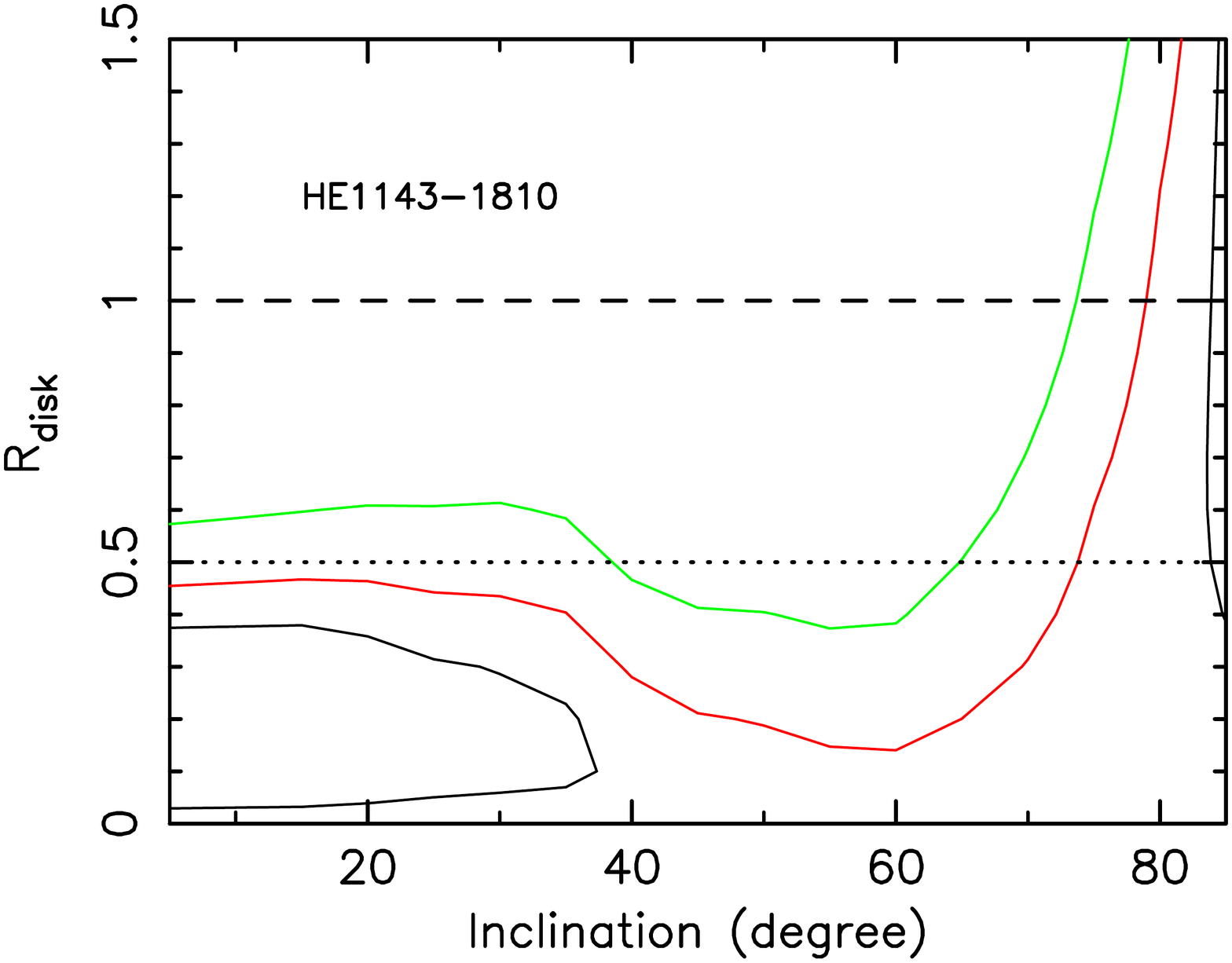}
\includegraphics[angle=0,width=58mm]{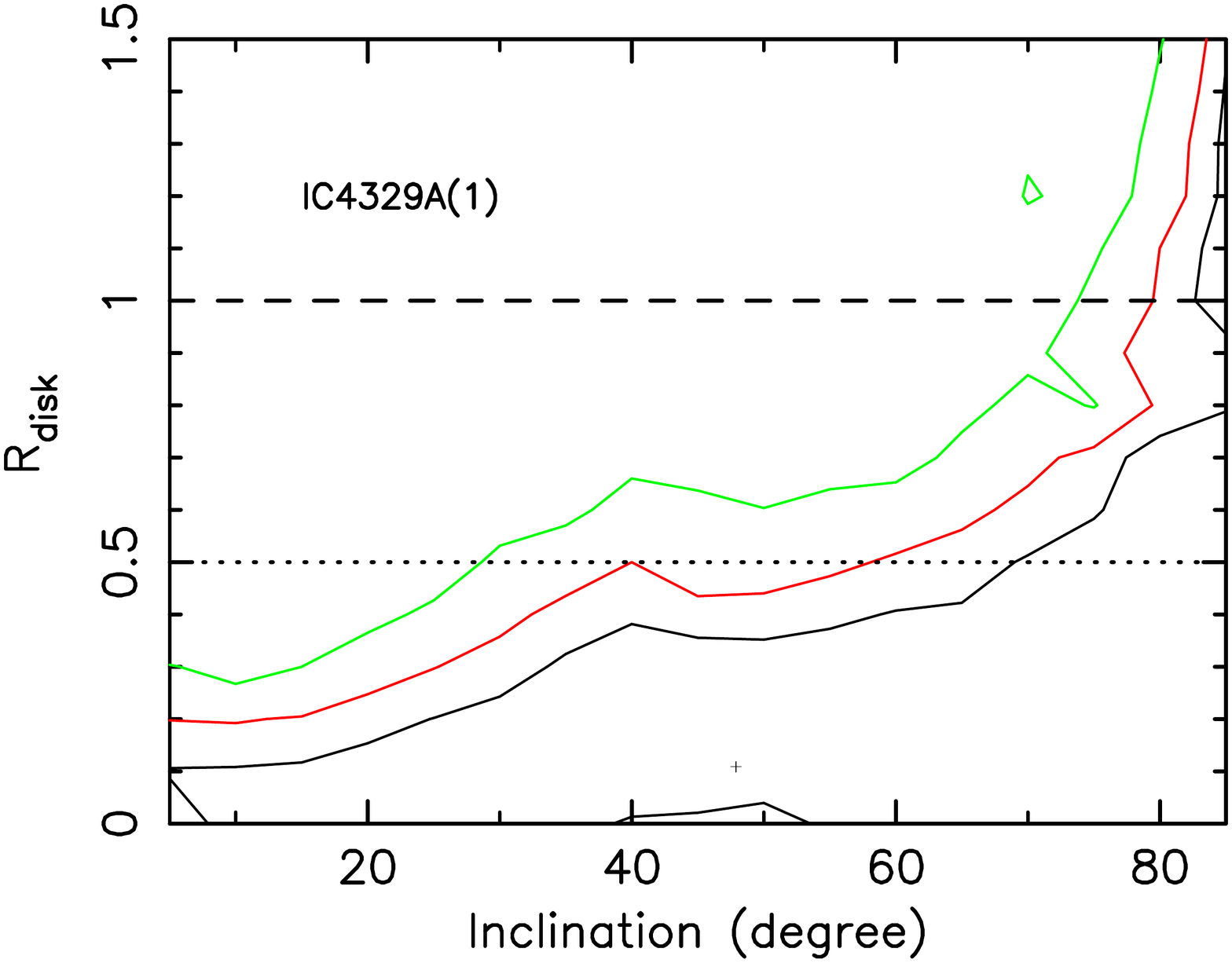}
\includegraphics[angle=0,width=58mm]{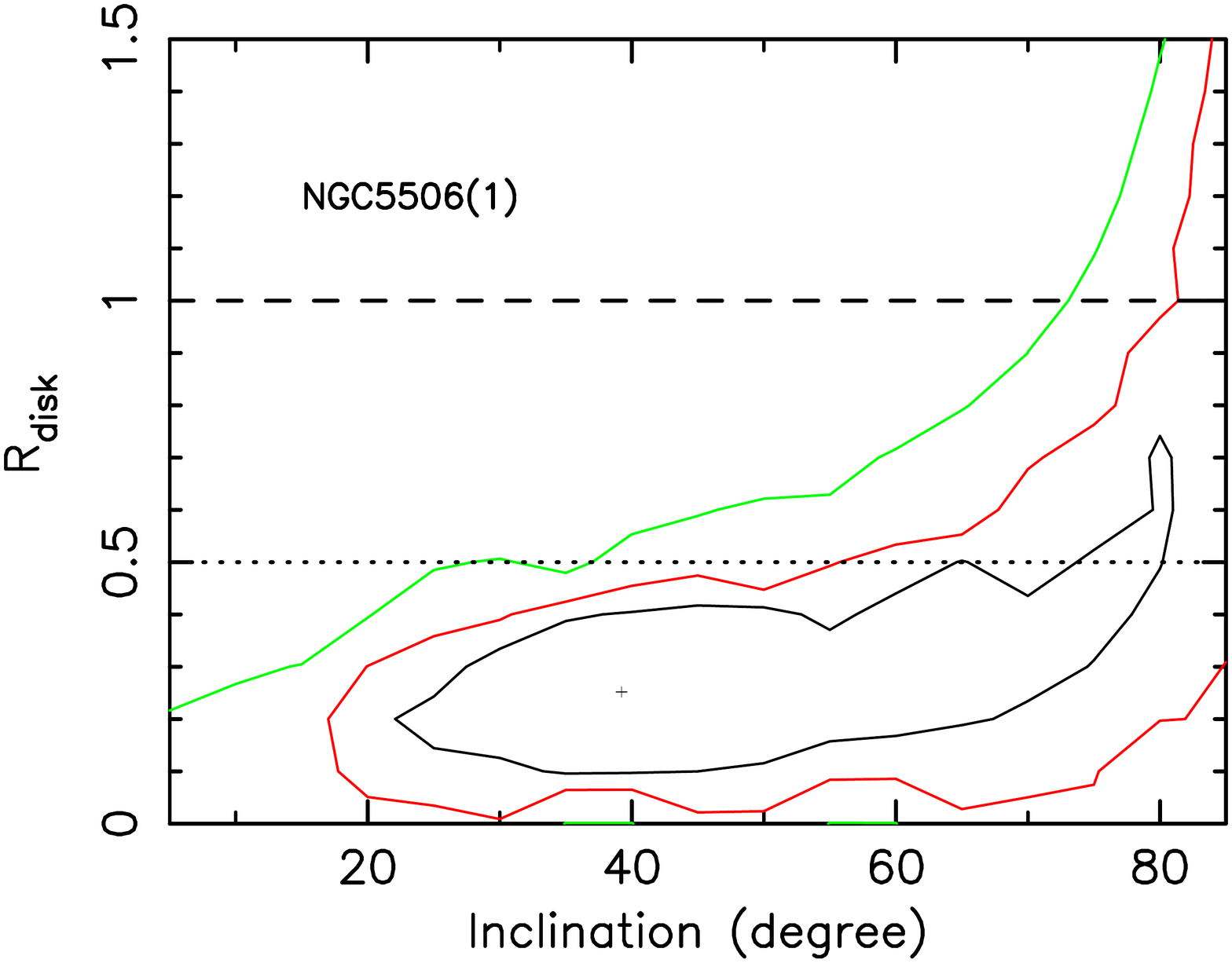}
\includegraphics[angle=0,width=58mm]{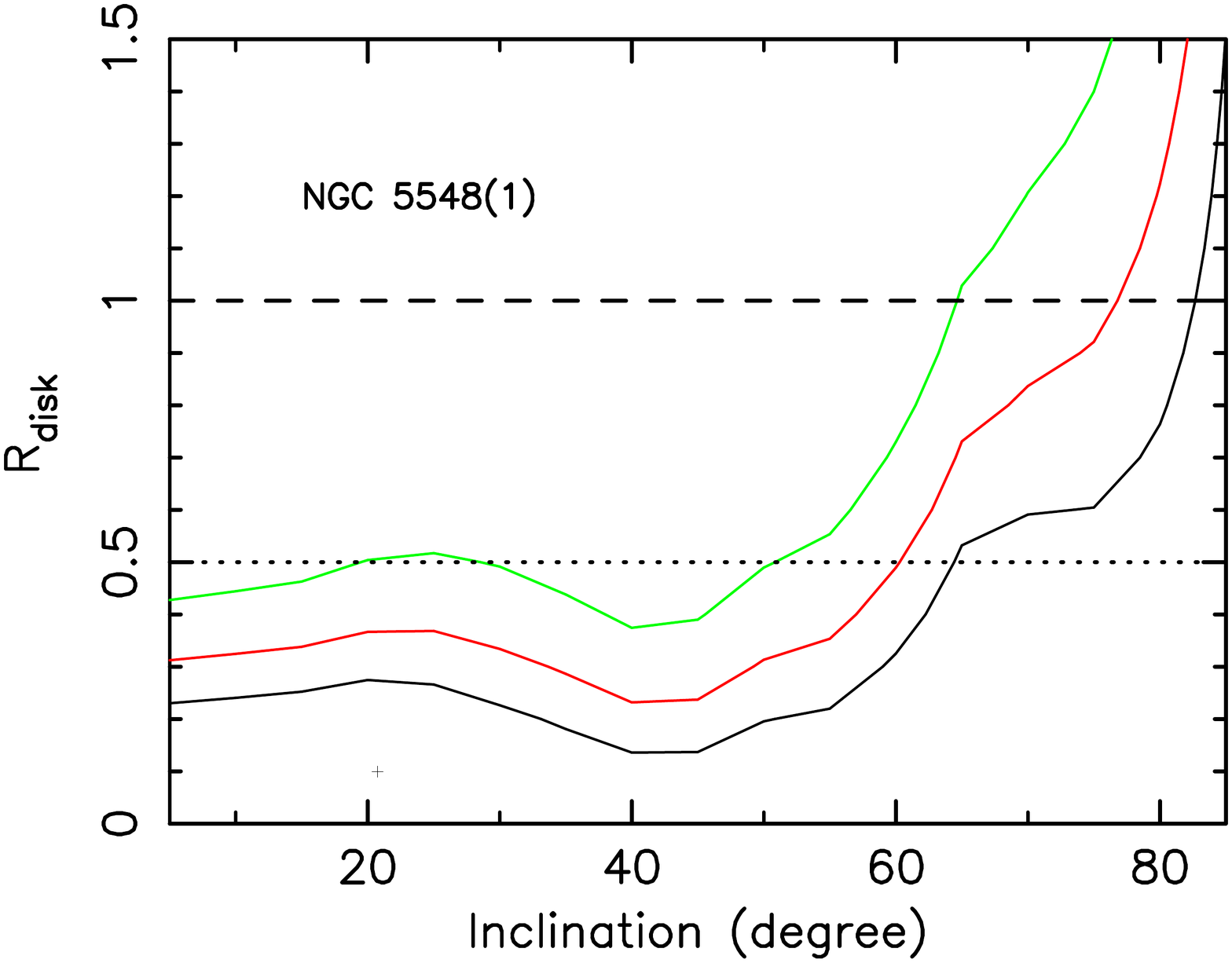}
\includegraphics[angle=0,width=58mm]{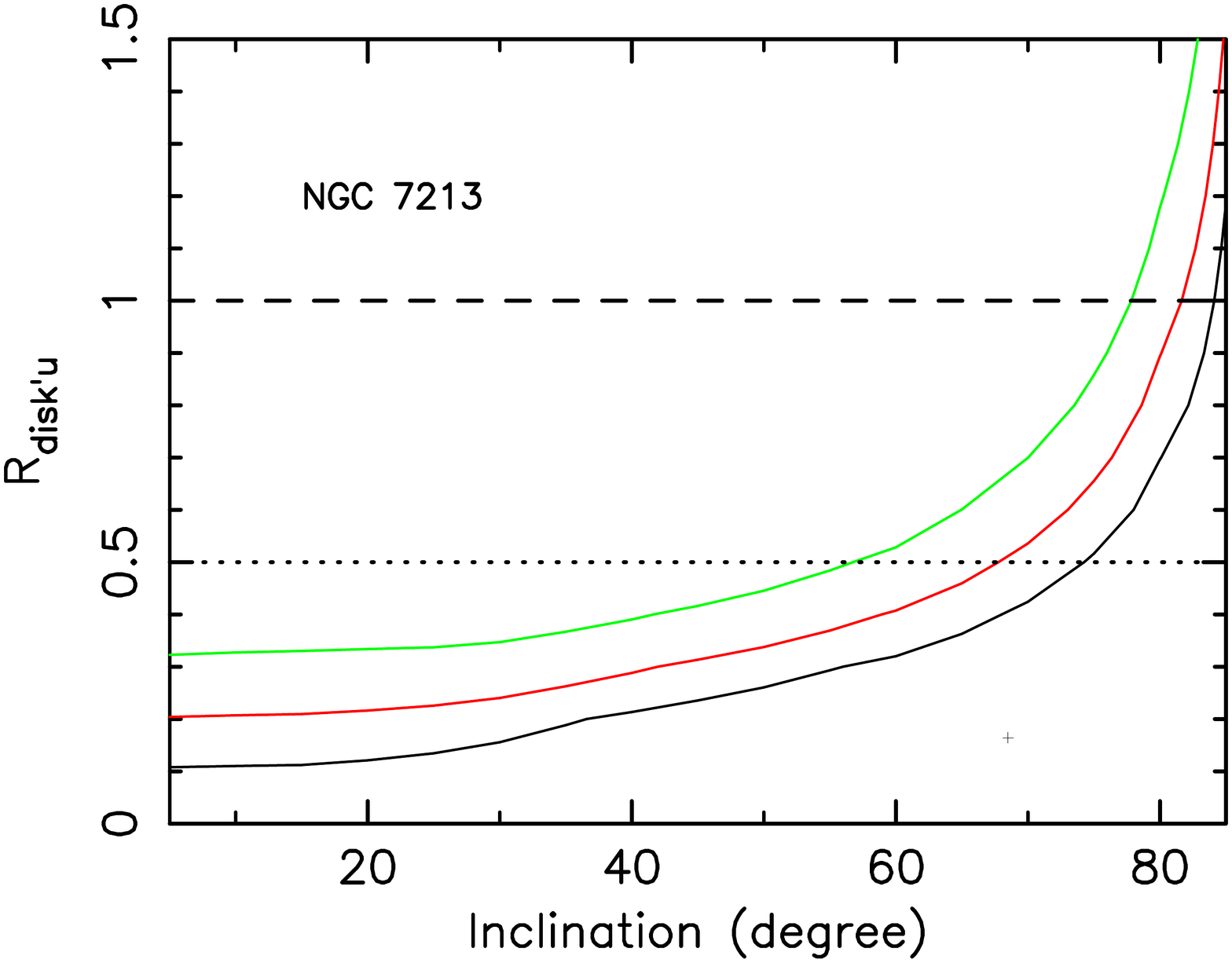}
\includegraphics[angle=0,width=58mm]{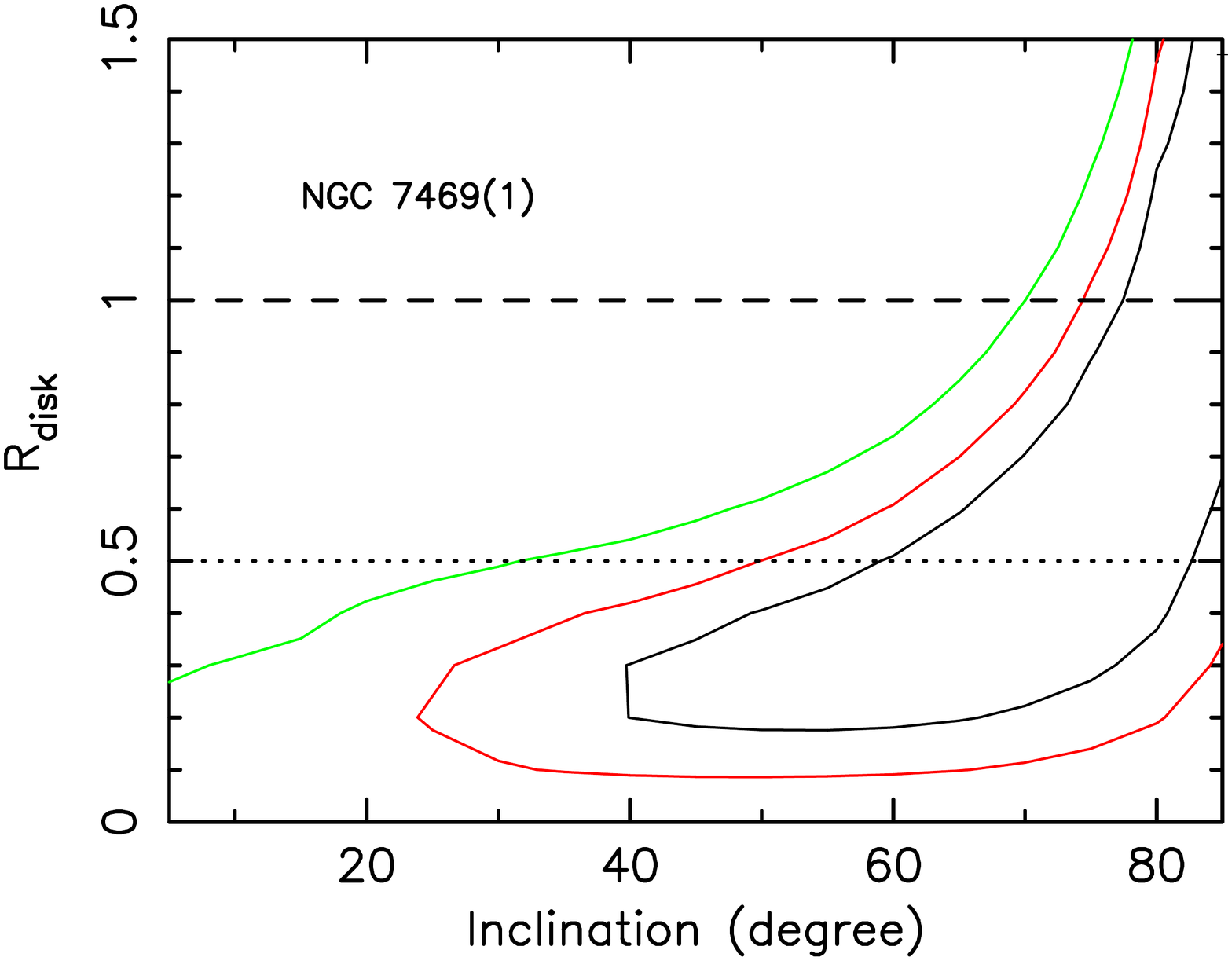}
}
\caption{Confidence contours of reflection fraction $R_{\rm disk}$ versus inclination for objects in which no broadened component of the line is significantly detected. In many objects, a relativistic line with parameters typical of the sample is easily allowed, and the lack of detection can be attributed to poor signal-to-noise ratio. In others, emission from the innter disk is only permitted if this disk inclination is very high.
\label{fig:nl_cont}}
\end{figure*}

This can be seen readily from the constraints listed in Tables~\ref{tab:kdpex} \& ~\ref{tab:kdbest}. 
Most of the observations in which the blurred reflection is undetected are compatible with $R_{\rm disk}=1$, the value expected for an accretion disk covering half the sky at the illuminating source. The strongest upper limits on $R_{\rm disk}$ are obtained for NGC 2110, HE1143-1810, and NGC 5548(1), which all have $R_{\rm disk} < 0.5$ at 68 per cent confidence. In order to produce more formal constraints on the presence or otherwise of a disk line in the undetected objects we have calculated confidence contours for $R_{\rm disk}$ versus $i$ for the 11 observations. These are shown in Figure~\ref{fig:nl_cont}.

This figure shows a variety of contour shapes, but as expected there is a general trend in which high values of the reflection fraction $R_{\rm disk}$ tend to be allowed at high inclinations, but not if the disk is seen relatively face-on. The horizontal dashed line shows $R=1$, the value expected for a point source illuminating a slab subtending $2\pi$ solid angle at the X-ray source. Typically, this value of $R_{\rm disk}$ is permitted only if the inclination of the accretion disk is very high. With a random distribution of inclinations, one does expect a relatively number of high inclination systems. For example, one quarter of randomly-oriented disks should be seen at inclinations $>75^{\circ}$. A value of $R=1$ is permitted in almost all individual observations in Fig.~\ref{fig:nl_cont} at 99~per cent confidence if $i>75^{\circ}$, so if the inclinations are indeed random there is no clear evidence from the sample as a whole that the broad reflector is ``missing" in any object.  This is even more the case when one considers that the typical reflection fraction for the sample is closer to $R=0.5$, i.e. the broad reflection appears to be a factor $\sim 2$ weaker than that expected for a $2\pi$ slab (and we note that this conclusion remains even if the objects in Fig.~\ref{fig:nl_cont} are excluded from the estimation of the mean $R_{\rm disk}$). With this smaller value of $R_{\rm disk}$, much lower disk inclinations are typically allowed so the overall inclination distribution is even more consistent with random orientations. 

On the other hand, as we have explicitly excluded Seyfert 2 galaxies from the sample, we should have a very strong selection against edge-on systems. For a typical torus half-opening angle of $\sim 60^{\circ}$, and assuming the torus and accretion disk axes are aligned, we would expect objects with $i>60$ to be seen as Seyfert 2s and hence be completely absent from the sample. If $R\sim 1$ then a large number of objects are inconsistent with $i<60$ at 99 per cent confidence. There is no observation in which we can exclude a disk inclination $i<60$ if the reflection component has $R=0.5$, though Mrk 110 and NGC 7213 come close.  We can therefore conclude very clearly that there are objects in which the disk line appears to be ``misssing", or that on average the reflection is significantly weaker than the expectation of $R_{\rm disk}=1$ based on our assumed geometry. 

\subsection{Evidence for black hole spin}

\begin{figure}
\includegraphics[angle=0,width=90mm]{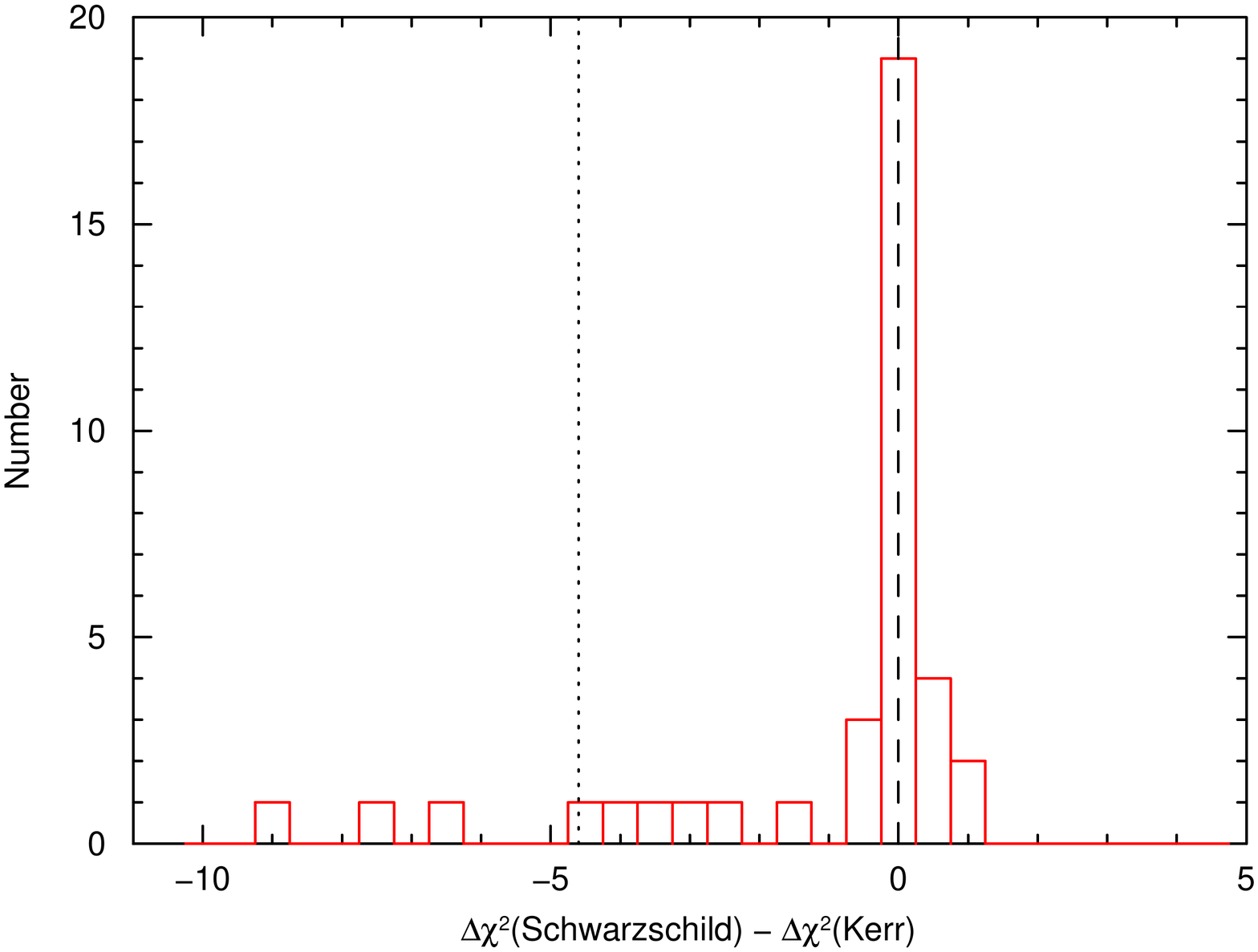}
\caption{Difference in $\chi^{2}$ for a model in which the inner disk radius is fixed to $6 r_{\rm g}$, appropriate for a Schwarzschild black holes, compared to one with $R_{\rm in} = 1.235 r_{\rm g}$, 
to represent a maxmally rotating Kerr hole. The $\Delta \chi^{2}$ distribution is strongly peaked around zero (vertical dashed line), showing that in the majority of cases it s not possible to distinguish between these two scenarios.  $\Delta \chi^{2}=-4.6$ (vertical dotted line) corresponds to an improvement at 99~per cent confidence according the the likelihood ratio test. Three observations show a substantial improvement in the Kerr geometry.}
\label{fig:kerr_delchi}
\end{figure}

Several workers have noted that the profiles of iron K$\alpha$ lines in AGN are one of the few ways to obtain constraints on the black hole spin, and evidence for rapidly rotating black holes has been presented in a number of cases, notably MCG-6-30-15 (Iwasawa et al. 1996; Wilms et al. 2001; Reynolds et al. 2004b, Brenneman \& Reynolds 2006). With our sample we can test this by allowing the inner radius in our model to take a smaller value, as might be expected in a Kerr geometry where stable orbits can approach the black hole's gravitational radius (Bardeen, Press \& Teukolsky 1972). For the simplest possible test, we have simply changed the model described in the previous section (i.e. model F) now fitting an inner radius of $1.24 r_{\rm g}$, close to the minimum value, and appropriate for a maximally-rotating Kerr black hole. 

A histogram showing the difference in $\chi^{2}$ between the Schwarzschild and Kerr fits is given in Fig.~\ref{fig:kerr_delchi}. It can be seen that typically the statistic changes very little when the inner radius is changed from $6$ to $1.24 r_{\rm g}$, showing that for most of the observations it is not possible to distinguish between rotating and non-rotating black holes. Based on a likelihood ratio test, however, a total of 3 observations do show a significant improvement for the smaller inner radius. These are MCG-6-30-15(1), NGC 3516(1), NGC 7469(2). The revised parameters for these three observations are shown in Table~\ref{tab:kdkerr}. 

\begin{table*}
\centering
\caption{Disk line parameters re-derived using for the 3 observations which show a significant improvement for $R_{\rm in}=1.24 r_{\rm g}$, expected for a rapidly rotating black hole.
Col.(1): Name and Observation number;
Col.(2): Reflection fraction of the blurred {\tt pexmon} component, where $R=1$ corresponds to a semi-infinite slab seen at an inclination $i$ subtending 2$\pi$ solid angle at the X-ray source; 
Col.(3): Inclination of the slab representing the blurred reflector (i.e. the accretion disk); 
Col.(4): Break radius for the emissivity law
Col.(5): Reflection fraction of the unblurred {\tt pexmon} component, where $R=1$ corresponds to a semi-infinite slab seen at an inclination of 60$^{\circ}$ and subtending 2$\pi$ solid angle at the X-ray source; 
Col. (6): $\chi^{2}$ and degrees of freedom including all components. 
Col. (7): Probability of obtaining $\chi^{2}$ by chance
Col. (8): Probability that the $\chi^{2}$ for a fit with $r_{\rm in }=1.24 r_{\rm g}$ would improve on that with $r_{\rm in }= 6 r_{\rm g}$ by chance, based on likelihood ratio test. 
\label{tab:kdkerr}}
\begin{center}
\begin{tabular}{llcllcccc}
\hline
Name &  $R_{\rm disk}$ & $i$ & $r_{\rm br}$ &  $R_{\rm tor}$  & $\chi^{2}$/dof & $P_{\chi^{2}}$ & $P_{lr}$ \\
 & ($\Omega/2\pi$) & (deg) & ($r_{\rm g}$) & ($\Omega/2\pi$) & & & \\
(1) & (2) & (3) & (4) & (5) & (6) & (7) & (8) \\
\hline
\hline
NGC 3516(1) & $2.0^{+1.4 }_{-0.9}$ & $34^{+4}_{-6}$ & $2.4^{+3.0}_{-1.2}$ & $0.9^{+0.3}_{-0.2}$  & 94.1/93 & 0.45 & $1.6 \times 10^{-3}$ \\
MCG-6-30-15(1) & $2.9^{+2.0}_{-1.7}$ & $37^{+2}_{-5}$ & $1.2^{+5.9}_{-0.0}$ & $0.5 ^{+0.4 }_{-0. 3}$  & 121.4/94 & 0.03 & $5.5 \times 10^{-4}$  \\
NGC 7469(2) & $4.5^{+2.4 }_{-2.1}$ & $85 ^{+0 }_{-2 }$ & $7.8^{+5.8}_{-4.4 }$ & $0.7^{+0.2 }_{-0.1}$ & 144.0/96 & 0.00 & $1.0 \times 10^{-4}$ \\
\hline
\end{tabular}
\end{center}
\end{table*}

\subsection{Non-solar iron abundance}
\label{sec:abun}

 It is also possible that the iron abundance in the reflecting material differs from the solar value, and we have therefore tested this possibility. Small deviations in the iron abundance away from our adopted value seem perfectly reasonable, and have been invoked in other objects (e.g. MCG-5-23-16 Reeves et al. 2007; MCG-6-30-15 Miniutti et al. 2007). The primary effect of the abundance variation will be to change the ratio of the iron line and edge strength in the reflected spectra to the strength of the Compton reflection hump. If iron is overabundant, the line will be relatively stronger, and vice versa. Because the reflection hump, which primarily affects the spectrum above 10 keV, is often not very well constrained with the XMM data, there will be a tendency for the fits to be driven by the strength of the iron K$\alpha$ line. So one explanation for the relatively low value of $R_{\rm disk} \sim 0.5$ found for the sample is that iron is in general underabundant relative to hydrogen and soft X-ray absorbing elements like oxygen. Some support for this has been found in the fits above, where we found evidence in a few objects for an additional ``hard tail" . 
  
With {\tt pexmon} we can model this self-consistently, and we have tested a variable abundance reflector against all the spectra. The abundance is allowed to vary in the range 0.1--10 times solar, with only iron changing, relative to all other elements. The abundance of both the distant and blurred reflector were allowed to vary, but they were tied together. Based on the F-test, six observations of four objects do show evidence for a significant departure from solar abundances at $>99$ per cent confidence, these being Akn120, NGC 4151(1,3), HE1143-1810 and NGC7469(1,2). In all cases a sub-solar abundance in the range $A_{\rm Fe} = 0.3-0.5$ is favoured.  We note that the observation designated NGC 4151(2), while not presenting a significant reduction in $\chi^{2}$ for a variable abundance model, is nonetheless consistent with the  low abundance derived from the other observations. The idea that the iron abundance in AGN might typically be less than the solar or cosmic value is intriguing, and provides a possible explanation for both the relatively weak disk reflection we see in the sample as a whole, and the absence of significant broadened lines in individual objects. At these point, however, we treat the improvements in the fit with low abundance models with great caution: in many cases they are achieved only with very steep underlying continuum slopes, and high reflection fractions $R>>1$. As we have already mentioned the \xmm\ EPIC data are relatively insensitive to the reflection continuum: the  confirmation of non-solar abundances therefore should await better high energy data. 

\section{DISCUSSION}

\subsection{Summary of main findings}

Our main motivation for studying the sample in this paper was to determine the iron K$\alpha$ line properties of Seyfert galaxies, and in particular assess the evidence for and robustness of broad emission lines which may arise from the inner regions. This study can be considered timely in the light of the fact that the paradigm established by {\it ASCA}, that accretion disk lines are very common, has been questioned by several authors in their analysis of \xmm\ data. We have used well-defined criteria for selecting our sample -- predominantly concentrating on those objects with the highest signal-to-noise ratio -- and in analysing the \xmm\ data we have taken a highly systematic approach to both the data analysis and spectral modeling. For the former, we have been relatively conservative in terms of data (and particularly background) screening. In regard to the latter, we have been careful to consider the physical origin and motivation for the various components we add to the spectra. An important advance in this regard is the use of the {\tt pexmon} model for the reflection continua, which enables modeling of the iron lines and reflection continuum in a self-consistent fashion. We have also taken careful account of ionized absorption in the spectra, as well as narrow emission and absorption features which may complicate the analysis of broad emission from the accretion disk. 

We first highlight a few simple, but important conclusions. Firstly, the continua of these Seyfert 1 galaxies outside the iron band is complex. Around half show a ``convex" shape in the 2.5-10 keV band, which we have interpreted as ionized absorption. A further set of objects show ``concave" spectra with hard tails, which are most easily interpreted in terms of Compton reflection. We can conclude, therefore, that even when the iron band is excluded well over half the sample show continua more complex than a power law. lt has long been appreciated that correct modeling of the continuum is critical in determining the emission line properties of AGN and our data are no exception. 

Once the continua are adequately modeled, we find that narrow emission lines associated with neutral or near-neutral iron K$\alpha$ are almost ubiquitous in Seyferts. However, these components cannot fully account for the complexities in the iron band. Just 5 of the 26 objects in our sample show acceptable fits which do not improve further if additional complexity is allowed. 

Excess emission in the iron band is the norm, with nearly 80~per cent of the spectra showing evidence for further complexity, and we initially modeled this with a Gaussian. This simple paramaterization provides a surprisingly good description of the spectra, and the typical parameters of the Gaussian show clearly that the excess emission is associated with iron. In about 2/3 of the sample observations this additional Gaussian is significantly broadened. The standard model for this broad emission is to invoke high velocities and gravitational redshift close to the central black hole.

Before proceeding to that conclusion, however, we have tested an alternative model against the data, consisting of up to two screens of ionized gas and a blend of narrow iron emission lines.  This model provides a reasonably good description of the majority of the spectra, particularly if emission form intermediate ionization states of iron in the range 6.4-6.7 keV is permitted. Comparing this model to the standard paradigm of blurred reflection from an accretion disk, however, we find that the accretion disk model is preferred for the sample in all senses, which is even more striking given this model has a smaller number of free parameters. This is a key conclusion of our work: in a fair and unbiased comparison we have found clearly that a model of broad emission from an accretion disk is statistically preferred to the leading alternative model, in which the broadening and complexity is due to ionized absorption and line blends. 

Despite this, narrow emission lines (either in absorption emission) and/or complex multi-zone absorption may be present in the spectra. Indeed, we have found evidence for such additional complexity in just under half the observations. Once these are accounted for our final model represents an excellent description of the vast majority of the spectra.  

The additional complexities make relatively little difference to the typical reflection parameters for the sample, however. These indicate that both the distant and nearby reflectors each subtend a solid angle $\sim \pi$ at the X-ray sources. Reflection from the inner regions is therefore weaker than expected from a standard flat accretion disk illuminated by a point source, although it is interesting to note that the {\it total} strength of the reflection (distant plus blurred)  is indicative on average of a $2\pi$ slab. The typical inclination for the inner reflecting slab is found to be $\sim 40^{\circ}$ and the emission comes from a characteristic radius of around $20 r_{\rm g}$. At this radius relativistic effects, including gravitational shifts are observed. The average properties of the sample give a somewhat misleading picture overall, however, as we find a significant dispersion in all of the relevant parameters of the reflection. For the simplest scenario, in which every source has the same structure in the central regions, the only parameter expected to have a significant spread is the disk inclination. Clearly, then, something more complex is occurring in the central regions of AGN. 

Investigating further, we find that the sample can be split into three general categories: those where significant relativistic broadening is seen ($<50 r_{\rm g}$), those with broad, but non-relativistic emission, and those without evidence for broad emission at all. There are a large number of objects ($\sim 40$~per cent the sample) in which relativistic effects -- probably from the accretion disk -- are robustly detected in the iron line emission. This fraction is an approximate agreement with the results of Guainazzi et al. (2006). These workers found a strong dependence of the apparent broad line fraction on the signal-to-noise ratio of the \xmm\ spectra. For the highest signal--to--noise spectra, as we consider here, these authors found evidence for broad emission in $\sim 50$~per cent of AGN. As we have also shown, very good statistics are required to deconvolve the broad emission from other spectral complexities. 

We have also investigated whether one can exclude the expected accretion disk contribution to the line emission in objects in which it is not significantly detected. In individual cases it is very difficult to exclude an accretion disk contribution, because the disk may be at  high inclination. We should have selected against the most edge-on systems in our sample, however, because we have excluded Seyfert 2 galaxies. Under these circumstances, it is clear that there are objects in which a blurred reflection component with $R_{\rm disk} =1 $ can genuinely be excluded. On the other hand, the average reflection fraction for the sample also appears to be lower than that from a flat disk. If this relatively weak inner disk reflection is the norm, it is much more difficult to detect when the signal-to-noise ratio is poor and/or the contrast with the continuum is weak, which it is for the broadest lines. An indication that this is occurring comes from the fact that for many of the observations in which no broad reflection is detected, reobservation of the same object with higher signal-to-noise ratio ends up revealing a broad emission line. 

We find evidence for a spinning black hole in three observations in the sample, but in no case is this very compelling and we conclude that the typical spectral quality of the data, in terms of signal-to-noise ratio, spectral resolution, and deblending from the continuum, is insufficient to determine the inner radius of the disk and hence the black hole spin. Tentative evidence is seen for sub-solar iron abundance in a few objects and this represents an additional possible explanation for the weakness or absence of broadened emission from the central regions. 

\subsection{Comparison with ASCA results}

The analysis of a number of the early \xmm\ spectra has raised a question of the consistency with the \asca\ data. The most obvious comparison work for the present study is with N97.  Their sample was smaller than ours, and excluded intermediate type Seyferts (e.g. 1.8 and 1.9).  Nonetheless, N97 found complex line emission in 14/18 objects (72 per cent), very similar to that found here. When modeled by a Gaussian, the parameters are also in excellent agreement.  For example N97 found $E_{\rm K \alpha}=6.34\pm 0.04$~keV, compared to $6.30\pm0.11$~keV found here. This agreement is all the more remarkable when one considers the fact that in the \xmm\ fits any narrow emission at 6.4 keV has been deconvolved, whereas in the comparable \asca\ fits it was not included. The agreement in the width is also good, with N97 quoting $0.43\pm 0.12$~keV compared to a value of $0.34 \pm  0.08$~keV found here. The main discrepancy is in the derived line equivalent width, where N97 found $EW_{\rm K \alpha} = 160\pm30$~eV, almost twice the value of 80 eV for the blurred reflector found here. This discrepancy can be attributed predominantly to the modeling of the narrow core as distant reflection in the present work. Based on the average value of $R_{\rm tor}$ in Table~\ref{tab:meanpars} this has $EW= 40\pm 7$ eV, and adding this to the average broad line EW this gives a total $EW=117 \pm17$~eV, consistent with the \asca\ value. A further reason for the observed discrepancy may be the fact that in the N97 work the effects of high ionization absorbers were ignored - this may have led to an overestimate of the blurred reflection component in some of the \asca\ fits. 

Nonetheless, when modeled by a simple Gaussian, the nature of the iron K$\alpha$ complexity seen in the \xmm\ data is both qualitatively and quantitavely very similar to that seen by N97 in the \asca\ spectra. We can therefore immediately dispense with the possibility that there are wholesale problems with the integrity of the data from either (or both) satellites. While there may be differences in the spectra of any given object (e.g. due to variability), any changes in the conclusions are due to differences in the modeling and interpretation. 

\subsection{Can complex absorption mimic the broad lines?}

For the majority of the objects in our sample, the interpretation of the spectra is relatively simple. In 14/26 objects (20/37 observations) the spectra are well described by a power law, possibly with a single-zone ionized absorber, plus distant and blurred reflection. The fits are acceptable, and improve no further when a variety of additional components are introduced. We have, however,  also found a very large number of additional components required to explain the remainder of the spectra. These include multiple zones of ionized absorption, and narrow emission and absorption components which can occur at various energies. All these phenomena have been reported previously for individual sources, and all appear to be required by some of the spectra in our sample. Following these procedures 34 of our 37 spectra give acceptable fits at the 99~per cent confidence level.  Despite the multiple complexities, the fits to MCG-6-30-15(1,2) and NGC 7469(2) remain unacceptable. It should be noted that these spectra have very high signal-to-noise ratio, so that any residual systematic uncertainty e.g. in the instrumental calibration may contribute to the poor fit. Also, while these spectra have formally poor fits, none is ``catastrophic" by our previous definition ($\chi_{\nu}^{2}>2.0$).

Perhaps more troubling for the present study is that there are several objects which show apparent evidence for strong gravitational effects which have very complex absorption, and in which it has been claimed in the literature that this absorption may reduce or obviate the need for broad emission. Specifically, these are NGC 3783, NGC 3516 and NGC 4151. We have, of course, attempted to account fully for complex absorption in our fits and still find evidence for strong gravity. We have taken a systematic approach to determining the nature and significance of these additional complexities but we have also restricted the range of allowed additional components, with an eye to physical self-consistency. This may bring our analysis into conflict with previous work. We therefore consider these objects as  ``case studies", typically the most complex spectra, and compare our best fit solutions with those presented previously in the literature. We thereby aim to address the issue of whether our systematic and semi-automated procedure for spectral fitting has come up with sensible results. Conversely, this also allows us to assess whether the more ad hoc approaches to spectral fitting which are typically followed for individual sources yield superior results to ours.  

Before discussing these individual cases, it is perhaps prudent to review the assumptions behind our modeling. First, we have considered only the time-averaged spectra in the present work. We have assumed that the narrow core of the line comes from very distant, optically thick material, such that it is narrow compared to the instrumental resolution and accompanied by Compton reflection. We also assumed that the reflection from the inner regions is from optically thick gas, in the form of an accretion disk, and hence it too is accompanied by a reflection component, and we have used a fixed emissivity law appropriate to a Newtonian geometry. In general we have assumed solar elemental abundances, although we did test the possibility that the reflector has non-solar abundances in \S\ref{sec:abun}. We have furthermore considered only a power law form for the underlying continuum. Finally, our XSTAR models are assumed to account adequately for spectral curvature induced by gas in the line--of--sight. We have furthermore made the assumption that the absorbers fully cover the line of sight. 

\subsubsection{NGC 3783}

NGC 3783 is a prime example of an object in which \asca\ previously reported a broad, accretion disk line (N97; George et al. 1998a), but where the evidence has been weakened based on higher signal-to-noise ratio \xmm\ data, which revealed a thick warm absorber components (Reeves et al. 2004). Ionized absorption unquestionably plays a key role in shaping the X-ray spectrum of this object, with high resolutions \chandra\ showing evidence for multiple zones of ionized gas (Kaspi et al. 2002; Netzer et al. 2003).  The Reeves et al. observation is the one we have designated NGC 3783(2). Our fitting procedure results in a very similar overall model to that found by those workers, in that we infer the existence of two zones of ionized gas as well as a strong H-like emission line. As noted by Reeves et al. (2004) there is also a strong He-like absorption feature in the spectrum, which is accounted for adequately in our {\it XSTAR} models. We also see evidence for an additional narrow emission feature at 6.52 keV. Regardless of whether that feature is included, we still find a very large improvement in the fit $\Delta\chi^{2}>100$ when a blurred Compton reflection component is included. Our conclusions are very similar to those of Reeves et al., in that we find that the disk reflection component is quite weak, with $R_{\rm disk}=0.5$. Comparing the two observations of NGC 3783, we find that both require blurred reflection. The best fit $r_{\rm br}$ for NGC 3783(1) is, however, non-relativistivistic ($\sim 250 r_{\rm g}$) while in NGC 3783(2) strong gravity is implied ($r_{\rm br} \sim 11 r_{\rm g}$; see Table~\ref{tab:kdbest}). Line profile variability between the two observations is a possibility. 

\subsubsection{NGC 3516}

Another similar example is NGC 3516, which also has a multi-zone ionized absorber in the line--of--sight (Kraemer et al. 2002; Netzer et al. 2002). Very strong evidence for a disk line was presented based on the \asca\ data, along with additional narrow features in the line complex (Nandra et al. 1999). The narrow features have been confirmed with \chandra\ and \xmm\ (Turner et al. 2002), but Turner et al. (2005) have suggested that very thick warm absorbers might be able to mimic some of the broader ``red wing" emission. Specifically, their suggestion was that a ``heavy" warm absorber component partially covering the X-ray source could account for the emission redward of the core. As we have not allowed for the possibility of partial covering of the warm absorber in our modeling this clearly accounts for the fact that we have not recovered this solution in our systematic fits. 

We have, however, tested such a model explicitly against the data for NGC 3516. For NGC 3516(1) we find that the ``heavy" partial covering warm absorber provides a similar quality fit to the spectrum, with $\chi^{2}=103.8$/93 d.o.f. compared to $\chi^{2}=100.5$/93 d.o.f for the blurred reflection model (both fits include two fully-covering warm absorbers and a distant reflector). For NGC 3516(2), we found an excellent fit ($\chi^{2} = 90.8$/94 d.o.f.) for the blurred reflector model E, with a single zone, fully covering warm absorber. This improved even further (to $\chi^{2}=76.5$) with the addition of a He-like absorption line. The ``heavy" partial covering model is substantially worse than these fits, with $\chi^{2}= 129.6$/92 d.o.f for the model favoured by Turner et al. (2005) which has two fully covering ionized absorbers in addition to the partial covering ionized absorber. Our simpler model, with just a single zone of ionized gas and a blurred reflector, is evidently a far superior fit to the second observation, at least over the limited bandpass considered in our work of 2.5-10 keV. A similar result was found by Turner et al. (2005), in that they found a highly significant improvement in the fit when adding a diskline to the ``heavy" partial covering model for this observation.

\subsubsection{NGC 4151}

A debate has existed for almost a decade over whether a relativistic line component is needed in this object. It was one of the earliest reported examples (Yaqoob et al. 1995), but due to the heavy and complex absorption in this object (e.g. Weaver et al. 1994; Kraemer et al. 2006) it has been very difficult to establish unambiguously whether it is present. In particular Schurch et al. (2003), who analysed the XMM-Newton observations we have designated NGC 4151(1) and NGC 4151(2), have suggested that the spectrum can be accounted for adequately with a non-relativistic line when Compton reflection is included, and the absorption is modeled with a combination of two fully covering absorbers: one neutral and one ionized.  
We also get an excellent fit formally with the Schurch et al. (2003) model to the NGC 4151(1) and NGC 4151(2) spectra (with $\chi^{2}=74.2/93$ dof and $\chi^{2}$=122.5/92 dof respectively). The former fit in particular is superior to our best fit (which has $\chi^{2}=93.7$/94 d.o.f.) and while we ultimately found a much better fit to the NGC 4151(2) (with $\chi^{2}=81.9$/85 dof) much of the improvement  can be attributed to the significant narrow line components in the source noted in Table~\ref{tab:addlines}. 

There are two substantial difference in the Schurch et al. interpretation compared to our analysis. Firstly, their model permits the core of the line to be somewhat broad. We find a strong requirement for this in our spectra, with FWHM $\sim 6,000$~km $s^{-1}$. While not relativistic, this width is clearly inconsistent with our assumption that the line core comes from very distant material in the torus. Additionally, in the Schurch et al. model the Compton reflection is decoupled from the 6.4 keV line emission. When taking this approach we find a requirement for extremely strong Compton reflection, with $R=6.9$ and $R=4.5$ respectively for NGC 4151(1) and NGC 4151(2). As we have tied the Compton reflection to the strength of the line to maintain physical consistency, we have failed to recover this solution in our modeling. 

The Schurch et al. (2003) model does not provide a good fit to the NGC 4151(3) observation (which was not included in their work), even if a significant core width is allowed and arbitrarily strong Compton reflection is included (in this case with $R=2$). The best fit to this model has $\chi^{2}=298.9$/93 d.o.f.  This can be compared, for example, to our model with two absorbers and a blurred disk reflector, which has $\chi^{2}=161.5/92$ d.o.f. While the latter is not formally acceptable (which we attribute to additional narrow components listed in Table~\ref{tab:addlines}, it is very much superior to the Schurch et al. (2003) model. It should also be noted that this observation has the highest signal--to--noise ratio of all the NGC 4151 observations, and hence should provide the strongest constraints.

Considered as a whole, our results are broadly in agreement with those of Schurch et al. (2003), in that we find no strong evidence for a relativistic line in the first two observations of NGC 4151, although they are consistent with having one. The final observation does require such effects and this time the data cannot be made consistent with the narrower line implied by observations 1 and 2. The main reason
for our conclusions is not the modeling of the absorption (in fact our absorption model is more complex than the one used by Schurch et al.) but the assumption that the core of the Fe K$\alpha$ line is narrow and that it is tied to the reflection continuum. 
 
\medskip

\noindent
Summarising the above case studies, while have always found evidence for broadened (and sometimes relativistic) lines in these objects with complex absorption, that conclusion depends quite strongly on the assumptions behind the modeling. This applies not only to the properties of the absorption (e.g. its ionization state, covering fraction and the number of zones), but also to the properties of the reflection. Thus, taking different approaches to the modeling of these phenomena can lead to different conclusions.  In addition, the curvature corresponding to the relativistic reflection is often relatively subtle (a good example being NGC 3783), making it all the more difficult to distinguish from other components, or small systematic effects in the data. 

An important caveat with our work is that we have restricted the bandpass of the spectral fits to a relatively small range, 2.5-10 keV. The lack of high energy data ($>10$~keV) may exacerbate the degeneracy between complex absorption and broad line emission, as an accurate determination of the direct and reflected continua is highly desirable when modeling the lines (e.g. Reynolds et al. 2004b). 
In this regard, it is interesting to consider recent, high signal to noise, broad-band X-ray spectroscopy of AGN with Suzaku, which has helped clarify our understanding of the origin of the broad iron line. While Suzaku has similar sensitivity as XMM-Newton at iron K, the high energy coverage of the Suzaku Hard X-ray Detector (HXD; Takahashi et al. 2007) provides a direct measurement of the strength of the reflection hump above 10 keV. Furthermore the extended hard X-ray bandpass out to $\sim100$\,keV gives tighter constraints on the intrinsic continuum form of the AGN, breaking the degeneracy between the continuum slope, complex absorption, reflection and the broad iron line inherent in data of more limited bandpass up to 10 keV (see Reeves et al. 2006 for a recent review).

The initial results from Suzaku appears to provide robust verification of the presence of relativistic iron lines in AGN, confirming our own results; examples include MCG -6-30-15, (Miniutti et al. 2007), MCG -5-23-16, (Reeves et al. 2007), 3C 120 (Kataoka et al. 2007), NGC 3516 (Markowitz et al. 2007). The broad iron line emission in these AGN appear to originate from a few to several tens of gravitational radii, confirming the findings in this work, while strong (R$\sim1$) reflection components are directly detected, likely arising from a combination of disk and torus emission. In the complex case of NGC 3516, the broad band Suzaku observations are able to distinguish between layers of high column density ionised absorption (with $N_{\rm H}\sim10^{23}$\,cm$^{-2}$, Markowitz et al. 2007) and the disk reflection component, showing that both complex absorption and blurred iron line emission are present in the AGN spectrum.

\subsection{The broad emission}
\label{sec:blines}

While we have established reasonably robust evidence for broadened emission in a large number of our sample objects, the details of this emission merit comparison with the simple physical picture on which we have based our spectral modeling. One assumption is our chosen emissivity law, which is appropriate for a semi-infinite, flat disk illuminated by a point like X-ray source in a Newtonian geometry, located above the disk on its axis of rotation, which we will describe as the ``lamppost" geometry. The emissivity law should nonetheless be a reasonably approximation to a disk illuminated by flares above its surface. We furthermore assume that the disk is in a neutral or near-neutral ionization state. The permitted free parameters in this model are only two, these being the height, $h$, of the illuminating source and the viewing inclination $i$. In reality we have also allowed the strength of the reflection to vary as a free parameter $R_{\rm disk}$. Furthermore, we have translated the height of the X-ray source into a characteristic radius $r_{\rm br}$ by making an approximation to the expected (Newtonian) emissivity law in this ``lampost" geometry. These two modifications to the simple lampost picture make the model somewhat more general, at the expense of strict physical self-consistency. In our parameterization the free parameters $R_{\rm disk}$ and $r_{\rm br}$  then encapsulate all the uncertainties about the precise geometrical arrangement of the disk and X-ray source. 

We have shown above that this model performs very well in terms of fitting the spectra, which must be considered very encouraging given the restricted number (3) of free parameters compared to standard disk line models often employed in the literature, which may introduce as many as 6 or 7 additional degrees of freedom. On the other hand, the assumptions in our model will surely be violated at some level in a real system. For example, the illumination pattern and emissivity of the disk will be affected by general relativistic effects in the vicinity of the black hole, the geometry of the X-ray source may differ from our assumptions above, the disk may well be significantly ionized, inhomogeneous etc. Nonetheless the typical parameters of the broad emission give important information regarding the nature of the central regions. 

Starting with the inclination $i$, the expectation value for the mean based on our best-fits to the spectra (i.e. models E+F) is $38 \pm 6$ degrees. This is similar to that found with \asca\ by N97, who interpreted this low average inclination (compared to a random distribution, where one expects $<i>=60^{\circ}$), as being due to selection effects. More specifically, because the N97 sample consisted only of Seyfert 1 galaxies, edge-on systems will be selected against. A similar effect should pertain here, in that we have excluded bona fide Seyfert 2s from our sample, although it ought to be less prevelant because we have at least allowed for intermediate type 1.8-1.9 Seyferts to remain in the sample. Some caution must be exercised in interpreting the mean inclination for the sample, however. Because the expected line emission in highly inclined disks is both weaker and broader than face-on disks, the inclination is likely to be poorly constrained. This will tend to push the average value to lower inclinations. 

The average break radius $r_{\rm br}$ in our best-fit models is found to be approximately $20 r_{\rm g}$, although we note that in only 8 observations is the break radius constrained to be $r_{\rm br}>6 r_{\rm g}$ (see Tables~\ref{tab:kdpex} and \ref{tab:kdbest}). While the precise value for $r_{\rm br}$ will depend on the true emissivity law and geometry of the disk, an advantage of our approach to the modeling is that, regardless of the true physical nature of the system, $r_{\rm br}$ should always represent approximately the characteristic radius at which the peak of the blurred reflection is emitted. At $20 r_{\rm g}$, strong gravitational effects are measurable, and hence our observations confirm the predictions originally envisioned in the theoretical models proposed by e.g. Fabian et al. (1989), Stella (1990) and Laor (1991). In other words, many of the spectra show evidence that blurred emission from very close to the black hole is being observed, enabling us to attempt to diagnose the structure and nature of the inner flow using the reflection signatures. As we have emphasized, this conclusion, at least for {\it some} of our sample sources is robust to the effects of distant reflection and ionized absorption. The conclusion should also remain regardless of whether the simple geometrical model we have assumed for the inner disk reflection is correct. 

A clear indication that the simple, Newtonian lamppost model does {\it not} adequately describe the disk-corona system in AGN is the fact that the derived reflection fraction $R_{\rm disk}$ is significantly lower than that expected in the lampost geometry. We find $\Omega/2 \pi \sim 0.5$, so the blurred reflection is approximately half the expected strength. The corresponding average equivalent width for the broad component is approximately $65 \pm 15$~eV. In addition, we find a large fraction of objects (Fig.~\ref{fig:nightmare}) with no discernible broadened reflection at all. That the accretion disk reflection is often weak, and sometimes absent is a key finding of our study, and the analysis of this phenomenon in a large sample places this conclusion on a robust statistical footing for the first time. As we have discussed previously, the finding of weak reflection in any individual object has little if any power to constrain models, as any individual case can be explained by a high inclination or variability (see below). Our observations (see Fig.~\ref{fig:nl_cont}) show clearly that if $R=1$ there would be too many very high inclination systems, considering that edge--on, type 2 Seyferts have been excluded from the sample. On the other hand, the observation of a weak average disk reflection component and the apparent absence of such a component in some cases may be related, in that Fig.~\ref{fig:nl_cont} shows that reflection with $R=0.5$ can almost never be excluded, even if the inclination is relatively low. 

\subsection{Why might the broad emission be weak or absent?}

A clear conclusion of our work is therefore that the very simple system we have assumed - that of a point source illuminating a flat, neutral accretion disk around a Schwarzschild black hole - does not adequately describe the sample spectra. Here we explore possible modifications to that simple picture which might account more fully for both the typical parameters of the broad reflection, and the range of spectral properties we see in the sample. 

\subsubsection{Geometrical effects}

The lampost model tends to maximise the observed reflection because in that geometry the disk sees the whole of the X-ray emission and furthermore the observer sees the whole of the disk. Other possible geometries include those with a central, spherical source and a cool disk surrounding it (e.g. Zdziarski, Lubinski \& Smith 1999). In that model the solid angle subtended by the disk at the X-ray source is less than $2\pi$, so the reflection is correspondingly weaker. Models where the X-ray source is embedded in a system of cloudlets also produce weaker iron K$\alpha$ emission for a given covering fraction, because the reflecting clouds can self-cover (e.g. Nandra \& George 1994). There would be a similar expectation if the hot, X-ray emitting plasma were embedded with the cool gas in an inhomogeneous accretion flow. 

\subsubsection{Abundance effects}  

An even simpler explanation for the apparent discrepancy between the expected and measured strength of the observed reflection is that our assumptions about the iron abundance are not correct. Because the most prominent feature in the reflection spectrum is the iron K$\alpha$ line, this can drive the fits and if the iron abundance is low, the derived $R_{\rm disk}$ value could be underestimated. We would still expect to see significant Compton reflection at high energies in the spectra, however, and possible examples of this in our sample are those which show ``concave" spectra. An excellent example is NGC 7469(2), which shows little evidence for broad emission around the iron K-complex, but which gives a very substantial improvement with a blurred reflector. The majority of the improvement in the fit is due to spectral curvature characteristic of reflection. We have also tested for abundance variations explicitly in our fits described in \S~\ref{sec:abun}, where we found apparent evidence for sub--solar iron abundance in a number of the spectra. A more general implication of these fits is that there are some objects in which we see Compton reflection without the expected iron K$\alpha$ strength associated with it.

\subsubsection{Extreme relativistic blurring} 

The confidence contours shown in Fig.~\ref{fig:nl_cont} indicate that, even in sources with no evident broad iron K$\alpha$ emission, a significant reflection component is nonetheless permitted if the accretion disk is highly inclined. This is because in an edge on disk the line is extremely broad, and is very difficult to distinguish from the continuum. A similar effect is expected in an extreme Kerr disk, where the emission is strongly concentrated in the inner regions. Once again, while the line emission from such a disk would be difficult to detect, reflection should still be present so this can be distinguished from some other possibilities with top quality high energy data. Compared to the low abundance case we expect the iron features --both emission line and absorption edge -- to be stronger, but broader. Distinguishing between these requires very high signal--to--noise ratio.  

\subsubsection{Anisotropy}

The predicted strength of the reflection depends on the observed continuum, with $R_{\rm disk}$ also representing the ratio of the observed power law emission to that seen by the disk. The simplest assumption is that the X-ray emission is isotropic, but this need not necessarily be the case. For example, Beloborodov (1999) has suggested that the direct emission might be beamed away from the disk by motion of hot plasma responsible for the X-rays. Strong gravitational effects are also expected in the central regions and these are expected to affect the incoming X-rays, as well as the outgoing reflection component. This has been discussed in detail by Miniutti \& Fabian (2004), who have explored several scenarios, concluding that the reflection can sometimes be weaker and sometimes stronger, than in the standard non-relativistic case. If the disk sees a different continuum than the observer, then it is very difficult to make clear predictions about the strength of the reflection continuum without knowing the detailed geometry of the inner regions. The weak relativistic reflection seen on average for our sample suggests that, if anything, the continuum emission is being more strongly beamed towards us than it is towards the disk. Specific models can be tested in principle, because in a given scenario the relativistic effects should predict the emissivity law in addition to the reflection strength. 

\subsubsection{Ionization} 

The intense illumination of the primary X-ray source will inevitably cause the disk surface to become ionized. In the most extreme case, the ionization could be sufficiently intense that iron becomes fully stripped, meaning that our primary diagnostic of the inner accretion disk (i.e. the broad iron K$\alpha$ line) is suppressed. Once again, in the default scenario where a ``lampost" X-ray source illuminates an $\alpha$-disk, the accreting material should be sufficiently dense that the disk surface can  plausibly be maintained in a low state of ionization, as has been assumed in this work. Numerous authors have pursued the idea that the disk might become significantly ionized, however (e.g. Ross \& Fabian 1993; Zycki et al. 1994). This might be expected in scenarios where the X-ray emission originates very close to the surface of the disk, such as magnetic flare models. In this case, even though the majority of the (optically thick) accretion disk maintains a low state of ionization, locally an ionized ``skin" can develop. The physical properties, in particular the ionization state of the this skin will control the observed reflection features. As discussed by e.g. Nayakshin \& Kallmann (2001), to maintain hydrostatic equilibrium, an instability exists such that arbitrary ionization states are unlikely. If the disk surface is ionized significantly beyond neutrality, it ought to be in the relatively stable Helium-like or Hydrogen-like states, or by ultra-ionized where iron is completely stripped. We have typically found that a low ionization disk provides an adequate description of our spectra, but where the broad iron line is weak or absent, a strongly ionized disk is a possibility. Again, Compton reflection should persist in the spectra, but if iron is fully stripped there will be no associated K--edge, nor any significant opacity below $\sim 10$~keV. Fully ionized reflection spectra would therefore show essentially no signatures below $\sim 10$~keV and are detectable only via subtle spectral curvature caused by Compton downscattering. 

\subsubsection{Hot inner accretion flow} 

A related scenario to that  just described is one in which the accretion flow is intrinsically hot. 
A basic assumption behind our modeling is that the black hole is fed by a standard accretion disk of the kind envisioned by, e.g., Shakura \& Sunyaev (1973), i.e. one in which the inner accretion flow is dense, optically thick, geometrically thin, relatively cool ($T_{\rm disk} \sim 10^{5-6}$~K) and radiatively efficient. Beginning with the work of Shapiro, Lightman \& Eardley (1976), the possibility that the inner flow might be hot ($T_{\rm disk} \sim 10^{9}$~K) has been raised, and this would clearly predict radically different reflection signatures than a standard ``$\alpha$-disk". This idea has been reinvigourated by Narayan \& Yi (1994), and subsequent work in which a hot and radiatively inefficient ``advective" flows can plausibly feed the back hole. Quite how relevant such scenarios are to the luminous AGN that are the subject of this study is not obvious. If, however, the inner accretion flow is indeed hot then we would not expect to see any Compton reflection from the inner regions at all.  This might account for the lack of obvious broad reflection features in some of the spectra, and hence the low average value for $R_{\rm disk}$. This model clearly cannot be applied generally to the sample because there are so many objects in which broad reflection {\it is} observed, unless a standard cold disk co-exists with the hot flow, or is present beyond a certain small transition radius. Effectively, then, this model becomes equivalent to one in which the geometry differs from the lampost model as, e.g., in the model of Zdziarski et al. (1999). 

\vspace{0.05in}

\noindent
The key to understanding which, if any, of the above models accounts for the lack of and/or weakness of the inner disk reflection in some cases is the comparison between the iron K$\alpha$ emission properties, and those of the reflected absorption edge and high energy Compton reflection continuum. The fact that at least some objects apparently show evidence for reflection stronger than -- or even without -- the associated line emission may hold the key to the interpretation of the weak and/or absent broad reflection. Of the options discussed above, the most likely options are sub-solar iron abundance, disk ionization, and geometrical variations. Unfortunately, due to the limited bandpass of the XMM-Newton spectra ($<10$~keV) it is in general difficult to constrain the strength of the Compton reflection continuum independently of the iron K$\alpha$ line. The possible models can only therefore be tested with high quality hard X-ray data. It is therefore interesting to note that recent data from {\it Suzaku}, which as we have mentioned cover a broader energy range including sensitivity to hard $\sim 20-30$~keV X-rays, show that at least one object in our sample, MCG-5-23-16, has evidence for a sub-solar iron abundance (Reeves et al. 2007). 

\subsection{Variability}

\begin{figure}
\includegraphics[angle=0,width=90mm]{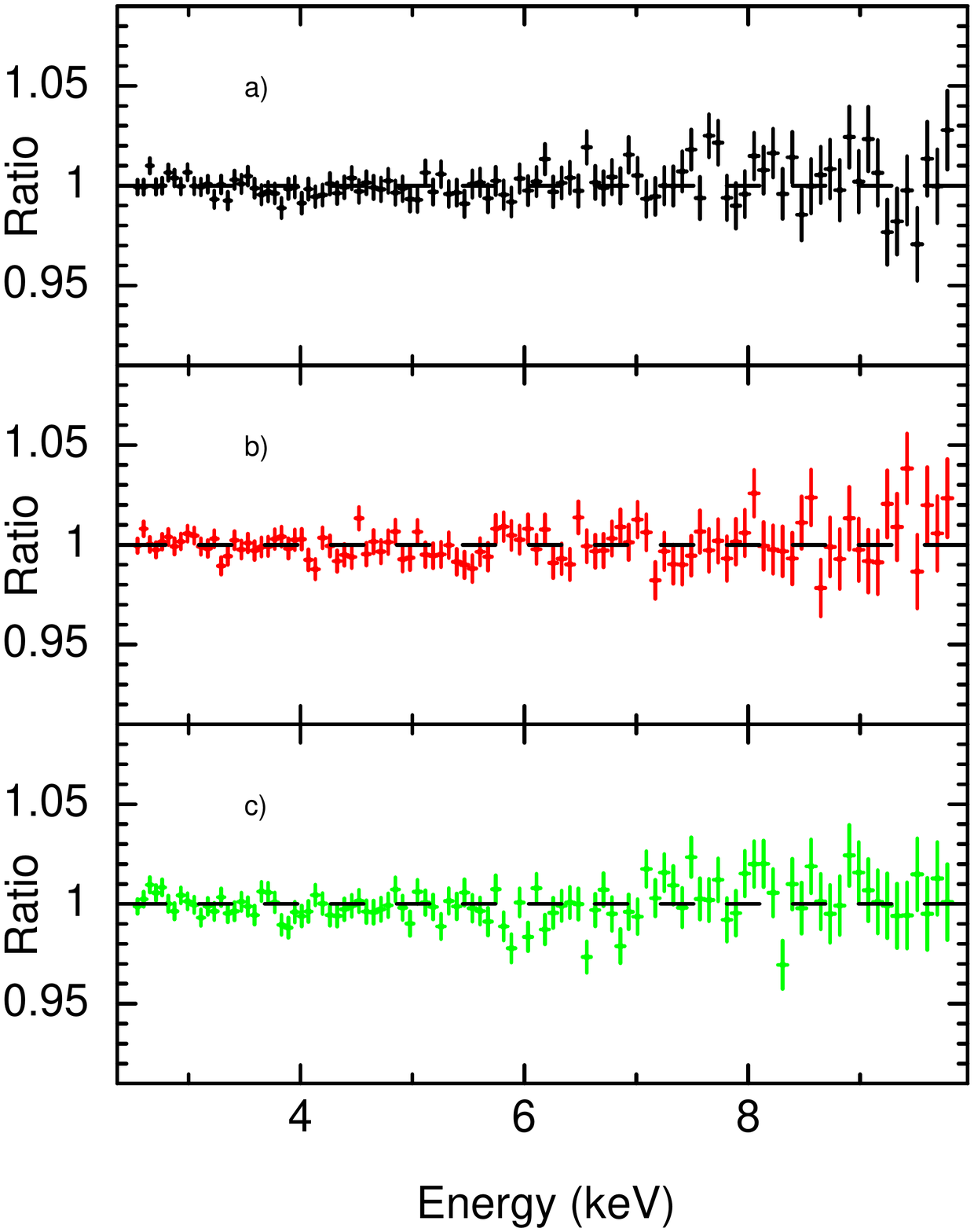}
\caption{Data/model ratios for simulated spectra with a variable power law index. To create the simulated spectra, the power law index is allowed to vary in the range $\Gamma = 1.7-2.2$ and the flux by a factor two. The simulated data were then fitted with a single power law, ignoring the range 4.5-7.5 keV, as in Fig. 1. Three cases are shown: a) where the index is independent of flux b) Where the index is positively correlated with the flux and c) where $\Gamma$ is negatively correlated with flux. In no case does a broad residual similar to a blurred reflection component appear falsely due to the spectral variability. 
}
\label{fig:varpl}
\end{figure}

Because we have only considered integrated spectra in our study, and because AGN are notoriously variable objects in the X-ray band, it is important to assess what effects variability might have on our results. We first dispense with a commonly--held misconception, that a variable power-law index might artificially induce broad residuals around the iron line. In Fig.~\ref{fig:varpl} we show residuals to synthetic spectra of very high signal-to-noise ratio created with a model of a simple power law, where the continuum varies in shape. We allow the index to vary in the range $\Gamma=1.7-2.2$ and the flux to vary by a factor $2$, and consider three cases: a) where the index varies independently of the flux b) where $\Gamma$ is positively correlated with the flux and covers the full range within the factor 2 variations c) as in b) but where $\Gamma$ is anti-correlated with the flux. In no case is there a broad residual observed in the spectraum of the kind characteristic of accretion disk reflection (c.f. Fig~1).  

Nonetheless, largely orthogonal constraints can be obtained from consideration of the variability of the emission lines in AGN. Several remarkable results have already been obtained by \xmm\ and previous satellites on line variability, but these have not revealed a clear or coherent picture. It is often stated that the variability data indicate that the inner disk reflection component in AGN is constant. In reality, this is a gross oversimplifcation based largely on the observations of MCG-6-30-15, which do indeed show this evidence that the variability of the inner disk reflection is weak (e.g. Fabian et al. 2002; Miniutti \& Fabian 2003). For other sources, even prior to \xmm, there were several reports of line flux and profile variations (e.g. Yaqoob et al 1996, Nandra et al. 1996, 1999, 2000), including in MCG-6-30-15 (Iwasawa et al. 1996, 1999; Vaughan \& Edelson 2001). \xmm\ has provided many further examples (e.g. Turner et al. 2002;  Miller et al. 2006). Of particular interest are cases where the ``red wing" of the iron line is seen to vary independently of the continuum (Vaughan \& Fabian 2004; Ponti et al. 2004; Tombesi et al. 2007), which provides robust evidence for an accretion disk origin for this feature and rules out absorption interpretations.  Another interesting case is that of Mrk 841, which has been demonstrated to show a narrow line at some epochs and a broad one at others (Petrucci et al. 2002; Longinotti et al. 2004; Petrucci et al. 2007). Were Mrk 841 in our sample, our conclusions about whether or not it possesses a broad reflector would then depend on which epoch we included in the analysis. 

Where we have multiple observations of the same source these seem to indicate that this kind of profile variability might be extremely common. Consider the observations shown in Fig.~\ref{fig:nightmare}, which show no evidence for blurred reflection. In all cases where there is a second observation of the source, that second observation {\it does} show evidence for velocity broadening of the reflection component. In addition, we find several cases where a broad, but non-relativistic reflection is required in one observation, but another shows evidence for relativistic or strong gravity effects. The trend in these multiple observations is clear, in that observations with higher signal--to--noise ratio always show stronger evidence for broad, or relativistic effects. One example is NGC 4151 where, as discussed above, observation 3 has very strong requirement for a relativistic line but observations 1 and 2, which have lower signal--to-noise ratio, do not. 

Another interesting example of similar nature is NGC 7469, where one observations shows a narrow line, but the second observation with higher signal--to--noise exhibits strongly blurred reflection, possbly even indicitive of a Kerr black hole.  We can explain the apparent lack of broad line in the lower S/N case by the fact that the reflection component is very strongly blurred and therefore very diffficult to detect. Indeed the reflection component in this case is apparently so strongly blurred that it is very difficult to tell whether it is truly reflection from the inner disk or a curved continuum. What is clear, however, is that data with ultra-high signal--to--noise ratio is required to detected such components. The component required in the second observation, which has 360,000 counts, is not detected in the first, which has 80,000, but the best--fit parameters of the reflection in the 80,000 count spectrum are entirely consistent.  Another interesting case is NGC 5506. This shows strong emission to the blue side of line. A question therefore arises as to whether the emission above 6.4 keV is due to Doppler blueshifting in an inclined disk, or a blend which includes helium--like and/or hydrogen--like components.  For the two observations, we find that a blend is prefered for observation 1, but an inclined disk is a better fit to observation 2. Here the observations have very similar signal--to--noise ratio, so it is difficult to say which interpretation should be preferred. We note, however, that when the line blend is allowed in observation 1, the parameters of the blurred reflector, while not signficant, are almost identical to those inferred for observation 2. So while, once again, there is no requirement for blurred reflection in the first observation, consistency between the two observations can be obtained if such a component is allowed. A final noteworthy example is NGC 5548, where the first, short observation shows only a narrow line. The much longer second observation reveals a broad, but non-relativistic line. It is interesting to note that an earlier \asca\ observation showed evidence for a relativistic emission line (N97; Yaqoob et al. 2001).

\medskip
\noindent
The apparent profile variability seen in a large fraction of the sample provides a natural explanation for the existence of observations in which no broad line is seen at a given epoch. The existence of such variability is not, however, predicted in the standard model and for the epochs without broad emission one must therefore rely on one of the explanations in Section~\ref{sec:blines}. Light-bending and/or anisotropic models are of particular interest in this regard, as they allow for flux and profile variability, in addition to departures from the simple assumed geometry, both of which we infer from our sample analysis. 

\subsection{The narrow 6.4 keV line}

In our modeling, we have associated the line core with very distant material, so that it is unresolved at the EPIC-pn resolution.   This assumption is supported by high resolution data from Chandra HETG, which show in general both that the line cores are unresolved, and they they are accompanied by a Compton shoulder  (e.g. Kaspi et al. 2001, 2002; Yaqoob \& Padhmanhaban 2004). Based on current data, an origin for the line core in, e.g. the optical broad line region - which would be both very mildly broadened, and not necessarily optically thick -- therefore seems unlikely (Nandra 2006) but it cannot be ruled out especially in individual cases. 

While the blurred reflection is found to be fairly common in the sample, narrow distant reflection is found to be almost ubiquitous. In the few cases where the distant reflection is not significantly detected, typically it is consistent with the average value of $R_{\rm tor}=0.4$. Taken at face value this $R$ value implies that the torus covers a solid angle of $\pi$ at the X-ray source. Assuming the disk covers half the sky (thus rendering the far side of the torus invisible) this implies a half opening angle for the torus of $\sim 50$~per cent.  For the assumed inclination of $60^{\circ}$, this reflection fraction can be translated into an approximate average equivalent width for the torus line of $40\pm 7$~eV. 

The detection of this narrow component offers strong support for standard AGN unification models (e.g. Lawrence \& Elvis 1982; Antonucci \& Miller 1985). In these, an optically thick material at a distance of $\sim 1 $~pc is  present in all AGN, including type Is. If the torus has a significant covering fraction then fluorescence emission should produce a line at 6.4 keV (Krolik \& Kallman 1987; Ghisellini, Haardt \& Matt 1994; Krolik, Madau \& Zycki 1994), with a velocity width of $\sim$few $\times 100$ km s$^{-1}$. This is unresolved at the EPIC resolution, and so is consistent with our data. The major alternative origin for the narrow core is in the optical broad-line region (BLR). Any iron K$\alpha$ emission from this region could (barely) be resolved with \xmm, but as has been mentioned a BLR origin for the Fe K$\alpha$ core is strongly disfavoured by the higher resolution Chandra HETG data (Yaqoob \& Padhmanhaban 2004; Nandra 2006).

The strength of the narrow emission line is at the low end of that expected in standard models, which predict the EW from the torus to be 50-100 eV for unobscured lines--of--sight and typical parameters. There are two possible explanations. First, as with the broader emission, the geometry could differ from that assumed in the models. In particular, the line EW is expected to be a strong function of the geomtry of the inner edge of the torus (Krolik et al. 1994). Alternatively, we might again be seeing evidence for a sub-solar iron abundance. Unless the metalicity of the torus is very low, our observations would tend to favour a relatively small covering fraction for the the torus of $\sim 50$~per cent. In this scenario, one would anticipate a ratio of type I to type II Seyferts of 1:1. This is reasonably consistent with new models for the synthesis of the X-ray background (Gilli, Comastri \& Hasinger 2007). At first glance, it may appear to be inconsistent with observations of local AGN, for which it is oft-quoted that the ratio for type 2 to type 1 AGN is 4:1, based on the work of Maiolino \& Rieke (1995). In fact, if one counts intermediate Seyfert types (e.g. types 1.8 and 1.9) as type 1, the ratio is close to 1:1, in agreement with our deduction from the iron K$\alpha$ line. 

\subsection{Other narrow line components}

\subsubsection{He- and H-like lines}

We have found evidence for narrow emission lines from Fe {\sc xxv} and {\sc xxvi} in a number of the spectra. One object, NGC 5506, shows an emission line from He-like iron and a further two NGC 3783 and NGC 4593 show emission from H-like iron. These features are also predicted in unification schemes (Krolik \& Kallmann 1987). The most likely scneario is that they are emitted by hot gas filling the torus, which is responsible for scattering nuclear BLR lines in to the line--of--sight in some Seyfert 2s (Antonucci \& Miller 1985). While we see the narrow 6.4 keV line in essentially all objects, however, these ionized lines appear to be much rarer. Overall, this implies either that the optical depth of the hot reflecting gas is quite small ($\tau < 0.01$) or that the gas is very highly ionized, so that even He and H-like species are absent. 

In three observations of two objects, NGC 3516(2) and MCG-6-30-15(1,2) we see evidence for {\it absorption} by the He-like species. This implies a very highly ionized zone of gas in the line--of--sight which is not well modeled by our XSTAR models. The inability of our absorption models to account for this feature in these two cases is perhaps unsurprising, given that the strength of absorption lines in the spectra is heavily dependent on the turbulent velocity of the gas. We have assumed this to be 100 km s$^{-1}$ in our models, but there is no reason to think that this might in reality be larger and hence the features stronger. Because we have not modeled the absorption features from the ionized gas in these observations, there will always be some doubt as to the exact broad iron line parameters we have derived, and this should be borne in mind when these parameters are interpreted. As we have mentioned, there is at least one object, NGC 3783, which shows a highly significant He-like absorption feature which {\it is} fully accounted for using out XSTAR models. We note that a local origin for these features  (e.g. McKernan, Yaqoob \& Reynolds 2005) is strongly disfavoured as the fits worsen substantially if the redshift is fixed to $z=0$, rather than the appropriate systemic value. 

\subsubsection{Intermediate Fe emission features}
\label{sec:intermed}

We see a total of 7 objects in which the fit shows a significant improvement when a narrow emission line {\it between} the neutral and He-like lines is introduced. This implies the existence of gas which is intermediate in ionization state. Such lines are unexpected for two reasons. Firstly, gas in these ionization state tends to be unstable. There should be a strong tendency to find gas either at the blackbody temperature, in which case iron will be close to the neutral state, or the Compton temperature in which it would be in the He-like state or even above.  Secondly, it has been suggested that the fluorescence lines would be resonantly trapped, and hence never observed to be strong (Ross \& Fabian 1993). Such lines are definitely detected in our sample, with a good example being Mrk 590 which has a prominent, narrow 6.5 keV line. In other cases, it is not clear that the intermediate lines are real or whether we are not quite modeling the disk lines correctly. Mismodeling of the broad line complex could occur very easily, especially as we have heavily restricted the allowed parameter space of models, and in particular have a very tight prescription for emissivity. Higher resolution spectra are needed to see if these lines are truly narrow or if they are parts of the disk line. A further complication is that in the XSTAR absorption models, narrow {\it absorption} lines around these energies are often seen. Because we do not know the precise parameters (e.g. turbulence velocity) of the ionized absorber, there may be cases in which these absorption lines are over-estimated (c.f. discussion of NGC 3516 and MCG-6-30-15 above). This, in turn, might lead to an overestimate of the strength and significance of any intermediate emission line, which can compensate for the excess absorption. 

\subsubsection{Redshifted absorption features}

Nandra et al. (1999) presented the first evidence for a redshifted absorption feature in the \asca\ X-ray spectrum of NGC 3516. These authors interpreted this as evidence for inflowing gas. Such absorption features represent a powerful potential probe of the inner regions of AGN and so, if confirmed, are very exciting.  A number of other recent reports now exist in the literature (e.g. E1821+643 Yaqoob et al. 2005; PG 1211+143 Reeves et al. 2005; Mrk 509 Dadina et al. 2005; Mrk 335 Longinotti et al. 2006). The detection significance  (2--3$\sigma$) of such features is usually quite marginal, however, leaving room for doubt as to whether they are real or not. We find just one  case in which we infer a significant redshifted absorption line -- Ark 120. The same feature was noted, but not considered signficant, by Vaughan et al. (2004). It should also be noted that the previously--reported absorption feature in the \asca\ spectrum of NGC 3516 (Nandra et al. 1999) is not detected in the higher quality spectra of the same source which we report here. The same comment applies to Mrk 509 (Dadina et al. 2005), with the absorption feature being detected only in the statistically-inferior BeppoSAX observation. While it is therefore tempting to interpret these feature as evidence for infalling material we must await confirmation of these results before opening a potentially new window on the study of accretion flows at small radii.

\subsubsection{Blueshifted absorption features}

We detect absorption features at around 7.5 keV in two objects,  IC4329A and NGC 4151. These features are difficult to interpret, as if there is only a single line the identification will always be ambiguous. Unless substantial velocity shifts are invoked, the most plausible identification with Fe K$\beta$, but this should always be accompanied by a much stronger K$\alpha$ features. As these are not typically observed they would need to be filled in with emission, which seems very contrived as one would have to fill in the K$\alpha$ line but not the K$\beta$. If, on the other hand, we identify this as a K$\alpha$ feature the most likely identification is with the H-like ion Fe {\sc xxvi}, as only a single feature is evident. In this case a very large blushift of order 0.1c must be invoked. Similar features have been reported previously, with the interpretation of a very fast outflow being preferred (Chartas, Brandt \& Gallagher 2003; Pounds et al. 2003; Reeves, O'Brien \& Ward 2003; Braito et al. 2007).  Taking our two cases, there is some ambiguity about the feature in NGC 4151, which may in reality be an iron K-edge, rather than an absorption line. Thus the only apparently robust case is  IC4329A, the feature in which has already been reported by Markowitz et al. (2005). As with the redshifted features, it would be very reassuring if these features were confirmed in repeat observations, before going too far in interpreting them. 

\subsubsection{Shifted emission features}

Turner et al. (2002) first reported evidence for shifted, narrow {\it emission} features, again in NGC 3516. These were identified with hot spots in the accretion disk, which if restricted to a narrow azimuthal range could be narrow but significantly Doppler and gravitationally shifted. There have been several additional reports (e.g. Porquet et al. 2004; Turner et al. 2004) and there have even been sugestions that the features move in energy in a periodic fashion (Iwasawa et al. 2004; Turner et al. 2006), providing the tell-tall signature of the rotation of the accretion disk. We find just one object in which a significant, shifted and narrow emission line is seen, this being NGC 4151. Two features were found in two separate observations, one at 3.7 keV in NGC 4151(2)and the other at 5.23 keV in NGC 4151(3). These observations have extremely high signal--to--noise ratio, so there is a danger that we may be interpreting small, systematic effects in the spectra. Furthermore, the 5.23 keV in particular may once again be associated with the broad accretion disk line, which has not been modeled adequately. Taken at face value, however, these features could be associated with hotspots in the disk of the kind envisaged by Turner et al. (2002). We note again, however, that many of the objects in which there have been previous reports for such phenonena, but which also form part of our sample (e.g. Mrk 766, NGC 3516), show no evidence for significant shifted features in our analysis.  This may in part be due to the more stringent criteria for statistical significance applied in the present work. As with the redshifted absorption features, however, another explanation is that the features are variable or transient. If so , they may not be revealed in a given observation, or in our integrated spectra at all, as if they are short lived they may be ``washed out" over the course of an entire observatation. If this is so then we may expect these narrow lines to integrate out into the broad emission line observed in so many of our spectra. 

We note finally that we find no evidence in any of our objects for narrow, blueshifted emission features of, e.g., the type reported by Yaqoob et al. (1999). 

\subsubsection{Underlying continuum}

We find a mean spectral index for the sample of $\Gamma=1.86\pm 0.06$. This can be compared to the values derived by Nandra \& Pounds (1994) from the {\it Ginga} observations of $\Gamma=1.95 \pm 0.05$ and N97 for the {\it ASCA} sample of $\Gamma = 1.91 \pm 0.07$ and is clearly in excellent agreement with both these values. Taking the weighted mean of all three samples we find $\Gamma = 1.91 \pm 0.03$, so $\Gamma=1.9$ can be taken as a useful representative value for the typical spectral index for nearby Seyfert galaxies. We find a significant intrinsic dispersion in the spectral indices of  $\sigma=0.22 \pm 0.05$, which can be compared to the {\it Ginga} and \asca\ values of $0.15\pm 0.04$ and $0.15\pm 0.05$, respectively and again entirely consistent. The fact that the derived $\Gamma$ values are consistent in all these three works gives us some confidence that we have not grossly mismodeled the underlying continuum, thus leading us to erroneous conclusions about the iron features. 

\section{Future Prospects}

Since the original prediction of accretion disk lines by Fabian et al. (1989), the hope has long been held that future missions might be able to exploit their diagnostic power to investigate the detailed working of the central engines of AGN.  In principle the detailed study of the lines can yield measurements of the black hole mass and spin, the geometry and physical conditions of the inner accretion flow, and of the X-ray source. In the most  optimistic case, the central regions of AGN might offer a useful laboratory in which to observe test particles moving in orbit close to the central black hole, which might provide an opportunity to test theories of gravity in the strong field. 

It is now clear that this potential has not yet been fully realised at least, by \xmm, which has revealed a level of complexity in the spectra that in many cases makes it extremely challenging to tease out the broad emission at all, let along determine the parameters of the accreting black hole system. On the other hand, our work has shown robust evidence for emission close to the central hole in a large fraction of our sample sources. With sufficiently high signal--to--noise ratio and spectral resolution from the next generation of X-ray observatories, further progress is possible, and indeed expected. Our work, and that of Guainazzi et al. (2006), has demonstrated that very high statistical quality is required for the data to be able to firmly rule in -- or out -- the presence of a broad, accretion disk reflection component. It is also clear that a bandpass which encompasses higher energies that \xmm, $>10$ keV, would be extremely helpful in disentangling the reflection components -- which apparently arise from both nearby and distant material -- from each other and from the broad iron K$\alpha$ line. Indeed, as we have just discussed, such observations arguably hold the key to determining why we see broad emission in some objects but not others, as they can distinguish between the various competing scenarios (e.g. geometry, ionization, abundance). 

Potentially the most exciting results are those already mentioned, where apparent periodic variability is seen (Iwasawa et al. 2004; Turner et al. 2006). Future variability observations can place all these results on a more sound footing via the use of more advanced statistical techniques, such as Fourier resolved spectroscopy (Gilfanov, Churazov \& Revnivtsev 2000; Papadakis et al. 2005, 2007) and principal component analysis (Milller et al. 2007), both of which are already yielding interesting results which are complementary to static spectral analysis such as that reported here.

Finally, a severe limitation of our present work is the relatively modest spectral resolution of the EPIC-pn cameras. In the presence of narrow features due to reflection, transmission and absorption, which we infer to be present in nearly all objects, higher resolution would be extremely helpful in devonvolving the broad emission. It is particularly needed in determining accurately the contribution from distant reflection, via observations of the narrow Fe K$\alpha$ core and the associated iron K-edge in the Compton reflected spectrum, and to correctly determine the absorption properties. With respect to the latter, the number of absorption screens covering the source can typically only be inferred with complete reliability if they can be separated in velocity. Subsequently, measurements of the absorption line strengths and widths for multiple ionization species can constrain the column density, ionization state and metallicity of the absorbing gas. Despite our best efforts in this work, without such observations one can never be completely sure about the presence and properties of the broad accretion disk emission. 

The kind of instrumental capabilities we need at iron-K for robust disk line studies is well within the grasp of current technology. Indeed, were it not for the tragic loss of the XRS calorimeters aboard the ASTRO-E and Suzaku missions, X-ray astronomy would have already entered the era of high throughput, high resolution spectroscopy at iron-K. The next generation of X-ray observatories, notably {\it Constellation-X} and {\it XEUS} will provide these capabilites, and indeed the potential of iron K$\alpha$ line studies provides the scientific motivation and drivers for the instrument requirement for these missions. The \xmm\ data have shown that extremely high data quality is required for meaningful broad iron line studies in AGN, but our work has also shown that if sufficiently high quality data are obtained, these lines can be revealed and the potential for X-ray spectroscopy to diagnose the strong gravity regime in AGN can be realised. 

\section{Conclusions}

Based on our study of very high signal-to-noise ratio spectra of AGN obtained using the \xmm-EPIC pn camera, we can conclude that:

\begin{itemize}
\item
Iron K$\alpha$ emission is universal in type I AGN, and complex emission beyond a single, narrow 6.4 keV line is extremely common, being seen in $\sim 80$~per cent of spectra.
\item
This complexity is poorly described by complex absorption phenomena and solely narrow line components. Nonetheless, at least half the sample require complex absorption in the form of a screen of ionized gas which affects the spectrum above 2.5 keV, and a smaller subset of observations show evidence for more than one such screen. Despite this, the very high signal--to--noise ratio of the EPIC data for our sample are able to break the degeneracy between complex absorption and blurred reflection, generally favouring the latter. 
\item
Around 2/3 of the sample show evidence for line emission which is broad compared to the instrumental resolution. The line broadening is very well explained by a model of reflection from an accretion disk around the black hole. In only about 30~per cent of the sample, however, is the broadening required to be relativistic ($<50 r_{\rm g}$).  
\item
The broad emission is, on average, weaker than that expected from a flat accretion disk illuminated by a point X-ray sources. In addition, several of the observations appear to exclude a significant accretion disk component, and this seem very unlikely to be explained solely by inclination effects. Possible scenarios to explain this include metallicity, geometrical and ionization effects. 
\item
Narrow emission at 6.4 keV is nearly universal in the spectra, and most likely arises from the obscuring torus envisaged in AGN unification schemes. The universality of the narrow line in Seyfert 1s provides strong support for these models, but again the line is somewhat weaker than expected, meaning that the covering fraction of the obscuring material is relatively small ($\sim 50$~per cent) and, possibly, that the metallicity is sub-solar. 
\item
A wide variety of additional emission and absorption features is seen in the spectra, including absorption and emission from He- and H-like iron, and emisison from intermediate states. The former imply significant columns of very hot and/or ionized gas in a few objects. 
\item
Possible velocity--shifted absorption and emission features are seen in a few cases. The examples are relatively few, and both the signficance and interpretation of these features are sufficiently ambiguous that further confirmation is required before firm conclusions can be drawn. 
\end{itemize}

\section*{Acknowledgements}

We are grateful to the anonymous referee for a several helpful comments on this manuscript, and to Jane Turner for numerous discussions. We thank the Leverhulme Trust (KN),  PPARC (PMO) and NASA (IMG, JNR) for financial support, Andy Fabian and Roderick Johnstone for providing the relativistic blurring codes, and Tim Kallman for extensive help with XSTAR model production. This work has made use of HEASARC and NED. We thank those we built and operate \xmm.


\end{document}